%% file: thesis.tex
\documentclass{scthesis}
\usepackage{psfig}
\usepackage[dvips]{epsfig}
\input{def}

\begin{document}

\submissionmonth{OCTOBER}
\submissionyear{2001}
\author{\bf {SANDHYA CHOUBEY} \\ SAHA INSTITUTE OF NUCLEAR PHYSICS \\ 
KOLKATA}
\title{NEUTRINOS AND THEIR FLAVOR MIXING IN NUCLEAR ASTROPHYSICS}
\maketitle

\include{ack}

%
%

\tableofcontents
\listoffigures
\listoftables

%
%

\include{chapter1}

\include{chapter2}

\include{chapter3}
\include{chapter4}

\include{chapter5}

\include{chapter6}

\include{chapter7}

\include{chapter8}

%
%

\include{list}

\end{document}

%% file: def.tex
\newcommand{\beq}{\begin{equation}}
\newcommand{\eeq}{\end{equation}}
\newcommand{\be}{\begin{eqnarray}}
\newcommand{\ee}{\end{eqnarray}}
\newcommand{\etal}{{\it et al.}}
\def\nue{{\nu_e}}
\def\anue{{\bar\nu_e}}
\def\numu{{\nu_{\mu}}}
\def\anumu{{\bar\nu_{\mu}}}
\def\nutau{{\nu_{\tau}}}
\def\anutau{{\bar\nu_{\tau}}}
\newcommand{\dm}{\mbox{$\Delta{m}^{2}$~}}

\newcommand{\tit}{\mbox{$\tan^{2}\theta$~}}
\newcommand{\st}{\mbox{$\sin^{2}2\theta$~}}

\newcommand{\chisq}{\mbox{$\chi^2$~}}
\newcommand{\chisqmin}{\mbox{$\chi^2_{\rm min}$~}}

\newcommand{\sabsq}{\mbox{$s_{12}^2$}}
\newcommand{\sacsq}{\mbox{$s_{13}^2$}}
\newcommand{\sbcsq}{\mbox{$s_{23}^2$}}

\newcommand{\sge}{\mbox{${\rm SG_e}$~}}
\newcommand{\mge}{\mbox{${\rm MG_e}$~}}
\newcommand{\sgmu}{\mbox{${\rm SG_\mu}$~}}
\newcommand{\mgmu}{\mbox{${\rm MG_\mu}$~}}

\newcommand{\br}{\mbox{$^{8}{B}~$}}
\newcommand{\ber}{\mbox{$^{7}{Be}$~}}
\newcommand{\cl}{\mbox{$^{37}{Cl}$~}}
\newcommand{\ga}{\mbox{$^{71}{Ga}$~}}
\newcommand{\chr}{\mbox{$\breve{\rm C}$erenkov~}}

%% file: ack.tex
\begin{acknowledgements}

{\it 
I am grateful to my thesis supervisor
Prof. Kamales Kar for all his help and guidance that he has extended 
to me during my research. He motivated me to take up neutrino 
physics and has incessantly   encouraged me to pursue my own line 
of thinking. I thank him for his incentive, patience and 
unremitting encouragement which has instilled in me a world of 
self confidence. 

I am indebted to Dr. Srubabati Goswami with whom I have done most of my 
research work. We share a perfect understanding which has helped us work 
together through thick and thin, particularly during the development 
of the solar and the atmospheric neutrino codes used extensively in 
this thesis. 
She has been a friend, teacher and guide and I thank her 
for all that she has done for me.

I would like to thank all my collaborators with whom I have 
worked at various stages of my research period. In particular, 
I would like to express my deep gratitude towards Prof. D.P. Roy 
with whom I had very stimulating discussions at numerous occasions 
which helped me understand the intricate details of 
neutrino physics. I also thank him for his constant encouragement 
which has been of immense help. I thank Prof. S.M. Chitre and 
Dr. H.M. Antia who introduced me to the subject of 
helioseismology and extended their help whenever needed. 
It was a pleasure working with them. 
I wish to thank Dr. Debasish Majumdar who I worked with during the 
initial stages of my work. Last but not the least, I thank 
Abhijit Bandyopadhyay for all his endurance and perseverance 
which helped us work together under the most difficult circumstances. 

Thanks are due to Prof. Amitava Raychaudhuri and Prof. Palash Baran 
Pal for many useful discussions and their unfailing encouragement. 

I wish to thank Serguey Petcov, Osamu Yasuda, Eligio Lisi and 
Carlos Penya Garay for their comments and suggestions at various 
stages which has been of immense importance. 

I thank all members of the Theory Group, SINP for creating a healthy and 
friendly atmosphere which made working here most enjoyable. Special 
thanks are due to Surasri, Ananda, Amit, Krishnendu, Subhasish, 
Rajen, Debrupa, Indrajit, Asmita, Sankha, Kaushik, Somdatta, Dipankar, 
Rajdeep, Sarmistha, Purnendu, Abhijit, Indranil, Tilak and Tanaya. 

Finally, I thank my family, my brother, my sister and particularly 
my parents. Without their support and encouragement this thesis 
would not have been possible. It is impossible to express my 
gratitude for them in words. }

\vskip 60pt
\noindent Kolkata, \hspace{9.4cm} Sandhya Choubey\\
\noindent October, 2001 \hspace{8.3cm} Theory Group, SINP

\end{acknowledgements}

%% file: chapter1.tex
\chapter{Introduction}

\section{Motivation}

The neutrino was proposed by Pauli in 1929 ``{\it ....as a desperate
remedy to save the principle of energy conservation in beta
decay}". This 
particle which can take part only in weak interactions 
was discovered experimentally by Reines and Cowan in 1956. Neutrinos can
come in different flavors in analogy to the different flavors of 
charged leptons. 
However, the decay width of Z$^0$ boson in the LEP2 experiment
restricts the number of {\it light active} neutrinos to three \cite{z1}. 
We now have experimental evidence for the existence of all the three 
types of neutrinos, $\nue$, $\numu$ and $\nutau$. 

The neutrino is known to be a neutral particle which carries a spin 1/2. 
But whether it has mass or not 
has been an intriguing issue ever since it was proposed. 
The direct upper limits on neutrino masses are quite poor. 
The limit on $\nue$ mass is obtained from tritium beta decay 
experiments and the best bound is $m_{\nue} < 5$ eV \cite{balasev}. 
The bounds on $\numu$ and $\nutau$ masses are much more weaker, 
$\numu < $ 190 keV and $\nutau < $ 18 MeV \cite{lab}. 
%
If the neutrinos are assumed to be Majorana particles then 
another experimental bound on neutrino mass comes from the 
neutrino-less double beta decay experiments ($\beta\beta 0\nu$ or 
$\beta\beta 0 \nu J$, where J is a Majoron). 
This is a lepton 
number violating process which will be possible only if the 
neutrinos are massive Majorana particles. From the non-observation 
of this process the most stringent bounds on the Majorana neutrino 
mass is 
$m_\nue < 0.35$ eV at 90\% C.L. \cite{bb1}. 

Even though these upper bounds, particularly the ones coming from direct 
mass searches are still weak (these limits on neutrino masses far exceed 
the cosmological bound which we will discuss later), they clearly indicate 
that the neutrino masses are {\it much} smaller than the mass of the 
corresponding charged leptons. This leads to an intra-familial 
hierarchy problem, a challenge for any model which can predict 
neutrino mass. 
In the Glashow-Weinberg-Salam standard model of particle physics,
which is consistent with all known experimental
data till date, 
the neutrinos are {\it assumed} to be massless.   
However there is no underlying gauge symmetry 
which forbids neutrino mass, unlike as in the 
case of the photons.  In most extensions of the standard 
model, the Grand Unified theories and the supersymmetric models, 
the neutrino is massive \cite{pbpal1,roulet1}.  

If one does allow for non-zero neutrino mass then the flavor eigenstates 
of the neutrino can be different from the mass eigenstates. 
One then encounters the quantum mechanical phenomenon of 
{\it neutrino flavor mixing} where one neutrino flavor {\it oscillates} 
into another flavor due to interference effects. 
The parameters involved in this process is the 
mass square difference between the two states which mix, \dm 
and the mixing angle between them, $\theta$. 
This mechanism can probe very small neutrino masses. Neutrino flavor 
oscillations was conjectured long ago as a plausible solution to 
the {\it solar neutrino problem} \cite{solar1} 
and the {\it atmospheric neutrino anomaly} \cite{sk1}. 

The thermonuclear fusion reactions responsible for
energy generation in the Sun release a huge flux of neutrinos. This
flux of pure $\nue$ arriving from the Sun have been measured for 
almost the last 40 years now by the 
Homestake, SAGE, GALLEX/GNO, Kamiokande, Super-Kamiokande and the 
SNO experiments. 
All these experiments have observed a
deficit of the solar neutrino flux predicted by the 
``standard solar model" (SSM) \cite{bp001}. 
This discrepancy between theory and experiment came to be known as 
the solar neutrino problem. Neutrino flavor mixing -- either
in vacuum \cite{gp1} or in solar matter \cite{msw1} -- 
can account for the solution to this apparent anomaly. The most 
favored solution to the global data 
demands a $\dm \sim 10^{-5}$ eV$^2$ and large 
values for the mixing angle. 

The atmospheric neutrinos are produced due to collision of the 
cosmic ray particles with the air nuclei. These neutrinos 
were detected in large water \chr experiments which reported a 
depletion of the $\numu$ flux compared to expectation. 
Solution to this 
atmospheric neutrino anomaly again called for neutrino flavor 
oscillations with $\dm \sim 10^{-3}$ eV$^2$ and maximal mixing to
reconcile data with predictions. The results from the Super-Kamiokande (SK) 
atmospheric neutrino experiment in 1998 \cite{sk1} became a hallmark in the 
history of particle physics when it finally confirmed that 
the atmospheric neutrinos do oscillate and are hence indeed massive. 
The observation of neutrino mass in the SK experiment (although indirect)
is the first and till date the
only evidence of physics beyond the standard model. 

There have been many terrestrial neutrino oscillation searches 
using both accelerators as well as reactors as neutrino sources 
\cite{bpet1,terr1}. But all of them except the LSND experiment 
at Los Alamos, have yielded negative results for neutrino oscillations. 
The LSND experiment has continued to give positive signal for oscillations 
since 1996 with $\dm \sim $ eV$^2$ \cite{lsndintro}. 
Thus we have three indications of neutrino oscillations. However since the 
three different hints demand three different 
values of \dm, it is difficult to explain all the experimental data 
in the framework of three neutrinos. 
There have been quite a few attempts in the literature to explain 
all the three experiments with three flavors but it is largely 
believed that if the LSND results are correct then one has to introduce 
a fourth sterile neutrino. 

Neutrinos 
are known to play a pivotal role in the supernova dynamics and
nucleosynthesis. 
A huge flux of neutrinos and antineutrinos are released during 
the thermal cooling phase of a core collapse supernova. 
These neutrinos can be detected in the 
terrestrial detectors. The detection of the neutrinos from SN1987A 
in the Kamiokande and IMB \cite{sn1987a1} gave birth to the 
subject of neutrino astronomy and heralded the beginning of a new era 
in neutrino physics. A careful study of the resultant neutrino 
signal from a galactic supernova 
in the terrestrial detectors, can throw light on not just the 
type-II supernova mechanism but also on the neutrino mass and 
mixing parameters. 

The neutrinos are the most abundant entities in the Universe after 
radiation. Hence even a small mass for the neutrinos can make a huge 
difference to the total mass of the Universe. The energy density of the 
neutrinos cannot exceed the total energy density of the Universe and 
this puts a strong upper bound on the sum of all the 
light neutrino species \cite{cowsik}, 
$\sum_i m_{\nu_i} < 46$ eV \cite{pbpal1}. Small neutrino masses can 
be an important component of the {\it dark matter}. 
Since neutrinos were relativistic at the time of structure formation 
they are called ``hot" dark matter. It is believed that a combination 
of hot+cold dark matter is required for a correct explanation 
of this problem. 
Neutrino properties and 
number of neutrino generations are also severely constrained by 
primordial nucleosynthesis arguments. 

Neutrinos play a crucial role in the understanding of stellar evolution, 
type-II supernova mechanism, nucleosynthesis and dark matter 
studies. A huge amount of theoretical and experimental 
effort has gone into the study of neutrino properties. 
Thus the importance of the subject warrants a detailed 
analysis of the neutrino mass and mixing parameters in the context of 
the current experimental data and a careful study of its implication for 
astrophysics.

\section{Plan of Thesis}

In this thesis we explore the implications of neutrino oscillations in the 
context of solar neutrino problem, 
atmospheric neutrino anomaly and the terrestrial 
accelerator/reactor neutrino oscillation experiments.
We perform detailed statistical analyses 
of the solar and atmospheric neutrino data and map out the regions of the 
parameters space consistent with the experiments. We investigate 
the effect of neutrino mass and mixing on the predicted neutrino signal 
from a galactic supernova in the current water \chr detectors. 

We begin in chapter 2 with the presentation of the basic aspects of 
neutrino oscillations in vacuum and in matter. We discuss the 
adiabatic and non-adiabatic propagation of neutrinos in a 
medium with varying density and present the expressions for the 
survival probability. 

In chapter 3 we give a detailed description 
of the current status of the neutrino oscillation experiments in terms 
of two flavor oscillations. 
For the solar neutrinos we present a brief review of the SSM predictions, 
summarize the main experimental results, work out the unified 
formalism for the solar neutrino survival probability, describe our solar 
neutrino code and perform a comprehensive \chisq analysis of the global solar 
data. Similarly for the atmospheric neutrinos we discuss the 
theoretical flux predictions, the SK experimental data, our 
atmospheric code and perform a \chisq fit with the $\numu-\nutau$ oscillation 
scenario. We briefly review the bounds from the most important 
accelerator/reactor experiments. 

We extend our study of the solar neutrinos in chapter 4 where we 
probe the potential of the energy independent solution 
in describing the global solar neutrino data. We investigate the 
signatures of this scenario in the future solar neutrino experiments. 

In chapter 5 we analyse the SK atmospheric neutrino data and the 
accelerator/reactor data in a three-generation framework. We 
present the results of the \chisq analysis of (1) only the SK data and 
(2) SK+CHOOZ data and compare the allowed regions with those 
obtained from the accelerator/reactor experiments. For the only SK 
analysis we indicate some new allowed regions with very small \dm 
which appear due to matter effects peculiar to the mass spectrum of the 
neutrinos that we have chosen here. 
We study the implications of this mass spectrum in the K2K experiment. 

In chapter 6 we work in a scheme where one of the components in the 
neutrino beam is unstable and explore the viability of this decay 
model as a solution to the atmospheric neutrino problem. 

In chapter 7 we make quantitative predictions for the number of events 
recorded in the current water \chr detectors due a galactic 
supernova. We examine the signatures of neutrino mass and mixing 
which can show up in the detectors and suggest various variables 
that can be used to study the effect of oscillations in the 
resultant signal. 

We present our conclusions in chapter 8 with a few comments on the 
physics potential of the most exciting future detectors.

%% file: chapter2.tex
\chapter{Neutrino Oscillations}

Neutrino oscillations in vacuum is analogous to $K^0-\bar{K^0}$ 
oscillations in its quantum mechanical nature. Neutrinos 
produced in their flavor eigenstates 
lack definite mass if they are massive and mixed. 
The states with definite mass are the 
mass eigenstates and neutrino propagation from the production 
to the detection point is governed by the equation of motion 
of the mass eigenstates. Finally the neutrinos are detected by weak
interactions involving the flavor states, leading to the 
phenomenon of neutrino oscillations, first suggested by 
Bruno Pontecorvo \cite{bruno} and later by Maki {\it et al} \cite{maki}. 
Some very good reviews on neutrino oscillations include 
\cite{bilpont,bilpet,kuo,bahcall,pbpal,kimpev}.

When neutrinos move through matter they interact with the 
ambient electrons, protons and neutrons. 
As a result of this they pick up an effective mass much the same 
way as photons acquire mass on moving through a medium. 
This phenomenon has non-trivial impact on the mixing scenario 
of the neutrinos. 
In section 2.1 we first develop the formalism for neutrino 
oscillations in vacuum. In section 2.2 we discuss in detail the 
effect of matter on the mass and mixing parameters of the neutrinos.

\section{Neutrino Oscillations in Vacuum}

The flavor eigenstate $|\nu_\alpha\rangle$ created in a weak 
interaction  process can be expressed as a linear superposition 
of the mass eigenstates $|\nu_i\rangle$
\be
|\nu_\alpha\rangle = \sum_{i=1}^NU_{\alpha i} |\nu_i\rangle
\label{super}
\ee
where $U$ is the unitary mixing matrix analogous to the 
Cabibo-Kobayashi-Maskawa matrix in the quark sector. We consider the 
general case of neutrinos with $N$ flavors and henceforth set $\hbar=c=1$.  
After time t, the initial $|\nu_\alpha\rangle$ evolves to 
\be                                                        
|\nu_\alpha(t)\rangle = \sum_{i=1}^Ne^{-iE_i t}U_{\alpha i}|\nu_i\rangle
\label{tsuper}
\ee
where $E_i$ is the energy of the $i^{th}$ 
mass eigenstate. For simplicity we assume 
that the 3-momentum {\bf p} of the different components of the neutrino beam 
are the same. The energy $E_i$ of the $i^{th}$ component is 
given by
\be
E_i = \sqrt{ {\bf p}^2 + m_i^2}
\ee
where $m_i$ is the mass of the $i^{th}$ mass eigenstate. 
However since the masses are non-degenerate, the $E_i$ are different 
for the different components and eq. (\ref{tsuper}) is a different 
superposition of $|\nu_i\rangle$ compared to eq. (\ref{super}). 
Hence one expects the presence of other flavor states in the resultant 
beam in addition to the original flavor. The amplitude of finding 
a flavor $\nu_\beta$ in the original $\nu_\alpha$ beam is 
\be
\langle\nu_\beta|\nu_\alpha(t)\rangle = \sum_{i=1}^N
e^{-iE_i t}U_{\alpha i}
U_{\beta i}^*
\ee 
Hence the corresponding probability is given by
\be
P_{\nu_\alpha \nu_\beta} = \sum_{i=1}^N |U_{\beta i}|^2 |U_{\alpha i}|^2 
+ {\rm Re}\sum_{i \ne j} U_{\beta i}^*U_{\beta j} U_{\alpha i} 
U_{\alpha j}^* ~e^{[-i(E_j - E_i)t]}
\ee
For ultra relativistic neutrinos with a common definite momentum {\bf p}, 
$E_i \approx {\bf |p|} + \frac{m_i^2}{2|{\bf p}|}$ and t can be replaced 
by the distance traveled L. Then we obtain
\be
P_{\nu_\alpha \nu_\beta} &=& \delta_{\alpha \beta} - 4~
\sum_{j > i}~ U_{\alpha i} U_{\beta i}^* U_{\alpha j}^* U_{\beta j}
\sin^{2}\left(\frac{\pi L}{\lambda_{ij}}\right)
\label{npr}
\\
\lambda_{ij} &=& (2.47 {\rm m}) \left(\frac{E}{MeV}\right)
\left(\frac{\rm {eV}^2}{\Delta_{ij}}\right)
\label{wv}
\ee
is the oscillation wavelength
which denotes the scale over which neutrino oscillation effects can be
significant and
$\Delta_{ij} = m_{i}^2 - m_{j}^2$.
The actual forms of the various survival and transition
probabilities depend on the neutrino mass spectrum assumed
and the choice of the mixing matrix $U$ relating the flavor
eigenstates to the mass eigenstates. The oscillatory character is embedded 
in the $\sin^{2}\left(\frac{\pi L}{\lambda_{ij}}\right)$ term. 
Depending on the value of $E$ and $\Delta_{ij}$, if the oscillation wavelength 
is such that $\lambda_{ij} \gg L$, 
$\sin^2 \left(\pi L/\lambda_{ij}\right) \rightarrow 0$, the 
oscillations do not get a chance to develop and 
the neutrino survival probability is $\sim 1$. 
On the other hand, $\lambda_{ij} \ll L$
implies a large number of oscillations, so that once the averaging over
energy and/or the distance traveled 
is done $\sin^{2}\left(\pi L/\lambda_{ij}\right)
\rightarrow 1/2$ and one encounters what is called average oscillations. 
But when $\lambda_{ij} \sim L$ then one has full oscillation effects and 
constraints on $\Delta_{ij}$ can be put from observations of the 
resultant neutrino beam.   

If we restrict ourselves to two-generations for 
simplicity then the mixing 
matrix takes a simple form
\be
{U\!=\!{\pmatrix{\cos\theta & \sin\theta
\cr -\sin\theta & \cos\theta \cr}}}
\label{2matrix}
\ee  
\be
\nue=\cos\theta\nu_1 + \sin\theta \nu_2\\
\numu=-\sin\theta \nu_1 + \cos\theta \nu_2
\ee
where $\theta$ is called the mixing angle. 
The expression of the transition probability 
from $\nu_\alpha$ to a {\it different} flavor $\nu_\beta$ reduces to 
\be
P_{\nu_\alpha \nu_\beta} = \sin^2 2\theta \sin^2 
\left(\frac{\pi L}{\lambda}\right )
\label{2genpr}
\ee
The survival probability of the original neutrino beam is simply
\be
P_{\nu_\alpha \nu_\alpha} &=& 1-P_{\nu_\alpha \nu_\beta} 
\nonumber\\
&=& 1 - \sin^2 2\theta \sin^2 \left(\frac{\pi L}{\lambda}\right )
\ee

\section{Neutrino Oscillations in Matter}

So far we have discussed the neutrino transition and survival 
probabilities in vacuum. 
When neutrinos move through a medium they interact with the ambient 
matter and this interaction modifies their effective masses and mixing. 
It was pointed out in \cite{wolfen,ms} that the patterns of neutrino
oscillations might be significantly affected if the neutrinos travel 
through a material medium. This is because normal matter has 
only electrons and so the $\nue$ experience both charged as well as 
neutral current interactions while the $\numu$ and $\nutau$ can 
participate in only neutral current processes.  

Since the flavor eigenstates are involved in weak interactions we 
work in the flavor basis for the time being and will introduce neutrino 
mixing later. All the expression for the interaction terms given 
below are for neutrinos. The 
corresponding expressions for antineutrinos are same with a -ve 
sign. 
The charged current 
scattering of $\nue$  with electrons gives the 
following contribution to the Lagrangian 
\be
{\cal L}_{eff} = -\frac{G_F}{\sqrt{2}}\{\bar e\gamma^\mu(1-\gamma_5)\nue\}
\{\bar \nue \gamma_\mu (1-\gamma_5) e\}  
\ee
where $G_F$ is the Fermi coupling constant. On rearranging the 
spinors using Fierz transformation one gets
\be
{\cal L}_{eff} = -\frac{G_F}{\sqrt{2}}
\{\bar\nue\gamma^\mu(1-\gamma_5)\nue\}
\{\bar e \gamma_\mu (1-\gamma_5) e\}  
\ee
For forward scattering of neutrinos off electrons, the 
neutrino momentum remains the same so that charged current 
contribution after averaging the electron field over the 
background becomes 
\be
-\frac{G_F}{\sqrt{2}}\{\bar\nue\gamma^\mu(1-\gamma_5)\nue\}
\langle \bar e \gamma_\mu (1-\gamma_5) e\rangle
\ee
The axial current part of 
$\langle \bar e \gamma_\mu (1-\gamma_5) e\rangle$ gives the 
spin in the non-relativistic limit while the spatial part of the 
vector component gives the average velocity, both of 
which are negligible for a non-relativistic collection of electrons. 
So the only non-vanishing contribution comes from the 
$\gamma_0$ component which gives the electron density.
\be
\langle \bar e \gamma_0 e\rangle = n_e
\ee
where $n_e$ is the ambient electron density. 
The extra contribution to the Lagrangian then reduces to 
$-\sqrt{2}G_Fn_e\bar\nu_{eL}\gamma^0\nu_{eL}$. 
Hence for unpolarized electrons 
at rest, the forward charged current scattering of neutrinos off 
electron  gives rise to an effective potential
\be
V_{cc} = \sqrt{2}G_Fn_e
\ee
The effect of this term is to change the 
effective energy of the $\nue$ in matter to
\be
E_{eff} &=& \sqrt{{\bf p}^2 + m^2} + \sqrt{2}G_F n_e
\nonumber\\
&\approx& |{\bf p}| + \frac{m^2}{2|{\bf p}|} + \sqrt{2}G_Fn_e\\
\nonumber
&=& |{\bf p}| + \frac{1}{2E}(m^2 + 2\sqrt{2}G_Fn_eE)
\ee
Thus the charged current interaction gives rise to an extra 
effective contribution to the $\nue$ mass square called $A$ 
or the Wolfenstein term \cite{wolfen} 
\be
A=2\sqrt{2}G_F n_e E
\ee
The neutral current 
contribution to the effective neutrino energies can be 
calculated in an identical manner and comes out to be
\be
V_{nc}=\sqrt{2}G_F\sum_fn_f\left[I_{3L}^f - 2\sin^2\theta_W Q^f\right ]
\ee
where $f$ stands for the electron, proton or neutron, 
$n_f$ is the density of $f$ in the surrounding matter, 
$Q^f$ is the 
charge of $f$ and $I_{3L}^f $ is the third component of weak isospin 
of the left chiral projection of $f$. Since for the proton 
$Q^f=1$ and $I_{3L}^f=1/2$, while for the electron 
$Q^f=-1$ and $I_{3L}^f=-1/2$ and since normal matter is charge neutral 
ensuring that $n_e=n_p$, the contributions due to electron and proton 
cancel each other exactly. Hence the only remaining contribution is 
due to the neutrons for which $Q^f=0$ and $I_{3L}^f=-1/2$ so that 
\be
V_{nc}=-\sqrt{2}G_Fn_n/2
\ee

In order to see the effect of these extra contributions due to 
interactions of the neutrino beam with the ambient matter, on neutrino 
mass and mixing parameters, let 
us look at the equation of motion for the neutrino states.  
We restrict ourselves to a two-generation scheme 
for simplicity ($\nue$ mixing with either $\numu$ or $\nutau$) 
and first write down the 
evolution equation for the mass eigenstates in vacuum which 
can be subsequently extended to include matter effects. The 
equation of motion for the mass eigenstates 
in vacuum can be written as
\be
i\frac{d}{dt}{\pmatrix {\nu_1 \cr \nu_2 \cr}}
=M^2{\pmatrix {\nu_1 \cr \nu_2 \cr}}
\ee
where $M^2$ is the mass matrix, diagonal in the mass basis
\be
M^2\!=\!{\pmatrix {E_1&0 \cr 0 & E_2 \cr}} \!\approx\! E \!+\!
{\pmatrix {m_1^2/2E & 0 \cr 0 & m_2^2/2E \cr}}
\ee

As the interaction terms are defined in the flavor basis 
we have to look at the evolution equation in the flavor basis. 
Since the mass eigenstates are related to the flavor eigenstates 
by the relation (\ref{super}) and since the mixing matrix in 
two-generations is given by (\ref{2matrix}), the equation of 
motion in terms of the flavor states can be written as
\be
i\frac{d}{dt}{\pmatrix {\nue \cr \numu \cr}}
=M_f^2{\pmatrix {\nue \cr \numu \cr}}
\label{eqofmot}
\ee
\be
M_f^2 &=& U~M~U^\dagger\\
          &=& E+\frac{m_1^2+m_2^2}{4E}+
\frac{\Delta}{4E}{\pmatrix {-\cos 2\theta &
\sin 2\theta \cr \sin 2\theta & \cos 2\theta \cr}}
\ee
where $M_f^2$ is the vacuum mass matrix in the flavor basis and 
$\Delta = m_2^2 - m_1^2$.  
We next include in the mass matrix the interaction terms in the mass matrix 
for neutrinos moving through matter. 
Once these extra contributions to the Hamiltonian 
due to interactions of the neutrino with matter are taken into account, 
the equation of motion for the neutrino beam 
in the flavor basis is given by eq. 
(\ref{eqofmot}) with the $M_f^2$ in vacuum 
replaced with $M_{fm}^2$ in matter given by
\be
M_{fm}^2&=& E+\frac{m_1^2+m_2^2}{4E}-\frac{1}{\sqrt{2}}G_Fn_n
\nonumber\\
&&+
\frac{1}{4E}{\pmatrix {4\sqrt{2}G_Fn_eE-\Delta \cos 2\theta &
\Delta \sin 2\theta \cr \Delta \sin 2\theta & \Delta \cos 2\theta \cr}}
\nonumber\\
&=&
E+\frac{m_1^2+m_2^2}{4E}-\frac{1}{\sqrt{2}}G_Fn_n + \frac{A}{4E}
\nonumber\\
&&+
\frac{1}{4E}{\pmatrix { A - \Delta \cos 2\theta &
\Delta \sin 2\theta \cr \Delta \sin 2\theta & -A + \Delta \cos 2\theta
\cr}}
\label{matrixm}
\ee
where $A=2\sqrt{2}G_Fn_eE$ as defined before. 
The terms proportional to the identity matrix play absolutely 
no role in flavor mixing and can be safely dropped. 
Hence we see that the neutral current term which affect all 
the flavors equally, falls out of the oscillation analysis and 
the only non-trivial contribution comes from the charged 
current interaction. 

The energy eigenvalues in matter are obtained by 
diagonalising the mass matrix $M_{fm}^2$. We define 
$E_i^m$ to be the mass eigenvalues and $U_m$ to 
be the mixing matrix in matter. In analogy with the vacuum 
case $U_m$ can be parametrized as 
\be
\pmatrix{\nue \cr \numu \cr} = U_m \pmatrix{\nu_1^m \cr \nu_2^m \cr}
\label{num}
\ee
\be
{U_m\!=\!{\pmatrix{\cos\theta_m & \sin\theta_m
\cr -\sin\theta_m & \cos\theta_m \cr}}}
\label{2matrixm}
\ee  
where $\theta_m$ is the mixing angle in matter. Then since 
\be
U_m^\dagger M_{fm}^2 U_m = \!{\pmatrix{ E_1^m & 0 \cr 
0& E_2^m \cr}}
\ee
the eigenvalues of the mass matrix are 
\be
E_i^m = E -\frac{1}{\sqrt{2}}G_Fn_n + \frac{M_i^2}{2E}
\label{eigenm} 
\ee
where
\be
M_{1,2}^2=\frac{1}{2}\left[(m_1^2+m_2^2+A)
\mp\sqrt{(-A+\Delta\cos2\theta)^2+(\Delta\sin2\theta)^2}\right]
\ee
Hence the mass squared difference $\Delta$ in vacuum is modified 
in presence of matter to
\be
\Delta_m = \left[ (-A + \Delta \cos 2\theta )^2 + 
(\Delta \sin 2\theta )^2
\right] ^{1/2}
\ee
while the mixing angle in matter is given by
\be
\tan 2\theta_m = \frac{\Delta \sin2\theta}{-A+\Delta \cos 2\theta}
\label{tanm}
\ee
or equivalently
\be
\sin^2 2\theta_m = \frac{(\Delta \sin2\theta)^2}{
(-A+\Delta \cos 2\theta)^2 + (\Delta \sin2\theta)^2}
\ee
The mass and the mixing angle in matter are changed substantially compared
to their vacuum values depending on the value of $A$ relative to $\Delta$. 
This change is most dramatic when the condition 
\be
A=\Delta \cos2\theta
\ee
is satisfied. If this condition is attained in matter then 
$\sin^22\theta_m = 1$ and the mixing angle in matter 
become maximal {\it irrespective} of the value of 
the mixing angle in vacuum. 
So even a small mixing in vacuum can be amplified to maximal mixing 
due to matter effects 
This is called the matter enhanced resonance 
effect or the 
the Mikhevey-Smirnov-Wolfenstein or the MSW effect
\cite{wolfen,ms,bethe} which is the favored solution to the
Solar Neutrino Problem, a subject we shall address later in great details.
Note that for $\Delta>0$ one encounters  
this resonance for neutrinos ($A>0$) only if $\theta < \pi/4$.
Also note that the expression (\ref{tanm}) can be rewritten as
\be
\tan 2\theta_m = \frac{\tan2\theta}{1-n_e^{pr}/n_e^{res}}
\ee
where $n_e^{pr}$ is the electron density at the point of 
neutrino production and $n_e^{res}$ is the corresponding 
density at resonance. So that for $n_e^{pr} \gg n_e^{res}$ 
the mixing angle in matter $\theta_m \rightarrow \pi/2$.

We next look for the electron neutrino survival probability. 
For a constant density medium the survival probability is given by
\be
P_{\nue\nue} &=& 1 - \sin^2 2\theta_m 
\sin^2\left(\frac{\pi L}{\lambda_m}\right)
\\
\lambda_m &=& (2.47 {\rm m}) \left(\frac{E}{MeV}\right)
\left(\frac{\rm {eV}^2}{\Delta_m}\right)
\label{wvm}
\ee
Hence for a constant density medium the expressions 
for transition and survival probabilities are exactly 
similar to ones for the vacuum case, 
with the mass squared difference and the mixing angle replaced by 
the corresponding terms in matter. But things get more complicated 
when the neutrinos move through a medium of varying density. 
We address this complex issue in the next section. 

\subsection{Neutrino Survival Probability in Medium with Varying 
Density}

In most situations encountered in nature the neutrinos propagate 
through medium with varying density. Since the eigenstates and 
the eigenvalues of the mass matrix both depend on the density 
of the medium, the mass eigenstates defined in eq. (\ref{num}) 
are no longer the stationary eigenstates. For varying density 
one can define the stationary eigenstates only for the 
Hamiltonian at a given point. Starting from the equation of motion 
in the flavor basis and dropping the irrelevant terms proportional 
to the identity matrix we have
\be
i\frac{d}{dx}{\pmatrix {\nue \cr \numu \cr}}&=&
\frac{1}{4E}{\pmatrix { A - \Delta \cos 2\theta &
\Delta \sin 2\theta \cr \Delta \sin 2\theta & -A + \Delta \cos 2\theta
\cr}}{\pmatrix {\nue \cr \numu \cr}}
\nonumber\\
&=&\frac{1}{2E}U_m{\pmatrix {M_1^2 & 0 \cr 0 & M_2^2 \cr}}
{\pmatrix {\nu_1^m \cr \nu_2^m \cr}}
\ee
where we have substituted $x$ for $t$. Keeping in mind that the mixing matrix 
is now $x$ dependent we obtain
\be
i\frac{d}{dx}{\pmatrix {\nu_1^m \cr \nu_2^m \cr}}&=&
\left[\frac{1}{2E}{\pmatrix {M_1^2 & 0 \cr 0 & M_2^2 \cr}} 
-U_m^\dagger i\frac{d}{dx}U_m \right ] 
{\pmatrix {\nu_1^m \cr \nu_2^m \cr}}
\nonumber\\
&=&
{\pmatrix {M_1^2/2E & -id\theta_m/dx \cr
id\theta_m/dx & M_2^2/2E \cr}}
{\pmatrix {\nu_1^m \cr \nu_2^m \cr}}
\label{off}
\ee  
The off diagonal terms in eq. (\ref{off}) mix the states $\nu_1^m$ and 
$\nu_2^m$ and the mass eigenstates in matter keep changing with $x$. 
Hence one has to solve the eq. (\ref{off}) to get the survival and 
transition probabilities. 
However the mass eigenstates travel approximately 
unchanged and unmixed as long as 
the off diagonal terms in (\ref{off}) are small compared to 
the diagonal terms. Since the terms proportional to the unit matrix 
do not affect oscillation probabilities, the only physical parameter 
in the diagonal elements is their difference. Thus the condition for 
the off diagonal term to be small can be written as
\be
\left|\frac{d\theta_m}{dx}\right | \ll \left | 
\frac{M_2^2 - M_1^2}{4E}\right | 
\label{adiabatic}
\ee
This is called the {\bf adiabatic condition}. Since $|d\theta_m/dx|$ 
has a maximum at resonance while $|M_2^2 - M_1^2|$ has a minimum, 
the adiabaticity condition becomes most stringent at the resonance. 
One can define an 
{\bf adiabaticity parameter} as 
\be
\gamma =\frac{\Delta \sin^22\theta}{2E \cos2\theta}\left
|\frac{d}{dx}ln n_e \right |_{x=x_{res}}^{-1}
\label{gamma}
\ee
The adiabatic condition then reduces to $\gamma \gg 1$. 
Depending on whether the adiabatic condition is satisfied or not 
neutrino propagation in matter may be of two types: {\it adiabatic} and 
{\it non-adiabatic}.

$\bullet${\bf The Adiabatic Case:} 
If one can neglect the off diagonal terms in 
eq. (\ref{off}) in comparison 
to the diagonal elements, $\nu_1^m$ and $\nu_2^m$ approximately 
become the 
eigenstate of the mass matrix. This is the adiabatic approximation.  
As long as the adiabatic condition is satisfied, the two mass states 
evolve independently and there is no mixing between them. The 
$\nu_i^m$ state created in matter then remains a $\nu_i^m$ always. 
This is the adiabatic propagation of neutrinos in matter.
Adiabatic propagation of neutrinos in matter has very interesting 
consequences. For example, consider the neutrinos produced at the center 
of the Sun or a supernova. 
Since the matter density at the point of 
production of these neutrinos is much higher that the resonance 
density, the mixing angle $\theta_m \approx \pi/2$ and since 
\be
\nue = \cos\theta_m\nu_1^m + \sin\theta_m\nu_2^m
\ee
the $\nue$ are created almost entirely in the 
heavier $\nu_2^m$ state. 
Thereafter the neutrino moves adiabatically towards lower densities, 
crosses the resonance when $A=\Delta\cos2\theta$ 
and finally comes out into the vacuum. 
Since the motion was adiabatic the heavier state remains the heavier state 
even after it comes out. Since in vacuum the heavier state is 
\be
\nu_2 = -\sin\theta\nue+ \cos\theta\numu
\ee
the survival probability of the electron neutrino is just 
$\sin^2\theta$. Hence for very small values of the mixing angle $\theta$ 
one may have 
almost {\it complete} conversion of the $\nue$ produced inside the 
sun. 

In the discussion above the 
survival probability is $\sin^2\theta$ for $\theta_m=\pi/2$. 
For any general angle $\theta_m$ at the production point of the 
neutrino, the survival probability is 
\be
P_{\nue\nue} = 
\frac{1}{2} + \frac{1}{2}\cos 2\theta \cos 2\theta_m
\label{probamsw}
\ee

$\bullet${\bf The Non-Adiabatic case:}
One may encounter cases where $\gamma \sim 1$ and the adiabatic 
condition (\ref{adiabatic}) is not satisfied. 
This breakdown of adiabaticity is 
most pronounced at the position of resonance as discussed in 
the previous section. This violation of adiabaticity at the resonance 
signals that the off diagonal terms in eq. (\ref{off}) become 
comparable to the diagonal terms and there is a finite probability 
of transition from one mass eigenstate to another. This transition 
probability between the mass eigenstates is called the 
{\it level crossing} or the {\it jump probability}. It is defined as
\be
P_J = \left |\langle \nu_2^m (x_+)| \nu_1^m (x_-)\rangle \right|^2
\label{pj}
\ee
where $x_\pm$ refer to two faraway points on either side of the 
resonance. $P_J$ can be determined by solving the equation of 
motion (\ref{off}) for a given matter density profile. For a linearly 
varying density, the jump probability is given by the 
Landau-Zener expression \cite{lz,parke}
\be
P_J = \exp(-\frac{\pi}{2}\gamma)
\ee
where $\gamma$ is the adiabaticity parameter given by eq. (\ref{gamma}). 
For the Sun the density profile is roughly exponential and we 
will address that issue in the next chapter. 

The electron neutrino survival probability, taking into account 
the finite level crossing between the mass eigenstates due to 
breakdown of adiabaticity is given by
\be
P_{\nue\nue}=
\frac{1}{2} + (\frac{1}{2}-P_J)\cos 2\theta \cos 2\theta_m
\label{probnamsw}
\ee
This can be easily computed if one knows the form of $P_J$.


\subsection{Matter Effects with Sterile Neutrinos}

In the entire discussion above on the effect of matter on 
neutrino mass and mixing we had tacitly assumed that both the 
flavors involved in oscillations were {\it active} flavors, that is, 
can take part in weak interactions. But one may consider situations 
where one of the states involved is {\it sterile}. 
Sterile states have no interactions with the 
surrounding medium, neither charged nor neutral. Thus for the 
$\nue-\nu_{\rm sterile}$ mixing, relevant for the solar neutrino problem, 
the mixing angle and the mass squared difference in matter are given by, 
\be
\Delta_m &=& \left[(-2\sqrt{2}G_Fn_eE+\sqrt{2}G_Fn_nE+\Delta\cos 2\theta)^2
+(\Delta\sin 2\theta)^2\right]^{1/2}
\label{delmste}
\\
\tan 2\theta_m &=& \frac{\Delta \sin 2\theta}{
-2\sqrt{2}G_Fn_eE+\sqrt{2}G_Fn_nE+\Delta\cos 2\theta}
\label{mixmste}
\ee
The survival probability $P_{\nue\nue}$ is still given by eqs. 
(\ref{probamsw}) and ({\ref{probnamsw}). 
For the $\numu-\nu_{\rm sterile}$ mixing, relevant for the 
atmospheric neutrino anomaly the corresponding expressions are 
\be
\Delta_m &=& \left[(\sqrt{2}G_Fn_nE+\Delta\cos 2\theta)^2 
+(\Delta\sin 2\theta)^2\right]^{1/2}
\label{delmst}
\\
\tan 2\theta_m &=& \frac{\Delta \sin 2\theta}{
\sqrt{2}G_Fn_nE+\Delta\cos 2\theta}
\label{mixst}
\ee
Note however that whether the neutrinos are oscillating into active 
or sterile species matter only when a material medium is involved. 
The oscillations in vacuum is the same for both the cases.


%% file: chapter3.tex
\chapter{Bounds on Two Flavor Neutrino Mixing Parameters}

The question whether neutrinos are massive or not has been 
answered. The Super-Kamiokande (SK) atmospheric neutrino data has 
quelled all apprehensions about the existence of neutrino flavor  
mixing and has propelled neutrino physics to the forefront of 
particle phenomenology. 
The other puzzle which has warranted neutrino 
oscillations as a plausible solution is the long standing solar neutrino 
problem. For the last four decades solar neutrino detectors have 
recorded a flux far less than that predicted by solar models. Though 
the theory of neutrino flavor oscillations 
can offer the best possible solution 
to this anomaly, 
there are still a lot of issues that have to be settled.
Finally there have been earnest searches for neutrino flavor 
mixing in the laboratory, using both reactors as well as accelerators 
as sources for neutrino beams. But all such quests have lead to 
negative results, with the exception of 
the LSND experiment at Los Alamos which has reiterated since 
1996 to have observed positive neutrino oscillation signals. 

In this chapter we survey the current status of the solar neutrino 
problem, the atmospheric neutrino anomaly and the laboratory 
experiments. We briefly discuss the incident neutrino fluxes, 
the detection techniques, present the main experimental results and 
perform detailed statistical analysis of the solar and the atmospheric 
neutrino data in terms of two-generation neutrino oscillations. 
We describe our solar and atmospheric neutrino code and 
discuss the method of \chisq analysis. 
We identify the best-fit solutions and display the allowed 
regions in the neutrino parameter space. We begin with the 
solar neutrino problem in section 3.1, 
next take up the case of the atmospheric neutrino anomaly 
in section 3.2 and finally move over to the description 
of the most stringent accelerator/reactor experiments in section 3.3.

\section{The Solar Neutrino Problem}

Neutrinos are an essential byproduct of the thermonuclear energy 
generation process inside the Sun 
whereby four proton nuclei are fused into 
an alpha particle. 
\be
4p~\longrightarrow~ ^4He + 2e^+ + 2\nue + 28~{\rm MeV}
\label{hb}
\ee
This process is called Hydrogen Burning and is responsible for the 
hydrostatic equilibrium of the Sun.
About 2-3\% of the total energy released in the process (\ref{hb}) is 
carried away by the neutrinos. The rest is in the form of 
electromagnetic radiation, which diffuses out from the core 
to the surface of the Sun, getting degraded in frequency 
to appear as sunlight. It takes millions of years for the photons  
to emerge from the Sun due to interactions with the solar matter. 
As a result of these repeated scatterings, the photons 
cannot give us any direct information about the core of the Sun. 
The neutrinos on the other hand have typical scattering cross sections 
of about $10^{-43}$ cm$^2$, which for a solar density of about 
100 g/cc gives a mean free path of the order of $10^{17}$ cm. This is 
much larger than the radius of the Sun. The neutrinos escape from 
the Sun unadulterated and bring along, all the information about the 
solar core imprinted on them. Thus a careful study of these neutrinos 
promises to provide detailed information about the solar interiors and 
the thermonuclear energy generation process inside the core and hence 
holds the potential to 
verify or refute the existing solar models. 
 
Keeping this in mind, Ray Davis began his pioneering experiment 
in which neutrinos are captured by the $^{37}Cl$ atoms \cite{davis}. 
The corresponding work on 
theoretical predictions for the solar neutrino fluxes were done  
by Bahcall and his collaborators \cite{bahcall1} 
in the framework of what is called 
the ``standard solar model". 
The first results of this experiment
were declared in 1968 \cite{davis}. This experiment has 
been running for almost four decades now \cite{cl} and the 
results have been remarkably robust; the experiment sees a 
depletion of the solar neutrino flux over the standard solar model 
(SSM) predictions. The solar neutrinos seem to vanish on their way 
from the Sun to the Earth and this mystery of missing solar neutrinos 
constitutes the {\it Solar Neutrino Problem} (SNP). Two decades later the 
Kamiokande water \chr experiment corroborated this observed 
deficit of solar neutrinos \cite{kam}. This conflict between the SSM
prediction and observation has been further substantiated by the 
results from the $^{71}Ga$ 
experiments -- the SAGE \cite{sage}, the GALLEX \cite{gallex} 
and now the GNO \cite{gno}, which is the upgraded 
version of the GALLEX. The advent of the Super-Kamiokande 
\cite{sksolar,sksolarspec,sksolardn}, the 
upgraded version the the original Kamiokande experiment, 
brought with it rich statistics, which provided further insight into 
the SNP in terms of the overall depletion of the solar flux 
\cite{sksolar}, the 
energy dependence of the suppression rate \cite{sksolarspec} 
and the presence of any 
difference in the observed rate at day and at night, a phenomenon 
called the day-night effect \cite{sksolardn}. 
The recently declared results on the electron scattering events (ES) 
from the Sudbury Neutrino Observatory (SNO) \cite{sno3} are consistent 
with the observations of SK while their charged current (CC) events when 
compared with the SK events signify the presence of a $\numu/\nutau$ 
component in the resultant neutrino beam at the $3.3\sigma$ level while 
the total solar flux is calculated to be $5.44\pm0.99 \times 
10^6 cm^{-2}s^{-1}$ which is in agreement with the SSM
predictions \cite{sno3}. 

Thus if the neutrinos are assumed
to be the standard particle predicted by the Glashow-Weinberg-Salam
standard model of particle physics then 
there seems to be an apparent discrepancy between the experimental 
observations and the SSM predictions. 
There have been several attempts in the literature 
to explain the 
experimental results by modifying the solar models, assuming neutrinos 
to be standard. These endevours are collectively called the 
{\it astrophysical solutions}. In the most general class of 
these solutions the solar neutrino fluxes are considered as free 
parameters, with the only requirement being the reproduction 
of the observed solar luminosity \cite{astrogen}. But all such 
analyses fail to explain the observations from all the three 
experiments, Cl, Ga and water \chr 
simultaneously. 
In fact the best-fit for these 
solutions predict a negative flux for the \ber neutrino which is 
an unphysical situation. The astrophysical solutions 
fail to explain even two of the solar neutrino results 
simultaneously. 
After the declaration of the SNO results the astrophysical solution 
has fallen into further disfavor \cite{bah1,beri}. In \cite{bah1} 
Bahcall has shown that while prior to SNO the astrophysical solution 
failed to fit the data at the $2.5\sigma$ level, 
after including the 
SNO data it is ruled out at $4\sigma$.  
Thus it is not the uncertainties in the solar models 
that are responsible for this anomaly. 
In fact the solar models have been remarkably 
refined in the last four decades and the flux uncertainties have 
been reduced to a large extent. Hence one needs to consider 
some non-standard property for the neutrino in order to be 
able to reconcile the data with the standard solar model predictions. 

Among the various particle physics 
solutions proposed till date, neutrino oscillations 
in vacuum \cite{bruno3} and/or in matter \cite{wolfen3,ms3}, 
have been the first and the most appealing solution 
\cite{bks,fl,gg,sg,postsno3}. 
Neutrino flavor mixing has 
the potential to explain not only the total suppression of the solar 
flux, but also the energy dependence of the suppression factor and 
the day-night asymmetry. 
The analysis of the solar neutrino problem in terms of 
neutrino mass and mixing gives four pockets of allowed area in the 
neutrino oscillation parameter space. One of them has $\dm \sim 10^{-6}$ 
eV$^2$ and mixing angles very small and is called the {\it small 
mixing angle} (SMA) region. Another has $\dm \sim 10^{-5}$ eV$^2$ with 
large mixing angles and is referred to as the {\it large mixing 
angle} (LMA) solution. A third allowed zone has \dm in the range 
$10^{-7}-10^{-9}$ eV$^2$ and mixing angle close to maximal. This 
is the LOW-QVO region, LOW stands for {\it low} \dm and QVO 
for {\it quasi vacuum oscillations}. The last region is the one 
associated with vacuum neutrino oscillations with $\dm \sim 10^{-10}$ 
eV$^2$ and mixing close to maximal. 
While doing an analysis of the solar data one has to consider 
$\nue$ oscillations to either an active flavor ($\numu$, $\nutau$) 
or some sterile species ($\nu_{\rm sterile}$). The latter case 
will be different, firstly because the sterile neutrinos will not 
show up in the detectors in the neutral current interactions 
even though $\numu$, $\nutau$ can. This 
feature affects both the MSW as well as the vacuum solutions. 
For the MSW solutions there is an extra difference as the 
sterile neutrinos do not have any interaction with the ambient matter,
both in the Sun and in the Earth. Thus the effect of matter 
on the mixing parameters will 
be different for the $\nu_{\rm sterile}$ as compared to 
$\nu_{\rm active}$, as discussed in the previous chapter. 

We perform detailed \chisq-fits for both the $\nue-\nu_{\rm active}$ 
and $\nue-\nu_{\rm sterile}$ transformations.  
We perform a global \chisq analysis of the solar neutrino 
data from all experiments in the framework of neutrino mass and  
mixing. We adopt an {\it unified} approach for the presentation 
of the MSW, the vacuum and quasi-vacuum oscillations 
solutions. We find the best-fit solution to the global data 
and show the allowed regions in the \dm-\tit parameter space. 
From the global analysis of all 
available solar neutrino data, one finds 
that the large mixing angle $\nue-\nu_{\rm active}$ 
MSW solution gives the best fit.

Section 3.1.1 is a brief discussion on the solar neutrino 
flux predictions in the standard solar models and the various 
uncertainties involved. In section 3.1.2 we briefly 
present the essential features of the solar neutrino experiments and 
summarize their main results. 
In section 3.1.3 we
discuss the solar neutrino code developed by us.
In section 3.1.4 we introduce 
neutrino flavor mixing and develop the unified formalism for the analysis 
of the SNP in the context of neutrino mixing. 
Finally in section 3.1.5 we present the results of a 
comprehensive $\chi^2$ analysis of the solar neutrino data. We 
identify the best fit solutions and give C.L. allowed areas in the 
neutrino parameter space.

\subsection{Neutrinos in Standard Solar Models}

\begin{figure}[h]
\vskip -4cm
    \centerline{\psfig{file=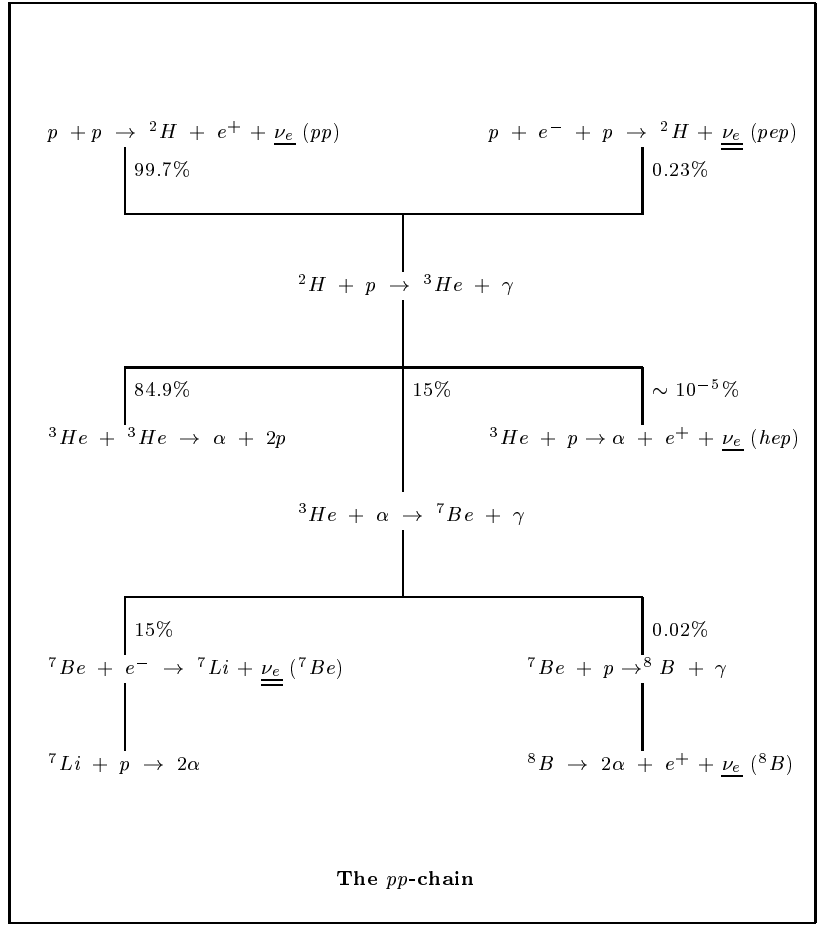,width=8.0in}}
\vskip -12cm
    \caption[The $pp$ chain.]{\label{ppchain}The reactions of the
$pp$-chain. The probability
of a particular
reaction is shown as a percentage. The neutrinos are shown underlined. 
Those with double underlines are monoenergetic. We indicate in 
parenthesis the popular names by which these various neutrinos 
are addressed.}
\end{figure}

``Standard Solar Model" is a solar model constructed with the best 
available physics and input data. 
Almost all solar models postulate the thermonuclear fusion of 
protons (\ref{hb}) to be the main energy generation process in the 
Sun \cite{bu88,bp92,tcl93,bp95,ds96,bp98,bp00}.
The eq. (\ref{hb}) is actually the compactified form for a 
chain of reactions in which four hydrogen nuclei are fused 
to form a helium nucleus. This chain of reactions, called the 
{\it pp chain} is shown in fig. \ref{ppchain}. 
The other series of nuclear reactions in the core of the 
Sun that release neutrinos is the $CNO$ cycle. 
However, since the $CNO$ cycle becomes important only above 
core temperature $T_c \sim 10^7 $ K, it produces only 
about 1.5\% of the total solar neutrinos released. 
Nevertheless the $CNO$ cycle is responsible for three 
sources of solar neutrinos and we call them the $^{13}N$, 
$^{15}O$ and $^{17}F$ neutrinos. So one has eight different 
types of solar neutrinos, five produced in the $pp$ chain and 
three in the $CNO$ cycle.

\begin{figure}[htb]
    \centerline{\psfig{file=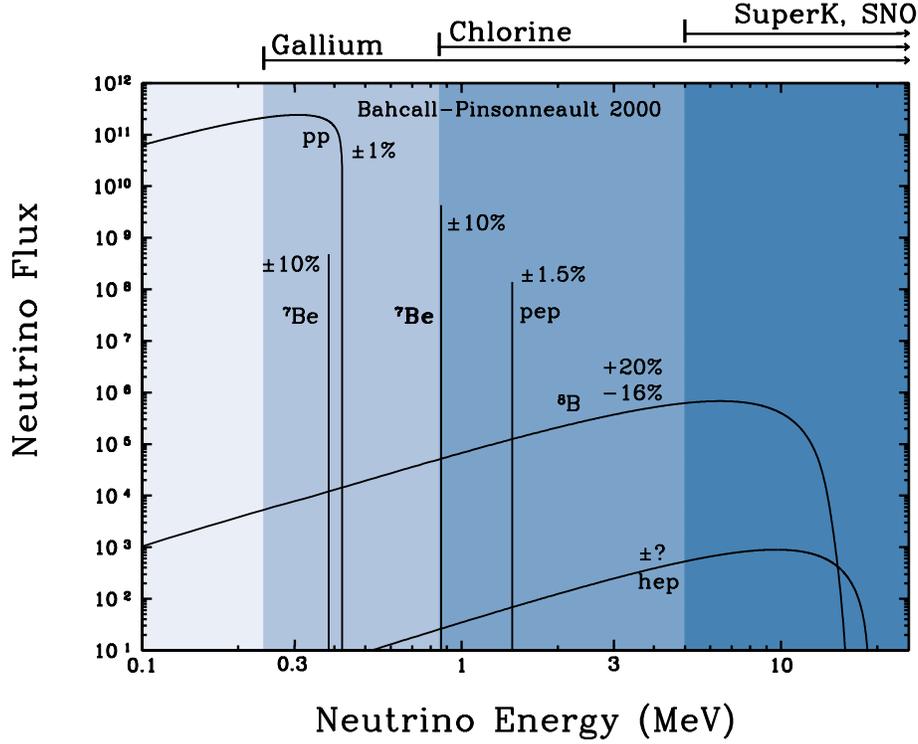,width=4.2in,angle=270}}
\vskip -1cm
    \caption[The solar neutrino spectrum.]{\label{nuspec} Solar neutrino 
spectrum in the standard solar model as a function of neutrino energy. 
The continuous spectra are in units of cm$^{-2}$MeV$^{-1}$s$^{-1}$, 
while the monoenergetic lines are in cm$^{-2}$s$^{-1}$. Figure 
shows the $\pm 1\sigma$ uncertainties in the model predictions of 
the various fluxes. Also shown are the energy ranges over which the 
solar neutrino experiments are sensitive.}

\end{figure}

From the observed solar luminosity and from the fact that 28 MeV 
energy are released per two electron neutrinos produced 
in eq. (\ref{hb}), one can 
make an order of magnitude estimate of the solar neutrino flux
\be
\Phi_\nue &=& \frac{\rm luminosity}{4\pi D^2 \times \frac{1}{2}(
\rm binding~energy~of~^4He)}
\nonumber\\
&\approx& \frac{4\times 10^{33} {\rm ergs/s}}{4\pi \times 
(1.5\times 10^{13} {\rm cm})^2 \times 14~{\rm MeV}}
\nonumber\\
&\approx& 6\times 10^{10} {\rm cm}^{-2} {\rm s}^{-1}
\label{estimate}
\ee
The exact solar neutrino flux calculations depend on a number 
of factors such as the nuclear reactions rates, the metallicity ($Z/X$), 
the age of the Sun and the opacity. Detailed solar neutrino flux 
calculations from the standard solar models are available 
\cite{bu88,bp92,tcl93,bp95,ds96,bp98,bp00} 
and agree with the rough estimate made in eq.   
(\ref{estimate}) in order of magnitude. 
The first two columns of Table \ref{bp00tab1} 
list the total solar neutrino fluxes along with their $\pm 1\sigma$ 
uncertainties, from the eight 
different reactions, as given by the year 2000 model of Bahcall, 
Pinsonneault and Basu, which we shall henceforth refer to as 
BPB00 \cite{bp00}. In fig. \ref{nuspec} we show the 
energy spectrum of
the neutrinos emitted in various reactions of the $pp$ chain 
in BPB00. Also shown are the $\pm 1\sigma$ uncertainties 
in the various fluxes\footnote{This figure has been taken from 
John Bahcall's homepage; www.sns.ias.edu/$\sim$jnb/}.

\begin{table}[htbp]
    \begin{center}
        \begin{tabular}{||c||c|c|c||} \hline\hline 
         {\rule[-3mm]{0mm}{8mm}
         source} & Flux & Cl & Ga \\
    [1ex]     & ($10^{10}$ cm$^{-2}$s$^{-1}$) & (SNU) & (SNU)\\ 
    [1ex]\hline\hline
         {\rule[-3mm]{0mm}{8mm}
         $pp$}  &  5.95($1.00^{+0.01}_{-0.01}$) 
          &  0.0 & 69.7 \\[1ex]
         $pep$ &  1.40$\times 10^{-2}$($1.00^{+0.015}_{-0.015}$) 
          & 0.22 & 2.8 \\[1ex]
         $hep$ &  9.3$\times 10^{-7}$ & 0.04 & 0.1\\[1ex]
         $^{7}Be$ & 4.77$\times 10^{-1}$($1.00^{+0.10}_{-0.10}$) 
          & 1.15 & 34.2 \\[1ex]
         $^8B$ & 5.05$\times 10^{-4}$($1.00^{+0.20}_{-0.16}$) 
          & 5.76 & 12.1 \\[1ex]
         $^{13}N$ & 5.48$\times 10^{-2}$($1.00^{+0.21}_{-0.17}$) 
          & 0.09 & 3.4 \\[1ex]
         $^{15}O$ & 4.80$\times 10^{-2}$($1.00^{+0.25}_{-0.19}$) 
          & 0.33 & 5.5 \\[1ex]
         $^{17}F$ & 5.63$\times 10^{-4}$($1.00^{+0.25}_{-0.25}$) 
          & 0.0 & 0.1 \\ [1ex]\hline\hline 
               &  & & \\
         \raisebox{1.5ex}[0pt] {Total} &  & 
         \raisebox{1.5ex}[0pt] {$7.6_{-1.1}^{+1.3}$}& 
         \raisebox{1.5ex}[0pt] { $128_{-7}^{+9}$} \\ 
         \hline\hline

         \end{tabular}
      \end{center}
\vskip -0.5cm
      \caption[The BPB00 predictions]{\label{bp00tab1}
    The BPB00 predictions for the solar neutrinos 
    fluxes and neutrino capture rates in the Cl and Ga detectors. 
    The expected \br flux in SK is 
    5.05$\times 10^{-6}{\rm cm}^{-2}{\rm s}^{-1}$.}
\end{table}  
Among the predictions for the neutrinos from different reactions 
in the Sun, the flux of the $^8B$ neutrino is the most uncertain. 
In fact the largest contribution to the differences in 
the predictions of the various standard solar models is 
their choice of different values for the $S_{17}$ which is the astrophysical 
$S$-factor for the reaction $(^7Be(p,\nu)^8B)$. 
While the SSM of Dar and Shaviv \cite{ds96} uses a value of 
$S_{17}$ as low as $17\pm 2$ eV barn, the earlier 1992 and 1995 
models of Bahcall and Pinsonneault \cite{bp92,bp95} had 
used $S_{17}=24 \pm 2$ eV barn. The value of $S_{17}$ used in 
BPB00 is $19^{+4}_{-2}$ eV barn which is the value accepted by 
the Institute of Nuclear Theory (INT) \cite{int}.

The solar fluxes are sensitive not just to the uncertainties in the  
nuclear reaction rates, but also to the value 
of the core temperature $T_c$. 
In fact the \br flux $\Phi_{^8B} \propto T_c^{18}$, the $pp$ flux 
$\Phi_{pp} \propto T_c^{-1.5}$ while the \ber flux $\Phi_{^7Be} 
\propto T_c^{8}$. 
Hence even a slight increase in the value of $T_c$ can seriously 
affect all the solar fluxes. 
In particular, it will sharply raise the \br and \ber fluxes and 
lower the $pp$ flux\footnote{This implies that just by adjusting the value
of $T_c$ alone one cannot solve the SNP since Cl and SK 
experiment will require a 
lowering of $T_c$ which will raise the $pp$ flux making the Ga results 
look all the more puzzling.}. 
The value of $T_c$ is sensitive to a number 
of factors including the value of the opacity of the solar core. 
If the value of the opacity is raised, 
it slows down the heat transport, leading to higher core temperatures. 
The value of the opacity in turn depends on the abundance of 
heavy elements in the Sun or the metallicity. Another very important 
ingredient in the solar models is the inclusion of element diffusion. 
Apart from convection, the two other mechanisms important for 
transporting solar matter are; (1) gravitational settling, which 
pulls heavier elements towards the center and (2) temperature gradient
diffusion, which results in pushing lighter elements outward. 
Both of these cause the inward diffusion of $^4He$ and outward 
diffusion of $^1H$. Hence 
diffusion increases the opacity, which results in a 
higher $T_c$ leading to an increase of $^8B$ and $^7Be$ fluxes 
and decrease of the $pp$ flux. 

In spite of the various uncertainties 
involved, only a fraction of which have been 
discussed above, 
it was shown in \cite{bp92,bp95} that if the same input physics is 
used, then all the standard solar models agree with one another to an 
accuracy of better than 10\%. Hence 
though we have presented the results on the total fluxes and the   
neutrino spectra from the standard solar model of Bahcall,      
Pinsonneault and Basu, predictions by almost all the standard models        
published so far are in reasonable agreement with each other.
For our analysis of the SNP in terms of neutrino mass and 
mixing, we have used the latest SSM predictions by 
Bahcall, Pinsonneault and Basu (BPB00) \cite{bp00}.

\subsection{The Solar Neutrino Experiments}

\subsubsection{The Cl Experiment (Homestake)}

This is the first and the longest running experiment on solar neutrinos 
started in the sixties by Davis and his collaborators with 615 
tons of $C_2Cl_4$ (perchloroethylene) in the Homestake Gold mine in 
South Dakota \cite{davis}. 
The neutrino detection process in this experiment is 
\be
\nue + \cl \rightarrow ^{37}Ar + e^-
\label{cl}
\ee
The $^{37}Ar$ atoms are extracted from the detectors at the end of 
a certain period of time and counted by detecting the Auger 
electron released when the $^{37}Ar$ decays by capturing a K-shell or 
an L-shell electron. 
The reaction (\ref{cl}) 
has a threshold of 0.814 MeV so that the Cl experiment 
predominantly detects the \ber and \br neutrinos. It misses out 
on the most abundant and least uncertain $pp$ neutrinos which have a 
maximum energy of only 0.42 MeV (cf. fig \ref{nuspec}). 
The third column of Table \ref{bp00tab1} 
shows the BPB00 predictions \cite{bp00} for the neutrino capture 
rates in the Cl experiment for the different neutrino sources.  
Also shown for the Cl detector are the total predicted rate and 
the $\pm 1 \sigma$ uncertainties 
in the model calculations. The numbers quoted are in a convenient unit 
called SNU, defined as, 
$1 ~{\rm SNU} = 10^{-36} {\rm events/target~atom/second}$. 
The observed rate of solar neutrinos in the experiment is 
\cite{cl}
\be
{\rm Observed~Rate}_{\rm Cl} = 2.56 \pm 0.23~{\rm SNU}
\ee
Compared to the BPB00 prediction of $7.6_{-1.1}^{+1.3}$ as in 
Table \ref{bp00tab1}, this gives 
a ratio of observed to expected SSM rate of
$0.335 \pm 0.029$.

\subsubsection{The Ga Experiments (SAGE, GALLEX, GNO)}

These are also radiochemical experiments that use \ga 
as their detector material. The \ga captures a $\nue$ to produce 
$^{71}Ge$ by the reaction
\be
\nue + \ga \rightarrow ^{71}Ge + e^-
\ee
This reaction has a threshold of only 0.233 MeV. Hence the advantage 
that this detector has is that it is capable 
of seeing the $pp$ neutrinos which are responsible for 98.5\% 
of the energy generation of the Sun. Hence the fact that the Ga 
detector could detect these neutrinos confirms the basic 
postulate of all the solar models, that the Sun generates its 
energy through thermonuclear burning. This itself was a very 
significant achievement of the Ga detectors. 

The SAGE ({\underline S}oviet {\underline A}merican 
{\underline G}allium {\underline E}xperiment) in Russia and GALLEX 
({\underline {Ga}}llium {\underline {Ex}}periment) in Italy 
are experiments that use this detection technique. The 
SAGE in Baksan Neutrino Observatory uses 
60 tons of metallic $Ga$ as the 
target. The $^{71}{Ge}$ produced is separated and counted. The 
observed rate is \cite{sage}
\be
{\rm Observed~Rate}_{SAGE} = 75.4\pm^{7.0}_{3.0}(stat.) \pm 
^{3.5}_{3.0}(syst.)
~{\rm SNU}
\ee
The GALLEX is located in the Gran Sasso laboratory in Italy and uses 
30 tons of $Ga$ in the form $GaCl_{3} - HCl$ solution. The observed 
neutrino rate in GALLEX is \cite{gallex}
\be
{\rm Observed~Rate}_{GALLEX} = 77.5 \pm ^{7.6}_{7.8}~{\rm SNU}
\ee
The GALLEX has now finished its run and has been upgraded to the 
GNO ({\underline G}allium {\underline N}eutrino {\underline O}bservatory) 
which has already given results \cite{gno}
\be
{\rm Observed~Rate}_{GNO} = 65.8 \pm ^{10.7}_{10.2}~{\rm SNU}
\ee
The combined SAGE and GALLEX+GNO results is 
\be
{\rm Observed~Rate}_{Ga} = 74.7 \pm 5.0 ~{\rm SNU}
\ee
which is more than 6$\sigma$ away from the SSM predicted rate of 
$128_{-7}^{+9}$ SNU \cite{bp00} (cf. Table \ref{bp00tab1}).

\subsubsection{The Water \chr Experiments (Kamiokande and
Super-Kamiokande)}

The water \chr detectors detect solar 
neutrinos by the forward scattering of electrons 
\be
\nue + e^- \rightarrow \nue + e^-
\label{nuescatt}
\ee
As it moves, the scattered electron emits \chr light,  
which is viewed by the huge number of photomultiplier tubes 
covering the entire detector volume. The water detector in general  
has a higher threshold so that it is sensitive to just the \br 
and the vanishingly small $hep$ 
neutrinos. But it has many other advantages. 
It is a real 
time experiment which has directional information. The reaction 
(\ref{nuescatt}) is forward peaked and the detector can 
reconstruct the direction of the incoming neutrino 
from the angle of the emitted \chr cone. Earlier the
Kamiokande \cite{kam} and now the Super-Kamiokande 
\cite{sksolar} have found an excess of events peaking broadly
in the solar direction and have thus confirmed that the 
observed neutrinos are indeed coming from the Sun. 
The water detector can observe not just the $\nue$ 
as in the radiochemical experiments, 
but neutrinos and antineutrinos of {\it all} flavors. Thus it can detect 
$\numu$ and $\nutau$ through neutral current electron scattering 
though the neutral current scattering cross-section is about $1/6^{th}$ 
of the charged current scattering cross-section. This 
is important if one wants to distinguish between neutrinos oscillating 
out into either $\numu/\nutau$ or to some $sterile$ species 
$\nu_{\rm sterile}$, 
which does not have any standard model interactions. 
But the real strength of this experiment lies in its ability to 
provide information about the incident neutrino energy spectrum 
from the observed recoil electron energy spectrum.
This piece of experimental observation tells us about the 
form of the energy dependence of the suppression rate which is 
extremely important in distinguishing between allowed solutions 
to the SNP. Also, since it is a real time experiment, 
it can divide its data set into day and night bins. Hence the detector 
can give information on the difference  
between the observed solar flux during day and night. 

\begin{figure}[htb]
    \centerline{\epsfig{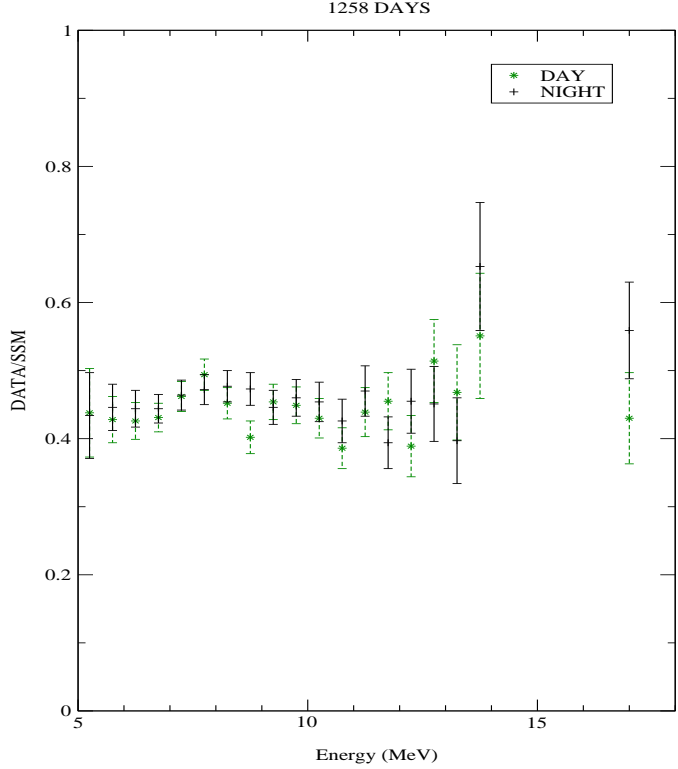}}
\vskip -0.5cm
    \caption[Super-K spectra at day and night.]{\label{dnspec} 
SK recoil energy spectra along with the $\pm 1\sigma$ errorbars 
for 1258 day data. The dotted errorbars are for the spectra at day 
while the solid errorbars are for the night bins.}

\end{figure}

The Kamiokande experiment, located in a deep mine at Mozumi, Japan, 
was a 4.5 ktons detector with a threshold of 7.5 MeV. 
The observed solar neutrino flux reported by this experiment is
\be
{\rm Observed~Rate}_{\rm Kamiokande} = (2.89 \pm 0.42) \times
10^{6} {\rm cm}^{-2}{\rm s}^{-1}
\ee   
Super-Kamiokande (SK) \cite{sksolar} is the upgraded 
version of Kamiokande. 
The first result of this experiment on solar neutrinos was released 
in 1998 \cite{sksolar}. 
The SK has now managed to reduce its threshold to 5.0 MeV 
\cite{sk1258} and the observed solar flux 
reported by SK after 1258 day of data taking is \cite{sk1258}
\be
{\rm Observed~Rate}_{\rm SK} = (2.32 \pm 0.08) \times
10^{6} {\rm cm}^{-2}{\rm s}^{-1}
\label{skobs}
\ee
The recoil electron energy spectrum \cite{sksolarspec}
released by the SK collaboration 
after 1258 day of data \cite{sk1258} 
is consistent with no spectral distortion. 
This means that the suppression rate observed is essentially 
energy independent. The 
SK gives not just the total recoil energy spectrum but also the 
spectrum at day and the spectrum at night. In fig \ref{dnspec} we show 
the day and the night spectra separately for the 1258 day SK data.
We can see that both the day and the night spectra are flat upto 
$1\sigma$. The night bins have {\it slightly} more events than the 
day bins. The degree of difference between the day and night event rates 
is conveniently measured by the day-night asymmetry, defined as 
${\cal A}=(\Phi_n - \Phi_d)/\Phi_{average}$ and  
$\Phi_{average}=\frac{1}{2}(\Phi_n + \Phi_d)$, where $\Phi_n(\Phi_d)$ 
is the observed flux during night(day). 
The SK reports \cite{sk1258}
\be
{\cal A}=0.033\pm 0.022({\rm stat.})_{-0.012}^{+0.013}({\rm sys.})
\label{dnasymm}
\ee   
This is just a 
$1.3\sigma$ effect which signifies almost no day-night asymmetry.

In order to study the day-night effect in greater details, the SK divide 
their data on total {\it rates} into 
a day and five night bins according to 
the zenith angle at which the neutrinos arrive \cite{sksolardn,sk1258}. 
They have now also 
divided their observed {\it spectrum} into zenith angle bins 
\cite{sk1258}. This helps 
to study the energy dependence of the suppression rate 
as well as the predicted day night asymmetry together in the 
most efficient manner. 

The Sun-Earth distance changes with the time of the year due to the 
eccentricity of the Earth's orbit. 
If the neutrinos oscillate in vacuum on their way from the Sun to 
Earth, then one should expect an extra modulation  
of the solar neutrino flux due to oscillations with 
the time of the year, as the survival probability depends 
very crucially on the distance that the neutrinos travel. 
The SK have reported the seasonal variation of the solar flux.
The data is consistent with the expected annual variation due to
orbital eccentricity of the Earth assuming no neutrino oscillations. 

\subsubsection {The Heavy Water Detector (SNO)}

The Sudbury Neutrino Observatory at Sudbury, Canada is the worlds first 
heavy water detector containing 1 kton of pure ${\rm D_2O}$
surrounded by 7 kton of pure ${\rm H_2O}$. The main 
detection process is the charged current (CC) breakup of the deuteron 
\be
\nu_e + d \rightarrow p + p +e^-
\label{nued}
\ee
It has now declared its first results on the observed \br flux \cite{sno3}
\be
\Phi_{CC}^{SNO} = 1.75 \pm 0.07(stat) ^{+0.12}
_{-0.11}(sys) \times 10^6 {\rm cm}^{-2} {\rm s}^{-1}
\label{ccrate}
\ee
The detector threshold for the kinetic energy of the observed electron 
released in (\ref{nued}) is 6.75 MeV. 
SNO has also released its observed $^{8}{B}$ flux measured 
by the electron scattering (ES) reaction (cf. eq. (\ref{nuescatt})) 
and they report \cite{sno3}
\be
\Phi_{ES}^{SNO}
= 2.39 \pm 0.34 (stat)^{+0.16}_{-0.14}(sys) \times 10^6 
{\rm cm}^{-2} {\rm s}^{-1}
\label{phiessno}
\ee
which agrees with the SK observation (\ref{skobs}), though the 
errors are still large. 
Apart from the total rates, SNO also gives the
recoil electron energy spectrum for the CC events and they do not report 
any 
significant distortion with energy. However the real strength of this detector 
is its ability to measure the flux of all the neutrino species with equal 
cross-section via the neutral current (NC) breakup of deuteron
\be
\nu_x + d \rightarrow n + p + \nu_x
\label{neutral}
\ee
The observed NC rate from SNO is being eagerly awaited. 
Implications of the NC rate for the mass and mixing parameters is 
discussed in detail in \cite{snopre}.

\subsubsection{Summary of the Experimental Results}

In order to summarize the main results available from all the 
solar neutrino experiments, we present in Table \ref{rates} the ratio 
of the observed to the expected total rates\footnote{From now 
onwards we shall call these ratios as the observed rates.}  
in the 
Ga, Cl, SK and SNO experiments\footnote{ We choose 
to neglect the Kamiokande data since the 
Kamiokande observations are consistent with the SK data which 
has much higher statistics.}. 
The corresponding rough estimates for the 
compositions of the observed flux is also shown. 
Since the ES rate in SK and SNO 
is sensitive to both $\nue$ as well as $\numu/\nutau$, 
we show in brackets separately the $\nue$ contribution to the observed 
rate assuming $\nue-\nu_{\rm active}$ oscillations. We note 
that the observed rate have a strong nonmonotonic dependence on the 
neutrino energy since the Ga experiments which see the lowest 
energy $\nue$ have the highest rate, the Cl experiment observes 
intermediate energy neutrinos and reports the lowest rate, while 
the SK and SNO 
which are sensitive to the highest energy neutrinos, have a rate that 
is intermediate between the Ga and Cl rates. 

\begin{table}[htbp]
    \begin{center}
\begin{tabular}{||c|c|c||} \hline\hline
{\rule[-3mm]{0mm}{8mm}
experiment} & $\frac{obsvd}{BPB00}$ & composition \\ \hline \hline 
Cl &
0.335 $\pm$ 0.029 & \br (75\%), \ber (15\%)
\\\hline
Ga & 0.584 $\pm$ 0.039 &$pp$ (55\%), \ber (25\%), \br (10\%)
\\\hline
& 0.459 $\pm$ 0.017  & 
\\
\raisebox{1.5ex}[0pt] {SK}
& (0.351 $\pm$0.017) &  \raisebox{1.5ex}[0pt] {\br (100\%)}
\\ \hline
SNO(CC) & 0.347 $\pm$ 0.027 &  \br (100\%)\\\hline 
& $0.473 \pm 0.074$ & \\
\raisebox{1.5ex}[0pt] {SNO(ES)}
&($0.368 \pm 0.074$)&\raisebox{1.5ex}[0pt] {\br (100\%)}\\
\hline\hline
\end{tabular}
      \end{center}
\vskip -0.3in
      \caption[The observed solar neutrino rates]{\label{rates}
   The ratio of the 
observed solar neutrino rates to the corresponding BPB00 SSM predictions. 
The rate due to $\nue$ events in SK and SNO(ES) 
assuming $\nue-\nu_{\rm active}$ 
oscillations 
is shown within parentheses. The Ga rate corresponds to the combined 
SAGE and GALLEX+GNO data. Also shown is the composition of the 
observed fluxes. }
\end{table}

In sharp contrast to the strong nonmonotonic energy dependence 
exhibited by the data on total rates, the recent SK data on the energy 
spectra at day and night show no evidence for any energy 
dependence. 
The SNO is also consistent with no spectral distortion, however the 
errorbars for the SNO spectrum is still high. 
Hence there is an apparent conflict between the 
total rates and the SK spectrum data since the former would prefer 
solutions with strong nonmonotonic energy dependence while the latter 
would favor solutions with relatively weak dependence on energy. 
In addition the SK data is consistent with little or no day-night 
asymmetry which we shall see rules out large parts of the 
parameter space which predict strong Earth matter effects.

\subsection{The Solar Neutrino Code}

We perform a dedicated analysis of the global solar neutrino data 
on the total observed rate
and the SK day-night recoil electron energy spectrum. 
This takes into account all available {\it independent} 
experimental features of the solar neutrino data. 
We take the rates from the 
Cl, Ga (SAGE and GALLEX+GNO combined), SK and SNO CC experiments. 
The SNO ES data is not incorporated as it 
has large error. We also leave out the SNO CC spectrum for 
the same reason.  
We do not incorporate the Kamiokande rate as discussed before.
We use the $\chi^2$ minimization 
technique to determine the best-fit parameters and draw the 
C.L. contours. For the statistical analysis for the total rates we 
define the $\chi^2$ function as 
\be
{\chi_R^2 = \sum_{i,j=1}^4 \left ( R_{i}^{th} - R_{i}^{exp}\right )
(\sigma_{ij}^{2})^{-1} \left ( R_{j}^{th} - R_{j}^{exp}\right )}
\label{chirate}
\ee
where $R_i^{th}$ is the theoretical prediction of the 
event rate for the $i^{th}$ 
experiment and $R_i^{exp}$ is the corresponding observed value 
shown in Table \ref{rates}. 
The error matrix $\sigma_{ij}^2$ contains 
the experimental errors, the theoretical errors and their correlations.
For the evaluation of the error matrix $\sigma_{ij}^2$ 
we have followed the procedure given in \cite{flerrorsolar}. 

The expected event rate for the radiochemical experiments Cl and Ga 
in presence of oscillations is
\be
R_i^{th} = \sum_{k=1}^8 \int_{E_{\nu}^{th}} \phi_k(E_\nu) \sigma_i(E_\nu) 
\langle P_{ee}(E_\nu)\rangle dE_\nu
\label{radiorate}
\ee
where $\sigma_i(E_\nu)$ is the capture cross section for the $i^{th}$ 
detector, $E_{\nu}^{th}$ is the detector threshold,  
$\langle P_{ee}(E_\nu)\rangle$ is the neutrino survival probability 
averaged over the distribution of the neutrino production region 
inside the Sun,
$\phi_k(E_\nu)$ is the neutrino spectrum from the 
$k^{th}$ source inside the Sun and the sum is over all the eight sources.
For the SK experiment the corresponding event rate is given by
\be
R_{SK}^{th}\!\! = \!\! \int_{E_A^{th}}\! dE_A \int
dE_T R(E_A,E_T)\int
dE_\nu\lambda_\nue (E_\nu)
\left[\frac{d\sigma_{\nue}}{dE_T}
\langle P_{ee}(E_\nu)\rangle+
\frac{d\sigma_{\nu_x}}{dE_T}
\langle P_{ex}(E_\nu)\rangle\right]
\label{skrate}
\ee
where
$\lambda_\nue$ is the normalized $^8B$ neutrino spectrum,
$E_T$ is the true and $E_A$ the apparent(measured) kinetic energy of
the recoil electrons, $E_A^{th}$ is the detector
threshold energy which is 5.0 MeV
and R($E_A$,$E_T$) is the energy resolution function which is
taken as \cite{bahlisi} 
\be
R(E_A,E_T)&=&\frac{1}{\sqrt{2\pi \sigma^2}}
\exp\left(-\frac{(E_T-E_A)^2}{2\sigma^2}\right)
\label{resol}
\\
\sigma &=& 1.6\sqrt{\frac{E_T}{10 {\rm MeV}}}
\ee
In eq. (\ref{skrate}) $\langle P_{ee}\rangle$
is the time averaged $\nu_e$ survival probability,
$\langle P_{ex}\rangle$ is the time averaged transition probability
from $\nue$ to $\nu_x$, where $\nu_x$ is either $\numu$ or $\nutau$,
$d\sigma_\nue/dE_T$ is the differential cross section for
($\nu_e-e$) scattering while $d\sigma_{\nu_x}/dE_T$ is the 
corresponding cross section for 
($\nu_x-e$) scattering. Note that if 
one has $\nue-\nu_{\rm sterile}$ transitions involved 
the second term will be absent 
and only the $\nue$ contribution to the 
scattering rate will survive. 

For the  $\nu_e-d$ CC event rate in SNO we use
\be
R_{CC}^{th} &=& \frac {\int dE_\nu \lambda_{\nu_e}(E_\nu)
\sigma_{CC}(E_\nu)\langle P_{ee}(E_\nu)\rangle} {\int dE_\nu
\lambda_{\nu_e} (E_\nu) \sigma_{CC}(E_\nu)} 
\\
\sigma_{CC} &=& \int_{E_{A}^{th}}dE_A\int_0^\infty dE_TR(E_A,E_T)
\frac{d\sigma_{\nu_e d} (E_T,E_\nu)}{dE_T} 
\label{ccosc}
\ee
For SNO $E_{A}^{th}=(6.75 + m_e)$ MeV, where $m_e$ is the mass of the 
electron and d$\sigma_{\nu_e d}$/dE$_T$ is the 
differential cross section of the $\nu_e-d$ interaction. One of the
major uncertainties in the SNO CC measurement stems from the
uncertainty in the $\nu_e -d$ cross-section. We use the
cross-sections from \cite{nakamura} which are in agreement with
\cite{butler}. Both calculations give an uncertainty of 3\% which
is also the value quoted in \cite{sno3}\footnote {It was recently
pointed out in \cite{beacom} that the calculation of both
\cite{nakamura} and \cite{butler} underestimate the total $\nu_e
-d$ cross-section by 6\%. We have not included this effect in our
calculation.}. $R(E_A,E_T)$ for SNO is given by the same functional form 
(\ref{resol}) with the $\sigma\equiv\sigma_{SNO}$ given as 
\cite{sno3}
\be
\sigma_{SNO}&=&(-0.462+0.547\sqrt{E_T}+0.008722E_T)
\ee

For the analysis of the day-night effect and the energy behavior 
of the suppression rate we define a $\chi^2$ function for the 
SK 1258 day day-night recoil electron energy spectra as 
\be
\chi_S^2 = \sum_{i,j=1}^{38} \left ( X_n S_{i}^{th} - S_{i}^{exp}\right )
(\sigma_{ij}^2)^{-1}\left (X_n S_{j}^{th} - S_{j}^{exp}\right )
\label{chispec}
\ee
where $S_{i}^{th}$ are the theoretically calculated predictions 
for the $i^{th}$ energy bin, normalized to BPB00, $S_{i}^{exp}$ 
are the corresponding observed values and the sum is over 19 day + 
19 night energy bins provided by SK. 
The error matrix for the spectrum analysis 
is defined as in \cite{ggerrorsolar}. 
In eq. (\ref{chispec}) $X_n$ is an overall normalization constant which is 
allowed to vary freely in the analysis. The SK provides information 
about three aspects of the solar neutrino flux suppression, 
(i) the overall suppression rate, 
(ii) the energy dependence of the suppression and 
(iii) the effect of Earth matter on the suppression rate. 
The information about the overall \br flux observed by SK 
is embodied both in the total rate and in the spectrum data. 
Since we have already accounted for this piece of information 
in the $\chi^2_R$ we avoid the double counting of the total 
suppression rate in $\chi^2_S$ by introducing this floating 
normalization $X_n$. Thus the SK 
day-night spectrum data provides information on 
only the presence of energy distortion, if any. It gives 
information on the the day-night asymmetry as well. 

For the global analysis we take into account the
data on total rates as well the SK day-night spectrum data and
define our total \chisq as $\chisq = \chi^2_R + \chi^2_S$.
If we assume no new property for the neutrino and use the flux 
predictions from BPB00, then the value of $\chisq=89.27$ which is 
definitely unacceptable. Even if the constraints on the solar models 
is relaxed, so that one allows the fluxes to take on any arbitrary 
value subject to the solar luminosity constraint, the fit is extremely 
poor if all the three experiments are considered 
together. The data 
cannot be explained by this approach, 
even if one takes only two experiments at a time. 
In fact as discussed in the introduction, all 
such fits predict ``missing \ber neutrinos". This happens 
because the Ga observed flux can be almost accounted for by the $pp$ 
and $pep$ fluxes alone, given the luminosity constraint. 
If simultaneously the observations of the water \chr experiments 
are to be accounted for, then 
there is an extra contribution from the \br flux in Ga 
leaving no room for the \ber flux. If on the other hand one 
considers Cl and SK together, then the expected \br flux in the former 
from the observation of the \br flux in the latter, more than compensates 
the observed rate in Cl, again demanding complete suppression 
of \ber \cite{bahcallfit}. 
With the advent of the SNO CC result the astrophysical solution 
gets comprehensively ruled out \cite{bah1}. 
Thus one has to invoke some new property for the neutrinos 
beyond the standard model of particle physics in order to solve the 
solar neutrino problem. 
We probe the viability of neutrino mass and 
flavor mixing as a possible explanation of this discrepancy. 

We first find the best-fit solution to the data on only the total 
rates by minimizing $\chi^2_R$. Next we take into account the 
global data on rates as well as the SK day-night spectrum data 
so that our total \chisq is $\chisq = \chi^2_R + \chi^2_S$. 
We minimize this \chisq for $\nue-\nu_{\rm active}$ 
oscillations keeping the \br flux normalization in the total rates 
fixed at the SSM prediction. We 
repeat the entire analysis for $\nue-\nu_{\rm sterile}$ oscillations. 
For both these neutrino flavor mixing 
analyses we adopt a unified approach to which we turn our
attention next.

\subsection{Unified Formalism for Analysis of Solar Data}

The general expression for the probability amplitude of survival for an
electron neutrino produced in the deep interior of the Sun, for
two neutrino flavors, is given by \cite{qvofl}
\be
A_{ee} = A_{e1}^{\odot}  A_{11}^{vac}
A_{1e}^{\oplus} + A_{e2}^{\odot}
A_{22}^{vac} A_{2e}^{\oplus}  
\ee 
where $A_{ek}^\odot (k=1,2)$ gives the probability amplitude of
$\nu_e \rightarrow \nu_k$
transition at the solar surface,
$A_{kk}^{vac}$ is the survival amplitude from the solar surface to the
surface of the Earth and
$A_{ke}^\oplus$ denotes the $\nu_k\rightarrow \nue$
transition amplitudes inside the Earth.
We can express
\be
A_{ek}^{\odot} = a_{ek}^{\odot}
e^{-i \phi^{\odot}_k}
\ee
where $\phi^{\odot}_k$ is the phase picked up by the neutrinos 
on their way from the production point in the central regions to the
surface of the Sun and
\be
{a_{e1}^{\odot}}^2 = \frac{1}{2} + (\frac{1}{2} - P_J)\cos2\theta_m
\label{ae1}
\ee
where 
$\theta_m$ is the mixing angle at the production point 
of the neutrino and is given by eq. (\ref{tanm}) for transitions 
to active and by eq. (\ref{mixmste}) for transitions to sterile 
neutrinos,  
$P_J$ is the non-adiabatic jump probability given by eq. (\ref{pj}) which
for the exponential density profile of the Sun can be conveniently
expressed as \cite{petcov}
\be
P_J &=& \frac{\exp(-\gamma_c \sin^2 \theta) -
\exp(-\gamma_c)}{1-\exp(-\gamma_c)}
\\
\gamma_c &=&\pi\frac{\Delta m^2}{E}\left| \frac{d~ln n_e}{dr}
\right |_{r=r_{res}}^{-1}
\ee
The survival amplitude
$A_{kk}^{vac}$ is given by
\be
A_{kk}^{vac} = e^{-i E_{k} (L - R_{\odot})}
\ee
where $E_k$ is the energy of the state $\nu_k$, $L$ is the distance 
between the center of the Sun and Earth and $R_\odot$ is the solar 
radius. 
For a two-slab model of the Earth --- a mantle and core with constant
densities of 4.5 and 11.5 gm cm$^{-3}$
respectively, the expression for $A_{2e}^{\oplus}$ can be written as
(assuming the flavor states to be continuous across the boundaries)
\cite{petearth}
\be
A_{2e}^{\oplus}&=&
\sum_{\stackrel{i,j,k,}{\alpha,\beta,\sigma}}
U_{ek}^M e^{-i\psi_k^M}
U_{\alpha k}^M U_{\alpha i}^C
e^{-i\psi_i^C}
U_{\beta i}^C U_{\beta j}^M
e^{-i\psi_j^M}
U_{\sigma j}^M U_{\sigma 2}
\ee
where ($i,j,k$) denotes mass eigenstates and
($\alpha,\beta,\sigma$) denotes flavor eigenstates, $U$,
$U^M$ and $U^C$ are the mixing matrices in vacuum, in the mantle and the
core respectively and $\psi^M$ and $\psi^C$ are the corresponding phases
picked up by the neutrinos as they travel through the mantle and the core
of the Earth.
The $\nu_e$ survival probability is given by
\begin{eqnarray}
P_{ee}  & = &  |A_{ee}|^2 \nonumber \\
        & = & {a_{e1}^\odot}^2 |A_{1e}^\oplus|^2
+ {a_{e2}^\odot}^2 |A_{2e}^\oplus|^2
                \nonumber \\
        &   & + 2 a_{e1}^\odot a_{e2}^\odot
               Re[A_{1e}^\oplus {A_{2e}^\oplus}^{*}
e^{i(E_{2} - E_{1})(L - R_{\odot})} e^{i(\phi_2^\odot - \phi_1^\odot)}]
\label{pr}
\end{eqnarray}
Identifying
$P_{\odot} = {a_{e1}^\odot}^2$ and
$P_{\oplus} = |A_{1e}^\oplus|^2$
eq. (\ref{pr}) can be expressed as \cite{qvofl,stp,murayama} 
\be
P_{ee}&=&P_{\odot}P_{\oplus} + (1-P_{\odot})
(1-P_{\oplus})\nonumber\\
&& + 2\sqrt{P_{\odot}(1-P_{\odot})
P_{\oplus}(1-P_{\oplus})}\cos\xi
\label{probtot}
\ee
where we have combined all the phases involved in the Sun, vacuum and
inside Earth in $\xi$.
This is the most general expression for survival probability for the
unified analysis of solar neutrino data.
Depending on the value of $\Delta m^2/E$ one recovers the well known
{\it Mikheyev-Smirnov-Wolfenstein} (MSW) \cite{wolfen3,ms3}
and vacuum oscillation (VO) \cite{bruno} limits:

$\bullet$ In the regime $\Delta m^2/E \stackrel{<}{\sim}
5\times 10^{-10}$ eV$^2$/MeV matter effects inside the Sun
suppress flavor transitions and
$\theta_m \approx \pi/2$. Therefore, from (\ref{ae1}), we obtain
$P_{\odot} \approx P_J \approx \cos^2 \theta$
as the propagation of neutrinos is extremely
non-adiabatic and likewise,
$P_{\oplus} = \cos^2\theta$ to give
\beq
P_{ee}^{vac} = 1 - \sin^2 2\theta \sin^2(\dm (L-R_\odot)/4E)
\label{probvac}
\eeq

$\bullet$ For $\Delta m^2/E \stackrel{>}{\sim} 10^{-8}$
eV$^2$/MeV, the total oscillation phase
becomes very large and the $\cos\xi$ term in eq. (\ref{probtot})
averages out to zero.
One then recovers the usual
MSW
survival probability
\be
P_{ee}^{MSW}=P_{\odot}P_{\oplus} + (1-P_{\odot})
(1-P_{\oplus})
\label{proball}
\ee

$\bullet$ In between the {\it pure} vacuum oscillation regime where the
matter effects can be safely neglected, and the {\it pure} MSW
zone where the coherence effects due to the phase $\xi$ can
be conveniently disregarded, is a region where both effects can
contribute. For
$5\times 10^{-10}$ eV$^2$/MeV $\stackrel{<}{\sim}
\Delta m^2/E \stackrel{<}{\sim} 10^{-8}$ eV$^2$/MeV, both matter effects
inside the Sun and coherent oscillation effects in the
vacuum become important. This is the {\it quasi vacuum oscillation}
(QVO) regime \cite{qvofl,qvo}. In this region, $P_\odot\approx P_J$ and
$P_\oplus=\cos^2\theta$ and the survival probability is given by
\cite{stp,flqvo}
\beq
P_{ee} = P_J\cos^2\theta + (1-P_J)\sin^2\theta
+\sin^22\theta \sqrt{P_J(1-P_J)}\cos\xi
\eeq
We calculate $P_J$ in this region using the
prescription given in \cite{flqvo}.

\subsubsection{Day-Night Effect}

For the the range of $\Delta m^2/E$ for which matter effects inside 
the Earth are important (the pure MSW regime), 
one expects a significant day-night 
asymmetry. 
During day time the neutrinos do not cross the Earth and 
$P_\oplus$ is simply the projection of the $\nu_1$ state onto the 
$\nue$ state. Hence the $\nue$ survival probability during day is simply 
\be
P_{ee}^D = \frac{1}{2} + (\frac{1}{2} - P_J) \cos2\theta\cos2\theta_m
\label{probday}
\ee
The probability during night 
is given by the full expression (\ref{proball}). 
If one factors out $P_{ee}^D$ 
in the complete expression (\ref{proball}) which includes the 
Earth matter effects then\footnote{Note that for the purpose of simplicity 
of presentation we show the expressions where the phases 
have been averaged out to zero. However for the actual 
calculation of the probabilities we use the unified expression 
(\ref{probtot}).}
\be
P_{ee} = P_{ee}^D +
\frac{(2P_{ee}^D
-1)(\sin^2\theta-P_{2e}^{\oplus})}
{\cos2\theta}
\label{probnight}
\ee
where $P_{2e}^{\oplus}=1-P_\oplus$. From eq. (\ref{probday}) and 
(\ref{probnight}) we see that the extra contribution coming 
due to the matter effects inside the Earth is
\be
P_{ee} - P_{ee}^D = \frac{(2P_{ee}^D                                       
-1)(\sin^2\theta-P_{2e}^{\oplus})}                      
{\cos2\theta}                                        
\label{regen1}
\ee
This is the total regeneration of $\nue$ inside the Earth which we 
shall call 
$R_E$. For convenience we shall define the {\it regeneration factor}
\be
{\rm f_{reg}} = P_{2e}^{\oplus} - \sin^2\theta
\label{freg}
\ee
In the absence of any Earth matter effects, 
$P_{2e}^{\oplus} = \sin^2\theta$ and so ${\rm f_{reg}}=0$ and we get
back $P_{ee}^D$ from eq. (\ref{probnight}). We also note that 
(with $P_{ei}^\odot = |a_{ei}^\odot|^2$)
\be
\frac{(2P_{ee}^D                                       
-1)}{\cos2\theta}
&=&(1-2P_J)\cos2\theta_m
\nonumber\\
&=& -(P_{e2}^\odot - P_{e1}^\odot)
\ee
In other words the above factor quantifies the amount of level 
crossing due to loss of adiabaticity at resonance. 
The Earth regeneration now can be conveniently expressed as
\be
R_E &=& (P_{e2}^\odot - P_{e1}^\odot){\rm f_{reg}} 
\label{regen3}\\
&=& -(1-2P_J)\cos2\theta_m{\rm f_{reg}}
\label{regen4}
\ee
Hence from eq. (\ref{regen3}) and (\ref{regen4}) we see 
that the $\nue$ regeneration inside the Earth depends on
\begin{enumerate}
\item {\it The adiabatic factor $P_J$:} 
$R_E$ is maximum for $P_J=0$, decreases with increasing
$P_J$, hits the minimum ($R_E=0$) for
$P_J=1/2$ and changes sign for $P_J>1/2$. Which means that 
for $P_J>1/2$ we 
have further depletion of $\nue$ as they pass through the Earth. 

\item {\it The value of $\cos2\theta_m$:} We can see from eq. (\ref{tanm})
that $\cos2\theta_m > 0$ if the resonance density is more than the 
density at which the neutrinos are produced, $\cos2\theta_m =0$ 
if the neutrinos are produced at the position of resonance and 
$\cos2\theta_m < 0$ if the resonance density becomes less than the 
production density. As the resonance density decreases 
$\cos2\theta_m$ decreases and reaches the value 
$\cos2\theta_m \approx -1$. 
As the value of \dm decreases the position of resonance 
for the solar neutrinos shifts further outward and the value of 
$\cos2\theta_m$ approaches $-1$. Thus for the LOW solution 
$\cos2\theta_m=-1$ for almost all neutrinos energies and one gets 
maximum regeneration for the LOW solution.

\item {\it The regeneration factor:} The net regeneration due to Earth 
matter effects depends crucially on the value of $f_{\rm reg}$ which 
determines quantitatively the actual effect of Earth matter.
\end{enumerate}

%
\begin{figure}[h]
    \centerline{\psfig{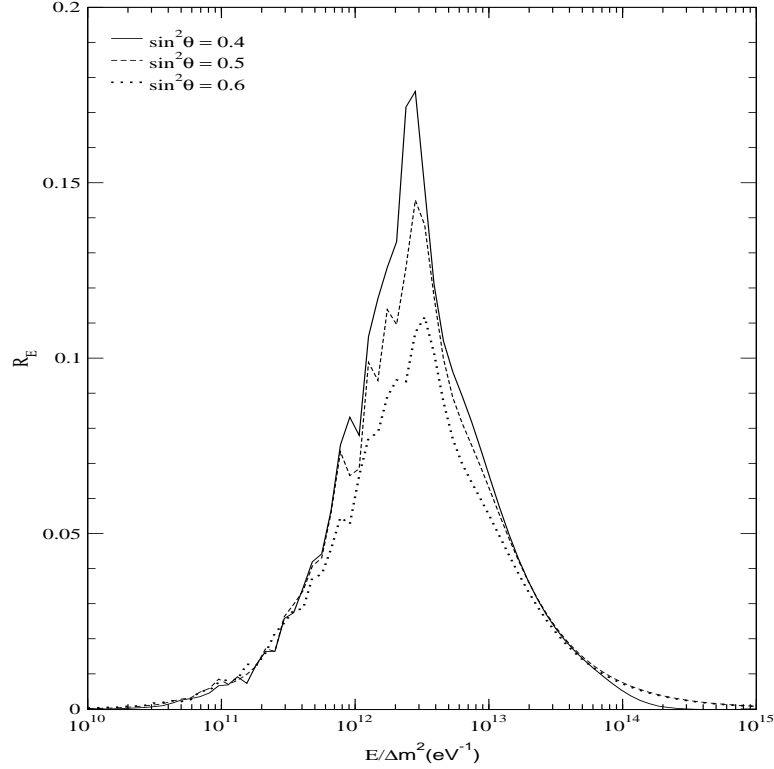}}
     \caption[Earth regeneration as a function of $\frac{E}{\dm}$]
     {\label{reg1}Earth regeneration of $\nue$ as a function of 
     $\frac{E}{\dm}$ for three different values of the mixing angle.}
\end{figure}
\begin{figure}[h]
    \centerline{\psfig{file=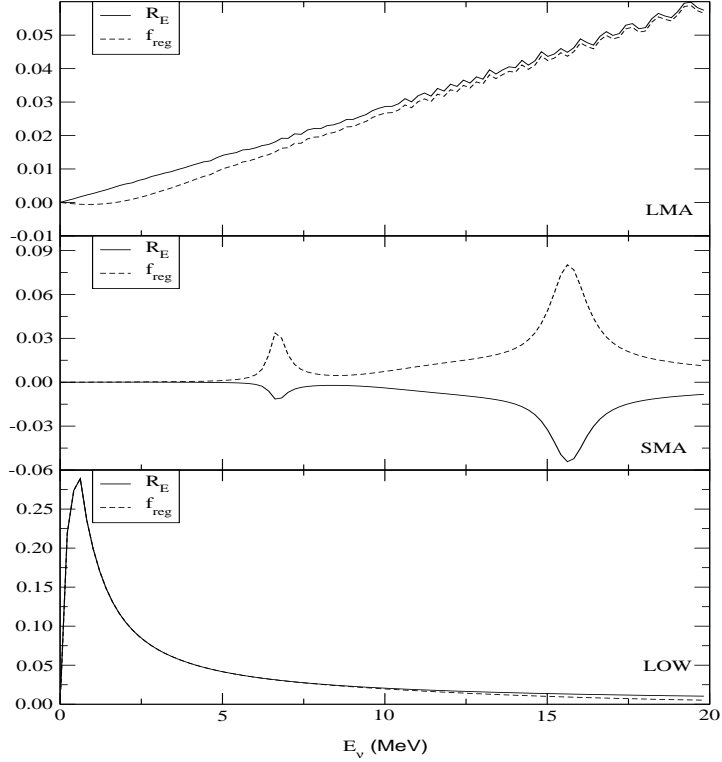,width=4.0in,height=10.5cm}}
\vskip -0.8cm
     \caption[The regeneration factor ${\rm f_{reg}}$
      as a function of neutrino energy]
     {\label{reg2}The regeneration factor ${\rm f_{reg}}$ 
and the net regeneration $R_E$ as a function of
     neutrino energy
     for typical values of the parameters in the SMA, LMA and LOW
     regions.}
\end{figure}

In fig \ref{reg1} we show the Earth regeneration $R_E$ as a function of
$\frac{E}{\dm}$ for three values of the mixing angle in the range 
$0<\theta<\frac{\pi}{2}$. We note that the Earth matter effects are important
only for $10^{11}$ eV$^{-1} \stackrel {<}{\sim} \frac{E}{\dm} 
\stackrel {<}{\sim} 10^{14}$ eV$^{-1}$ which is the ``pure" MSW 
regime and peaks at $\frac{E}{\dm} \sim 3\times 10^{12}$ eV$^{-1}$. 
Since around $E=5-15$ MeV the SK 
day-night data allows for very small day-night 
asymmetry, most of regions around $(2-8)\times 10^{-6}$ eV$^2$ for 
large mixing angles are disfavored. 
In fig \ref{reg2} we show the regeneration factor 
${\rm f_{reg}}$ and the total Earth regeneration $R_E$ vs energy at the 
SK latitude 
for typical values of the parameters 
in the SMA, LMA and LOW-QVO regimes. 
Since the latitude of the other detectors are not very different
we do not expect $f_{reg}$ and $R_{E}$ to be very different for them. 
Noteworthy point is that while the {\it 
regeneration factor} ${\rm f_{reg}}$ is positive for all the three 
cases considered, $R_E$ turns out to be negative for the SMA case. 
This is because for the SMA solution $P_J > 1/2$, or in other words 
$P_{e2}^\odot - P_{e1}^\odot < 1$, signifying large 
level crossing from the $\nu_2$ to the $\nu_1$ state in the solar 
matter at resonance. 
On the other hand for the LMA and LOW solutions the neutrino 
moves adiabatically inside the Sun, the $\nue$ produced in the 
$\nu_2$ state remains in a $\nu_2$ state throughout 
and $P_{e2}^\odot - P_{e1}^\odot 
\sim 1$, so that $R_E \approx {\rm f_{reg}}$. Thus for both 
LMA and LOW solutions one has positive regeneration of $\nue$ inside 
the Earth, the effect being more for the latter since for low \dm 
all the neutrinos resonate far away from the production zone and   
$\cos2\theta_m$ is closer to -1. Also note that for the LOW solution 
the regeneration is important at low energies while LMA has 
more regeneration for higher energy neutrinos. 

\begin{figure}
\vskip -0.5cm
    \centerline{\psfig{file=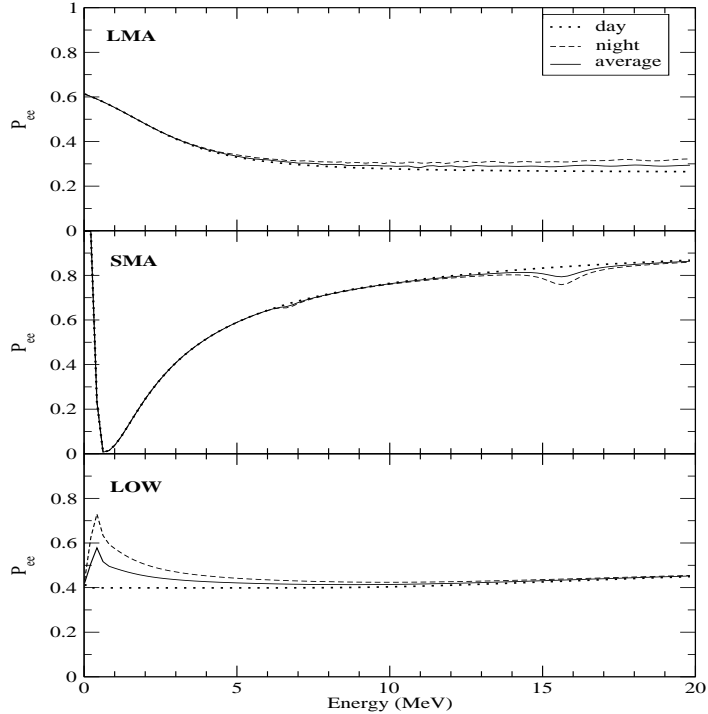,width=4.0in,height=10.5cm}}
\vskip -1cm
      \caption[The neutrino survival probability vs. energy]
     {\label{prob}The neutrino survival probability $P_{ee}$ 
     as a function of neutrino energy for the LMA, SMA and 
     LOW solutions.}
\end{figure}

We finally present in fig \ref{prob} the actual survival probability 
$P_{ee}$ vs energy 
during day (shown by dotted lines), during night (shown 
by dashed lines) and the day-night average (shown by solid lines) for 
SMA, LMA and LOW case. In order to understand the nature of the 
probabilities we call $\cos2\theta \equiv \epsilon$ and note that:

$\bullet$ For the SMA region $\epsilon \approx 1$ and from fig. \ref{reg2} 
we observe
that $f_{reg}$ is very small excepting for two peaks at
E $\approx$ 6 MeV and E $\approx$ 15 MeV corresponding to
strong enhancement of the earth regeneration effect for the neutrinos
passing through the core \cite{akh,petearth}. Hence
\beq
P_{ee}^{SMA} \approx P_{ee}^D
\label{prsma}
\eeq
In this region
for low energy ($pp$) neutrinos, resonance is not encountered
(resonance density $\gg$ maximum solar density) and hence $P_J\approx 0$
and $\cos2\theta_{m} \approx 1$ giving $P_{ee}^{SMA} \approx 1$.
For intermediate energy ($^{7}{Be}$) neutrinos
$\cos2\theta_{m} \approx -1$ (resonance density $\ll$ production density)
and $P_{ee}^{SMA} \approx P_{J} \approx 0$ for these energies. 
For high energy ($^8B$) neutrinos also, $\cos2\theta_{m} \approx -1$ and
$P_{ee}^{SMA} \approx P_{J}$, with $P_J$ rising with energy.

$\bullet$ For the LMA solution 
the motion of the
neutrino in the solar matter is adiabatic for almost all neutrino
energies and $P_J \approx 0$.
For low energy neutrinos the matter effects are weak both
inside the Sun and in Earth giving $f_{reg} \approx 0$ and
$\cos 2\theta_{m} \approx \epsilon$ so that for Ga energies
\cite{concha}
\beq
P_{ee}^{LMA} \approx \frac{1}{2}(1 + \epsilon^2)
\label{prgalma}
\eeq
At SK and SNO energies matter effects result in $\cos2\theta_{m}\approx -1$
while $f_{reg}$ is small
but non-zero
($\approx$ 0.03 at 10 MeV as seen from fig. \ref{reg2}) 
giving
\begin{eqnarray}
P_{ee}^{LMA} &\approx& \frac{1}{2}(1 - \epsilon) + f_{reg}\nonumber\\
&=& \sin^2\theta + f_{reg}
\label{prsklma}
\end{eqnarray}

$\bullet$
In the LOW region $\cos2\theta_m \approx -1$ for all neutrino energies
and  $P_J\approx 0$ (except for very high energy neutrinos) and thus
for all neutrino energies 
\beq
P_{ee}^{LOW} = \frac{1}{2}(1 - \epsilon) + f_{reg}
\label{prlow}
\eeq
where 
$f_{reg}$ is small for high energy neutrinos and large for
low energy neutrinos (cf. fig. \ref{reg2}).

\subsection{Results and Discussions}

We present in Table \ref{totchi} the results of the $\chi^2$ analysis 
for $\nue-\nu_{\rm active}$ oscillations, 
using data from the Cl, Ga, SK  and SNO\footnote{We 
incorporate only the SNO CC rate as the SNO ES rate and the 
SNO CC spectrum still have large errors.} 
experiments. We use the total rates given in 
Table \ref{rates} and the 1258 day SK 
recoil electron energy spectrum at day and night. 
We show the best-fit values of the parameters 
\dm and \tit, $\chi^2_{\rm min}$ and the goodness of fit (GOF) 
for the SMA, LMA, LOW-QVO, VO and Just So$^2$ \cite{ragh} solutions. 

The best-fit for the only 
rates analysis comes in the VO region which is favored at 28.79\%. 
Prior to SNO the SMA solution could explain the nonmonotonic energy 
dependence of the survival probability from the Cl, Ga and SK experiments 
well and was the best-fit solution. But with the advent of SNO it falls 
into disfavor and is allowed at only 6.59\%. 
For the LMA solution on the other hand the survival probability is 
given by eqs. (\ref{prgalma}) and (\ref{prsklma}) at Ga and SK/SNO 
energies respectively and for the values of $\epsilon$ from Table \ref{totchi} 
and $f_{\rm reg}$ given in fig. \ref{reg2}, it 
approximately reproduces the rates of Table \ref{rates}. 
LMA is allowed at 18.27\% while LOW-QVO is barely allowed at 1.55\%. 
In fact the 
LOW solution gets allowed only due to the strong Earth regeneration 
effects at low energies which 
helps the LOW solution to 
explain the Ga data better.
In addition to these four solutions we have a fifth solution called the 
Just So$^2$ solution \cite{ragh} at $\dm \sim 5.38\times 10^{-12}$ eV$^2$. 
For these \dm one gets a very small survival
probability for the $^{7}{Be}$ neutrinos while for the \br
neutrinos the survival probability is close to 1.0 \cite{justso2}.
Therefore it cannot explain the total rates data.

\begin{table}[htb]
    \begin{center}
       \begin{tabular}{||c||c|c|c|c|c||} \hline\hline
        &Nature of & $\Delta m^2$ &
         $\tan^2\theta$&$\chi^2_{min}$& Goodness\\
            &Solution & in eV$^2$&  & & of fit\\
          \hline\hline
 &SMA & $7.71\times 10^{-6}$&$1.44 \times 10^{-3}$ & 5.44 & 6.59\% \\
 &LMA & $2.59 \times 10^{-5}$ & 0.34 & 3.40 & 18.27\% \\
rates &LOW-QVO & $ 1.46 \times 10^{-7}$ & 0.67 & 8.34 & 1.55\%\\
 &VO& $7.73\times 10^{-11}$& 0.27 & 2.49 &28.79\%\\
&Just So$^2$& $5.38\times 10^{-12}$ & 1.29 & 19.26 &$6.57\times10^{-3}$\%\\
\hline
   &SMA & $5.28 \times 10^{-6}$&$3.75 \times 10^{-4}$ &
   51.14 & 9.22\%  \\
rates &LMA & $4.70 \times 10^{-5}$ & 0.38 &
   33.42 &  72.18\% \\
 +  &LOW-QVO & $ 1.76 \times 10^{-7}$ & 0.67 & 39.00 & 46.99\%
   \\
 spectrum   &VO& $4.64\times 10^{-10}$& 0.57 & 38.28 & 50.25\%\\
    &Just So$^2$&$5.37\times 10^{-12}$& 0.77&51.90&8.10\%\\
         \hline\hline
        \end{tabular}
      \end{center}
\vskip -0.5cm
      \caption[\chisq fits to the solar data for  $\nue-\nu_{\rm active}$]
      {\label{totchi}The best-fit values of the parameters,
        \chisqmin, and the goodness of fit from the
       global analysis of
       rates and rates+spectrum data for MSW oscillations involving
       two active neutrino flavors.}
\end{table}

We next perform a complete global analysis 
of the solar neutrino data taking the four total rates 
and the 1258 day SK day-night recoil electron spectrum data. 
We present in Table \ref{totchi} the results obtained by minimizing 
$\chi^2 = \chi^2_R + \chi^2_S$. 
We get five allowed solutions LMA, VO, LOW, SMA and Just So$^2$ in 
order of decreasing GOF. 
The LMA solution can approximately reproduce the rates as discussed above 
and since the survival probability 
(\ref{prsklma}) for SK 
is approximately energy independent it can account for the flat recoil
electron energy spectrum. LMA thus gives the best-fit being allowed at 
72.18\%. The LOW solution with best-fit 
$\epsilon=0.2$ and 
$f_{reg}\sim 0.2$ for Ga and $\sim 0.025$ for SK energies 
(cf. fig. \ref{reg2}) can just about reconcile the 
Ga and SK rates. However it provides a very good description of the 
flat SK spectrum and is allowed at 46.99\%. 
The VO solution at $\dm\sim 4.64 \times 10^{-10}$ eV$^2$ gives a 
very low \chisq for the spectrum data and hence the overall fit for 
VO is very good. However for the SMA solution there is a mismatch between 
the parameters that give the minimum \chisq for the rates data and the 
spectrum data. The spectrum data prefers value of \tit which are one 
order of magnitude lower than those preferred by the rates data. Thus 
the overall fit in the SMA region suffers and it is allowed at only 
9.22\%. The Just So$^2$ solution is very bad for the rates 
but since it gives a
flat probability for the \br
neutrinos the spectrum shape can be accounted for and the global analysis gives
a GOF of 8.1\%.

\begin{figure}[t]        
\vskip -3cm             
  \centerline{\psfig{file=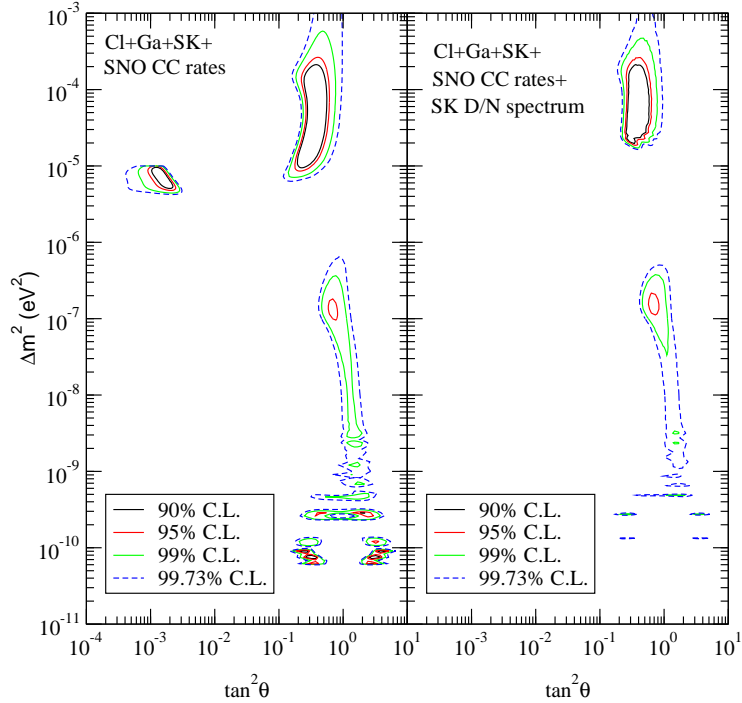,width=4.0in}}
\vskip -1in
   \caption[The allowed areas from
   global analysis of solar data]{\label{contsolar}
   The 90\%, 95\%, 99\% and  99.73\% C.L. allowed areas from the
   analysis of the total rates (left panel) and global analysis of 
the rates
   and the 1258 day SK recoil electron spectrum at day and night 
(right panel),
   assuming MSW conversions to sequential neutrinos.}
\end{figure}

In fig. \ref{contsolar} we show the 90\%, 95\%, 99\% and 99.73\% C.L. 
allowed areas from the analysis of the data on total rates (shown in the 
left hand panel) and the combined data on rates and the SK spectrum 
(shown in the right hand panel). 
For the only rates case we have allowed areas in the SMA, LMA, LOW-QVO 
and VO regions. For the global analysis we get allowed zones in 
the LMA and the LOW-QVO zones. In the VO region we get just two 
small areas which are allowed. But the most significant feature is the 
disappearance of the SMA solution 
from the global fit even at 99.73\% C.L. ($3\sigma$). 

We have repeated the entire analysis for the $\nue-\nu_{\rm sterile}$ 
case and have reported the results of the fit in Table \ref{chisterile}. 
We find that after the inclusion of the SNO data, all the solutions are 
disfavored with a probability of more than 99\% from the total rates 
analysis while for the global analysis the GOF of these become 
much worse. 
\begin{table}
    \begin{center}
       \begin{tabular}{||c||c|c|c|c|c||} \hline\hline
        &Nature of & $\Delta m^2$ &
         $\tan^2\theta$&$\chi^2_{min}$& Goodness\\
            &Solution & in eV$^2$&  & & of fit\\
          \hline\hline
&SMA & $4.18 \times 10^{-6}$&$5.72 \times 10^{-4}$ &
   17.24& $1.80 \times 10^{-2}$
   \%  \\
& LMA & $4.98 \times 10^{-5}$ & 0.54 &
   23.96 & 6.27$\times 10^{-4}$\% \\
rates&LOW-QVO & $1.00\times10^{-7}$ & 0.94 & 24.26 & 5.40$\times10^{-4}$\%
   \\
&VO& $1.07 \times 10^{-10}$& 0.27 &15.71 & $3.88 \times 10^{-2}$\%\\
 &Just So$^2$& $5.37 \times 10^{-12}$& 1.28 &19.40 & $6.13 \times 10^{-3}$\%\\
 \hline
    & SMA & 5.59 $\times 10^{-6}$&$2.83 \times 10^{-4}$ &
   54.21 & 5.35\%  \\
rates  &LMA & $6.13 \times 10^{-5}$ & 0.50 &
   52.93 & 6.75\% \\
+&LOW-QVO & $ 2.93 \times 10^{-8}$ & 1.00 & 53.18 & 6.45\%\\
spectrum    &VO& $4.67\times 10^{-10}$& 0.37 & 46.28 & 19.70\%\\
 &Just So$^2$& $5.37\times 10^{-12}$& 0.77 & 52.09 & 7.83\%\\
         \hline\hline
        \end{tabular}
      \end{center}
\vskip -0.5cm
      \caption[\chisq fits to the solar data for $\nue-\nu_{\rm sterile}$]
      {\label{chisterile}The best-fit values of the parameters,
        \chisqmin, and the goodness of fit from the
       global analysis of
       rates and rates+spectrum data for MSW oscillations involving
       $\nue-\nu_{\rm sterile}$ oscillations.}
\end{table}

\section{The Atmospheric Neutrino Anomaly}

The atmospheric neutrinos are produced due to the collision of
cosmic rays ($N_{cr}$) with the nuclei in the atmosphere
($N_{air}$) resulting in a chain of reactions which culminates in 
the production of neutrinos and antineutrinos with expected flavor 
ratio of roughly $\nue:\numu:\nutau \approx 1:2:0$.
\be
N_{cr} + N_{air} &\longrightarrow& \pi^{\pm},~K^{\pm},~K^0,......
\label{atmchain}
\\
\pi^{\pm}(~K^{\pm}) &\longrightarrow& \mu^{\pm} + \numu(\anumu)
\nonumber\\
\mu^{\pm} &\longrightarrow& e^{\pm} + \nue(\anue)+\anumu(\numu)
\label{decaychain}
\ee
The first pioneering measurement of events induced by these 
neutrinos were made in the Kolar Gold Fields in India \cite{kgf} 
and almost simultaneously in South Africa \cite{sa}. 
But the interpretation 
of the results from these experiments remained ambiguous due to 
both theoretical as well as experimental uncertainties. 
It was only since late eighties that people earnestly started pursuing 
this problem \cite{kamatm1,kamatm,imb,frejus,nusex,skatm1} 
and around 1988 Kamiokande, which was originally designed to detect 
proton decay,   
declared its results on atmospheric neutrino measurements 
which showed a marked deficit of observed to expected $\numu/\nue$ 
ratio \cite{kamatm1}. This embodies the {\it Atmospheric Neutrino 
Anomaly}. This was the second evidence of missing neutrinos after 
the solar neutrino problem and again called for the existence 
of neutrino mass and mixing for the correct interpretation of 
the experimental results.

However due to the large theoretical and experimental uncertainties 
involved in the atmospheric neutrino measurement, it failed 
to prove itself as a compelling evidence of neutrino oscillations 
until the arrival of the Super-Kamiokande atmospheric results in 
1998 \cite{skatm1}. The SK atmospheric neutrino data 
\cite{skatm1,skatm2,skatm3}  not 
only confirmed the suppression of the muon type neutrinos, it 
also firmly established the fact that the observed deficit 
has a zenith angle dependence. That is, the neutrinos coming 
at larger zenith angles and hence traveling distances of 
the order of the diameter of the Earth were suppressed more 
compared to neutrinos arriving directly from the top of the 
detector. This differential depletion of the atmospheric $\numu$ 
flux is referred to as the observed {\it up-down asymmetry}. 
The $\nue$ events on the other hand are reported to be 
consistent with theoretical expectations. 

The most convincing particle physics scenario which can explain 
{\it all} aspects of the SK data is $\numu-\nutau$ oscillations 
with $\dm \sim 10^{-3}$ eV$^2$ and $\st \sim 1$ 
\cite{lisi,fl4atm,G_G,yasuda}. 
Thus the theory of  
neutrino mass and mixing, which was conjectured in the late 
sixties 
as a plausible solution to the 
solar neutrino problem, was established as an accepted reality 
by the SK atmospheric neutrino 
results in 1998. This observation of neutrino mass by the SK 
(although indirect) is the {\it first} and till date the {\it only} 
evidence of physics beyond 
the standard model of particle physics. 

The observed depletion of atmospheric neutrinos was also reported 
by the IMB \cite{imb} earlier and more recently by the 
Soudan2 \cite{soudan2} and MACRO \cite{macro} 
collaborations\footnote{The Frejus \cite{frejus} and Nusex \cite{nusex} 
experiments though were consistent with no deficit of atmospheric 
neutrinos.}. In order to do justice to all these experiments one should
perform a global analysis taking all the experimental data into 
account. However due to its overwhelming statistics, the SK data 
dictates the atmospheric neutrino 
analysis, with the other experiments having little impact on the 
fit and the allowed parameter regions. Hence considering the rich 
statistics and low systematics of the SK, 
we choose to work with just the SK data in our 
analysis of the atmospheric neutrino anomaly and 
perform dedicated \chisq-fits to the SK 1144 day data. 
We describe two different established methods of \chisq analysis 
of the atmospheric data and discuss their merits and demerits, present 
the best-fit \dm and \st and display the 90\% and 99\% C.L. allowed 
zones in the \dm-\st parameter space for two-generation $\numu-\nutau$
oscillations.

In section 3.2.1 we give an essence of the atmospheric neutrino flux 
calculations and the theoretical uncertainties associated with it. 
In section 3.2.2 we discuss the various features of the SK atmospheric 
experiment and state the main experimental results. In section 
3.2.3 we describe our atmospheric neutrino code and discuss 
the advantages and disadvantages of the two methods of \chisq 
analysis that we have used. Finally in section 3.2.4 we present the results
of our \chisq fits.

\subsection{Atmospheric Neutrino Flux Predictions}

Atmospheric neutrinos are the result of a cascade of reactions 
following the interaction of the cosmic rays with the 
air nuclei (cf. eq. (\ref{atmchain}), (\ref{decaychain})). 
In the decay chain 
(\ref{decaychain}), for every electron neutrino (or antineutrino) created 
one has two muon type neutrinos (or antineutrinos)\footnote{For the 
purpose of discussion we shall henceforth 
use the term neutrino to mean both 
neutrino as well as antineutrino though for the actual calculation 
they are treated differently.} produced. 
Thus one may naively guess that the ratio of fluxes 
\be
{\cal R} &=& \frac{\numu + \anumu}{\nue + \anue}
\nonumber \\
&\approx & 2
\label{ratio}
\ee
But for a 
realistic prediction for the atmospheric neutrino fluxes one needs to
fold in the cosmic ray fluxes with the hadronic interaction model, which 
takes into account the complete reaction chain and perform a 
comprehensive calculation. Two such widely used 
atmospheric flux calculations are by Honda \etal \cite{honda} and 
Agarwal \etal \cite{bartol}. 

The primary cosmic rays which are the main ingredient of the atmospheric 
neutrino flux calculations have 
large uncertainties in their over all normalization, composition, 
as well as in their spectrum, all of which are extremely 
crucial in the atmospheric neutrino flux calculations. Though the 
cosmic ray fluxes are relatively well known at $E\stackrel{<}{\sim} 
30$ GeV (these give rise to the sub-GeV atmospheric neutrino fluxes), 
there are few experimental measurements for the 
higher energy cosmic rays ($E\stackrel{<}{\sim} 1000$ 
GeV which result in multi-GeV neutrino fluxes) 
and hence this regime is plagued with large uncertainties. 

The resultant atmospheric neutrino fluxes also depend crucially on the
hadronic interaction model. In their calculation, Honda 
\etal use a full Monte Carlo method for the sub-GeV neutrinos 
while for the multi-GeV fluxes they employ a ``hybrid model" 
\cite{honda}. They make their atmospheric neutrino flux calculations 
using a one-dimensional approximation where they assume that all the 
secondary particles including the mesons and the leptons 
are collinear with the incident primary cosmic ray. Though this 
approximation works well for the higher energy calculations, 
it fails for the low energy sub-GeV fluxes, particularly 
for the horizontal zenith angles and therefore should be replaced 
by a full three-dimensional calculation\cite{3dhonda}. 

Let us next discuss some of the most important factors which affect 
the atmospheric neutrino flux predictions.
\begin{enumerate}
\item {\it The solar activity}: The effect of the solar wind 
on the cosmic rays entering the solar sphere of influence is to 
deviate their trajectories away from the Earth. This affects the low energy
cosmic rays more than the higher energy ones and is known as the solar 
modulation of the cosmic ray fluxes. Since the solar wind depends on 
the solar activity, the fluxes are maximum at the solar minimum and 
minimum at the solar maximum. 
%
\item {\it The geomagnetic field}: Since the cosmic rays consist of 
charged particles they experience a repulsion due to the 
Earth's magnetic field and may be deflected away. 
Hence only particles with momentum above a certain threshold 
can break this barrier and enter the Earth's atmosphere. This 
threshold is called the {\it rigidity cut off} and is defined as, 
rigidity=momentum/charge. 
This effect results in cutting off the lower 
energy cosmic rays which affects the low energy neutrino fluxes. 
The geomagnetic field bends the cosmic rays and thus 
introduces a directionality into the atmospheric fluxes, the 
effect obviously being more for lower energy neutrinos. 
This effect of the geomagnetic field results in predicting 
more neutrino flux from the west than from the east and is 
known as the {\it east-west effect}.  
SK has made observations of this predicted east-west anisotropy 
and has confirmed this estimated directional behavior \cite{skeastwest}. 
\item {\it The density structure of the atmosphere:} Once 
the pions and kaons are produced in the air, they may either decay creating
muons and neutrinos or may interact with other particles. Which of 
these processes dominate is dictated by the energy of the meson and 
decay becomes comparable to interaction when
\be
{\rm c}\frac{E}{{\rm mc}^2}\tau \sim \frac{1}{\sigma n},~~{\rm or} 
~~E\sim \frac{{\rm mc}^2}{{\rm c}\tau\sigma n}
\label{samerate}
\ee
where $\tau$ is the rest frame lifetime of the meson of energy $E$ 
and mass m, $\sigma$ is the interaction cross-section with air and 
$n$ is the number density of the air nuclei. 
Since the interaction rate depends inversely on the 
density of the atmosphere, and since the cosmic rays moving 
along the horizontal zenith angle 
interact with the air nuclei at a higher altitude 
(and hence lower density), the decay probability for these is 
more compared to interaction. 
Hence the resultant neutrino fluxes are maximum along the 
horizontal and minimum along the vertical directions.  
This gives rise to the 
zenith angle variation of the neutrino flux. Again, because the 
interaction rate 
of the meson is determined by its energy,  
the zenith angle dependence of the neutrino flux increases with energy.
\item {\it Decay lifetime of the mesons:} The ratio 
${\cal R}\approx 2$ only for $\pi$ decays at rest 
for which the decay (\ref{decaychain}) is complete. 
However above $E\stackrel {>}{\sim} 
5$ GeV the decay lifetime of the produced muon becomes so large 
that they fail to decay in air. This results 
in raising the value of ${\cal R}$ from 2. 
Again, as the effective atmospheric 
depth along  
the horizontal direction is more than that along the vertical, the decay 
chain (\ref{decaychain}) has comparatively 
a better chance of completion along the 
horizontal direction and this results in endowing ${\cal R}$ with a 
zenith angle dependence. 
\end{enumerate}

\begin{figure}
    \centerline{\psfig{file=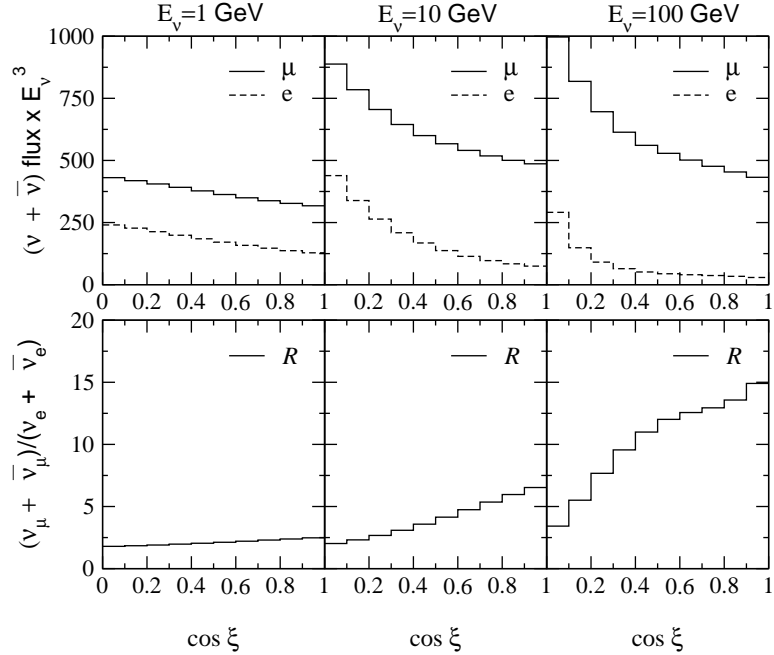,width=4.0in}}
\vskip -0.5cm
    \caption[The atmospheric fluxes.]{\label{atmflux} The 
atmospheric neutrino fluxes as a function of the neutrino 
zenith angle for three different fixed energies. Shown are 
the absolute fluxes as well as the ratio ${\cal R}$.}
\end{figure}

Though the uncertainty due to primary cosmic rays for the sub-GeV 
neutrinos is small, they are significantly modulated by the solar activity 
and the geomagnetic field. Being low in energy, they have less 
zenith angle dependence. The multi-GeV fluxes on the other hand have less 
dependence on solar activity and geomagnetic field, but have 
large zenith angle variation and have to contend with huge 
cosmic ray uncertainties.
We show in fig \ref{atmflux} the atmospheric neutrino flux predictions by
Honda \etal \cite{honda} as a function of the zenith angle for three
typical values of energies. Also shown in the lower panels are the 
corresponding values of ${\cal R}$. The fluxes shown are without the geomagnetic
effects so that they are symmetric about the horizontal and hence the 
sign of $\cos\xi$ is not important. We note that both $\nue$ as well 
as $\numu$ decrease along the vertical, displaying the zenith angle 
behavior discussed above, the effect being more for the 
$\nue$. Thus the ratio ${\cal R}$ gets raised from the often quoted 
value of 2 along the vertical. The above effect is seen to intensify with 
energy. 

Among the major uncertainties involved in the atmospheric flux 
calculations, the most important contribution comes from the lack 
of correct estimates for the cosmic ray flux normalization, 
composition and spectrum. This results in about $10\%$ error 
in the sub-GeV fluxes and 20\% error in the multi-GeV flux range. 
The hadronic interaction model brings about 10\% uncertainty above 
300 MeV. The other sources of uncertainties include the method 
of calculation employed and the use of one-dimensional approximation. 
All these factors sum up to result in total uncertainty of about 
20--30\% in the absolute values of the fluxes \cite{honda}. 
For the ratio ${\cal R}$ 
on the other hand 
the uncertainties due to the primary cosmic rays and the error in the 
calculation scheme cancel out and one has only about 5\% error 
for the sub-GeV and about 10\% error for the 
multi-GeV neutrinos \cite{honda}.

\subsection{Results from the Super-Kamiokande}

\begin{figure}[h]
\vskip -2cm
    \centerline{\psfig{file=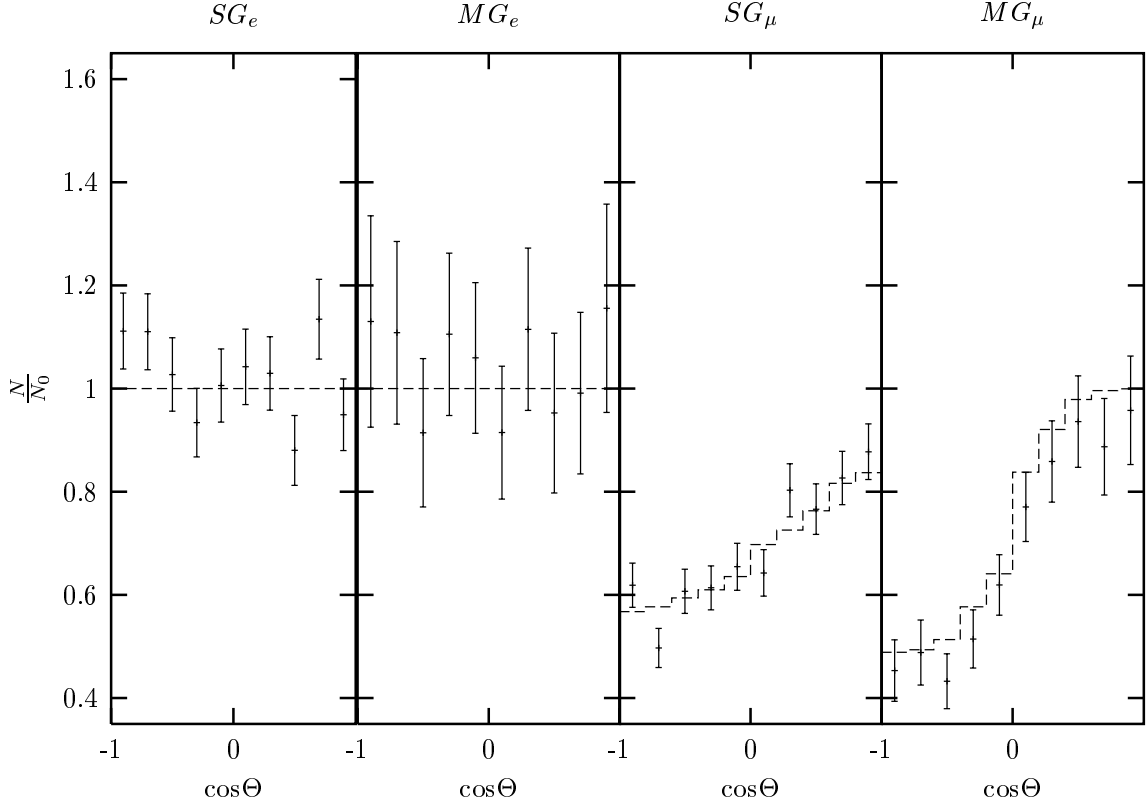,width=7.0in}}
\vskip -2cm
    \caption[The zenith angle distribution of the 
Super-K atmospheric neutrino events.]{\label{skatmdata}The 
sub-GeV and multi-GeV event distributions vs.
zenith angle. $N$ is the number of observed $e$ type or $\mu$ type 
events and $N_{0}$ is the Monte Carlo expectation. The errorbars show 
the $\pm 1\sigma$ uncertainties. Also shown by the dashed lines 
are the zenith angle 
distributions for the various data samples predicted by the two-generation 
$\numu-\nutau$ oscillation best-fit solution.}
\end{figure}

The Super-Kamiokande as described before 
is a large water \chr detector. 
It contains 50 kton of ultra pure water, divided 
into two concentric cylinders. The inner detector  
with a fiducial volume of 22.5 ktons, has its surface 
lined by 11,146 inward facing 50cm photomultiplier tubes and  
1885 outward facing 20cm photomultiplier tubes. 
The outward facing phototubes view the 
outer detector volume which serves both as a particle shield 
as well as a detector. 
The {\it main} process by which it detects the atmospheric neutrinos 
is the charged current quasi-elastic interaction
\be
\nu_l + N \rightarrow l + N^\prime
\ee
where the flavor of the incident neutrino is tagged onto 
the flavor of the released lepton. As this lepton moves 
in water it emits a \chr cone which can be viewed by the 
phototubes lining the detector walls. The SK 
classifies its data into 
\begin{itemize}
%
\item {\it $e$ type or $\mu$ type events}:
The single-ring events can be classified into 
$e$ or $\mu$ type depending on whether the observed \chr ring is 
fuzzy or sharp respectively.  
\item {\it contained events:} Events for which the products 
of neutrino interaction are contained within the detector volume 
are called contained events. If the products are fully 
contained within the inner detector volume itself, then 
these events are termed {\it fully-contained events} (FC). On the other 
hand if the produced leptons exit the inner detector volume and 
stop in the outer detector, then these events are called 
{\it partially-contained events} (PC). Only muons are penetrating enough 
to be able enter the outer detector and so only $\numu$ have PC events. 
\item {\it sub-GeV and multi-GeV events:} Depending on the value of 
the visible energy of the released lepton ($E_{vis}$) the FC events 
are categorized as sub-GeV (if $E_{vis} < 1.33$ GeV) or 
multi-GeV (if $E_{vis} > 1.33$ GeV).
\item {\it upward-going muons:}  If the muons are produced in the 
rock below the detector and penetrate into the detector volume, then such 
events are grouped together to be called upward-going muons \cite{skup}. 
If these muons traverse the entire detector volume and escape 
out then they are termed as {\it upward through-going muons}. 
If on the other hand they stop within the detector then one has 
{\it upward stopping muons}. These upward-going muons are 
recorded for zenith angle $>90^o$ only. The neutrino induced 
muons with zenith angle less than 
this are not taken into account due the huge contamination resulting 
from the muons in the cosmic ray fluxes themselves. 
\item {\it zenith angle bins:}  
At the detector, the neutrino flux come from all directions.
Thus, the total path length between the
production point in the atmosphere and the detector varies
from about 10 km to 13,000 km depending on the zenith angle.
Neutrinos with zenith angle less than $90^{\rm o}$ ({\it
downward neutrinos}) travel a distance of $\sim$ 10 -- 100 km from their
production point in the atmosphere to the detector while the neutrinos
with larger zenith angles ({\it upward neutrinos}) cross a distance of
up to $\sim$ 13,000 km to reach the detector. 
The SK can see the zenith angle 
of the lepton that emits the \chr cone. The lepton zenith angle 
$\Theta$ is related to the incoming neutrino zenith angle 
$\xi$ through the relation
\be
\cos \Theta = \cos \xi \cos \psi + \sin \xi \cos \phi \sin \psi
\label{zenith}
\ee
where $\psi$ is the angle
between the incoming neutrino $\nu_l$ and the scattered lepton $l$ 
and $\phi$ is the azimuthal angle
corresponding to the incident neutrino direction. The SK divides 
its events into bins corresponding to the zenith angle of the 
observed lepton $\cos\Theta$.
\end{itemize}
After 1144 day of data taking \cite{sk1144} 
the SK divide their observed number of sub-GeV and multi-GeV  
$e$ type and $\mu$ type events into 10 zenith angle bins each. 
Instead of showing the actual 
number of events reported by SK, for the sake of clarity we present 
in fig \ref{skatmdata} the ratio of the number of observed 
($N$) to the number of expected ($N_0$) events, for the 
sub-GeV $e$-type (\sge), multi-GeV $e$-type (\mge), sub-GeV $\mu$-type 
(\sgmu) and multi-GeV $\mu$-type (\mgmu) data samples.
We have included the PC events along with the FC \mgmu events 
in the panel for \mgmu. Henceforth we will follow this procedure for 
representation of the PC and \mgmu data samples. The ratio 
$N/N_0$ are shown for the ten zenith angle bins along 
with the $\pm1 \sigma$ errorbars. 
Had there been no anomaly 
this ratio would have been one for all the data bins. However we 
observe that 
\begin{enumerate}
\item The $e$-type events, both \sge and \mge, 
are consistent with theoretical expectations.
\item The observed $\mu$-type events on the other hand show depletion
compared to expectations.
\item The deficit of $\mu$-type events has a zenith angle distribution. 
The $\numu$ depletion being most in the first zenith angle 
bin (upward) and least in the last one (downward), for both the \sgmu and 
\mgmu data samples. Thus the data has a marked up-down asymmetry. 
\item In addition the $\numu$ deficit has an energy dependence. 
The up-down asymmetry being more for the multi-GeV  
than for the sub-GeV neutrinos. 
\end{enumerate}
Since the uncertainties in the absolute fluxes are immense 
($\approx 30\%$), one may 
instead use the double ratio,
\be
R = \frac{(N_{\mu}/N_e)|_{\rm obs}}{(N_{\mu}/N_e)|_{MC}}
\label{ratm}
\ee
where $(N_{\mu}/N_e)|_{\rm obs}$ are the ratio of the total number 
of observed $\mu$-type to $e$-type events in the detector and 
$(N_{\mu}/N_e)|_{MC}$ is the corresponding Monte Carlo expectation. 
Different atmospheric flux calculations agree to within better 
than 5\% on the magnitude of this quantity.
The up-down asymmetry can be then expressed in terms of an 
up-down asymmetry parameter $Y_l$ defined as \cite{yasuda}
\be
Y_{l} \equiv {(N_{l}^{-0.2}/N_{l}^{+0.2})|_{obs}
\over (N_{l}^{-0.2}/N_{l}^{+0.2})|_{MC}}
\label{yatm}
\ee
Here $N_{l}^{-0.2}$ denotes the number of $l$-type events
produced in
the detector with zenith angle $\cos \Theta < -0.2$, {\it i.e.} the upward
neutrino events while
$N_{l}^{+0.2}$ denotes the number of $l$-type events for $\cos \Theta >
0.2$ {\it i.e.} events coming from downward neutrinos.
The central bin has contributions from both upward and downward
neutrinos and is not useful for studying the
up-down asymmetry.
For the 848 day and 535 day data the $R$ and $Y_l$ are given in Table 
\ref{skrydata}. The $Y_e$ is close to 1 but the $Y_\mu$ and 
$R$ are less than unity for both the sub-GeV as well multi-GeV 
cases. 
\begin{table}[htbp]
    \begin{center}
         \begin{tabular}{||c||c|c||c|c||} \hline\hline
{} & \multicolumn {2}{c||} {\rm 848 ~ day ~ data} &
\multicolumn {2}{c||} {\rm 535 ~ day ~ data} \\ \cline {2-5}
{\rm Quantity} & {\rm Sub-GeV} & {\rm Multi-GeV}
& {\rm Sub-GeV} & {\rm Multi-GeV} \\ \hline\hline
{$R$} & {$0.69\pm 0.05$} & {$0.68\pm 0.09$} & {$0.63\pm 0.06$} 
& {$0.65\pm 0.09$} \\ 
{$Y_\mu$} & {$0.74\pm 0.04$} & {$0.53\pm 0.05$} & {$0.76\pm 0.05$} 
& {$0.55\pm 0.06$} \\ 
{$Y_e$} & {$1.03\pm 0.06$} & {$0.95\pm 0.11$} & {$1.14\pm 0.08$} 
& {$0.91\pm 0.13$} \\ \hline \hline

        \end{tabular}
     \end{center}
\vskip -0.5cm
     \caption[The $R$ and $Y_l$ from the SK atmospheric data]
     {\label{skrydata}
     The double ratio $R$ and up-down asymmetry parameter 
     $Y_l$.}
\end{table}
%

\subsection{The Atmospheric Neutrino Code}

The expected number of $l$ (e or $\mu$) like 1 ring events recorded
in the detector in presence of oscillations is given by
\begin{eqnarray}
N_l
& = & n_T
\int^{\infty}_{0} dE
\int^{(E_l)_{\rm max}}_{(E_l)_{\rm min}} dE_l
\int_{-1}^{+1} d\cos \psi
\int_{-1}^{+1} d\cos \xi\
{1 \over 2\pi}
\int_{0}^{2\pi} d\phi
\nonumber\\
&\times&
{d^2F_l (E,\xi) \over dE~d\cos\xi}
\cdot{ d^2\sigma_l (E,E_l,\cos\psi) \over dE_l~d\cos\psi }
\epsilon(E_l)
\cdot
{\ }P_{\nu_l \nu_l} (E, \xi).
\label{rate}
\end{eqnarray}
where 
$n_T$ denotes the number of target nucleons, $E$ is the neutrino energy,
$E_l$ is the energy of the final charged lepton, $\psi$ is the angle
between the incoming neutrino $\nu_l$ and the scattered lepton $l$,
$\xi$ is the zenith angle of the neutrino and $\phi$ is the azimuthal angle
corresponding to the incident neutrino direction (the azimuthal angle
relative to the $\psi$ has been integrated out). The zenith angle of the
charged lepton in terms of $\psi$, $\xi$ and $\phi$ is given by eq. 
(\ref{zenith}). $d^2F_l /dE d\cos\xi$ is the differential flux of
atmospheric
neutrinos of type $\nu_l$,
$d^2\sigma_l/dE_l d\cos\psi$ is the differential cross section
for $\nu_l N \rightarrow l X$ scattering and $\epsilon(E_l)$
is the detection efficiency for the 1 ring events in the detector.
The efficiencies that were available to us are not the detection
efficiencies
of the charged leptons but some function which we call $\epsilon^\prime(E)$
defined as \cite{private}
\be
\epsilon^\prime(E) = \frac{ \int{ \frac{d\sigma}{dE_l} \epsilon(E_l)
dE_l}}
{\int{ \frac{d\sigma}{dE_l} dE_l}}
\ee
$P_{\nu_l \nu_l}$ is the survival probability of a
neutrino flavor $l$ after traveling a distance $L$ given by,
\be
L= \sqrt{(R_e+h)^2 - {R_e}^2 \sin^2 \xi} - R_e \cos \xi
\ee
$R_e$ being the radius of the Earth and $h$ is the height of the atmosphere
where the neutrinos are produced.
We use the atmospheric neutrino fluxes from \cite{honda}.
For the sub-GeV events the dominant process is the charged current
quasi-elastic scattering from free or bound nucleons.
We use the cross-sections given in \cite{gaisser}.
The events in multi-GeV range have contributions coming from
quasi-elastic scattering, single pion production and multi pion production
and
we have used the cross-sections given in \cite{lipary}. For the multi-GeV
events we assume that the lepton direction $\Theta$ is the same as the
incoming neutrino direction $\xi$. But actually they are slightly
different. We simulate this difference in the zenith angles by
smearing the angular distribution of the number of events with a
Gaussian distribution having a one sigma width of $15^{\rm o}$ for
$\mu$ type events and $25^{\rm o}$ for the $e$ type events \cite{G_G}.
For the sub-GeV
events, difference in direction between the charged lepton and the neutrinos
are exactly taken care of according to eq. (\ref{rate}) and (\ref{zenith}).

In order to do a statistical analysis of the data we define a 
\chisq function as \cite{yasuda}
\be
\chi^2 = \sum_{i} \left[\left({R}^{exp} - R^{th} \over \Delta
R^{exp} \right)^2
+ \left({Y^{exp}_{\mu} - Y^{th}_{\mu} \over \Delta Y^{exp}_{\mu}}\right)^2
+ \left({Y^{exp}_{e} - Y^{th}_{e} \over \Delta Y^{exp}_{e}}\right)^2
\right]
\label{chiry}
\ee
where the sum is over the sub-GeV and multi-GeV cases. 
The experimentally observed rates
are denoted by the superscript ``exp"
and the theoretical predictions for the quantities are labeled by ``th".
$\Delta R^{exp}$ is the
error in $R$ obtained by combining the statistical and
systematic errors in quadrature.
$\Delta Y^{exp}$ corresponds to the error in $Y$. For this we take only
the statistical errors since these are much larger compared to the
systematic errors.
We include both the $e$-like and the $\mu$-like up-down asymmetries in the
fit so that we have 4 degrees of freedom (6 experimental data - 2
parameters) for the oscillation analysis in the two parameters
$\Delta m^2$ and $\sin^2 2\theta$.

The use of these type of ratios for the $\chi^2$ analysis test has been
questioned in \cite{lisi1} because the error distribution of these ratios is
non-Gaussian in nature. However as it has been shown in
\cite{yasuda} the use of the $R$'s and $Y$'s as defined above is
justified within the 3$\sigma$ region around the best-fit point for
a high statistics experiment like SK and provides an alternative way
of doing the $\chi^2$-analysis.
A comparison of the results of \cite{yasuda} with those obtained in
\cite{G_G,lisi} shows that the best-fit points and the allowed regions
obtained do not differ significantly in the two approaches of data
fitting.
The advantage of using the ratios is that they are relatively
insensitive to the uncertainties in the neutrino fluxes and
cross-sections as the overall normalization factor gets canceled out
in the ratio. We have included the $Y_e$ in our analysis because to
justify the $\nu_\mu - \nu_\tau$ oscillation scenario, it is necessary to
check that $\chi^2$ including the data on electron events  gives a low value
and hence it is the standard practice to
include these in the $\chi^2$-analysis \cite{yasuda,G_G,lisi}.
  
Even though the method of \chisq analysis defined above is acceptable, 
a better method is to use the absolute number of
e or $\mu$ type events taking into account the errors and their
correlations properly \cite{lisi,G_G}. Following this method we 
define our \chisq as \cite{lisi,G_G}
\be
\chi^2 =
\sum_{i,j=1,40} \left(N_i^{th} -
N_i^{exp}\right)
(\sigma_{ij}^2)^{-1} \left(N_j^{th} - N_j^{exp}\right)
\label{chilisi}
\ee
where the sum is over the sub-GeV and multi-GeV electron and muon bins.
The experimentally observed number of events
are denoted by the superscript ``exp"
and the theoretical predictions for the quantities are labeled by ``th".
The element of the error matrix $\sigma_{ij}$ is calculated as in
\cite{lisi}, including the correlations between the different bins.
For two-generation 
analysis of the SK atmospheric neutrino data we employ this 
second more widely used method of \chisq analysis and present our 
results in the next section. 

\subsection{Results and Discussions}

Since the $e$ type events are consistent with the Monte Carlo 
expectations, the only possible two-generation oscillation schemes 
that one should consider to be responsible for the $\numu$ 
deficit are the $\numu-\nutau$ and $\numu-\nu_{\rm sterile}$ oscillation
modes\footnote{In more general four generation schemes the 
$\numu$ can oscillate into a mixture of $\nutau$ and 
$\nu_{\rm sterile}$ \cite{fl4atm}.}. The two cases differ from 
each other in the Earth matter effects on the mass and mixing parameters 
for the upward neutrinos. While for $\numu-\nutau$ oscillations there 
are no effects of the Earth matter, the $\numu-\nu_{\rm sterile}$ 
has its \dm and \st modulated in presence of matter since the 
sterile neutrinos do not have any interaction with 
the ambient matter. 
For the $\numu-\nu_{\rm sterile}$  case 
the mass squared difference and mixing angle in matter are 
given by eq. (\ref{delmst}) and eq. (\ref{mixst}) of chapter 2. 
Since for the atmospheric neutrino anomaly the mixing angle in 
vacuum is close to maximal, the effect of matter in this case is 
mostly to reduce the mixing. Thus the $\numu-\nu_{\rm sterile}$ 
oscillations predict a lower suppression of the upward neutrinos 
than demanded by the data and fail to reproduce the correct 
zenith angle distribution, the effect being more for the higher 
energy neutrinos. Thus the $\numu-\nu_{\rm sterile}$  solution gets 
disfavored by the zenith angle 
data. In addition, since the sterile neutrinos 
do not induce any neutral current event in the detector while both the 
$\mu$ and $\tau$ neutrinos do, for $\numu-\nu_{\rm sterile}$ mixing 
one expects a reduction of the neutral current events while for 
$\numu-\nutau$ oscillations they remain unchanged. 
In their paper \cite{sksterile}, the SK collaboration have shown 
that their neutral current sample is consistent with expectations, 
there is no depletion of the events and hence transition of $\numu$ to 
$\nu_{\rm sterile}$ are disfavored. 
The SK collaboration have given 99\% C.L. {\it exclusion plots} 
for a combined 
analysis of the neutral current enriched multi-ring events, the 
high energy partially contained events and the upward-going 
muon data sets, in terms of 
$\numu-\nu_{\rm sterile}$ oscillations. They superimpose the 
99\% C.L. {\it allowed region} that they obtain from the \chisq analysis 
of the FC data over these exclusion plots and show that there 
is no region of overlap between the two. They thus conclude that 
pure two-generation $\numu-\nu_{\rm sterile}$ oscillations are 
disfavored by the data and hence the pure 
$\numu-\nutau$ mixing, which can provide 
excellent fit to all the atmospheric 
data samples without any inconsistency 
is the favored alternative \cite{sksterile}.  The latest 1289 day data 
sample on the upward-going events are consistent with $\nutau$ appearance 
at the $2\sigma$ level \cite{sk12893} thereby further disfavoring the 
sterile option. We therefore concentrate on the 
$\numu-\nutau$ oscillation scenario. 

\begin{figure}[t]
   \centerline{\hskip -2cm\psfig{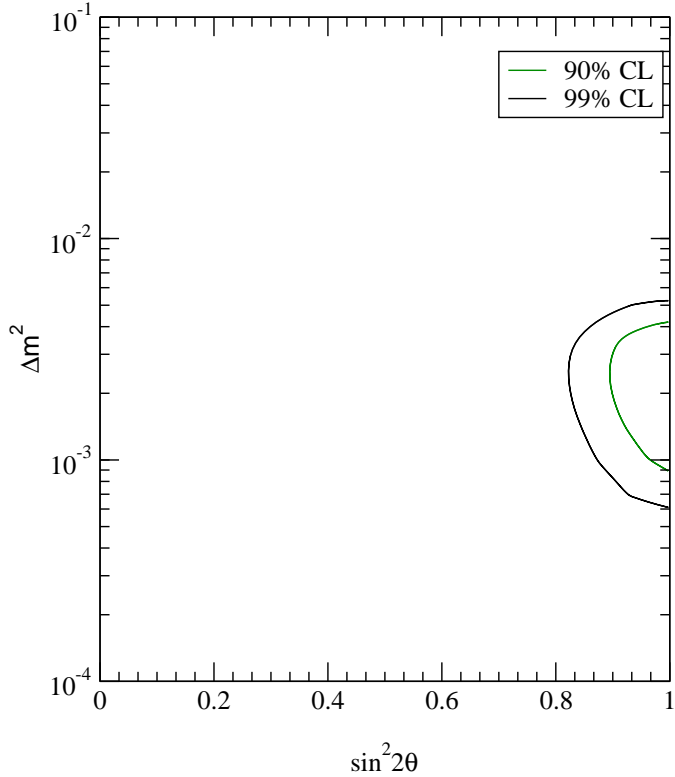}}
\vskip -0.5cm
    \caption[The C.L. contours for $\numu-\nutau$ 
    oscillations of the atmospheric neutrinos]
    {\label{c2atm}
    Region of the parameter space allowed by the $\numu-\nutau$ oscillation
    solution to the 1144 day SK atmospheric neutrino
    data. Shown are the 90\% and 99\% C.L. allowed zones.}
\end{figure}

In the two-flavor $\numu-\nutau$ picture the survival probability 
of an initial $\nu_{\mu}$
of energy $E$ after traveling a distance $L$ in vacuum is
\be
P_{\nu_{\mu}\nu_{\mu}} =1- \sin{^2}2\theta \sin{^{2}}(\pi L/\lambda_{osc})
\label{p2nu}
\ee
where $\theta$ is the mixing
angle between the two neutrino states in vacuum and
$\lambda_{osc}$ is the oscillation wavelength defined as,
\be
\lambda_{osc} = (2.47~ km)~ \frac{E}{GeV} \frac{eV^2}{\Delta m^2}
\label{lo}
\ee
where $\Delta m^2$ denotes the mass squared difference between the two
mass eigenstates. If we insert eq. (\ref{p2nu}) in eq. (\ref{rate}) 
and minimize the \chisq function defined by (\ref{chilisi}), then 
for the 1144 day of SK data we get the following best-fits and 
\chisqmin
\begin{itemize}
\item{
$\chi^2_{min} = 36.23$,
$\Delta m^2$ = 0.0027 eV$^2$, $\sin^2 2\theta$ = 1.0}
\end{itemize}
For 38 degrees of freedom ($40-2$) this corresponds to a 
goodness of fit of 55.14\%. We present in fig. \ref{c2atm} 
the 90\% and 99\% C.L. allowed area 
in the neutrino parameter
space for two-generation $\numu-\nutau$ oscillations. 
In fig \ref{skatmdata} we show with dashed lines 
the histograms for the best-fit value for
two-generation $\numu-\nutau$ oscillations. Both the absolute number 
of events as well as the zenith angle 
distributions of the SK data are seen to be reproduced well.
%
%

\section{Bounds from Accelerator and Reactor Experiments}

Neutrino oscillations experiments can be broadly classified into two 
types:\\
$\bullet$ {\it Disappearance experiments}: In these one starts with  
a neutrino beam of a definite flavor $\nu_\alpha$ and looks 
for a reduction in the original flux, after the 
beam has traveled a certain distance.\\
$\bullet$ {\it Appearance experiments}: These experiments search for a 
new flavor $\nu_\beta$ in the original $\nu_\alpha$ beam, after it 
has been made to travel a distance L.

Both the {\it solar neutrino problem} and the 
{\it atmospheric neutrino anomaly} that we have discussed above 
are disappearance experiments. 
But both these experiments have the intrinsic defect of 
large uncertainties in model predictions of neutrino fluxes. 
The neutrino beams in the  
laboratory experiments on the other hand have well controlled 
spectrum as well as path length and hence are better suited for 
precise study of neutrino mass and mixing parameters. The laboratory 
accelerator and reactor 
experiments may operate in either the disappearance or appearance modes.  
While the appearance experiments have the advantage of being apparently 
easier, since one is looking for a new flavor absent in the initial 
beam, the disappearance experiments have the strength  
to probe the oscillations of the initial $\nu_\alpha$ to all 
channels including $\nu_\alpha-\nu_{\rm sterile}$ which can never 
be seen in an appearance experiments. 

In two-generations the conversion probability of $\nu_\alpha$ to another 
flavor $\nu_\beta$ after traveling distance $L$ is given by 
eq. (\ref{2genpr})
\be
P_{\nu_\alpha \nu_\beta} &=& \sin^2 2\theta \sin^2 (\frac{\pi L}{\lambda})
\nonumber \\
&=& \sin^2 2\theta \sin^2 \left(1.27\frac{\dm L}{E}\right)
\ee
$\lambda$ being the oscillation wavelength. Sensitivity of a 
given experiment can be ascertained in terms of the minimum 
\dm and \st that the experiment can probe. 
For \dm very large 
$\lambda \ll L$, and $\sin^2 (\frac{\pi L}{\lambda}) \rightarrow 
\frac{1}{2}$. For the limiting value of $P_{\nu_\alpha \nu_\beta}^0$ 
seen by the experiment
\be
\sin^22\theta_{\rm min} = 2P_{\nu_\alpha \nu_\beta}^0
\label{sinmin}
\ee
On the other hand if \dm is such that $\left(\frac{\dm L}{E}\right)$ 
is very small then we can replace the sine function by its argument.
\be
P_{\nu_\alpha \nu_\beta}^0 = \sin^2 2\theta \left(1.27\frac{\dm L}{E}\right)^2
\ee
Hence the minimum value of \dm that the experiment can constraint 
corresponds to the case where $\st \rightarrow 1$, that is 
\be
\Delta m^2_{\rm min} = \frac{P_{\nu_\alpha \nu_\beta}^0}{1.27L/E}
\label{delmin}
\ee

Using the above argument the terrestrial experiments give exclusion 
plots in the \dm--\st plane.
All accelerator and reactor experiment till date barring the LSND 
in Los Alamos have failed to observe any positive signal of 
neutrino oscillations\footnote{The preliminary results from the 
K2K experiment in Japan also gives a positive indication of neutrino 
oscillations.} and give exclusion plots in the \dm--\st plane. 
We give below a list of the most prominent experiments and their 
main features.

$\bullet$ {\bf Reactor Experiments}\\
The neutrinos in these experiments are produced in 
$\beta$ decay of fission products in the reactors. Thus 
typically one has an initial $\anue$ beam with energies of the order a 
few MeV. Since the energies are so small these experiments operate 
in the disappearance mode. 
These $\anue$ are detected by liquid scintillation detectors 
through the reaction 
\be
\anue +p\rightarrow e^+ + n
\ee
Since these are necessarily disappearance experiments, the incident 
neutrino flux should be very accurately known to reduce the 
systematic uncertainties. Due to these high systematic uncertainties 
the reactor experiments fail to probe very low values of \st. 
To overcome this 
problem to some extent and to reduce the uncertainties, most 
experiments use two or more detectors placed at different distances 
and what is used in the analysis 
is the ratio of the neutrino flux observed 
at these different detectors. 
The characteristics of the most important reactor experiments 
and their limits on the oscillation parameters are 
listed in Table \ref{artable}.
Among the reactor experiments the most stringent bounds in the low 
\dm range are 
obtained from the CHOOZ long-baseline experiment in France 
\cite{chooz}. This experiment has a baseline of 1 km 
so that it has an average $L/E \sim 3$ ($E\sim 3$ MeV) and 
hence is sensitive to \dm as low as $10^{-3}$ eV$^2$. It observes 
no neutrino oscillations in the $\anue$ disappearance mode and 
rules out $\st > 0.1$ for $\dm > 7\times 10^{-4}$ eV$^2$ \cite{chooz}. 

The Bugey experiment \cite{bugey} also in France is most constraining 
in the higher \dm region. It rules out large parts of the 
LSND allowed zone so that only $\dm > 0.2$ eV$^2$ of the allowed 
parameter space remains valid. 
\begin{table}[t]
   \begin{center}
     \begin{tabular}{||c|c||c|c|l|l||} \hline \hline
      &  & probes & signal & & \\
      & experiment & oscillation & for & \dm in eV$^2$ & \st \\
      &           &  channel & oscillation & & \\ \hline\hline
      & G\"{O}SGEN \cite{gosgen}& 
          $\nue \rightarrow \nu_x$ & negative & $<0.02$ & $<0.02$ \\
      & Krasnoyarsk \cite{krasno} 
        & $\nue \rightarrow \nu_x$ & negative & $<0.014$ & $<0.14$ \\
      $I$
      & BUGEY\cite{bugey} 
        & $\nue \rightarrow \nu_x$ & negative & $<0.01$ & $<0.2$ \\
      & CHOOZ \cite{chooz}
        & $\nue \rightarrow \nu_x$ & negative & $<0.002$ &$<0.1$  \\
      & Palo Verde\cite{palo} 
        & $\nue \rightarrow \nu_x$ & negative & $<0.002$ & $<0.6$
      \\ \hline \hline
      & CDHSW \cite{cdhsw} 
        &$\numu \rightarrow \nu_x$ & negative & $<0.23$ & $<0.02$ \\
      & CHARM \cite{charm} 
        &$\numu \rightarrow \nu_x$ & negative & $<0.29$ & $<0.2$ \\
      & CCFR \cite{ccfr2}
        &$\anumu \rightarrow \bar\nu_x$ & negative & $<15$ & $<0.02$ \\
      & E776\cite{e776}
        &$\anumu \rightarrow \bar\nu_x$ & negative & $<0.075$ 
        & $<0.003$ \\
      & E531 \cite{e531}
        &$\numu \rightarrow \nutau$ & negative & $<0.9$ & $<0.004$ \\
      $II$
      & E531 \cite{e531}
        &$\nue \rightarrow \nutau$ & negative & $<9$ & $<0.12$ \\
      & LSND \cite{lsnd1} &$\anumu \rightarrow \anue$ & 
         \bf{positive} & & \\
      & LSND \cite{lsnd2} &$\numu \rightarrow \nue$ & 
         \bf{positive} & \raisebox{1.5ex}[0pt]
         {\bigg\} 1.2}& \raisebox{1.5ex}[0pt]{\bigg\}0.003}\\
      & KARMEN2 \cite{karmen2} &$\anumu \rightarrow \anue$ & negative & 
         $<0.007$&$<0.0021$ \\
      & K2K \cite{k2k} & $\numu \rightarrow \nu_x$& 
         \bf{positive} & awaited & awaited\\ \hline \hline

      \end{tabular}
     \end{center}
\vskip -0.5cm
     \caption[The reactor and accelerator experiments]
     {\label{artable}The reactor (I) and accelerator (II) experiments 
that are running/completed. We display the minimum value 
of \dm (cf. eq. (\ref{delmin})) and minimum value of \st 
(cf. eq. (\ref{sinmin})) that are excluded by the experiments 
that observe null oscillations. For LSND we give the best-fit 
\dm and \st. For K2K an oscillation analysis is still awaited.}
\end{table}

$\bullet$ {\bf Accelerator Experiments}\\
The neutrino beam in these experiments consist of mainly 
$\numu$, $\anumu$, $\nutau$, $\anutau$, produced from decay of 
pions and kaons in accelerators. These experiments have much higher 
energy neutrinos and may operate either in the disappearance or the 
appearance mode, depending on the energy of the initial neutrino 
beam. Table \ref{artable} reviews the characteristics of 
the main accelerator experiments. 

The {\underline L}iquid
{\underline S}cintillator 
{\underline N}eutrino {\underline D}etector at the Los Alamos Meson 
Physics Facility has claimed to have seen neutrino
mass and mixing \cite{lsnd1,lsnd2}. In this experiment a high energy 
proton beam is made to impinge on a target thereby 
predominately producing $\pi^+$. The small 
number of $\pi^-$ created are absorbed 
so that the dominant decay chain is $\pi^+ \rightarrow \mu^+\numu$, 
and $\mu^+ \rightarrow e^+\nue\anumu$. Thus this beam has a paucity  
of $\anue$ and in their first experiment, the LSND 
made a search for $\anumu-\anue$ oscillations by looking for 
$\anue$ appearance by recording the 
$\anue p\rightarrow e^+ n$ events. They reported a 
detected transition probability of $P_{\anumu\anue} = 
(0.31_{-0.1}^{+0.11} \pm 0.05)\%$ \cite{lsnd1}. In this case 
the $\pi^+$ and $\mu^+$ decay at rest (DAR). 
On the other hand if they are made to decay in flight (DIF) (the energy 
for this has to be much higher) then 
the $\mu^+$ do not decay much and the 
flux of $\nue$ in the beam is small making it possible to 
probe the $\numu-\nue$ oscillation channel 
as well. The LSND again claims to 
have seen a positive signal with a 
transition probability of $P_{\numu\nue}=
(0.26\pm1.0\pm0.5)\%$ \cite{lsnd2}. They have recently done a 
reanalysis of their entire data sample collected between 
1993-1998 for both $\anumu-\anue$ and $\numu-\nue$ oscillations 
with a common selection criteria and use a novel event reconstruction 
\cite{lsnd}. 
The final transition probability that they report to have 
observed is $(0.264\pm 0.067\pm 0.045)\%$. The best-fit value of the 
oscillation parameters \dm and \st that they give are shown in the 
Table \ref{artable}. 

On the other hand KARMEN, the \underline{K}arlsruhe-\underline{R}utherford 
\underline{M}edium \underline{E}nergy \underline{N}eutrino 
experiment, at the ISIS spallation neutron facility, is an 
experiment almost similar to LSND which uses $\numu$, $\anumu$ and 
$\nue$ beam from $\pi^+$ and $\mu^+$ decays at rest and looks for 
$\numu-\nue$ and $\anumu-\anue$ oscillations. This experiment 
has been consistently inconsistent with the positive neutrino 
oscillation observed at LSND \cite{karmen2}. It gives an exclusion 
plot in the \dm--\st plane which excludes the entire LSND allowed 
region above $\dm \geq 2$ eV$^2$ and part of the zones allowed for 
lower \dm.  

The other experiment which has reported to have seen a depletion 
of expected flux due to oscillations is the still running 
K2K long-baseline experiment in Japan. It sends a pure and intense 
$\numu$ beam from KEK to SK with a baseline of 250 km 
\cite{k2k} and is the first 
long-baseline accelerator experiment. It has an average $E\approx 1.3$ 
GeV which gives $L/E \approx 200$ and has two detectors. The near one  
at a distance of about 300m from the source 
is a one kton water \chr detector, 
similar in technology to the SK. The far detector, 250 km away, is the 
SK itself. The experiment can work both in the 
$\numu$ disappearance as well as in  
the $\nue$ appearance modes. 
This experiment promises to probe the region of the parameter 
space allowed by the SK atmospheric data and thus holds the 
potential to pin down the exact values of the \dm and \st 
responsible for the atmospheric neutrino anomaly.  
The K2K collaboration has already released their first results 
on muon events. They detect 28 events in SK against an 
expectation of $37.8_{-3.8}^{+3.5}$ which is based on the interactions 
of the incident beam at the near detector \cite{k2k}. 
Thus there is a suppression 
of the $\numu$ flux, though the discrepancy is still 
within statistical errors. Even the neutrino energy spectrum that 
they observe is as of yet very poor in statistics. 
Robust results 
and oscillation analyses, on both the total fluxes as well as the 
observed energy spectrum from this experiment is eagerly awaited.

%


%% file: chapter4.tex
\chapter{The Energy Independent Solution to the 
Solar Neutrino Problem}

In the previous chapter we have seen that 
the suppression of the solar neutrino flux has been confirmed now by a
number of experiments \cite{gno7,cl7,sk7,sno7}, 
covering a neutrino energy range of $0.2
- 20$ MeV.  However there is still considerable uncertainty regarding
its energy dependence.  An energy independent suppression factor has
been advocated by several groups over the years \cite{eind7}.  This was shown
to be disfavored however by the combined data on the total
suppression rates \cite{eind27}.  More recently the weight of experimental
evidence changed in favour of an energy independent solution following
the SK data on the day/night spectrum, showing practically no energy
dependence nor any perceptible day/night effect \cite{sk7}.

In this chapter we evaluate the
experimental status of the energy independent solution {\it vis a vis} the
MSW solutions after the inclusion of the SNO CC data.  
For the $\nue$ survival probability $P_{\nue\nue}$ we use the full expression 
given by eq. (\ref{probtot}) in chapter 3, including the coherent term 
containing the phase $\xi$ which allows us to probe down the entire 
parameter space down to $\dm \sim 10^{-11}$ eV$^2$. 
For the composition of the neutrino 
fluxes seen in various experiments and the 
observed rates we refer to the Table \ref{rates} presented in 
chapter 3 where all the survival rates are shown
relative to the standard solar model (SSM) prediction of BPB00 \cite{bp7}.

The apparent energy dependence in the survival rates of Table \ref{rates} is
conventionally explained in terms of the vacuum oscillation (VO),
small and large angle MSW (SMA and LMA) as well as the LOW solutions
\cite{bks7}. The VO and SMA solutions show strong nonmonotonic
energy dependence and SMA is essentially ruled out now with the 
inclusion of the SNO CC data and the SK day/night
spectrum \cite{sk7} 
data. On the other hand the spectrum data is compatible with
the LMA and LOW solutions, which predict modest and monotonic decrease
of the survival rate with energy. We will study the potential of 
an energy independent solution as a possible explanation of the global 
solar neutrino data. 
We shall see that with reasonable allowance for the
renormalisations of the Cl rate and the \br neutrino flux the data are
described well by the energy independent solution.  These
renormalisations shall also be seen to improve the quality of fits with
the oscillation solutions and enlarge the region of their validity in
the parameter space substantially.  Moreover we shall see that most of
this enlarged region of parameter space shows weak matter effect on
$P_{\nu_e \nu_e}$, implying practically energy independent suppression
of the solar neutrino flux.  Thus the energy independent solution can
be looked upon as an effective  parameterization of the oscillation
solutions over this region.

We shall perform a \chisq 
fit of the combined data on the survival rates of Cl, Ga, SK and 
SNO (CC) and the SK spectrum at day and night 
with the energy independent solution as well as the four
traditional solutions mentioned above.  
We have discussed the \chisq function, the error matrix and the method 
of analysis in chapter 3. 
In order to reconcile the
energy independence of the spectrum with the apparent energy
dependence in the rates of Table \ref{rates}, 
we shall consider the following
variations in the Cl rate and \br neutrino flux, since the Cl
experiment \cite{cl7} has 
not been calibrated while the \br neutrino flux is
very sensitive to the underlying solar model.

\begin{enumerate}
\item[{i)}] 
Since the Cl experiment \cite{cl7} has not been
calibrated, there are several fits in the literature disregarding this
rate \cite{gp7,gpns7,gfm7}. 
We shall consider an upward renormalisations of the Cl
rate by 20\% (i.e. $2\sigma$), which will push it marginally above the
SK and SNO rates.  This is favored not only by the energy independent
solution but also the LMA and LOW solutions, showing monotonic energy
dependence. 

\item[{ii)}] 
The \br neutrino flux is very
sensitive to the solar core temperature and hence to the underlying
solar model. 
We shall consider a downward variation of the \br
neutrino flux of BPB00 \cite{bp7},
\be
f_B = 5.15 \times 10^6/{\rm cm}^2/{\rm sec} \left(1.0 \matrix{+.20 \cr
\cr -.16}\right)
\label{one}
\ee
by upto $2\sigma$.  A downward renormalisation of this flux is
favored by the energy independent solution and to a lesser extent by
LOW, though not by the LMA solution.  It is also favored by 
some helioseismic models, e.g. the model of \cite{antia7} 
giving $f_B = (4.16 \pm 0.76) \times 10^6/{\rm cm}^2/{\rm sec}$.
It may also be noted here that using 1$\sigma$
lower limits of the appropriate nuclear reaction rates Brun,
Turck-Chi\`{e}ze and Morel \cite{tc7} have obtained a relatively low value of
\br neutrino flux, $f_B = 3.21 \times 10^6/{\rm cm}^2/\sec$.

\end{enumerate}

In section 4.1 we present the results of the \chisq fit with the energy 
independent solution. In section 4.2 we identify the regions of the 
parameter space with $\stackrel{<}{\sim}10\%$ energy dependence and 
call these the quasi-energy independent regions. In section 4.3 we 
present the MSW solutions with renormalized Cl and $X_B$ (the 
\br flux normalization factor) and show 
the C.L. contours in the $\dm-\tan^2\theta$ plane. Most of the 
these allowed regions are seen to overlap with the quasi-energy 
independent zones. We end in section 4.4 with some discussions on the 
potential of future experimental measurements in distinguishing the 
energy independent solution with the MSW solutions. 

\section{Energy Independent Solution}

Table \ref{eindtab} 
summarizes the results of fitting the global rate + spectrum
data with an energy independent survival probability
\be
P_{\nu_e \nu_e} = 1 - {1 \over 2} \sin^2 2\theta.
\label{three}
\ee
We fit the data for Cl observed and Cl renormalized and for 
both the cases present the results for $X_B$ fixed at SSM and 
$X_B$ varying freely. 
It shows that even without any readjustment to the Cl rate or the
\br neutrino flux the energy independent solution gives an acceptable
goodness of fit 
(g.o.f.) of 24\%.  An upward renormalisation of the Cl rate by 20\%
improves the g.o.f. to 42\%.  And varying the \br neutrino flux
downwards improves it further to 49\%, which corresponds to a
renormalisation factor $X_B = 0.7$ for the \br neutrino flux.  Note
however that the g.o.f. and the best-fit value of the mixing angle for
$X_B = 1$ are very close to those for $X_B = 0.7$.  This is because
the large error in the \br neutrino flux of eq. (\ref{one}) is already
incorporated into the error matrix involved in the \chisq analysis 
for $X_B=1$ case. 

\begin{table}[htb]
\begin{center}
\begin{tabular}{||c|c|c|c|c||}
\hline\hline
&$X_{B}$&$\sin^22\theta\left(\matrix{\tan^2\theta \cr {\rm
or} \cr \cot^2 \theta}\right)$& $\chi^2_{\min}$ & g.o.f   \\
\hline \hline
Chlorine& 1.0&0.93(0.57)& 46.06 &23.58\%\\
Observed&0.72 & 0.94(0.60) & 44.86 & 27.54\%\\
\hline
Chlorine&1.0 &0.87(0.47) &41.19& 41.83\%\\
Renormalized&0.70&0.88(0.48)&38.63 & 48.66\%\\
\hline \hline
\end{tabular}
\end{center}
\caption[The energy independent fit]{\label{eindtab}
The best-fit value of the parameter, the $\chi^2_{\min}$
and the g.o.f from a combined analysis of rate and spectrum with the
energy independent solution given by eq. (\ref{three}).}
\end{table}

Traditionally the energy independent solution (\ref{three}) is
associated with the averaged vacuum oscillation probability at
distances much larger than the oscillation wave-length, as originally
suggested by Gribov and Pontecorvo \cite{gribov7}.  As we shall see below
however an effectively energy independent solution holds around the
maximal mixing region over a wide range of $\Delta m^2$ even after
including all the matter effects in Sun and Earth.

\section{Regions of Energy Independent Solution in the $\Delta m^2 -
\tan^2 \theta$ Plane}

The energy dependence of the survival probability arises from
different sources in different regions of the parameter space.

\begin{enumerate}
\item[{i)}] For $\Delta m^2/E < 10^{-14} \ {\rm eV}$ the Earth
regeneration effect can be safely neglected.  Then the survival
probability 
\be
P_{\nu_e \nu_e} = P_1 \cos^2 \theta + P_2 \sin^2 \theta + 2 \sqrt{P_1
P_2} \sin\theta \cos\theta \left({\Delta m^2 L \over E}\right),
\label{four}
\ee
where $L$ is the distance between the neutrino production point at the
solar core and its detection point on Earth; and $P_2 \ (= 1 - P_1)$
is the probability of the produced $\nu_e$ emerging from the Sun as
the heavier mass eigen-state $\nu_2$
The coherent interference term, represented by the last term of
eq. (\ref{four}), is responsible for the nonmonotonic energy
dependence of the VO and QVO solutions.

\item[{ii)}] For $\Delta m^2/E \sim 10^{-14} - 10^{-11} \ {\rm eV}$,
the coherent term is negligible but the Earth regeneration
contribution can be significant leading to energy dependence of 
the survival probability. Besides over a large part of this
region, represented by the MSW triangle, the $\nu_e$ is adiabatically
converted into $\nu_2$ in the Sun, i.e. $P_2 = 1$.  The the day/night
averaged probability around these adiabatic zones is \cite{gpns7} 
\be
\bar P_{\nu_e \nu_e} = \sin^2\theta + {\eta_E \sin^2 2\theta \over 4(1
- 2\eta_E \cos 2\theta + \eta^2_E)},
\label{six}
\ee
where
\be
\eta_E = 0.66 \left({\Delta m^2/E \over 10^{-13} \ {\rm eV}}\right)
\left({g/{\rm cm}^2 \over \rho Y_e}\right).
\label{seven}
\ee
Here $\rho$ is the matter density in the Earth\footnote{Though in 
eqs. (\ref{six}) and (\ref{seven}) we 
have shown the regeneration for a constant density Earth, 
we have used the two slab model for actual calculations.} 
and $Y_e$ the average
number of electrons per nucleon. The regeneration contribution
is always positive and peaks around $\eta_E \sim 1$, i.e. $\Delta
m^2/E \sim 3 \times 10^{-13} \ {\rm eV}$ (cf. fig. \ref{reg1}). 
Regions of the parameter space which give non-adiabatic neutrino 
propagation in the Sun pick up a strong energy dependence. 

\item[{iii)}] Finally for $\Delta m^2/E > 10^{-11} \ {\rm eV}$ the
survival probability can be approximated by the average vacuum
oscillation probability of eq. (\ref{three}).  The MSW solutions (LMA
and SMA) lie on the boundary of the regions ii) and iii), i.e. $\Delta
m^2 \sim 10^{-5} \ {\rm eV}^2$ for $E \sim 1$ MeV. The survival
probability $P_{\nu_e \nu_e}$ goes down from $1 - {1\over2} \sin^2
2\theta \ (> 0.5)$ to $\sin^2\theta \ (< 0.5)$ in going up from Ga
to SK (SNO) energy.
\end{enumerate}

All the solar neutrino rates except that of SK have 
$\stackrel{>}{\sim} 10\%$ error,
which is also true for the SK energy spectrum.  The SK normalization
has at least similar uncertainty from the \br neutrino flux.
Therefore we shall 
treat solutions, which predict survival probability $P_{\nu_e \nu_e}$
within 10\% of eq. (\ref{three}) over the Ga to SK energy range, as
effectively energy independent solutions. Moreover the predicted
Ga, Cl and SNO rates will be averaged over the respective energy
spectra, while the predicted SK rates will be averaged over 0.5 MeV
bins, corresponding to the SK spectral data, since experimental
information is available for these averaged quantities only.

\begin{figure}[t]
\vskip -0.8cm
    \centerline{\psfig{file=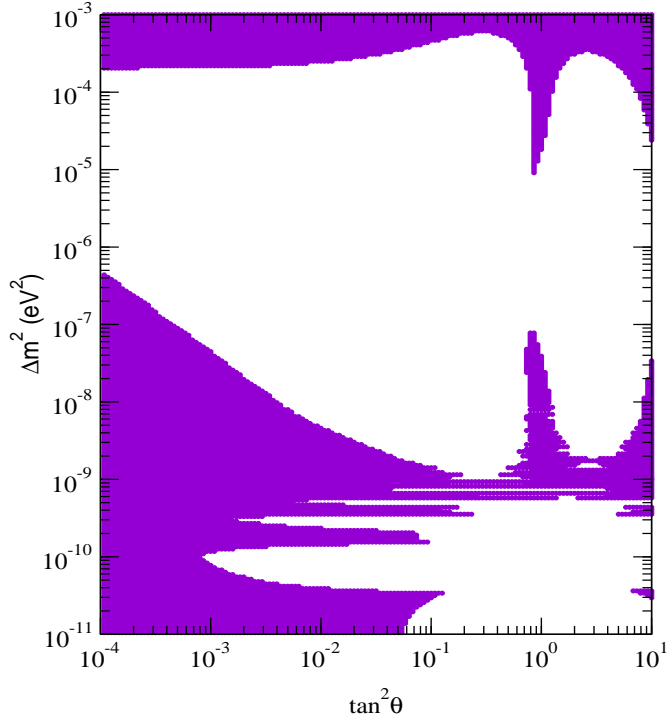,width=4.0in,height=4.0in}}
\vskip -0.5in
    \caption[The effective energy independent regions]{\label{eindfig}
The quasi-energy independent regions 
in $\Delta m^2-\tan^2 \theta$ parameter
space where the solar neutrino survival probability agrees with
eq. (\ref{three}) to within 10\% over the range of Ga to SK energies.}
\end{figure}

The shaded parts of fig. \ref{eindfig} 
shows the regions of effective energy independent solution as
per the above definition. 
The parameter space has been restricted to
$\Delta m^2 < 10^{-3} \ {\rm eV}^2$ in view of the constraint from the
CHOOZ experiment \cite{chooz7}. One sees that the energy independent solution
(\ref{three}) is effectively valid over the two quasi-vacuum
oscillation regions lying above and below the MSW range.  Moreover it
is valid over a much larger range of $\Delta m^2$ around the maximal
mixing region, since the solar matter effect does not affect $P_{\nu_e
\nu_e}$ at $\tan^2\theta = 1$.  It is this near-maximal mixing strip
that is relevant for the observed survival probability, $P_{\nu_e
\nu_e} \sim 1/2$.  The upper strip $(\Delta m^2 = 10^{-3} - 10^{-5} \
{\rm eV}^2)$ spans the regions iii) and part of ii) till it is cut off
by the Earth regeneration effect.  The lower strip $(\Delta m^2 =
10^{-7} - 5 \times 10^{-10} \ {\rm eV}^2)$ spans parts of region ii)
and i) till it is cut off by the coherent term contribution.  Note
that this near-maximal mixing strip represents a very important region
of the parameter space, which is favored by the socalled bimaximal
mixing models of solar and atmospheric neutrino oscillations 
\cite{barger7,bimax7}. 

One can easily check that averaging over the SK energy bins of 0.5 MeV
has the effect of washing out the coherent term contribution for
$\Delta m^2 > 2 \times 10^{-9} \ {\rm eV}^2$.  
But including this term enables us 
to trace the contour down to its lower limit.  It was claimed in
ref. \cite{bgp7} that the lower strip disappears when one includes the
coherent term contribution.  This may be due to the fact that their
predicted rate in the SK energy range was not integrated over the
corresponding bin widths of 0.5 MeV.

\begin{figure}
\vskip -0.5in
    \centerline{\psfig{file=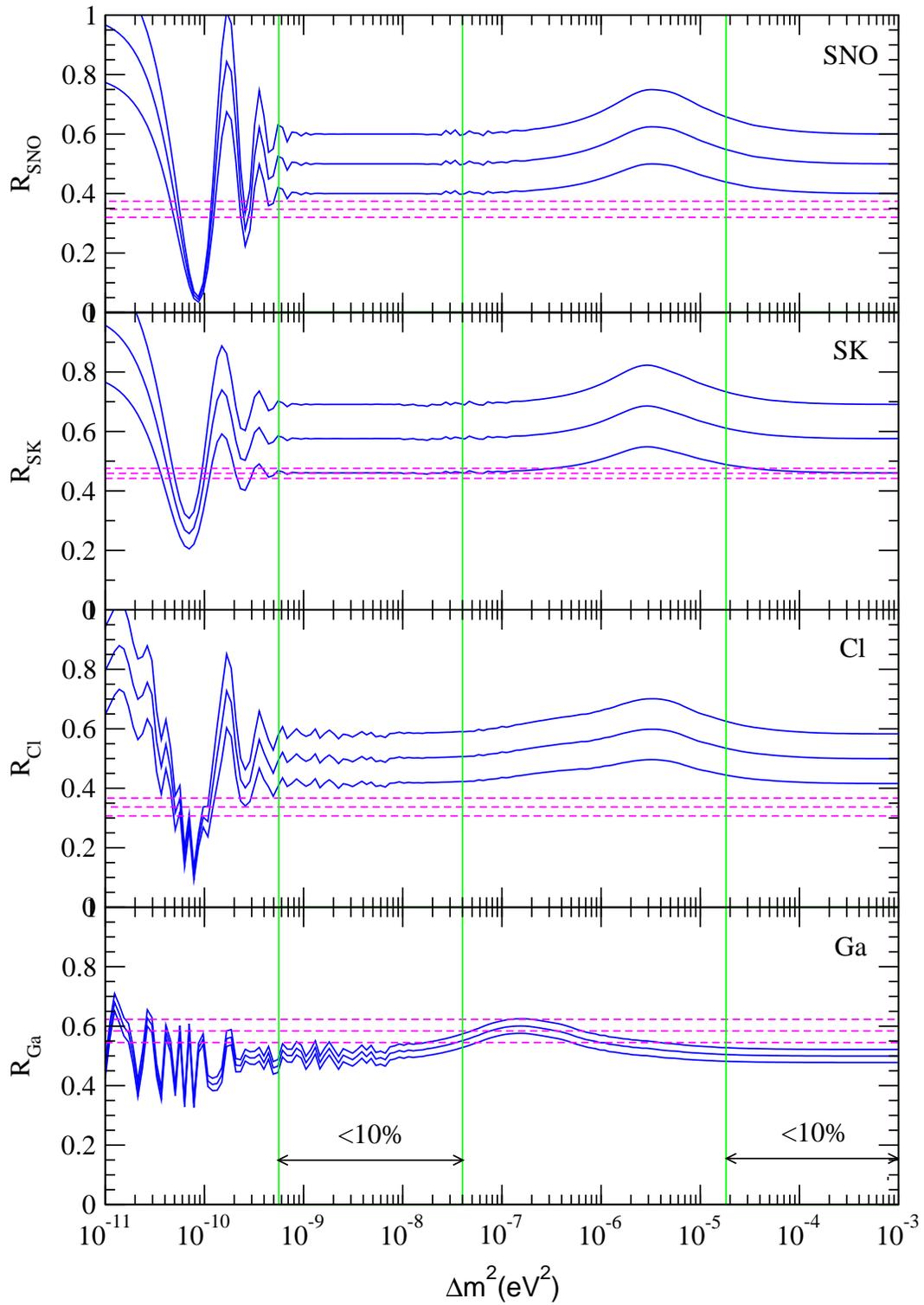,width=6.0in}}
\vskip -0.5in
    \caption[The solar neutrino rates at maximal mixing]{\label{maxrate}
The predicted survival rates at maximal mixing
against $\Delta m^2$ for Cl, Ga, SK and SNO 
experiments (See text for details.)}
\end{figure}

To get further insight into the oscillation phenomenon in the maximal
mixing region we have plotted in fig. \ref{maxrate} 
the predicted survival rates
at maximal mixing against $\Delta m^2$ for the Ga, Cl, SK and SNO
experiments.  In each case the rate has been averaged over the
corresponding energy spectrum.  This is similar to the fig. 7 of
Gonzalez-Garcia, Pena-Garay, Nir and Smirnov \cite{gpns7}. 
As in \cite{gpns7} the
predictions have been shown relative to the central value of the various 
neutrino fluxes of BPB00 along with those differing by $\pm 1\sigma$ from
the central value, e.g. for the \br flux 
$X_B = 1 \pm 0.2$.  We have found that these
curves are in good agreement with the corresponding ones of \cite{gpns7}. 
The two regions of $< 10\%$ energy dependence for maximal mixing 
are indicated by vertical
lines.  One can easily check that in these regions the central curves
lie within 10\% of the energy independent prediction $R = 0.5$ (note 
that the SK rate apparently looks higher 
due to the neutral current contribution).  One
can clearly see the energy dependence due to the 
violent oscillations in the VO region (on the left)
and the Earth regeneration peak in the LOW region (in the middle), 
particularly for  the Ga experiment. 
It should be
noted that the gap between the two energy independent regions due to
the Earth regeneration effect in fig. \ref{eindfig} 
is a little narrower than
here.  This is due to a cancellation between the positive contribution
from the Earth regeneration effect and the negative contribution from
the solar matter effect at $\tan^2\theta < 1$.  It ensures that the
resulting survival rate agrees with the energy independent solution
(\ref{three}) over a somewhat larger range of $\Delta m^2$ slightly
below the maximal mixing angle.

The observed rates from the Ga, Cl, SK and SNO experiments are shown
in fig. \ref{maxrate} 
as horizontal lines along with their $1\sigma$ errors.  With
20\% downward renormalisation of the \br neutrino flux $(X_B = 0.8)$
the energy independent prediction is seen to agree with the SK rate and
also approximately with SNO.  It is higher than the Cl rate by about
$2\sigma$.  The agreement with the Ga rate can be improved by going
to a little smaller $\theta$ and compensating the resulting deviation
from the other rates by a somewhat smaller $X_B$ as in Table \ref{eindtab}.
Nonetheless the maximal mixing solution for $X_B = 0.8$, shown here,
is in reasonable agreement with the observed rates over the energy
independent regions.  The Earth regeneration effect can be seen to
improve the agreement with the Ga experiment for the LOW solution.

\section{The SMA, LMA, LOW and VO Solutions}

\begin{figure}[t]
    \centerline{\psfig{file=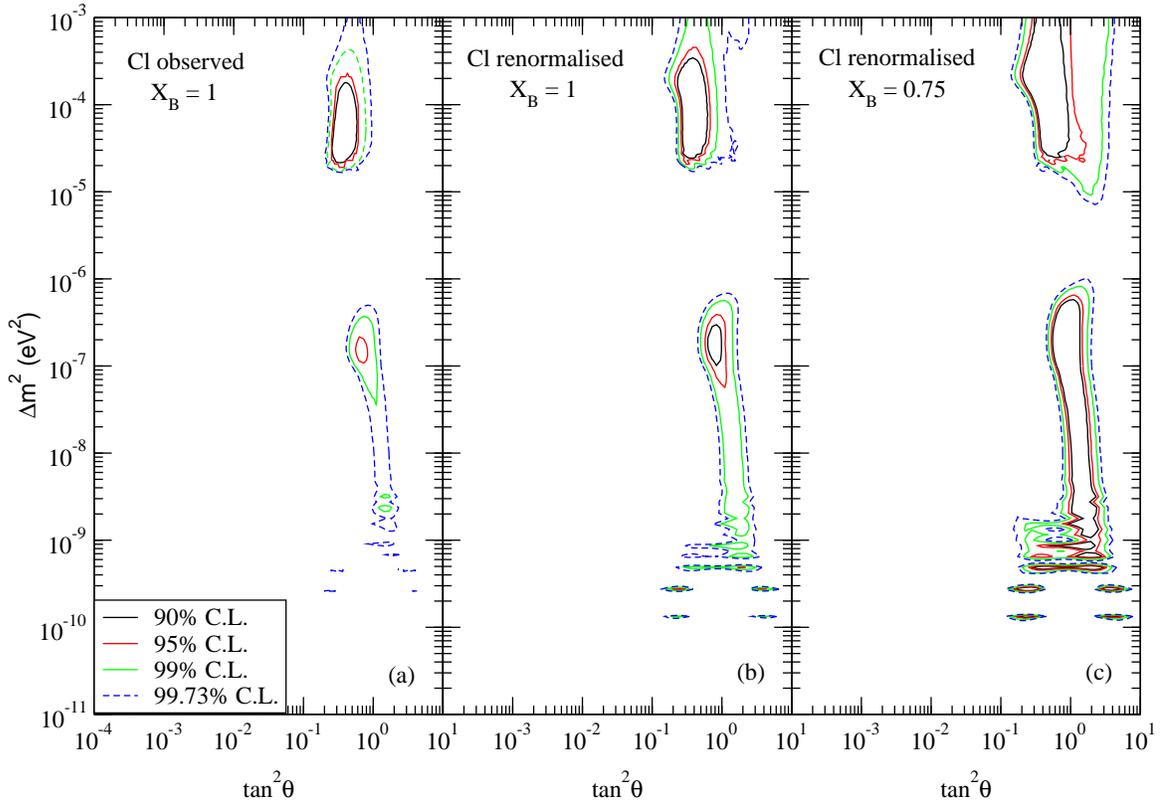,width=7.0in}}
    \caption[Allowed contours for renormalized Cl and $X_B$]
{\label{cont4}
The 90\%, 95\%, 99\% and 99.73\% C.L. allowed area from the
global analysis of the total rates from Cl (observed  and
20\% renormalized), Ga, SK and SNO (CC) detectors
and the 1258 days SK recoil electron spectrum at day and night,
assuming conversions to active neutrinos.}
\end{figure}

Tables \ref{totchi7} and 
\ref{totchi0.75} summarise the results of fits to the global rates +
spectrum data in terms of the conventional oscillation solutions.
Table \ref{totchi7} shows solutions to the data with observed and renormalized
Cl rate with the neutrino flux of BPB00 $(X_B = 1)$, while 
Table \ref{totchi0.75}
shows the effects of renormalizing this \br neutrino flux downwards by
25\% $(X_B = 0.75)$.  The corresponding 90\%, 95\%, 99\% and 99.73\%
$(3\sigma)$ contours are shown in fig. \ref{cont4}.
\begin{table}
\begin{center}
\begin{tabular}{||c|c|c|c|c|c||}
\hline\hline& Nature of & $\Delta m^2$ &
& & \\
& Solution & in eV$^2$& \raisebox{1.5ex}[0pt] {$\tan^2\theta$} &
\raisebox{1.5ex}[0pt] {$\chi^2_{min}$}&\raisebox{1.5ex}[0pt] {g.o.f}\\
\hline \hline
& SMA & $5.28 \times 10^{-6}$&$3.75\times 10^{-4}$
&51.14 &9.22\%  \\
Cl& LMA & $4.70\times 10^{-5}$ &0.38&
33.42 & 72.18\%  \\
Obsvd.& LOW & $1.76\times 10^{-7}$ & 0.67 &39.00&46.99\%  \\
&VO&$4.64\times 10^{-10}$ & 0.57 & 38.28 & 50.25\% \\
\hline
& SMA & $4.94 \times 10^{-6}$& $2.33 \times 10^{-4}$
& 50.94& 9.54\%  \\
Cl& LMA & $4.70\times 10^{-5}$ &0.38&30.59& 82.99\%\\
Renorm.& LOW & $1.99 \times 10^{-7}$ &0.77& 34.26& 68.57\% \\
&VO&$4.61\times 10^{-10}$ & 0.59 & 32.14 & 77.36\% \\
\hline\hline
\end{tabular}
\end{center}
\caption[Results of MSW fits with $X_B=1.0$]
{\label{totchi7}
The best-fit values of the parameters, the $\chi^2_{\min}$
and the g.o.f from a combined analysis of the Cl, Ga, SK and SNO CC
rates and the SK day-night spectrum in terms
of $\nu_e$ oscillation into an active neutrino, including the
matter effects. $X_{B}$ is kept fixed at the SSM value (=1.0). }
\end{table}
\begin{table}
\begin{center}
\begin{tabular}{||c|c|c|c|c|c||}
\hline\hline
& Nature of & $\Delta m^2$ &
& & \\
& Solution & in eV$^2$& \raisebox{1.5ex}[0pt] {$\tan^2\theta$} &
\raisebox{1.5ex}[0pt] {$\chi^2_{min}$}&\raisebox{1.5ex}[0pt] {g.o.f}\\
\hline \hline
& SMA & $5.28 \times 10^{-6}$&$3.75\times 10^{-4}$
&48.39 &14.40\%  \\
Cl& LMA & $4.65\times 10^{-5}$ &0.49&
38.90 & 47.44\%  \\
Obsvd.& LOW & $1.74\times 10^{-7}$ & 0.71 &39.91&42.95\%  \\
& VO & $4.55\times 10^{-10}$ & 0.44 & 37.17 & 55.36\%\\
\hline
& SMA & $8.49 \times 10^{-6}$& $1.78 \times 10^{-4}$
& 50.77& 9.82\%  \\
Cl& LMA & $4.64\times 10^{-5}$ &0.51&34.48& 67.61\%\\
enorm.&LOW & $2.09 \times 10^{-7}$ &0.81& 33.47& 71.97\% \\
&VO & $4.59\times 10^{-10}$&0.53&30.63& 82.86\%\\
\hline\hline
\end{tabular}
\end{center}
\caption[Results of MSW fits with $X_B=0.75$]{\label{totchi0.75}
Best fits to the combined rates and SK day-night spectrum data in
terms of $\nu_e$ oscillation into active neutrino with a
fixed $X_B = 0.75$.}
\end{table}

As we see from these tables and fig. \ref{cont4} the SMA solution gives rather
poor fit in each case, with no allowed contour at $3\sigma$ level.
This result agrees with the recent fits of \cite{postsno7} 
to the global data
including SNO.  The fit of \cite{bgg7} to these data shows a small allowed
region for SMA solution at $3\sigma$ level due to a slightly different
method of treating the normalization in the SK spectrum data, as
explained there.

The LMA and LOW solutions give good fits to the original data set,
which improve further with the upward renormalisation of the Cl rate
by $2\sigma$.  This is because the monotonic decrease of rate with
energy, implied by these solutions, favors the Cl rate to be
marginally higher than the SNO and SK rates as mentioned earlier.
For the renormalized Cl case, downward renormalisation of 
the \br neutrino flux by 25\% is seen to give
a modest increase (decrease) of g.o.f. for the best LOW (LMA) solution.
On the other hand the allowed ranges increase in both cases as we see
from fig. \ref{cont4}b and c.  
Combined together they imply that the downward
renormalisation of the \br neutrino flux modestly favors the LOW
solution but makes little difference to the LMA.  Its main effect on
these two solutions is increasing their allowed ranges in the
parameter space.  Comparing fig. \ref{cont4}c 
with fig. \ref{eindfig} shows that much of
these enlarged ranges of validity correspond to effectively energy
independent solution.
It is interesting to note that
the best-fit values of parameters in the LMA region are 
same for Cl observed and Cl renormalized cases while the $\chi^2_{min}$ 
improves for the latter. 
This shows that the  best-fit already chose
a probability at Cl energy, which is a little higher than 
that at SK/SNO energy.
Renormalizing the Cl rate  brought that point up to the fitted curve
without changing the best-fit parameters. 

While the best vacuum solution seems to show remarkably high
g.o.f. particularly for renormalized Cl rate and \br neutrino flux,
its regions of validity are two miniscule islands just below the lower
energy independent strip of fig. \ref{cont4}b,c. This solution has also been
obtained in the global fits of ref. \cite{postsno7,bgg7} 
as well as the SK fit to
their rate and spectrum data \cite{skosc7}. The position and size of its range
of validity suggest this to be a downward fluctuation of the
effectively energy independent quasi vacuum oscillation rather than a
genuine VO solution of Just-So type.  To get further insight into this
solution we have analysed the resulting energy dependence.  It shows
practically no energy dependence below 5 MeV, but a 15\% fall over the
$5-12$ MeV range.  The latter seems to follow the SK spectral points
rather closely amidst large fluctuation.  To check the stability of
this trend we have repeated the fit to the SK spectral data points,
plotted over 8 broad energy bins shown in \cite{skosc7}, which show much less
fluctuation than the $2 \times 19$ points sample.  The solution
completely disappears from this fit.  This confirms that the above VO
solution is simply an artifact of the sampling of the SK spectral
data.

For completeness we summarise in Table \ref{totchixb} 
the best fits of the above
solutions with free \br neutrino flux normalization.  The SMA solution
favors a very low \br neutrino flux $(X_B \simeq 0.5)$, which raises
the SK and SNO rates more than the Cl, thus accentuating the
nonmonotonic energy dependence of Table \ref{rates}. Still the g.o.f. of the
SMA solution is rather marginal.  On the other hand the LMA solution
favors $X_B > 1$, which suppresses the SK and SNO rates more than the
Cl, resulting in a monotonic decrease of rate with energy.  But the
corresponding g.o.f. are no better than those of Table \ref{totchi7}.  
The results
of the LOW and VO fits are similar to those of Table \ref{totchi0.75}.
\begin{table}
\begin{center}
\begin{tabular}{||c|c|c|c|c|c|c||}
\hline
& Nature of & &$\Delta m^2$ &
& & \\
& Solution & \raisebox{1.5ex}[0pt] {$X_B$}&
in eV$^2$& \raisebox{1.5ex}[0pt] {$\tan^2\theta$} &
\raisebox{1.5ex}[0pt] {$\chi^2_{min}$}&\raisebox{1.5ex}[0pt] {g.o.f}\\
\hline \hline
&SMA & 0.51& $5.25 \times 10^{-6}$&$3.44\times 10^{-4}$
&46.83 &15.41\%  \\
Cl& LMA & 1.18& $4.73\times 10^{-5}$ &0.33&
32.32 & 72.89\%  \\
Obsvd.& LOW &0.88&  $1.75\times 10^{-7}$ & 0.67 &38.75&43.57\%  \\
&VO& 0.70&$4.55\times 10^{-10}$& 0.44 & 37.24 & 50.44\%\\
\hline
& SMA & 0.48 & $4.66 \times 10^{-6}$& $2.32 \times 10^{-4}$
& 46.18& 17.01\%  \\
Cl& LMA & 1.15 & $4.71\times 10^{-5}$ &0.36&30.32& 80.80\%\\
Renorm.&LOW & 0.83 & $2.03 \times 10^{-7}$ &0.79& 33.18& 69.17\% \\
&VO&0.75&$4.63\times 10^{-10}$ & 0.55& 30.56 & 79.92\%\\
\hline\hline
\end{tabular}
\end{center}
\caption[Results of MSW fits with $X_B$ free]{\label{totchixb}
Best fits to the combined rates and SK day-night spectrum data in
terms of $\nu_e$ oscillation into active neutrino with
$X_B$ free.}
\end{table}

\section{Comparisons and Discussions}

Let us conclude by briefly discussing whether some of the forthcoming
neutrino experiments will be able to discriminate between the energy
independent and the MSW solutions. In particular the SNO experiment 
\cite{snonc7} 
is expected to provide both the charged current and neutral current scattering
rates over roughly the same energy range as SK. Thus the \br neutrino flux
can be factored out from their ratio, CC/NC. For oscillations into active
neutrino the corresponding double ratio
$R_{SNO}^{CC}/R_{SNO}^{NC}$
is predicted to be larger than 0.5 for the energy
independent solution and smaller than 0.5 for the LMA and LOW
solutions. 
In the absence of the neutral current data from SNO one can try
to make a similar comparison with the ratio of SK elastic and SNO
charged current scattering rates, 
\be
R^{el}_{SK} = X_B P_{\nu_e \nu_e} + r(1 - P_{\nu_e \nu_e}) X_B, \ r =
\sigma_{nc}/\sigma_{cc} \simeq 0.17,
\label{eight}
\ee
\be
R^{cc}_{SNO} = X_B P_{\nu_e \nu_e},
\label{nine}
\ee
where we have assumed a common survival rate neglecting the small
difference between the SK and SNO energy spectra \cite{snonc7}. One can
eliminate $P_{\nu_e \nu_e}$ from the two rates; and the resulting \br 
neutrino flux can be seen to be in good agreement with the BPB00
estimate \cite{sno7}. Alternatively one can factor out the flux from the
ratio 
\be
R^{es}_{SK}/R^{cc}_{SNO} = 1 - r + r/P_{\nu_e \nu_e}.
\label{ten}
\ee

Table \ref{ccnc} 
shows the best fit values of the above ratio for the LMA, LOW
and the energy independent solutions along with the corresponding
predictions for $R^{cc}_{SNO}/R^{nc}_{SNO} = P_{\nu_e \nu_e}$.  The
predictions of the maximal mixing solution is also shown for
comparison.  While the LMA and LOW 
predictions for the $CC/NC$ ratio differ by
$\sim 50\%$ they differ by only about $\sim 15\%$ in the case of the
$ES/CC$ ratio.  The observed ratio $R^{es}_{SK}/R^{cc}_{SNO}$ is seen
to favour the LMA over the LOW and energy independent solutions; but
even the largest discrepancy is only $\sim 1.5\sigma$.  With the
expected sample of several thousand $CC$ and $NC$ events from SNO one
expects to reduce the $1\sigma$ error for each of these ratios to
about 5\%.  Then one will be able to discriminate between the three
solutions meaningfully, particularly with the help of the $CC/NC$
ratio from SNO.
On the other hand the LOW solution predicts a large
Day-Night asymmetry of $>$ 10\% for the \ber neutrino 
\cite{gpns7,gfm7} at the Borexino \cite{borex7} 
and the KamLAND \cite{kaml7} 
experiments. This will be able to distinguish the LOW
from the LMA and the energy independent solutions.
Lastly it should be noted
that the reactor neutrino data at KamLAND is expected to show oscillatory
behavior for the LMA solution \cite{kam27},
which will help to distinguish it from the LOW or a generic energy
independent solution. 
\begin{table}
\begin{center}
\begin{tabular}{||c|c|c|c|c|c||}
\hline\hline
& $\Delta m^2$ & $\tan^2\theta$ &  $R^{CC}_{SNO}/R^{NC}_{SNO}$ &
$R^{ES}_{SK}/R^{CC}_{SNO}$ & Expt. value of \\ 
& & & & & $R^{ES}_{SK}/R^{CC}_{SNO}$ \\ \hline \hline
LMA & $4.7 \times 10^{-5}$ & 0.38 & 0.30 & 1.36 & \\
LOW & $1.99 \times 10^{-7}$ & 0.77 & 0.45 & 1.19 &  \\
energy-independent & - & 0.47 & 0.56 & 1.13 &  \raisebox{1.2ex}[0pt]{1.33 $\pm$ 0.13} \\
maximal mixing & - & 1.0 & 0.5 & 1.15 & \\
\hline\hline
\end{tabular}
\end{center}
\caption[$R^{CC}_{SNO}/R^{NC}_{SNO}$ at the best-fit values]{\label{ccnc}
The values of the ratios $R^{ES}_{SK}/R^{CC}_{SNO}$ and
$R^{CC}_{SNO}/R^{NC}_{SNO}$ at the best-fit values for the
LMA, LOW and energy independent  solutions for the
renormalized Cl and $X_{B} =1.0$ case.
Also shown are the predictions for the  maximal mixing ($P_{ee}=0.5$)
solution and  the experimental value of
$R^{ES}_{SK}/R^{CC}_{SNO}$.}
\end{table}

In summary the recent SK data on day/night spectrum is in potential
conflict with the apparent energy dependence in the suppression rates
observed in Ga, Cl, SK  and SNO 
experiments. Including matter effects one can get acceptable
oscillation solutions to both rates and spectrum data only
over limited regions of mass and mixing parameters.  However
an upward renormalisation of the Cl rate by 20\% $(2\sigma)$ results
in substantial improvement of the quality of fit.  Moreover a downward
renormalisation of the \br neutrino flux by 25\% $(1.5\sigma)$ as
suggested by the helioseismic model enlarges the allowed region of the
parameter space substantially. Over most of this enlarged region the
energy dependence resulting from the matter effects is too weak to be
discernible at the present level of experimental accuracy. Hence with
these renormalisations of the Cl rate and the \br neutrino flux the
data can be described very well by an energy independent solution.

%% file: chapter5.tex
\chapter{A Three Generation Oscillation Analysis of the Super-Kamiokande
Atmospheric Neutrino Data Beyond One Mass Scale Dominance Approximation}

In chapter 3 we have made detailed analysis of the 
solar and atmospheric neutrino data and presented our results in the 
two generation framework. 
In real world there are three active flavors of neutrinos 
\cite{lep4}. 
If one wants to do an analysis of the SK atmospheric neutrino 
data \cite{sk} 
in a three-generation framework, 
then there are two possibilities.   
The most popular three-generation picture 
in the context of the SK data is the one 
where one of the mass squared differences is in the solar neutrino
range and the other is suitable for atmospheric neutrino oscillations
\cite{lisi4,yasuda3,bar4,3gennew4,ccv4,lisiglob4}. 
In such a scheme one mass scale dominance (OMSD) applies for atmospheric
neutrinos and the relevant probabilities are functions of two of the 
mixing angles and one mass squared difference.  
This picture however cannot explain the LSND results \cite{lsnd4}.  
In this chapter we perform a three flavor $\chi^2$-analysis of the SK 
atmospheric neutrino data assuming a mass pattern with $\Delta_{12} 
\simeq \Delta_{13}$ fixed in the eV$^2$ range 
(henceforth $\Delta m_{ij}^2\equiv \Delta_{ij}$) and 
allowing the other mass scale to vary arbitrarily.
We also incorporate the CHOOZ reactor results \cite{chooz4} into the 
analysis. 
Apart from being suitable to explain the SK atmospheric neutrino data 
this spectrum is also interesting for the laboratory based 
neutrino oscillation experiments as the higher mass scale is explorable  
in the short base line experiments, whereas the lower mass scale 
can be probed in the long base line experiments. 
In this scheme to a good approximation,
neutrino oscillation in the short-base line 
accelerators and reactors will be governed
by one (the higher) mass scale 
\cite{one,fogli94} -- and only two of the mixing
angles appear in the expressions for the oscillation probabilities.
For the atmospheric and the long baseline experiments 
the characteristic energy and length
scales are such that in general both mass differences are of relevance 
and the probabilities involve all the three mixing angles. 
However the higher mass scale gives rise to 
\dm independent average oscillations 
and it does not enter the $\chi^2$ fit directly. 
We determine the best-fit values of 
$\Delta_{23}$ and the three mixing angles by performing a $\chi^2$
analysis of the  
\begin{itemize}
\item SK atmospheric neutrino data
\item SK atmospheric and CHOOZ data
\end{itemize}
We do a bin by bin analysis of the data taking into account the errors 
and their correlations and use 
the \chisq function described in eq. (\ref{chilisi}) of  
chapter 3.
Finally we compare the allowed values of the mixing angles as obtained
from the above analysis with those allowed by the other accelerator and
reactor neutrino oscillation data including LSND and KARMEN2.

The mass scheme of this paper was first considered in \cite{minakata,ska}
after the declaration of the LSND result. 
These papers performed a combined three
generation analysis of accelerator and reactor results as well as the 
Kamiokande atmospheric neutrino data. 
Three-generation picture 
with the higher mass difference in the eV$^2$ range and the lower 
mass difference in the atmospheric range has also been considered in  
\cite{ap,flms97} (pre-SK) and \cite{thun,ol1,flms99,ol2} (post-SK). 
These papers attempted to explain both solar and 
atmospheric neutrino anomalies mainly by
maximal $\nu_{\mu} \leftrightarrow \nu_e$  oscillations driven by
$\Delta_{ATM} \sim 10^{-3}$ eV$^2$. 
Although it was claimed in \cite{thun,ol1} that this scenario can provide a
good fit to all the available data on neutrino oscillations, it was shown in
\cite{flms99} and also later in \cite{ol2} that this scenario cannot reproduce
the zenith angle dependence of the SK atmospheric neutrino data.   

In this chapter our aim is to
determine the allowed oscillation parameter ranges consistent 
with SK atmospheric, CHOOZ, LSND and other accelerator and 
reactor experiments. 
The solar neutrino problem 
can be explained by invoking a sterile neutrino.  
We discuss in the conclusions how the solar neutrino flux
suppression can be explained 
in our scenario.

The plan of the chapter is as follows. 
In section 5.1.1 we present the 
formalism for three-generation oscillation
analysis in vacuum. 
We calculate the required probabilities including the 
earth matter effects in section 5.1.2. We use this expression 
of the probability for the actual calculation of the number of events. 
In section 5.2 we present the three-generation $\chi^2$ analysis 
of only SK atmospheric neutrino data using the computer code described in 
detail in chapter 3. In section 5.3 we present the
combined $\chi^2$ analysis of SK and CHOOZ data. 
In section 5.4 we compare the allowed values of mixing angles from the
above analyses with those allowed by the other accelerator and
reactor data including the latest results from LSND and KARMEN2. 
In section 5.5 we discuss the implications of 
our results for the future long baseline experiments 
and end in section 5.6 with some discussions and conclusions.

\section{Three-Flavor Analysis}

\subsection{The Vacuum Oscillation Probabilities}
The general expression for the probability that an initial
$\nu_{\alpha}$ of energy $E$ gets converted to a $\nu_{\beta}$
after traveling  a distance $L$ in vacuum is given by eq. (\ref{npr}) 
in chapter 2. 
The actual forms of the various survival and transition
probabilities depend on the neutrino mass spectrum assumed
and the choice of the mixing matrix $U$ relating the flavor
eigenstates to the mass eigenstates. We choose the flavor states
$\alpha =$ 1,2, and 3 to correspond to e, $\mu $ and $\tau $
respectively. The most suitable
parameterization of $U$ for the mass spectrum chosen by us is $U =
R_{13} R_{12} R_{23}$ where $R_{ij}$ denotes the rotation matrix in
the $ij$-plane. This yields:
\begin{equation}
U = {\pmatrix {c_{12}c_{13} & s_{12}c_{13}c_{23} - s_{13}s_{23}
& c_{13}s_{12}s_{23} + s_{13}c_{23} \cr
-s_{12} & c_{12}c_{23} &c_{12}s_{23} \cr
-s_{13}c_{12} & -s_{13}s_{12}c_{23} - c_{13}s_{23} &
-s_{12}s_{13}s_{23} + c_{13}c_{23}\cr}}
\label{um}
\end{equation}
where $c_{ij} =\cos{{\theta}_{ij}}$ and $s_{ij}
=\sin{{\theta}_{ij}}$ here and everywhere else in this chapter. 
We have assumed CP-invariance so that $U$ is
real. The above choice of $U$ has the advantage that ${\theta}_{23}$
does not appear in the expressions for the probabilities for the
laboratory experiments \cite{ska}. 

\noindent
The probabilities relevant for atmospheric neutrinos are
\be
P_{\nu_e \nu_e}&=&1 - 2c_{13}^{2}c_{12}^2 + 2c_{13}^{4}c_{12}^4 - 4
(c_{13}s_{12}c_{23} - s_{13}s_{23})^{2} (c_{13}s_{12}s_{23} +
s_{13}c_{23})^{2}
~\rm S_{23}
\label{pnueatm} 
\ee
\be
P_{\nu_\mu \nu_e}&=&2c_{13}^2c_{12}^2s_{12}^2 - 4c_{12}^2c_{23}s_{23}
(c_{13}s_{12}c_{23} - s_{13}s_{23})(c_{13}s_{12}s_{23} +
s_{13}c_{23}) 
~\rm S_{23}
\label{pmueatm}
\ee
\be
P_{\nu_\mu \nu_\mu}&=&1 - 2c_{12}^{2}s_{12}^2 -
4c_{12}^4c_{23}^2s_{23}^2
~\rm S_{23}
\label{pnumuatm}
\ee 
\noindent 
where $\rm S_{23} = \sin^2(\pi L/\lambda_{23})$, where $\lambda_{ij}$ 
is given by eq. (\ref{wv}). 
Apart from the most general three generation regime, the following 
limits are of interest, as we will see later in the context of the SK 
data:
\begin{enumerate}
\item  {\underline {The two-generation limits}}\\
Because of the presence of more parameters as compared to the one mass
scale dominance picture there are twelve possible two-generation 
limits \cite{sg4} with the oscillations driven by either $\Delta_{LSND}$
or $\Delta_{ATM}$. Below we list these limits specifying the mass scales that 
drive the oscillations:
\begin{itemize}
\item{
$s_{12} \rightarrow 0, s_{13} \rightarrow 0~~ 
(\nu_\mu-\nu_\tau,~~ \Delta_{ATM}$), 
$~~~s_{12} \rightarrow 0, s_{13} \rightarrow 1~~ 
(\nu_\mu-\nu_e,~~ \Delta_{ATM}$)\\
$s_{12} \rightarrow 1, s_{13} \rightarrow 0~~ 
(\nu_e -\nu_\tau,~~ \Delta_{ATM}$),
$~~~s_{12} \rightarrow 1, s_{13} \rightarrow 1~~ 
(\nu_e-\nu_\tau,~~ \Delta_{ATM}$)}

\item{
$s_{13} \rightarrow 0, s_{23} \rightarrow 0~~
(\nu_\mu - \nu_e,~~ \Delta_{LSND}$),
$~~~s_{13} \rightarrow 1, s_{23} \rightarrow 0~~
(\nu_\mu - \nu_\tau,~~ \Delta_{LSND}$)\\
$s_{13} \rightarrow 0, s_{23} \rightarrow 1~~
(\nu_\mu - \nu_e,~~ \Delta_{LSND}$),
$~~~s_{13} \rightarrow 1, s_{23} \rightarrow 1~~
(\nu_\mu - \nu_\tau,~~ \Delta_{LSND}$)}

\item{
$s_{12} \rightarrow 0, s_{23} \rightarrow 0~~
(\nu_e-\nu_\tau,~~ \Delta_{LSND}$),
$~~s_{12} \rightarrow 1, s_{23} \rightarrow 0~~
(\nu_e-\nu_\tau,~~ \Delta_{ATM}$)\\
$s_{12} \rightarrow 0, s_{23} \rightarrow 1~~
(\nu_e - \nu_\tau,~~ \Delta_{LSND}$),
$~~s_{12} \rightarrow 1, s_{23} \rightarrow 1~~
(\nu_e - \nu_\tau, ~~\Delta_{ATM}$)}

\end{itemize} 

\item  
\underline{$s_{12}^2$ = 0.0} \\
In this limit the relevant probabilities become
\be
P_{\nue \nue} &=& 1 - 2 c_{13}^2 s_{13}^2 +
4 s_{13}^2 c_{23}^2 s_{23}^2 S_{23} 
\label{pnues12}
\ee
\be
P_{\nue \numu} &=& 4 s_{13}^2 s_{23}^2 c_{23}^2 S_{23}
\label{pmues12}
\ee
\be
P_{\numu \numu} &=& 1 - 4 c_{23}^2 s_{23}^2 S_{23}
\label{pnumus12}
\ee
Thus $P_{\numu \numu}$ is the same as the two generation limit, 
$P_{\numu \nue}$ is governed by two of the mixing angles and
one mass scale and $P_{\nue \nue}$ is governed by two mixing angles and both
mass scales. 

\item 
\underline{$s_{13}^2$ = 0.0} \\
For this case the probabilities take the form 
\be
P_{\nue \nue} &=& 1 - 2 c_{12}^2 s_{12}^2 -
4 s_{12}^4 c_{23}^2 s_{23}^2 S_{23}
\label{pnues13}
\ee
\be
P_{\nue \numu} &=& 2 c_{12}^2 s_{12}^2 - 4 c_{12}^2 s_{12}^2 c_{23}^2 s_{23}^2
S_{23}
\label{pmues13}
\ee
\be
P_{\numu \numu} &=& 1 - 2c_{12}^2 s_{12}^2 - 
4 c_{12}^4 c_{23}^2 s_{23}^2 S_{23}
\label{pnumus13}
\ee
\noindent
In this case the probabilities are governed by two mass scales and two mixing angles. 

\end{enumerate}
We note that for cases (2) and (3) the probabilities are symmetric
under the transformation $\theta_{23} \rightarrow \pi/2 - \theta_{23}$.
The probabilities for these cases are functions of at most two mixing angles 
as in the OMSD case \cite{lisi4} but they are governed by both mass scales
making these limits different from the OMSD limit.  

\subsection {Earth Matter Effects}

Since on their way to the detector the upward going neutrinos pass through the 
earth, it is important in general to include the matter effect in 
the atmospheric neutrino analysis. 
The matter contribution to the effective squared mass of the electron 
neutrinos:
\begin{eqnarray}
A = 2{\sqrt 2}~ G_F~ E~n_e 
\end{eqnarray}
where $E$ is the neutrino energy and $n_e$ is the ambient electron 
density. Assuming a typical
density of 5 gm/cc and $E$ = 10 GeV, the matter potential $A 
\simeq  3.65 \times 10^{-3}$ eV$^2$ and since this is of the 
same order as $\Delta_{23}$ in our case, 
matter effects should be studied carefully.

\noindent
The mass matrix in the flavor basis in presence of matter is given by
\begin{eqnarray}
M_F ^2 = U~M^2~U^\dagger ~+~ M_A
\end{eqnarray}
where $M^2$ is the mass matrix in the mass eigenbasis, $U$ is the 
mixing matrix and 
\begin{eqnarray}
M_A = {\pmatrix {A & 0 & 0 \cr
0 & 0 & 0 \cr
0 & 0 & 0 \cr}}
\end{eqnarray}
Since $\Delta_{12} \sim \Delta_{13} \gg \Delta_{23} \sim A$, one can 
solve the eigenvalue problem
using the degenerate perturbation theory, where the $\Delta_{23}$ 
and $A$ terms are treated as a perturbation to the dominant 
$\Delta_{12}$ and $\Delta_{13}$ dependent terms. The mixing angle in matter 
is then given by
\begin{eqnarray}
\tan 2\theta_{23}^M = \frac{\Delta_{23}\sin2\theta_{23} -
As_{12}\sin2\theta_{13}}{\Delta_{23} \cos 2\theta_{23} -
A(s_{13}^2 - c_{13}^2 s_{12}^2)}
\label{theta23m}
\end{eqnarray}
while the mass squared difference in matter turns out to be
\begin{eqnarray}
\Delta_{23}^M = {\left[(\Delta_{23} \cos 2\theta_{23} - 
A(s_{13}^2 - c_{13}^2 s_{12}^2))^2 + (\Delta_{23}\sin2\theta_{23}
- As_{12}\sin2\theta_{13})^2\right]}^{1/2}
\label{del23m}
\end{eqnarray}
The mixing angles $\theta_{12}$ and $\theta_{13}$ as well as the 
larger mass squared difference $\Delta_{12}$ remain unaltered 
in matter. From eq. (\ref{theta23m}) and (\ref{del23m}) 
we note the following 
\begin{itemize}
\item In the limit of both $s_{12}\rightarrow 0$ 
and $s_{13}$ $\rightarrow 0$, 
the matter effect vanishes and we recover the two-generation 
$\nu_{\mu} - \nu_{\tau}$ limit. 
\item The resonance condition now becomes 
$\Delta_{23}\cos 2\theta_{23} = A (s_{13}^2 - c_{13}^2 s_{12}^2)$.
So that for $\Delta_{23} > 0$, one can have resonance for both 
neutrinos -- if 
$s_{13}^2 > c_{13}^2 s_{12}^2$ -- as well as for antineutrinos -- if 
$s_{13}^2 < c_{13}^2 s_{12}^2$.
This is different from the OMSD picture where for 
$\Delta m^2 > 0$ only neutrinos can resonate.

\item In the limit of $s_{12} \rightarrow 0 $ 
\begin{equation}
\tan 2\theta_{23}^{M} = \frac{\Delta_{23} \sin2\theta_{23}}{\Delta_{23} 
\cos 2\theta_{23} - A s_{13}^2}
\label{l1}
\end{equation} 
Here one gets resonance for neutrinos only (if $\Delta_{23}>0$).  

\item In the limit $s_{13} \rightarrow 0$ 
\begin{equation} 
\tan 2\theta_{23}^{M} = \frac{\Delta_{23} \sin 2\theta_{23}}{\Delta_{23} 
\cos 2\theta_{23} + A s_{12}^2}
\label{l2}
\end{equation} 
For this case for $\Delta_{23}>0$, there is no resonance for 
neutrinos but antineutrinos can resonate. 

\item In the limit where $\Delta_{23}\rightarrow 0$ 
\begin{eqnarray}
\tan 2\theta_{23}^M = \frac{s_{12} \sin 2\theta_{13}}
{s_{13}^2 - c_{13}^2 s_{12}^2},~~~ 
\Delta_{23}^M = A(s_{13}^2 + c_{13}^2 s_{12}^2)
\label{lowdelmsq}
\end{eqnarray}
Thus even for small values of $\Delta_{23} < 10^{-4}$ 
the mass squared difference in matter is $\sim A$ and one may 
still hope to see oscillations for the upward neutrinos due to matter
effects. The other point to note is that the mixing angle in matter 
$\theta_{23}^M$ depends only on $\theta_{12}$ and
$\theta_{13}$ and is independent of the vacuum mixing angle 
$\theta_{23}$ and $\Delta_{23}$. 
Contrast this with the OMSD 
and the two-generation 
$\nu_\mu - \nu_e$ oscillations.
For both the two-generation $\nu_\mu - \nu_e$
as well as the three-generation OMSD case, 
for $\dm \rightarrow 0$, the mixing angle in matter 
$\rightarrow 0$, 
but for the mass spectrum considered in this paper the 
$\tan 2\theta_{23}^M$ maybe large depending on the values 
of $\sabsq$ and $\sacsq$.
Hence we see that the {\it demixing effect} which
gives the lower bound on allowed values of \dm in
the two generation $\nu_\mu - \nu_e$ or the three-generation
OMSD case, does not arise here and we hope to get allowed regions 
even for very low values of $\Delta_{23}$. 
On the other hand even small values
of $\theta_{23}$ in vacuum can get enhanced in matter.
This special case where $\Delta_{23} \sim 0$ was 
considered in an earlier paper \cite{rujula}.


\item In the limit of $\sbcsq \rightarrow 0$ 
\begin{eqnarray}
\tan 2 \theta_{23}^M = \frac{-A s_{12} \sin 2\theta_{13}}
{\Delta_{23} -A (s_{13}^2 - c_{13}^2 s_{12}^2)}
\label{lowtheta1}
\end{eqnarray}
\item While for $\sbcsq \rightarrow 1$
\begin{eqnarray}
\tan 2 \theta_{23}^M = \frac{-A s_{12} \sin 2\theta_{13}}
{-\Delta_{23} -A (s_{13}^2 - c_{13}^2 s_{12}^2)}
\label{lowtheta2}
\end{eqnarray}
For the last two cases, corresponding to $\sin^2 2\theta_{23} \rightarrow 0$, 
again the mixing angle $\theta_{23}$ in matter is
independent of its corresponding value in vacuum and hence for
appropriate choices of the other three parameters, $\Delta_{23}$,
$\sabsq$ and $\sacsq$, one can get large values for $\sin^2 2\theta_{23}^M$
even though the vacuum mixing angle is zero.
  
\end{itemize}

\noindent
The amplitude that an initial $\nu_\alpha$ of energy $E$ is 
detected as $\nu_\beta$ after traveling through the earth is
\begin{eqnarray}
A(\nu_\alpha,t_0,\nu_\beta,t) = \sum_{\sigma,\lambda,\rho}
\sum_{i,j,k,l}&& [(U_{\beta l}^{M_m} e^{-iE_l^{M_m}(t-t_3)}
U_{\sigma l}^{M_m})(U_{\sigma k}^{M_m} e^{-iE_k^{M_c}(t_3-t_2)}
U_{\lambda k}^{M_c}) \times
\nonumber \\
&& (U_{\lambda j}^{M_m} e^{-iE_j^{M_m}(t_2-t_1)}U_{\rho j}^{M_m})
(U_{\rho i}e^{-iE_i (t_1-t_0)}U_{\alpha i})]
\label{amp}
\end{eqnarray}
where we have considered the earth to be made of two slabs, a mantle 
and a core with constant densities of 4.5 gm/cc and 11.5 gm/cc 
respectively and include the non-adiabatic effects at the boundaries. 
The mixing matrix in the mantle and the core are given 
by $U^{M_m}$ and $U^{M_c}$ respectively. $E_i^{X} 
\approx m_{iX}^2/2E$, $X$ = core(mantle) and
$m_{iX}$ is the mass of the 
$i^{th}$ neutrino state in the core(mantle).
The neutrino is produced at time $t_0$, hits the earth mantle at 
$t_1$, hits the core at $t_2$, leaves the core at $t_3$ and 
finally hits the detector at time $t$. The Greek indices 
($\sigma,\lambda,\rho$) denote the flavor eigenstates 
while the Latin indices ($i,j,k,l$) give the 
mass eigenstates.
The corresponding 
expression for the probability is given by
\begin{eqnarray}
P(\nu_\alpha,t_0,\nu_\beta,t) = |A(\nu_\alpha,t_0,\nu_\beta,t)|^2
\label{prmatter}
\end{eqnarray}
For our calculations of the number of events we have used the full 
expression given by eq.(\ref{amp}) and (\ref{prmatter}).

\section{$\chi^2$-analysis of the SK Data}

\begin{figure}[t]
    \centerline{\psfig{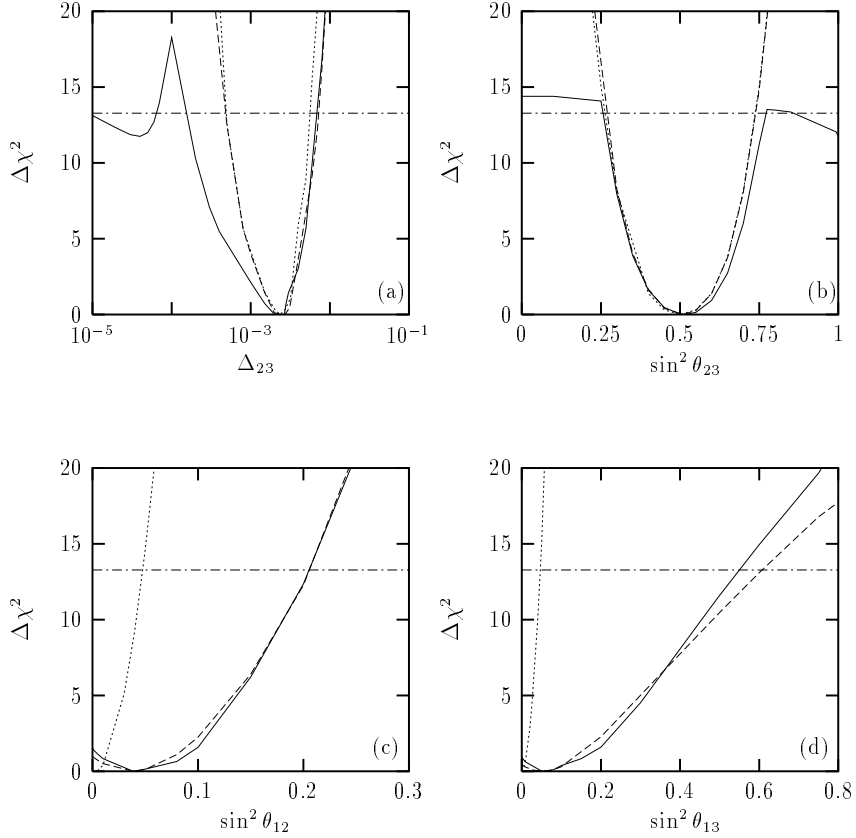}}
\vskip -1cm
   \caption[The variation of $\Delta \chi^2$ with mixing parameters]
{\label{delchi3gen}
The variation  of $\Delta \chi^2 = \chi^2 - \chi^2_{min}$
with one of the parameters keeping the other three unconstrained.
The solid (dashed) line corresponds to
only SK data when matter effects are included (excluded)
while the dotted curve gives the same for
SK+CHOOZ. The dashed-dotted line shows the 99\% C.L. limit for 4 parameters.}
\end{figure}

We do a bin by bin \chisq analysis of the 1144 days SK atmospheric 
neutrino data \cite{sk11444} following the procedure of statistical 
analysis given in \cite{lisi4,G_G4}.
We minimize the $\chi^2$ function defined in chapter 3, 
eq. (\ref{chilisi}).
For contained events there are  forty experimental data points.
The probabilities for
the atmospheric neutrinos are explicit 
functions of one mass-squared difference and
three mixing angles making the number of degrees of freedom (d.o.f) 36. 
The other mass squared difference gives rise to \dm independent 
average oscillations and hence does not enter the fit as an 
independent parameter. 


\noindent
For the three-generation scheme considered here 
the $\chi^2_{min}$ and the best-fit values of parameters that we get are
\begin{itemize}
\item
$\chi^2_{min}/d.o.f. = 34.65/36$, $\Delta_{23} = 0.0027$ eV$^2$,
$s^2_{23} = 0.51$,$s_{12}^2 = 0.04$ and
$s_{13}^2 =0.06 $
\end{itemize}
This solution is allowed at 53.28\% probability.

The  solid(dashed) lines in fig. \ref{delchi3gen} present 
the variation of the $\Delta \chi^2 = \chi^2 - \chi^2_{min}$ 
for the SK data, with respect to 
one of the parameters keeping the other three unconstrained, 
when we include(exclude) the matter effect.  
In fig. \ref{delchi3gen}(a) as we go towards smaller values of 
$\Delta_{23}$  around $10^{-3}$ eV$^2$
the effect of matter starts becoming important   
as the matter 
term is now comparable to the mass term. 
If matter effects are not there then for values of $\Delta_{23}$ 
$\stackrel{<}{\sim} 10^{-4} $ eV$^2$ the 
$S_{23}$ term in eq. (\ref{pnumuatm})
is very small and there is no up-down asymmetry
resulting in very high values of $\chi^2$ as is evident 
from the dashed curve. 
If the matter effects are included, then in the limit of very 
low $\Delta_{23}$ the matter term dominates and 
$\Delta_{23}^M$ is given by eq.(\ref{lowdelmsq}).
Since this term $\sim 10^{-3}$ eV$^2$ there can be 
depletion of the 
neutrinos passing through the earth causing an updown asymmetry. 
For $\Delta_{23}$ around $10^{-4}$ eV$^2$, there is cancellation 
between the two comparable terms in the numerator of 
eq. (\ref{theta23m}) and 
the mixing angle becomes very small and hence the $\chi^2$ around 
these values of $\Delta_{23}$ comes out to be very high. 

Fig \ref{delchi3gen}(b) 
illustrates the corresponding variation of $\Delta \chi^2$ with
$\sbcsq$ while the other three parameters are allowed to vary 
arbitrarily. 
For small and large values of $s_{23}^2$  
the inclusion of matter effect makes a difference. 
For $s_{23}^2$ either very small or large ($\sin^2 2\theta_{23} 
\rightarrow 0$) the overall suppression of the 
$\nu_\mu$ flux is less than that required by the data if vacuum oscillation 
is operative and so it is ruled out. If we include matter effects then    
in the limit of $s_{23}^2 =0$ and $s_{23}^2 =1$ the matter mixing angle is 
given by eqs. (\ref{lowtheta1}) and (\ref{lowtheta2}), which can 
be large for suitable values of $\sacsq$ and $\sabsq$ and hence 
one gets lower $\chi^2$ even for these values of $s_{23}^2$. 

In figs \ref{delchi3gen}(c) and \ref{delchi3gen}(d) 
we show the effect of $\sabsq$ and $\sacsq$
respectively on $\Delta \chi^2$. From the solid and the dashed lines 
it is clear that matter effects do not vary much the allowed ranges of 
$\sabsq$ and $\sacsq$.    

The dashed-dotted line in the figure 
shows the 99\% C.L. (= 13.28 for 4 parameters) 
limit. In Table \ref{allowed} we give the allowed ranges of the mixing 
parameters, inferred from fig. \ref{delchi3gen} 
at 99\% C.L. for the 1144 day SK atmospheric data, with 
and without matter effects. 

\begin{table}[htbp]
    \begin{center}
\begin{tabular}{||c|c|c|c|c||}\hline\hline
{} & {$\Delta_{23}$ in eV$^2$}&{$\sbcsq$}&{$\sabsq$}&{$\sacsq$}\\ \hline
{with} & {$\!\!1.6\!\!\times \!10^{-4}\!\leq\! 
\Delta_{23}\! \leq \!7.0\!\!\times\!\! 10^{-3}\!\!$} & 
{$0.26 \leq \sbcsq \leq 0.77$} & {$\sabsq \leq 0.21$} & 
{$\sacsq \leq 0.55$} \\
{$\!\!$matter effects$\!\!$} & {$\Delta_{23} \leq 6.5\times 10^{-5}$}
& {$\sbcsq \geq 0.85$} & {} & {} \\ \hline
{without}&{}
&{}&{}&
{}\\
{$\!\!$matter effects$\!\!$}&\raisebox{1.5ex}[0pt]{$\!\!5\!\!\times\!\!
10^{-4}\!\leq \!\Delta_{23}\!
\leq 7.0\! \times \!\!10^{-3}\!\!$}&\raisebox{1.5ex}[0pt]{$0.27 \leq \sbcsq
\leq 0.74$}&\raisebox{1.5ex}[0pt] {$\sabsq \leq 0.21$}&
\raisebox{1.5ex}[0pt] {$\sacsq \leq 0.6$}
\\ \hline\hline
\end{tabular}
\end{center}
\caption[Allowed range of parameters of the 3-gen analysis]
{\label{allowed}
The allowed ranges of parameters for the 1144 day SK data.}
\end{table}

\begin{figure}
    \centerline{\psfig{file=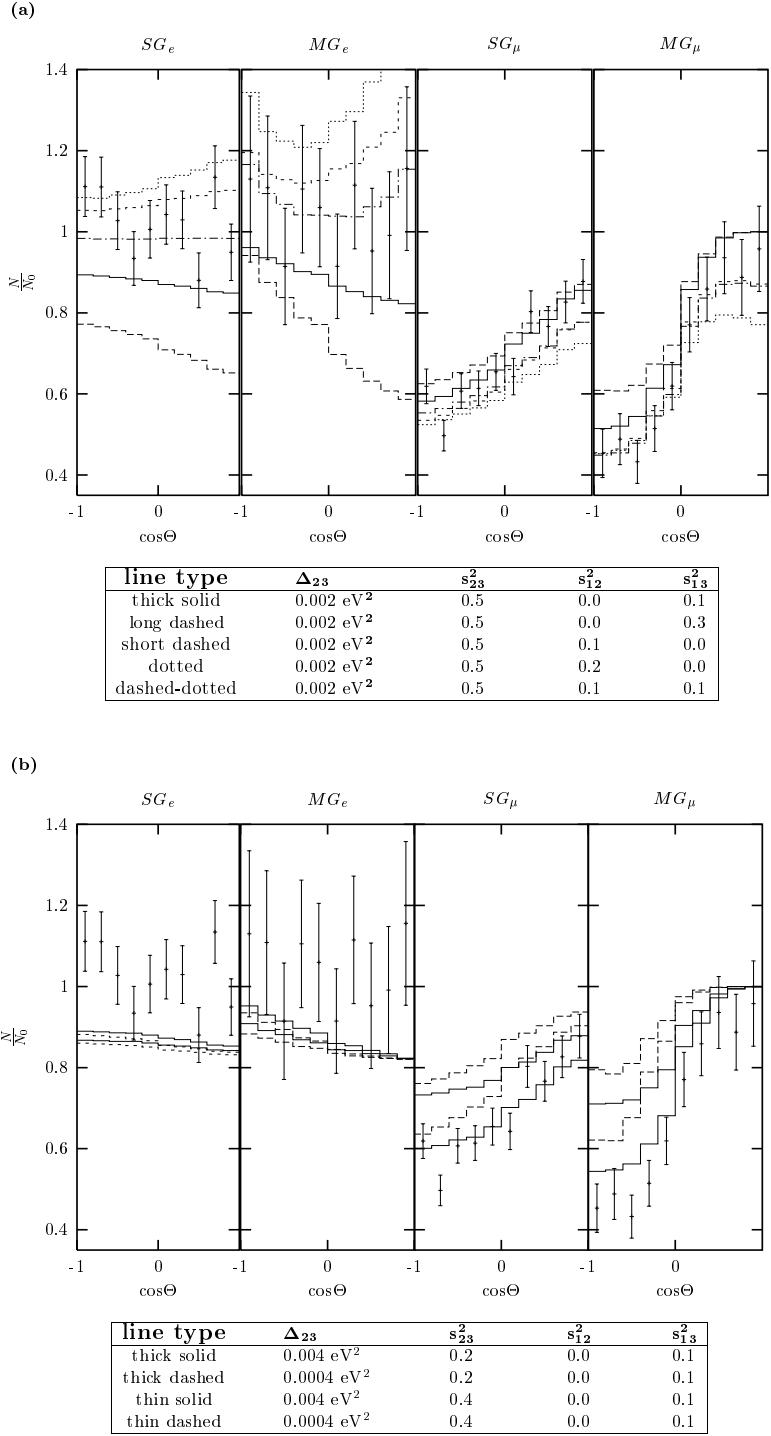,width=5.0in}}
\vskip -2.0cm
\caption[Zenith angle distribution of the lepton events]
{\label{3a}
(a)The zenith angle distribution of the lepton events
with $\Delta_{23}=0.002$ eV$^2$ and $\sbcsq$
= 0.5 for various combinations of
$\sabsq$ and $\sacsq$.
$N$ is the number of events as given by eq. (\ref{rate}) 
and $N_0$ is the corresponding number with survival probability 1.
The panels labelled $SG_{\alpha}$ and $MG_{\alpha}$ ($\alpha$ can be e or
$\mu$) give the
histograms for the sub-GeV and multi-GeV $\alpha$-events respectively.
Also shown are the SK experimental data points with $\pm$ 1$\sigma$ error
bars.
(b)Same as in (a) for fixed $\sabsq=0.1$ and $\sacsq=0.0$
varying $\Delta_{23}$ and $\sbcsq$.}

\end{figure}

\begin{figure}
    \centerline{\psfig{file=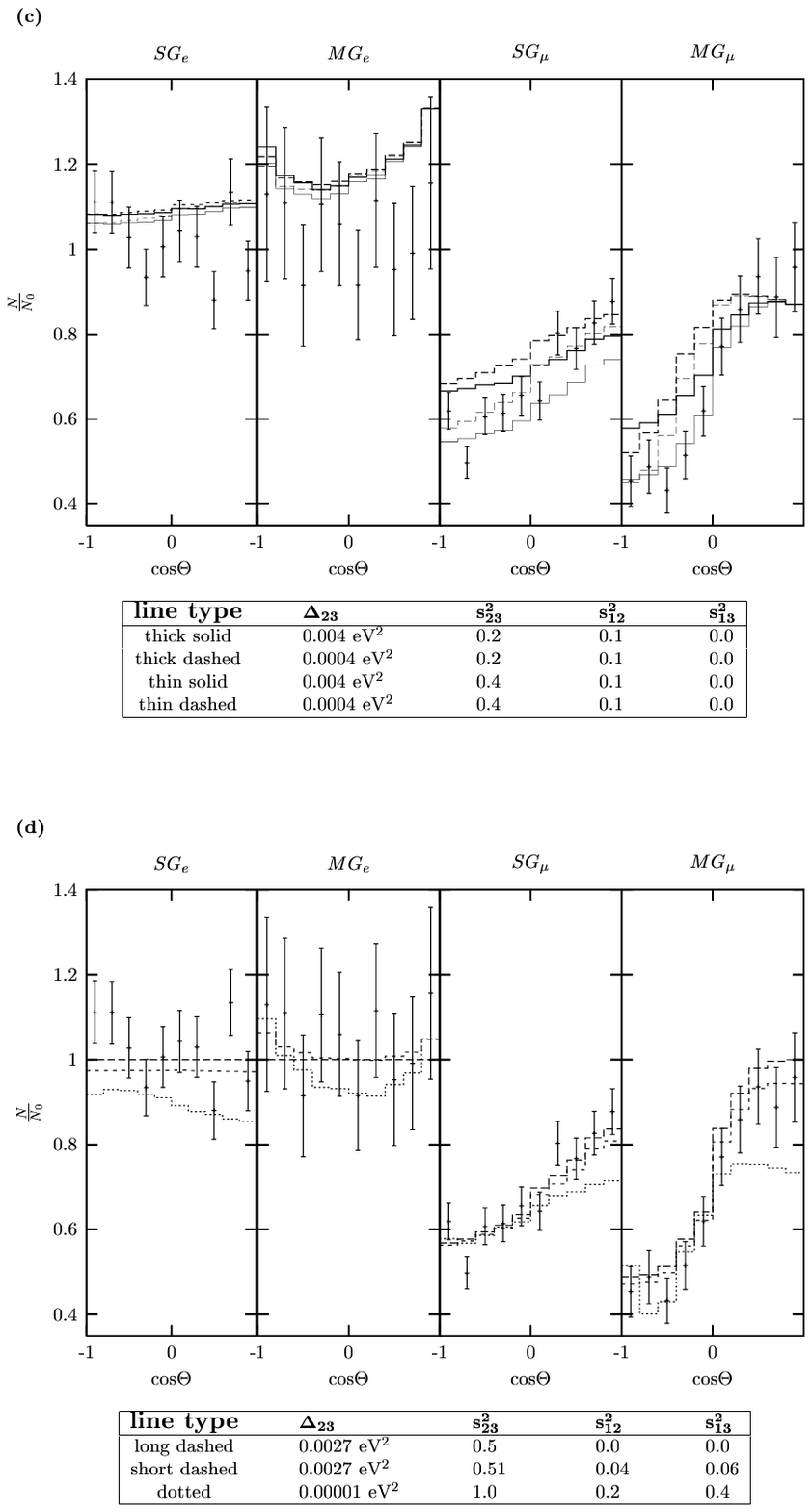,width=5.0in}}
\vskip -2cm
\caption[Zenith angle distribution of the lepton events]
{\label{3b}
(c)Same as in (a) fixing $\sabsq=0.0$ and $\sacsq=0.1$ for different
$\Delta_{23}$ and $\sbcsq$ values.
(d)The long-dashed (short-dashed) line gives the
zenith angle distribution of the lepton events for the
best-fit cases of the two-generation (three-generation)
oscillation solutions for SK. The dotted line gives
the corresponding distribution for $\Delta_{23}=10^{-5}$ eV$^2$,
$\sabsq=0.2$, $\sacsq=0.4$ and $\sbcsq=1.0$.}

\end{figure}

%
%

\subsection{Zenith-Angle Distribution}

Since the probabilities in our case are in general 
governed by two mass scales and
all  three
mixing  angles it is difficult to understand the allowed regions.  
To facilitate the qualitative understanding we present 
in figs. \ref{3a} and \ref{3b} 
the histograms which describe the zenith angle distribution.  
The event distributions in these histograms are approximately given by, 
\begin{equation}
\frac{N_\mu}{N_{\mu_{0}}} \approx P_{\numu \numu} +
\frac{N_{e_{0}}}{N_{\mu_{0}}} 
P_{\nue \numu}
\label{muno}
\end{equation}
\begin{equation}
\frac{N_e}{N_{e_{0}}} \approx P_{\nue \nue} + \frac{N_{\mu_{0}}}{N_{e_{0}}} 
P_{\numu \nue}
\label{eno}
\end{equation}
where the quantities with suffix 0 indicates the no-oscillation values. 
For the sub-GeV data $N_{\mu_{0}}/N_{e_{0}} \approx 2$ to a good approximation 
however for the multi-GeV data this varies in the range 2 (for  
$\cos\Theta$ =0) to 3 (for $\cos\Theta = \pm $1) \cite{lisi4}.
 

In fig. \ref{3a}(a) we study the effect of varying $s_{12}^2$ and $s_{13}^2$
for fixed values of $\Delta_{23}$ = 0.002 eV$^2$ 
and $s_{23}^2$ = 0.5.  
From eq. (\ref{theta23m}), (\ref{del23m}) and from fig. \ref{delchi3gen} 
we see that for the values of the $\Delta_{23}$ and $s_{23}^2$ 
considered in this figure 
the matter effects are small and we can understand the 
histograms from the vacuum oscillation probabilities.  
The thick solid line shows the event distribution for $s_{12}^2 = 0$
and $s_{13}^2 = 0.1$. 
As $s_{13}^2$ increases from 0, keeping $s_{12}^2$ as 0, from eqs. 
(\ref{pnues12}) and (\ref{pmues12}) 
$P_{\nue \nue}$ decreases from 1 and 
$P_{\nue \numu}$ increases from zero  resulting in a net
electron  depletion according to eq. (\ref{eno}). 
The long dashed line corresponds to $s_{13}^2$ = 0.3  
for which the electron depletion is 
too high as compared to data. 
The muon events are also affected as $P_{\numu \nue}$ increases with 
increasing $s_{13}^2$ even though $P_{\numu \numu}$ 
is independent of $s_{13}^2$. 
On the other hand for $s_{13}^2$ = 0.0, the effect of  
increasing $s_{12}^2$ is to increase the number of electron events 
and decrease the number of muon events according to eqs. 
(\ref{pnues13}), (\ref{pmues13}), (\ref{pnumus13}), (\ref{eno}) 
and (\ref{muno}). 
This is shown by the short-dashed and dotted lines in fig. \ref{3a}(a). 
For $s_{12}^2$ = 0.2 the electron excess and muon 
depletion both becomes too high as compared to the data.  
For the case when both
$s_{12}^2$ and $s_{13}^2$ are 0.1 the electron depletion caused by
increasing $s_{12}^2$ and the excess caused by increasing $s_{13}^2$ 
gets balanced and the 
event distributions are reproduced quite well, 
shown by the dashed-dotted line. 

 
In fig. \ref{3a}(b) 
we study the effect of varying $s_{23}^2$ and $\Delta_{23}$ 
in the limit of $s_{12}^2=0$ with $s_{13}^2$ fixed at 0.1. 
Although we use the full probabilities including the 
matter effect, for 0.004 eV$^2$ this is not so important and one can 
understand the histograms from the vacuum oscillation probabilities. 
For fixed $\Delta_{23}$ as $s_{23}^2$ increases, $P_{\numu \numu}$ decreases, 
making the muon depletion higher. This is shown in the figure 
for two representative values of $\Delta_{23}$.
The electron events are not affected much by change of $\sbcsq$. 
The slight increase with $\sbcsq$ is due to increase of  
both $P_{\nue \nue}$ and $P_{\numu \nue}$. 
To understand the dependence on $\Delta_{23}$ we note that   
for $\sbcsq=0.2$, if one looks at the vacuum oscillation probabilities, 
$N_{\mu}/N_{\mu 0} \approx 1 - 0.65 S_{23}$.
For 0.004 eV$^2$
the contribution of $S_{23}$ is more resulting in a lower number of 
muon events. 
For the electron events however the behavior with $\Delta_{23}$ is 
opposite, with $N_e/N_{e0} = 0.82 +
0.12 S_{23}$. Thus with increasing $\Delta_{23}$ the number of electron 
events increase. Also note that 
since the contribution of $S_{23}$ comes with 
opposite sign 
the zenith-angle distribution for a fixed $\Delta_{23}$ is opposite for 
the muon and the electron events. 

In fig. \ref{3b}(c) 
we show the histograms in the limit of $s_{13}^2$ = 0.0,
keeping $s_{12}^2$ as 0.1 and varying $\Delta_{23}$ and $s_{23}^2$. 
As $s_{23}^2$ increases all the relevant probabilities 
decrease
and therefore both $N_{\mu}/N_{\mu 0}$ and 
$N_{e}/N_{e 0}$ 
decrease giving less number of events for both. 
For this case the $S_{23}$ term comes with the same sign (negative) in both 
$N_{\mu}/N_{\mu 0}$ and $N_{e}/N_{e 0}$.
Therefore the depletion is more for higher $\Delta_{23}$  
for both muon and electron events.

Finally, the long dashed line in fig. \ref{3b}(d) represent the
histograms for the best-fit value for 
two-generation $\numu-\nutau$ oscillations, for which $P_{\nue\nue}=1$.
The short dashed line gives the histograms for the
three-generation best-fit values. Both give comparable explanation 
for the zenith angle distribution of the data. The dotted line 
gives the event distribution for $\Delta_{23} = 10^{-5}$ eV$^2$. 
As discussed earlier even 
for such low value of $\Delta_{23}$, we find that due to the 
unique feature of the beyond OMSD neutrino mass spectrum, 
earth matter effects ensure that  
both the sub-GeV as well as the multi-GeV upward muon events are 
very well reproduced, as are the electron events. But since $\sabsq$ 
is high, the downward $\numu$ are depleted more than the data requires.

\subsection{Allowed Parameter Region}

\begin{figure}[t]
\vskip -0.8cm
    \centerline{\psfig{file=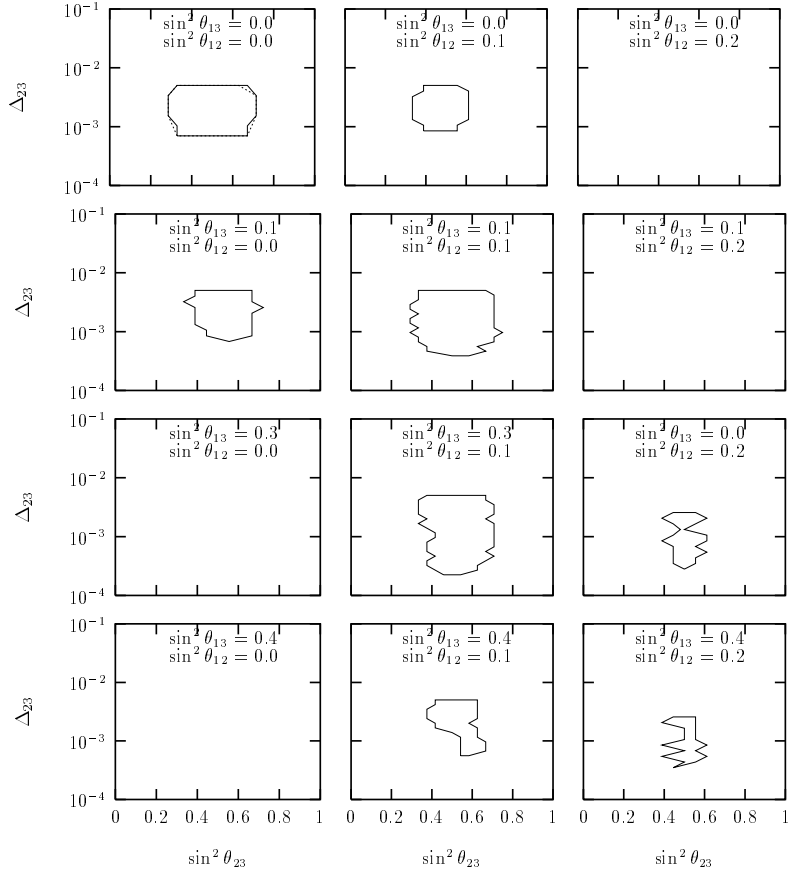,width=4.5in}}
\vskip -1.5cm
\caption[Allowed regions in $\Delta_{23}-\sin^2\theta_{23}$ plane]
{\label{c1}
The allowed parameter regions in the $\Delta_{23} -
\sbcsq$ plane for various fixed values of $\sabsq$ and
$\sacsq$, shown at the top of each panel. The solid lines
corresponds to the 99\% C.L. contours from the SK data alone, while
the dotted line
gives the 99\% contour from the
combined analysis of the SK+CHOOZ data.}
\end{figure}

In fig. \ref{c1} the solid lines give the 99\% C.L. allowed area 
from SK data in the 
$\Delta_{23}$-$s^2_{23}$ plane keeping the values of
$s^2_{13}$ and $s^2_{12}$ fixed in the 
allowed range from fig. \ref{delchi3gen} and Table \ref{allowed}. 
The first panel represents the two-generation $\numu - \nutau$ oscillation 
limit modulo the difference in the definition of the C.L. limit as the 
number of parameters are different. 
We have seen from the histograms in fig. \ref{3a}(a) that raising $\sabsq$ 
results in electron excess and muon depletion. 
On the other hand 
increase in $\sacsq$ causes electron depletion. 
The above features are reflected in the shrinking 
and disappearance of the 
allowed regions in the first row and column. In the panels where both 
$\sabsq$ and $\sacsq$ are nonzero one may get allowed regions only when the 
electron depletion due to increasing $\sacsq$ is replenished by the 
increase in $\sabsq$.  

\begin{figure}[t]
\vskip -0.8cm
    \centerline{\psfig{file=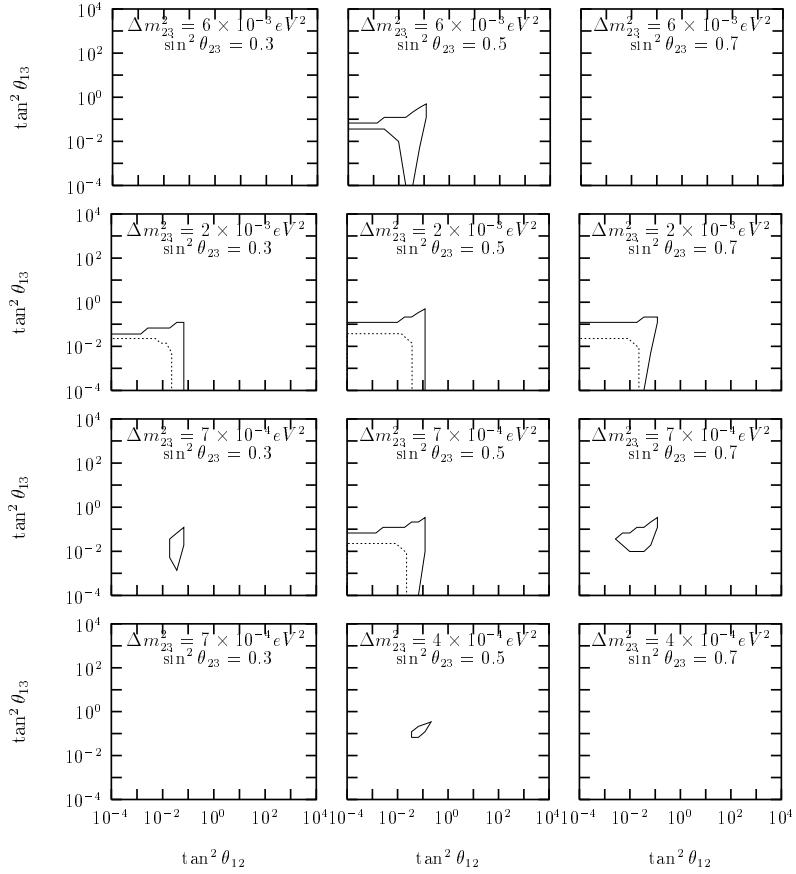,width=4.5in}}
\vskip -1.5cm
\caption[Allowed regions in $\tan^2\theta_{12}-\tan^2\theta_{13}$ plane]
{\label{c2}
Same as fig. \ref{c1} but in the bilogarithmic $\tan^2\theta_{12}-
\tan^2\theta_{13}$ plane for
fixed values of $\Delta_{23}$ and $s_{23}^2$.}
\end{figure}

In fig. \ref{c2} we present the 99\% C.L.  
allowed areas 
in the  bilogarithmic $\tan^2\theta_{12} - \tan^2\theta_{13}$ plane for various
fixed values of the parameters $\Delta_{23}$ and $s^2_{23}$.
We use the $\log(\tan)$ representation which enlarges the allowed regions
at the corners and the clarity is enhanced.  
The four corners in this plot refer to the two-generation limits discussed
before.
The extreme left corner ($\theta_{12} \rightarrow 0, \theta_{13}
\rightarrow 0$) correspond to 
the two generation $\nu_\mu - \nu_\tau$ oscillation limit.
As we move up increasing $\theta_{13}$,  
one has $\nu_e - \nu_\mu$ and $\nu_e - \nu_\tau$ mixing 
in addition and for $\sacsq \rightarrow 1 $ one
goes to the two generation 
$\nu_\mu - \nu_e$ oscillation region.
For the best-fit values of $\Delta_{23}$ and $\sbcsq$
if we take $\sabsq$ and $\sacsq$ to be 0 and $1$ respectively, 
then the $\chi^2_{min}$ is 66.92 which is therefore ruled out. 
Both the right hand corners in all the panels 
refer to pure $\nu_e - \nu_\tau$ 
oscillations and therefore there are no allowed regions in these zones.
For the panels in the first row, $\Delta_{23}=0.006$ eV$^2$ and  
the 2-generation $\numu-\nutau$ oscillation limit is 
just disallowed. The small area allowed for the middle 
panel of first row (between the solid lines) 
is due to the fact that for non-zero 
$\sabsq$ and $\sacsq$ the electron events are better 
reproduced, while $\sbcsq=0.5$ takes care of the muon events. 
Hence for this case
slight mixture of $\numu-\nue$ and $\nue-\nutau$ oscillations
is favoured. 
This feature was also reflected in the fact that 
in the fig. \ref{c1}, the panel for $\sabsq=0.1$ and $\sacsq=0.3$ has 
more allowed range for $\Delta_{23}$ than 
the panel for the 2-generation $\numu-\nutau$ limit. 
 For the panels with $\Delta_{23}=0.002$ eV$^2$, both
the pure $\numu-\nutau$ limit as well as full three-generation 
oscillations, give good fit.
For the last two rows with $\Delta_{23}= 0.0007$ eV$^2$ and $0.0004$ eV$^2$ 
the matter effects are important in controlling the 
shape of the allowed regions. 
Infact the allowed region that one gets for $0.0004$ eV$^2$ and 
$s_{23}^2$ = 0.5 is the hallmark of the matter effect in this 
particular three-generation scheme. As can be seen 
from fig. \ref{delchi3gen}(a) and 
Table \ref{allowed},
if one does not include the matter effect, then there are no allowed 
regions below $\Delta_{23}$ = 0.0005 eV$^2$ for any arbitrary 
combination of the other three parameters. 
Even for the first and the last panels with $\Delta_{23}=0.0007$ eV$^2$, 
one gets allowed areas solely due to matter effects.

\begin{figure}[t]
\vskip -0.8cm
    \centerline{\psfig{file=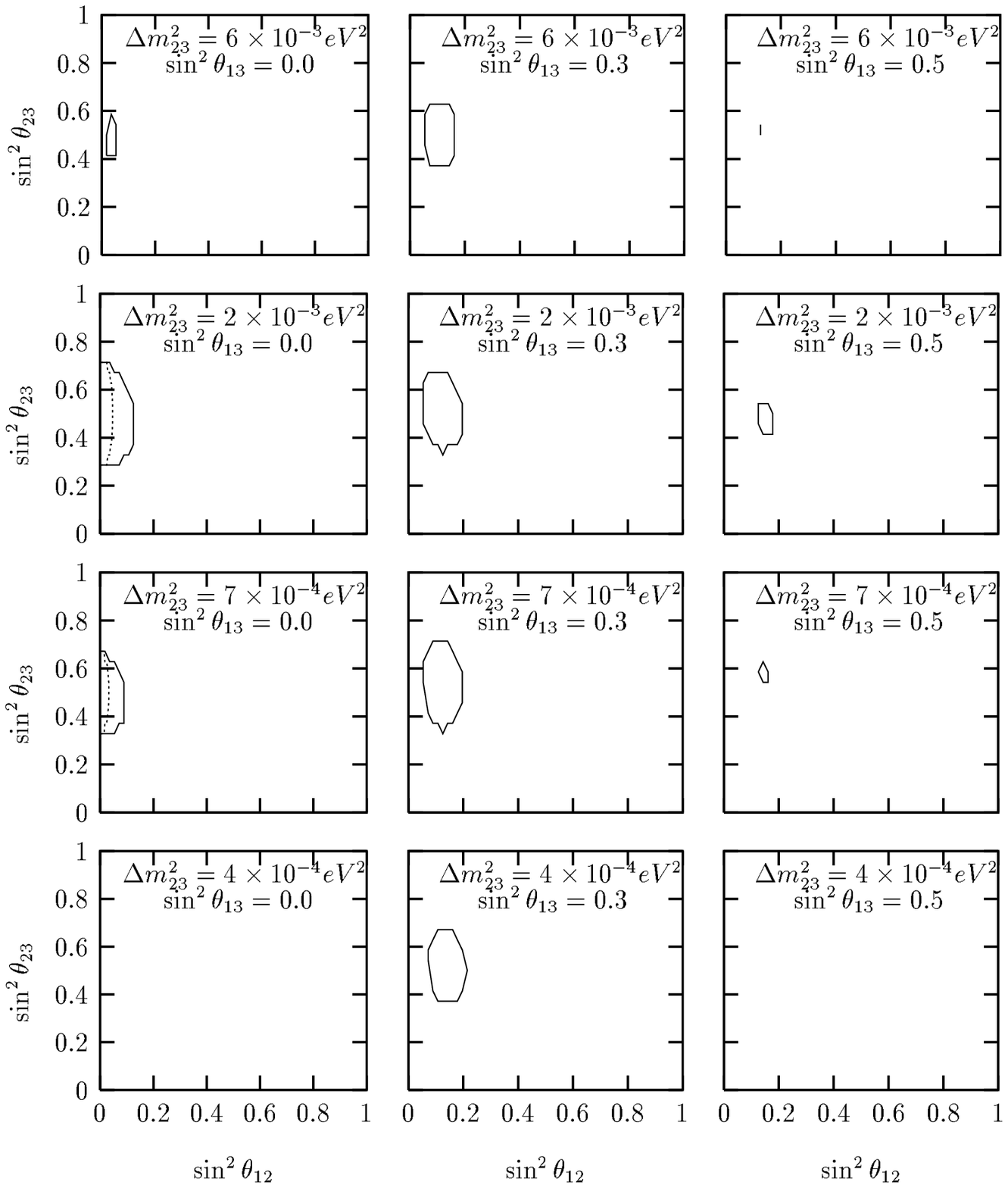,width=4.5in}}               
\vskip -1.5cm
\caption[Allowed regions in $\sabsq-\sbcsq$ plane] 
{\label{c3}
Same as fig. \ref{c1} but in the $\sabsq-\sbcsq$ plane for fixed values
of $\sacsq$ and $\Delta_{23}$.}
\end{figure}

In fig. \ref{c3} the solid lines show the 99\% C.L. allowed regions 
from SK data in the 
$s^2_{23} - s^2_{12}$ plane for fixed values
of $\Delta_{23}$ and $s^2_{13}$.
In contrast to the previous figure, 
here (and in the next figure)  we use the $\sin-\sin$ representation 
because the allowed regions are around $\theta_{23} =\pi/4$ and this
region gets compressed in the $\log(\tan)-\log(\tan)$ representation. 
For explaining the various allowed regions we separate the figures in two sets

$\bullet$ For $s^2_{13}$ = 0.0, 
the four corners of the panels represent the no-oscillation limits  
inconsistent with the data. Also as discussed in the earlier sections 
for $s_{23}^2$ = 0.0 or 1.0 one goes to the limit of 
pure $\nu_\mu-\nu_e$ conversions driven by
$\Delta_{LSND}$, which is not consistent with data. One obtains 
allowed regions only when 
$s^2_{23}$ is close to 0.5 with $s_{12}^2$ small, so that
$\numu-\nutau$ conversions are dominant.  
The allowed range of $\sabsq$ is controlled mainly by the electron 
excess as has been 
discussed before while the allowed range of $\sbcsq$ is determined 
mostly by the muon depletion.  

$\bullet$ For $s_{13}^2 \neq 0$, the four corners represent
the two-generation $\nu_e - \nu_\tau$  oscillation 
limit and hence these corners are not allowed. 
For $s_{23}^2=0.0$ or 1.0 and $s_{12}^2 \neq $ 0 or 1 
one has $\Delta_{LSND}$ driven $\numu-\nue$ and $\numu-\nutau$ conversion and 
$\Delta_{ATM}$ driven $\nu_e-\nu_\tau$ conversions. 
This scenario is not allowed as it gives excess of electron events
and also fails to reproduce the correct zenith angle dependence.
For a fixed $\Delta_{23}$ as $s_{13}^2$ increases the
electron depletion increases 
which can be balanced by increasing $s_{12}^2$ which increases the number
of electron events. Hence for a fixed $\Delta_{23}$ 
the allowed regions shift towards higher $s_{12}^2$ values. 

As in fig. \ref{c2} the allowed area in the middle panel of the last row is 
due to the inclusion of the matter effect. 

\begin{figure}[t]
\vskip -0.8cm
    \centerline{\psfig{file=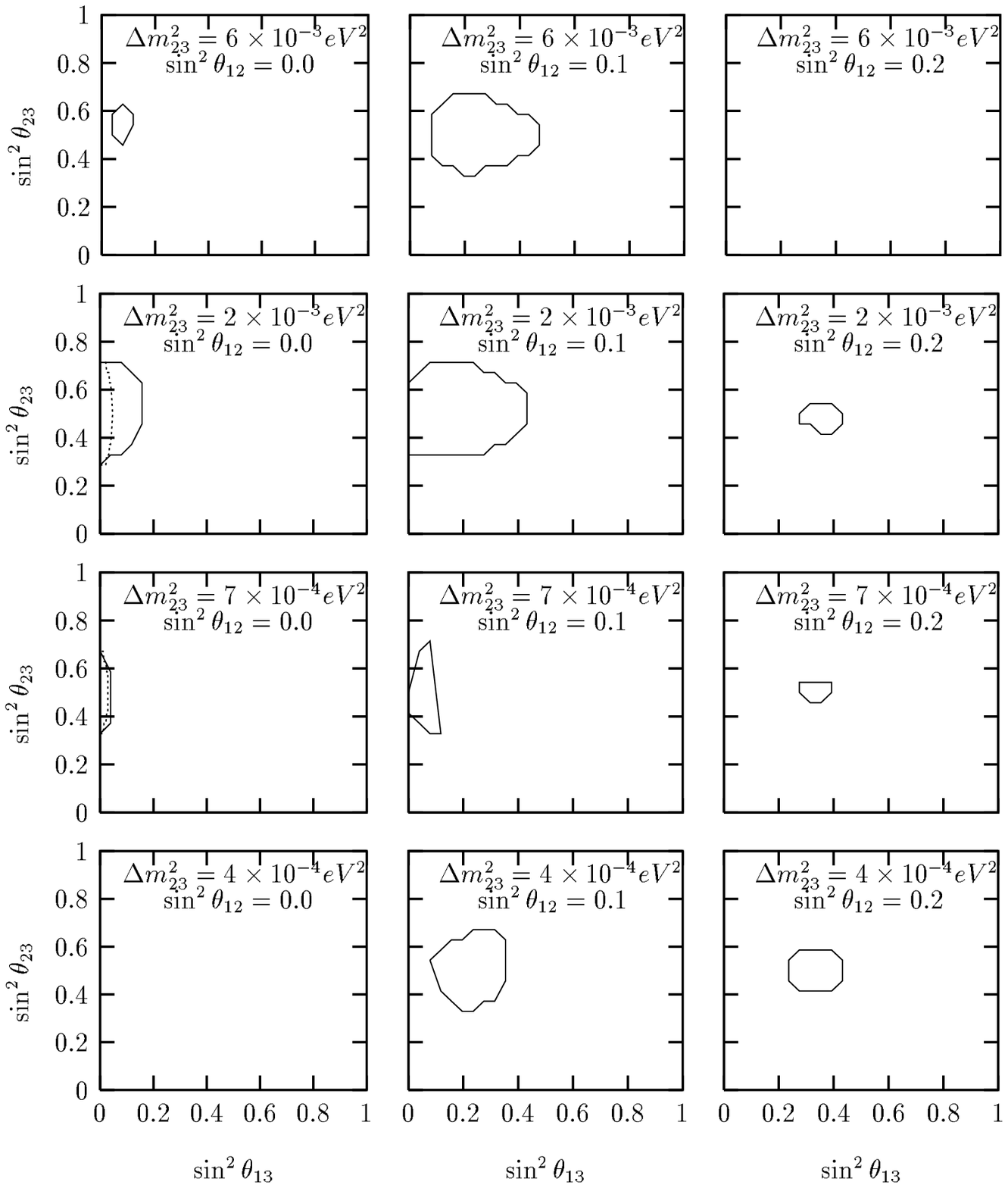,width=4.5in}}               
\vskip -1.5cm
\caption[Allowed regions in $\sacsq-\sbcsq$ plane] 
{\label{c4}
Same as fig. \ref{c1} but in the $\sacsq-\sbcsq$ plane for various fixed
values of $\sabsq$ and $\Delta_{23}$.}
\end{figure}
In fig. \ref{c4} the solid contours refer to the  99\% C.L. allowed areas
from SK atmospheric neutrino data 
in the  $s^2_{13} - s^2_{23}$ plane for various values
of $\Delta_{23}$ and $s^2_{12}$. 

$\bullet$ For $s_{12}^2$ = 0.0 
the corners represent no oscillation limits.
In the limit $\sbcsq \rightarrow 0$ or 1, one gets $\nue-\nutau$ 
oscillation driven by $\Delta_{LSND}$ which is also not allowed. 
For $s_{13}^2$ = 0.0 and $\sbcsq \sim 0.5$ one has maximal two-flavour 
$\numu-\nutau$ oscillation limit which is therefore allowed (not 
allowed for $\Delta_{23} = 0.006$ eV$^2$ as discussed before).    
As $s_{13}^2$ increases the electron depletion 
becomes higher and that restricts 
higher $s_{13}^2$ values. 

$\bullet$ For $s_{12}^2 \neq 0$, 
the four corners represent two-generation limits driven by
$\Delta_{LSND}$. This is the regime of average oscillations
and cannot explain the zenith angle dependence of the data.
For a fixed $\Delta_{23}$ the allowed region first expands and then  
shrinks in size and also shifts towards higher $s_{13}^2$ values as
$s_{12}^2$ increases just as in fig. \ref{c3}.

Matter effect is important for the last two rows and the increase 
in the allowed areas for the last two panels of $\Delta_{23}=0.0004$ 
eV$^2$ are typical signatures of matter effect. 


\begin{figure}[t]
    \centerline{\psfig{file=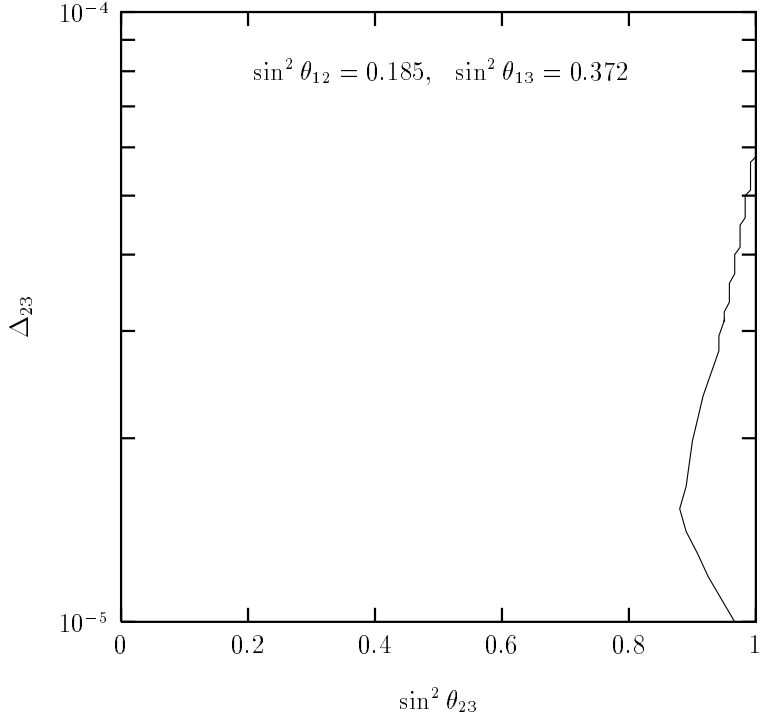,width=4.5in}}               
\vskip -1.0cm
\caption[Allowed regions in $\Delta_{23}-\sbcsq$ plane for 
$\Delta_{23}\sim 10^{-5}$ eV$^2$] 
{\label{c5}
The allowed parameter space in the $\Delta_{23}-\sbcsq$ plane
with $\Delta_{23}$ in the range $10^{-5}-10^{-4}$ eV$^2$ and with
fixed values of $\sabsq=0.185$ and $\sacsq=0.372$.}
\end{figure}
In fig. \ref{c5} we present the allowed range in the $\Delta_{23}-\sbcsq$ 
plane with $\Delta_{23}$ in the $10^{-5} - 10^{-4}$ eV$^2$ range 
and $\sabsq$, $\sacsq$ fixed at 0.185 and 0.372 respectively. We get 
allowed regions in this range of small $\Delta_{23}$ and small 
mixing due to matter effects -- a feature unique to the mass spectrum 
considered in this chapter.
 
\section{$\chi^2$ analysis of the SK + CHOOZ Data}

The CHOOZ experiment can probe upto $10^{-3}$ eV$^2$ and hence it can be
important to cross-check the atmospheric neutrino results.
In particular a two-generation analysis shows that CHOOZ data disfavours
the $\nu_\mu - \nu_e$ solution to the atmospheric neutrino problem.
The general expression for the survival probability of the electron neutrino
in presence of three flavours is
\begin{equation}
P_{\nu_e \nu_e} = 1 - 4 U_{e1}^2 ( 1 - U_{e1}^2){\sin^2}
({\pi L /\lambda_{12})} - 4 U_{e2}^2 U_{e3}^2
{\sin^2}({\pi L/\lambda_{23}})
\label{pnue3}
\end{equation}
This is the most general expression without the one mass scale dominance
approximation.
We now minimize the $\chi^2$ defined as 
\begin{equation}
\chi^2 = \chi^2_{ATM} + \chi^2_{CHOOZ}
\label{comb}
\end{equation}
where $\chi^2_{ATM}$ is calculated as before using eq. (\ref{chilisi}) 
and we define
$\chi^2{_{CHOOZ}}$ as \cite{yasuda4}
\begin{equation}
\chi^2_{CHOOZ} = \sum_{j=1,15}(\frac{x_{j} - y_{j}}{\Delta x_{j}})^2
\end{equation}
where $x_{j}$ are the experimental values, $y_{j}$ are the corresponding
theoretical predictions and the sum is over 15 energy bins of data of the
CHOOZ experiment \cite{chooz4}. 
For the CHOOZ experiment the $\sin^2(\pi L/\lambda_{12})$ term does not 
always average out to 0.5 (for SK this term always averages to 0.5)
and one has to do the energy integration properly.
For our analysis we keep the $\Delta_{12}$ fixed at 0.5 eV$^2$ and do
a four parameter fit as in SK. 
The $\chi^2_{min}$ and the best-fit values of parameters that we get are
\begin{itemize}
\item
$\chi^2_{min}/d.o.f. = 42.22/51$, 
$\Delta_{23}$ = 0.0023 eV$^2$, $s^2_{23} = 0.5$, $s_{12}^2 = 0.0022$
and $s_{13}^2 = 0.0$. 
\end{itemize}
Thus the best-fit values shift towards the two-generation
limit when we include the CHOOZ result.  
This provides a very good fit to the data 
being allowed at 80.45\% probability. 

\begin{figure}[t]
\vskip -0.8cm
    \centerline{\psfig{file=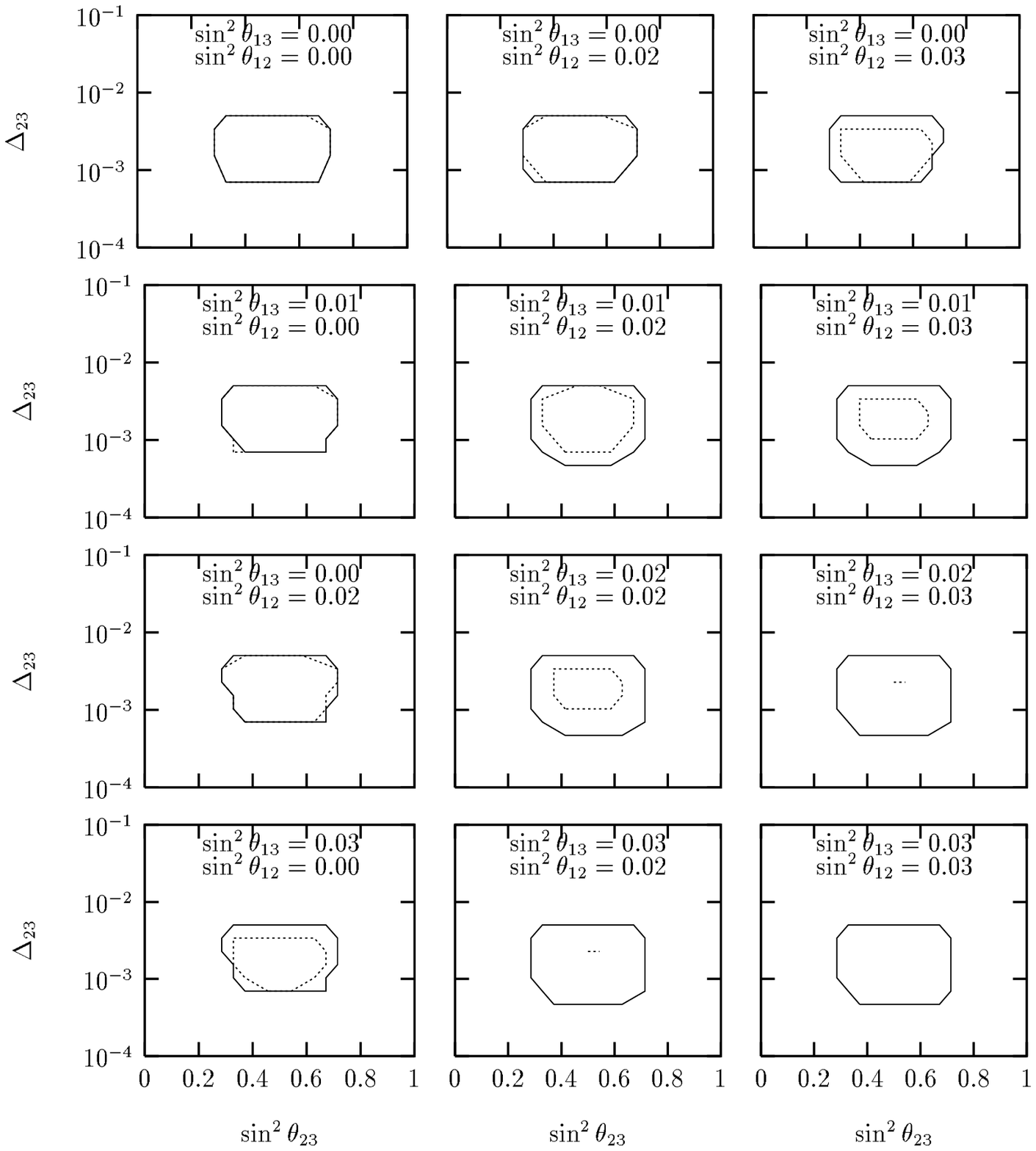,width=4.5in}}               
\vskip -1.5cm
\caption[Allowed regions in $\Delta_{23}-\sbcsq$ plane for $\sabsq$ and 
$\sacsq$ small] 
{\label{c6}
Same as fig. \ref{c1} but for smaller values of $\sabsq$ and
$\sacsq$, chosen from the range determined by the
SK+CHOOZ dashed line in fig. \ref{delchi3gen}.}
\end{figure}

The dotted lines in fig. \ref{delchi3gen} give 
the combined SK+CHOOZ $\Delta \chi^2 (= \chi^2 - \chi^2_{min})$ 
given by eq. (\ref{comb}), as a function of 
one of the parameters, keeping the other three unconstrained. 
We find that the CHOOZ data severely restricts the allowed ranges 
for the parameters $\sabsq$ and $\sacsq$ to values 
$\stackrel {<}{\sim} 0.047$, while $\Delta_{23}$ and 
$\sbcsq$ are left almost unaffected. 
Since CHOOZ is consistent with no oscillation one requires 
$P_{\nue\nue}$ close to 1. So the second and the third terms in eq. 
(\ref{pnue3}) should separately be very small. The second term 
implies $U_{e1}^2$ to be close to either 0 or 1. $U_{e1}^2$ close 
to zero implies either $\sabsq$ or $\sacsq$ close to 1 which 
is not consistent with SK. Therefore $U_{e1}^2$ is close to 1. 
Then from unitarity both $U_{e2}^2$ and $U_{e3}^2$ are close to 0 and 
so the third term goes to zero irrespective of the value of 
$\Delta_{23}$ and $\sbcsq$. Hence contrary to expectations, CHOOZ 
puts {\it almost} no restriction on the 
allowed values of $\sbcsq$ and $\Delta_{23}$, 
although $\Delta_{23} \sim 10^{-3}$ eV$^2$ -- in the regime in which 
CHOOZ is sensitive. On the other hand it puts severe constraints 
on the allowed values of $\sabsq$ and $\sacsq$ in order to 
suppress the average oscillations driven by $\Delta_{12}$.
Because of such low values of
$\sabsq$ and $\sacsq$ the matter effects for the atmospheric
neutrinos are not important and the additional allowed area with
low $\Delta_{23}$ and high $\sbcsq$ obtained in the SK analysis
due to matter effects are no longer allowed.
The 99\% C.L. regions allowed by 
a combined analysis of SK and CHOOZ data is shown by the dotted lines
in figs. \ref{c1}-\ref{c4}. 
It is seen that most of the regions allowed by the three-flavour analysis
of the SK data is ruled out when we include the CHOOZ result. 
None of the allowed regions shown in fig. \ref{c1} are allowed excepting the 
two-generation $\numu-\nutau$ oscillation 
limit because CHOOZ does not allow such high values of
either $s_{13}^2$ or $s_{12}^2$. Hence we present again 
in fig. \ref{c6} the allowed regions in the $\Delta_{23}-
\sbcsq$ plane for various fixed values of $\sabsq$ and 
$\sacsq$, determined from the dotted lines in fig. \ref{delchi3gen}. 
The solid lines in fig. \ref{c6} give the 99\% C.L. area allowed
by the SK data while the dotted lines give the corresponding allowed 
region from the combined analysis of SK+CHOOZ.  
We find that for the combined analysis 
we get allowed regions in this plane only for much smaller values
of $s_{12}^2$ and $s_{13}^2$, which ensures that the electron events are 
neither less nor more than expectations.

\section{Combined Allowed Area from Short Baseline Accelerator 
and Reactor Experiments}

\begin{figure}[t]
    \centerline{\psfig{file=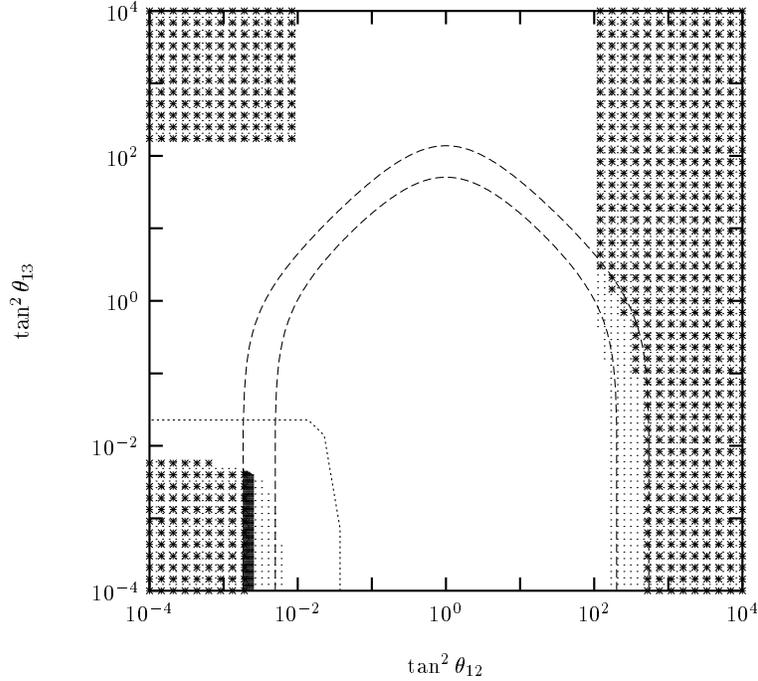,width=4.5in}}               
\vskip -1.0cm
\caption[Combined allowed 
regions in $\tan^2\theta_{12}-\tan^2\theta_{13}$ plane] 
{\label{combfig}
The area between the dashed lines is the 90\% C.L. region
allowed by LSND while the
light shaded zone gives the 90\% C.L. allowed region
from the non-observance of neutrino oscillation in the other
short baseline accelerator and reactor experiments except KARMEN2.
The corresponding area which includes KARMEN2 as well is
shown by the region shaded by asterix.
The 90\% C.L. allowed region from
SK+CHOOZ analysis is within the dotted line.
The dark shaded area corresponds to the combined allowed region.}
\end{figure}

As mentioned earlier the higher mass scale of this scenario 
can be explored in the accelerator based
neutrino oscillation search experiments. 
For the mass-pattern considered the most constraining 
accelerator experiments 
are LSND \cite{lsnd4}, CDHSW \cite{cdhs4}, 
E531 \cite{e5314} and KARMEN \cite{karmen4}.
Among these only LSND reported positive
evidence of oscillation. Other experiments are consistent with 
no-oscillation hypothesis. 
Also important in this mass range are the constraints from the reactor
experiment Bugey \cite{bugey4}. 
The relevant probabilities are \cite{ska}
\begin{itemize}
\item Bugey
\begin{equation}
P_{\overline{\nu}_e\overline{\nu}_e} = 1 - 4c_{13}^2c_{12}^2
{\sin^2}({\pi L/\lambda_{12}})  + 
4c_{13}^4c_{12}^4 {\sin^2}({\pi L/\lambda_{12}})
\label{bugey} 
\end{equation}    
\item CDHSW
\begin{equation}
P_{\overline{\nu}_{\mu}\overline{\nu}_{\mu}} = 1 -
4 c_{12}^2 s_{12}^2 {\sin^2}({\pi L/\lambda_{12}}) 
\end{equation}
\item LSND and KARMEN
\begin{equation}
P_{\overline{\nu}_{\mu}\overline{\nu}_e} = 4 c_{12}^2 s_{12}^2
c_{13}^2 {\sin^2}({\pi L/\lambda_{12}})
\label{pnumul}
\end{equation}
\item{E531}
\begin{equation}
P_{\nu_{\mu} \nu_{\tau}} = 4 c_{12}^2 s_{12}^2 s_{13}^2 \sin^2({\pi
L/\lambda_{13}}) 
\label{pe531}
\end{equation}
\end{itemize}
We note that the probabilities are functions of one of the mass scales
and two mixing angles. Thus the one mass scale dominance approximation
applies. There are many analyses in the literature of the accelerator
and reactor data including LSND under this one mass scale dominance 
assumption \cite{ska,fogli95}.  
These analyses showed that when one considers the results from
the previous (prior to LSND) 
accelerator and reactor experiments there are three
allowed regions in the $\theta_{12} - \theta_{13}$ plane \cite{ska,fogli95}
\begin{itemize}
\item low $\theta_{12}$ - low $\theta_{13}$
\item low $\theta_{12}$ - high $\theta_{13}$
\item high $\theta_{12}$ - $\theta_{13}$ unconstrained
\end{itemize}
When the LSND result was combined with these results then only the
first and the third 
zones remained allowed in the mass range $0.5 \leq \Delta_{12} \leq 2$ eV$^2$.
In these earlier analyses of the accelerator and 
reactor data \cite{ska,fogli95}
E776 \cite{e7764} was more constraining than KARMEN. But with the new 
data KARMEN2 gives stronger constraint than E776. Also the 
results from the KARMEN2 experiment now rule out most of the region 
allowed by the LSND experiment above 1 eV$^2$ \cite{karmen4}. The LSND          
collaboration has also now done a reanalysis of their entire data sample 
and report their final transition probability in \cite{lsndnew4}. 
We have repeated the analysis with the latest LSND and KARMEN results 
for one representative value of $\Delta_{12}=0.5$ eV$^2$ 
and present the allowed region in fig. \ref{combfig}.

The light-shaded area in fig. \ref{combfig} 
shows the 90\% C.L. allowed area in the 
bilogarithmic $\tan^2\theta_{12}-\tan^2\theta_{13}$ 
plane from the observance of 
no-oscillation in all the other 
above mentioned accelerator and reactor experiments
except KARMEN2. The inclusion of the KARMEN2 results as well gives the 
90\% C.L. region shown by the area shaded by asterix. 
The 90\% allowed region by the LSND experiment is within the 
dashed lines. The KARMEN2 data severely restricts the LSND allowed regions.
The dotted line shows the 90\% C.L. ($\chi^2 \leq \chi^2_{min} + 7.78$) 
region allowed by the combined 
$\chi^2$ analysis of the SK+CHOOZ data keeping $\Delta_{23}$ and $s_{23}^2$
at 0.002 eV$^2$ and 0.5 respectively. 
The combined SK  
atmospheric and the CHOOZ reactor data rule out 
the third zone (high $\theta_{12}$ with $\theta_{13}$
unconstrained ) allowed from LSND and other accelerator and reactor
experiments. 
Thus if one takes into account constraints from all experiments 
only a small region in the first zone (small $\theta_{12},\theta_{13}$) 
remains allowed. This common 
allowed region is shown as a dark-shaded area in the fig. \ref{combfig}. 
As evident from the expression of the probabilities for the 
accelerator and reactor experiments the 
combined allowed area of all the accelerator reactor 
experiments remains the 
same irrespective of the value of $\Delta_{23}$ and $\sbcsq$.  
Even though the combined area in fig. \ref{combfig} 
shows that in the 
first zone (small $\theta_{12},\theta_{13}$), SK+CHOOZ data 
allows more area in the $\theta_{12}-\theta_{13}$ plane for 
$\Delta_{23} = 0.002$ eV$^2$ and $\sbcsq =0.5$, 
from fig. \ref{c2} we see that for some other combinations of 
$\Delta_{23}$ and $\sbcsq$ one does not find any allowed zones 
from the SK+CHOOZ analysis, even at 
99\% C.L.. For those sets of values of $\Delta_{23}$ and $\sbcsq$ 
the SK+CHOOZ analysis is more restrictive than the LSND and other 
accelerator reactor data. 

\section{Implications}

From our analysis of the SK atmospheric data the explicit 
form for the $3\times3$ mixing matrix $U$ at the best-fit values of parameters
is 

\begin{equation}
U = {\pmatrix {0.95 & -0.039 & 0.31 \cr
-0.2 & 0.686 & 0.7 \cr
-0.24 & -0.727 & 0.644 \cr}}
\end{equation}

From the combined SK+CHOOZ analysis the mixing matrix at the best-fit 
values of the parameters is 

\begin{equation}
U = {\pmatrix {0.999  & 0.033 & 0.033 \cr
-0.047 & 0.706 & 0.706 \cr
-0.0 & -0.707 & 0.707 \cr}}
\end{equation}

From the combined allowed area of fig. \ref{combfig} the mixing matrix at 
$\Delta_{12}$ = 0.5 eV$^2$, $\Delta_{23} = 0.0028$ eV$^2$,
$s_{12}^2 = 0.005$, $s_{13}^2$ = 0.001 and
$s_{23}^2$ = 0.5, is 
\begin{equation}
U = {\pmatrix {0.997 & 0.028 & 0.072 \cr
-0.071 & 0.705 & 0.705 \cr
-0.032 & -0.708 & 0.705 \cr}}
\end{equation}

Thus the allowed scenario corresponds to 
the one where $\langle \nu_{1}|\nu_{e} \rangle$ is close to 1 
while the states $\nu_{2}$ and $\nu_{3}$ are combinations of
nearly maximally mixed $\nu_{\mu}$ and $\nu_{\tau}$ 
\footnote{ Thus this scenario is the same as the one termed 3a 
in Table VI in the pre-SK analysis
of \cite{flms97}. In their notation the states 2 and 3 were 1 and 2. 
It was disfavoured from solar neutrino results.}.  

\begin{figure}[t]
\vskip -0.8cm
    \centerline{\psfig{file=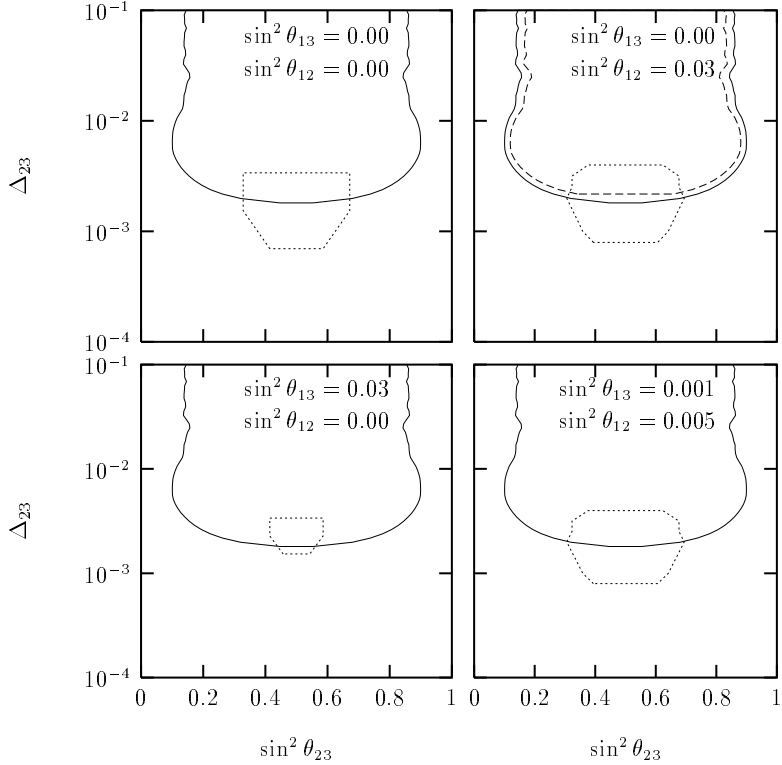,width=4.5in}}               
\vskip -1.5cm
\caption[Sensitivity of K2K in $\Delta_{23}-\sbcsq$ plane] 
{\label{k2k1}
90\% C.L. regions in the $\Delta_{23}-\sbcsq$ plane
that can be explored
by the $\numu-\numu$ (solid line) and $\numu-\nue$ (dashed line)
oscillation channels in the K2K experiment. The area inside the dotted line
shows the 90\% C.L. region allowed by SK+CHOOZ. The curves are presented
for fixed values of $\sabsq$ and $\sacsq$ with $\Delta_{12}=0.5$ eV$^2$.}
\end{figure}

Long baseline (LBL) experiments
can be useful to confirm if the atmospheric neutrino anomaly is
indeed due to neutrino oscillations, using well monitored 
accelerator neutrino beams.  
Some of the important LBL experiments are 
K2K (KEK to SK, L $\approx$ 250 km)\cite{k2k4}
\footnote{K2K has already presented some preliminary 
results \cite{k2knew4}.},
MINOS (Fermilab to Soudan, L $\approx$ 730 km ) \cite{minos} 
and the proposed CERN to Gran Sasso experiments (L $\approx$ 730 km)
\cite{cgs}. 
In this section we explore the sensitivity of the LBL experiment 
K2K in probing the parameter spaces allowed by the SK+CHOOZ and other 
accelerator and reactor experiments including LSND.
K2K will look for $\nu_\mu$ 
disappearance as well as $\nu_e$ appearance.
In fig. \ref{k2k1} we show the regions in the $\Delta_{23} - \sbcsq$ 
plane that can be probed by K2K using their 
projected sensitivity from \cite{k2k4}.
The top left panel is for the two-generation $\nu_\mu - \nu_\tau$ limit. 
The other panels are for different fixed values of $\sabsq$ and $\sacsq$ 
while $\Delta_{12}$ is fixed at 0.5 eV$^2$. 
For LBL experiments the term containing 
$\Delta_{12}$ averages to 0.5 as in the atmospheric case. 
The solid lines in the panels show the region that can be probed by 
K2K using the $\numu$ disappearance channel while the dotted lines 
give the 90\% C.L. contours allowed by SK+CHOOZ. One finds that for  
for $\Delta_{23} \geq 2 \times 10^{-3}$ eV$^2$, the whole region allowed
by SK+CHOOZ can be probed by the $\nu_\mu$ disappearance channel in 
K2K. The dashed lines show the 90\% C.L. area that K2K can probe 
by the $\nue$ appearance mode. 
As $s_{12}^2$ increases the constraint from the $P_{\numu \nue}$ channel 
becomes important as is seen in the top right panel of fig. \ref{k2k1}.   
However such high values of $s_{12}^2$, although allowed by SK+CHOOZ, is
not favoured when one combines LSND and other accelerator and reactor results. 
For lower $s_{12}^2$ values allowed by all the accelerator, reactor and
SK atmospheric neutrino experiment the projected sensitivity in the
$\numu-\nue$ channel of K2K is not enough to probe the allowed regions
in the $\Delta_{23} - \sbcsq$ plane as is shown by the absence of the 
dashed curves in the lower panels. 

\begin{figure}[t]
\vskip -0.8cm
    \centerline{\psfig{file=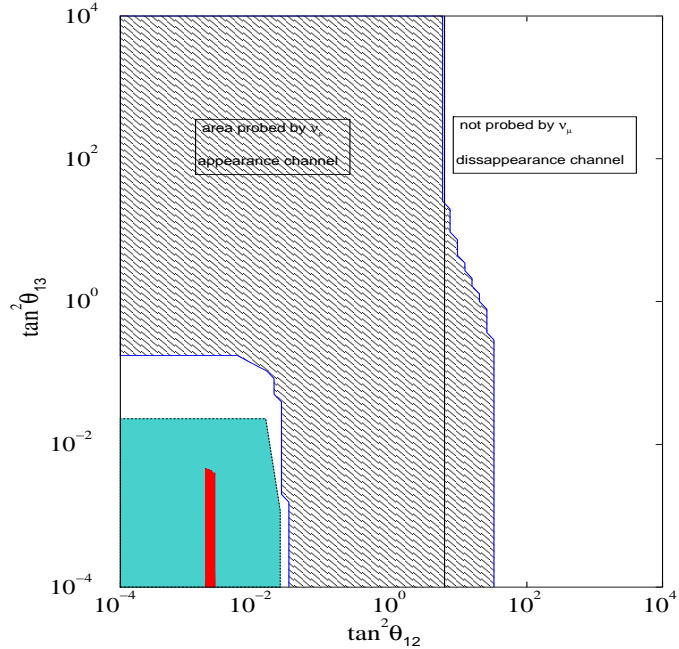,width=3.0in,height=2.5in}}               
\vskip 2.5cm
\caption[Sensitivity of K2K in $\tan^2\theta_{12}-\tan^2\theta_{13}$ plane] 
{\label{k2k2}
Sensitivity of the K2K experiment in the $\tan^2\theta_{12}-
\tan^2\theta_{13}$ plane for $\Delta_{23}=0.002$ eV$^2$,
$s_{23}^2=0.5$ and $\Delta_{12}=0.5$ eV$^2$.
The area that
can be explored by the $\numu - \numu$ (left of solid line)
and $\numu - \nue$ (hatched area)
channels in K2K at 90\% C.L. is shown.
The light-shaded area is allowed by SK+CHOOZ
and the dark-shaded region is the combined area allowed by all
accelerator and reactor data at 90\% C.L..}
\end{figure}

In fig. \ref{k2k2} 
we show the regions in the bilogarithmic $\tan^2 \theta_{12} - 
\tan^2 \theta_{13}$ plane which can be probed by K2K.
For drawing these curves we fix $\Delta_{23}=0.002$ eV$^2$,  
$s_{23}^2=0.5$ and $\Delta_{12}=0.5$ eV$^2$. 
Shown is the area that 
can be explored by the $\numu - \numu$ (left of the solid line)
and $\numu - \nue$ (hatched area) 
channels in K2K at 90\% C.L.. 
The light-shaded area is allowed by SK+CHOOZ
and the dark shaded area is allowed by the combination of all the  
accelerator, reactor and SK atmospheric neutrino data at 90\% C.L.. 
It is clear from the figure that even though the 
sensitivity of the $\nue$ appearance channel is not 
enough, K2K can still probe the 
combined allowed region in the $\theta_{12}-\theta_{13}$ plane 
from $\numu$ disappearance.

The projected sensitivities of MINOS and the CERN to ICARUS proposals 
are lower than K2K and it will be interesting to check if one can
probe the regions allowed in this picture better in these experiments. 
However since in our case the OMSD approximation is not applicable 
one has to do the energy averaging properly 
to get the corresponding contours in the three-generation parameters space, 
and one cannot merely scale the allowed regions 
from the two-generation plots. For K2K we could use the
fig. 5 of \cite{k2k4} to circumvent this problem. However since the 
analogous information for MINOS and CERN-Gran Sasso proposals is not 
available to us we cannot check this explicitly. 

An important question in this context is whether one can 
distinguish between the OMSD three generation and this mass scheme. 
In both pictures the SK atmospheric neutrino data can be explained
by the dominant $\nu_\mu-\nu_\tau$ oscillations  mixed with little amount of 
$\nu_e - \nu_\mu (\nu_\tau)$  transition.
However the mixing matrix $U$ is 
different. 
A distinction                               
can be done if one can measure the mixing angles very accurately.

What is the prospect in LBL experiments to distinguish between these
pictures?
We give below a very preliminary and qualitative  discussion on this.
If we take $s_{12}^2$ = 0.02, $s_{13}^2$ = 0.02 and $s_{23}^2$ = 0.5,  
$P_{\numu \nue}$ would be  (0.038 + 0.0004
$\langle S_{23} \rangle$).
As the second term is negligible one has average oscillations. 
This is different from the OMSD limit where $P_{\numu \nue} = 4 U_{\mu3}^2 
U_{e3}^2 S_{23}$ is energy dependent. 
If one combines the other accelerator and reactor 
experiments including LSND then the allowed values of   
of $s_{12}^2$ and $s_{13}^2$ are even less 
and choosing $s_{12}^2$ = 0.005, $s_{13}^2$ = 0.001 and $s_{23}^2$ = 
0.5 we get $P_{\nue \numu} = 0.01 - 0.004  \langle S_{23} \rangle$. 
Here also the term involving $\langle S_{23} \rangle$  
is one order of magnitude smaller and the oscillations will be averaged. 
Thus this channel has different predictions for the OMSD limit 
and beyond the OMSD limit.

\section{Discussions and Conclusions}

In this chapter we have done a detailed $\chi^2$ analysis of the SK
atmospheric
neutrino data going beyond the OMSD approximation. 
The mass spectrum chosen is such that $\Delta_{12} = \Delta_{13} \sim$
eV$^2$ to explain the LSND data and $\Delta_{23}$ is
in the range suitable for the atmospheric neutrino problem. 
We study in details the implications of the earth matter 
effects and bring out the essential differences of our mass pattern 
with the OMSD scenario and the two-generation limits.

We first examine in detail 
what are the constraints obtained from only SK data considering
its overwhelming statistics. 
The allowed regions include
\begin{itemize}
\item the two-generation $\nu_\mu - \nu_\tau$ limit (both $s_{12}^2$ and 
$s_{13}^2$ zero) 
\item regions where either $s_{12}^2$ or $s_{13}^2$ is zero;
in this limit the probabilities are functions in general of two
mixing angles and two mass scales.
\item  
the three-generation 
regions with all three mixing angles non-zero and the probabilities 
governed by both mass scales.

The last two cases correspond to dominant $\numu-\nutau$ oscillation 
with small admixture of $\numu-\nue$ and $\nue-\nutau$ oscillation. 
\item
regions with very low $\Delta_{23}$ ($<10^{-4}$ eV$^2$) 
and $\sbcsq$ close to 1, for which 
the earth matter effects enhance the oscillations of the upward 
neutrinos and cause an up-down flux asymmetry. This region is 
peculiar to the mass spectrum considered by us and is 
absent in the two-generation and the OMSD pictures. 
\end{itemize} 
We present the zenith angle distributions of the events in these cases.
With the inclusion of the CHOOZ result the allowed ranges of the mixing angles
$s_{12}^2$ and $s_{13}^2$ is constrained more ($\stackrel{<}{\sim} 0.047$),  
however the allowed ranges of
$\Delta_{23}$ and $s_{23}^2$ do not change much (see 
fig. \ref{delchi3gen}) except 
that the low $\Delta_{23}$ region allowed by SK due to matter effects 
is now disallowed. 
The inclusion of the constraints from LSND and other accelerator and
reactor experiments may restrict the allowed area in the 
$\theta_{12}-\theta_{13}$ plane for certain values of $\Delta_{23}$ and 
$\sbcsq$, but for some other combinations of $\Delta_{23}$ and 
$\sbcsq$, SK+CHOOZ turns out to be more constraining. 
We have included the latest results from LSND and KARMEN2 in our 
analysis.

In order to explain the solar neutrino problem in this picture one has to add 
an extra light sterile neutrino. 
With the new LSND results the allowed 4 neutrino scenarios are 
\begin{itemize}
\item
the (2+2) picture where two degenerate mass states
are separated by the LSND gap \cite{sg4,grimus,garcia,barger}.
\item
the (3+1) scheme with three neutrino states closely degenerate in mass and
the fourth one separated from these by the LSND gap 
\cite{barger,giunti4,vallenew4}.
\end{itemize} 
Our scenario can be easily extended to the (2+2) scheme which is still 
allowed after the inclusion of the SNO results \cite{ggnew4}. The 
(3+1) scheme is however shown to be ruled out from the atmospheric data 
in \cite{vallenew4}. 

 
To conclude, one can get allowed regions from the SK atmospheric 
neutrino data 
where both the mass scales and all the three mixing angles are 
relevant. The beyond one mass scale dominance spectrum considered 
in this chapter allows new regions in the low mass -- low mixing 
regime due to the earth matter effects. 
With the inclusion of the CHOOZ, LSND and other accelerator 
reactor results, the allowed regions are constrained severely. 
It is, in principle, possible to 
get some signatures in the LBL experiments to 
distinguish this picture from the OMSD limit.

%% file: chapter6.tex
\chapter{Decay of Atmospheric Neutrinos}


In the previous chapters we have seen that neutrino flavor oscillations
in vacuum, both in two as well as three flavors,   
gives an excellent fit to the SK atmospheric neutrino data. 
However there are other possibilities \cite{vep5,vli5,tor5,fcnc5,
decoh5,led5,val5} that 
have the potential to explain the data, though may be not as well as 
$\numu-\nutau$ oscllations do.  
Nevertheless one has to {\it rule out} all these possibilities before 
coming to any definite conclusion about the fate of the
atmospheric neutrinos. 
One among these
is neutrino decay \cite{lipari5,pak5}.

One aspect peculiar to the oscillation hypothesis is the periodicity 
of the resultant neutrino beam at the detector. But so far one 
has no direct experimental evidence for this periodicity. The other 
way to remove any ambiguity is to directly observe the number of 
$\nutau$ predicted by the $\numu-\nutau$ oscillation scenario. The latest  
1289 day SK data for the upward-going sample \cite{sk12895} is consistent 
with $\nutau$ appearance at the $2 \sigma$ level. But one needs more 
statistics before coming to any final conclusion. 

If one assumes the existence of a characteristic wavelength $\lambda$ 
for the vanishing $\numu$ such that $\lambda$ has a power law 
dependence on energy ($\lambda \propto E^n$) then the $\numu-\nutau$ 
conversion probability can be assumed to be parameterized as 
\cite{lisilbye5,sobel5} 
\be
P_{\numu\nutau} = \alpha \sin^2\left(\beta LE^n\right)
\ee
where $\alpha$, $\beta$ and $n$ are free parameters. If then one   
performs a most general \chisq analysis, keeping the energy exponent 
$n$ a free parameter, the best-fit comes out for $n\approx-1$ 
\cite{lisilbye5,sobel5}. In ref. \cite{lisilbye5} Fogli {\it et al.} further 
corroborate this feature by showing that 
only an $L/E$ distribution can fit the muon zenith angle data 
\cite{lisifig5}. 
Hence it can be conclusively said that the 
data strongly prefers theoretical models which predict a $L/E$ 
dependence for the $\numu$ survival probability. Oscillations 
in vacuum are hugely favored since they predict this $L/E$ behavior. 
But so does decay and it will be interesting to check if neutrino decay can 
really offer a decent fit to the observed atmospheric neutrino deficit. 

In ref. \cite{lipari5} it was shown that 
neutrino decay gives a poor fit to the
data. However they considered neutrinos with zero mixing. 
Barger {\it et al.} considered the situation 
of neutrino decay in the general case of neutrinos with non-zero mixing angle
\cite{pak5}. 
They showed that the neutrino decay fits the $L/E$ distribution of the SK
data well. The $\Delta m^2$ taken by them was $>$  0.1 $eV^2$ so that the 
$\Delta m^2$ dependent term in the expression for the neutrino survival 
probability averages out to zero. 
As pointed out in \cite{pak5} such a constraint on \dm is valid when the 
unstable  state decays into some other state with which it mixes. If however
the unstable state decays into a sterile state with which it does not mix
then there is no reason to assume \dm $>0.1$ eV$^2$. 

In this chapter we present our results of the 
neutrino decay solutions to the atmospheric neutrino
problem by  
doing $\chi^2$-fit to the 848 day of 
sub-GeV and multi-GeV Super-Kamiokande data \cite{kate5}.
We also present the results of $\chi^2$-fit to the 
535 day SK data and compare it with the results for the 
848 day data.  For the \chisq we use the definition given by 
eq. (\ref{chiry}) in chapter 3, where we use the double ratio
$R$ and the up-down asymmetry parameter $Y_\alpha$ ($\alpha = e,\mu$) 
\cite{fl5,yasuda5}. 
We first present the results of this \chisq fit for the two-generation 
$\numu-\nutau$ oscillations.  
For the neutrino decay analysis we take the most general case of
neutrinos with non-zero mixing and consider two pictures 
\begin{itemize}
\item \dm $>$ 0.1 $eV^2$ (scenario (a))
\item \dm unconstrained  (scenario (b))
\end{itemize}
We also explicitly 
demonstrate the behavior of the up-down asymmetry parameters 
in both scenarios.
 
Our analysis shows that scenario (a) 
is ruled out at 100\%(99.99\%) C.L. by the 848(535) day of SK data. 
However if we remove the constraint on $\Delta m^2$ 
and consider the possibility of decay into a sterile 
state
then one can get an acceptable fit for $\Delta m^2$ $\sim 0.001 eV^2$ and
\st large.

In section 6.1 we present the most general expression for the 
$\numu$ survival and transition probabilities with unstable 
neutrinos. We first consider neutrinos to be stable and 
display our results 
for two-generation $\nu_\mu-\nu_\tau$ oscillation analysis in section 6.2.1. 
In section 6.2.2 we present our results for the neutrino decay
solution constraining \dm to be $> 0.1 eV^2$. 
In section 6.2.3 we do a three parameter 
$\chi^2$ analysis by removing the constraint on
$\Delta m^2$.
In section 6.3 we conclude by performing a 
comparative study of the three cases and
indicate how one can distinguish experimentally between the scenario
(b) and the $\nu_\mu - \nu_\tau$ oscillation case though both give almost
identical zenith-angle distribution.

\section{Neutrino Survival Probabilities}

Neutrinos are assumed to be stable in the standard model of particle
physics. But if one allows for neutrino decay then the
the analysis of neutrino oscillation experiments become quite
different. We will here assume that the only unstable component 
is $\nu_2$ which decays into 
some other lighter state $\nu_j$ which may be an 
active or sterile species. Since radiative decays of neutrinos are severely
constrained \cite{fukugita5} we consider the two possible non-radiative 
decay modes discussed in the literature.
\begin{itemize}
\item Model 1: If neutrinos are Dirac particles
one has the decay channel $\nu_2 \rightarrow \bar{\nu}_{jR} + \phi$,
where ${\bar{\nu}}_{jR}$ is a right handed singlet and $\phi$ is an
iso-singlet scaler.
Thus all the final state particles
for this model are sterile and there is no distinct signature
of this decay apart from in disappearance experiments.
This model is discussed in \cite{app5}. In this model a light scalar boson
$\phi$ with lepton number -2 and a singlet right handed
neutrino is added to the
standard model. The neutrino coupling to this scalar boson is given by
$g_{2j} \nu_{R_j}^T C^{-1} \nu_{R_2}$,
$C$ being the charge conjugation operator.

\item Model 2: If neutrinos are Majorana particles,
the decay mode is $\nu_2 \rightarrow \bar{\nu}_j + J$,
where J is a Majoron, produced as a result of spontaneous
breaking of a global $U(1)_{L_e - L_\mu}$ symmetry \cite{anjan5}.
In this model the neutrino masses are generated by extending
the higgs sector of the standard model.
\end{itemize}
In both the decay
scenarios the rest frame lifetime of $\nu_{2}$ is given by \cite{app5}
\begin{equation}
\tau_{0} = \frac{16 \pi}{g^2} \frac{m_{2} (1 + m_{j}/m_{2})^{-2}}
{\Delta_d}
\label{tau0}
\end{equation}
where $g$ is the coupling constant, $m_i$ is the $\nu_i$ mass and
$\Delta_d  = m_2^2 - m_j^2$, the mass squared difference between the 
states that are involved in decay.
Assuming
$m_{2} >> m_{j}$  the equation (\ref{tau0}) can be written as
\be
g^2 \Delta_d \sim 16 \pi \alpha
\label{galpha}
\ee
where $\alpha$ is the decay constant related to $\tau_0$ as
$\alpha = m_{2}/\tau_{0}$.
We assume a scenario where 
\be
\nue \approx \nu_1
\ee
\be
\nu_\mu \approx \nu_2 \cos\theta + \nu_3 \sin\theta
\label{nm}
\ee
From eq. (\ref{nm}) the survival and transition
probabilities of the $\numu$ of energy E,
with an unstable component $\nu_2$,
after traveling a distance $L$ is given by,
\be
P_{\numu\numu} &=& \sin^4\theta + \cos^4\theta \exp(-4 \pi
L/\lambda_d)\nonumber\\
&& {}+ 2\sin^2\theta \cos^2\theta \exp(-2 \pi L/\lambda_d) \cos(2 \pi
L/\lambda_{osc})\,,
\label{pmumudo}
\ee
\be
P_{\numu\nutau} = \frac{1}{4} \sin^2 2\theta\{1+\exp(-4\pi L/\lambda_d)
-2 \exp(-2\pi L/\lambda_d) \cos(2\pi L/\lambda_{osc})\}
\label{ntb}
\ee
where $\lambda_d$ is the decay length
(analogous to the oscillation wavelength $\lambda_{osc}$ 
given by eq. (\ref{wv})) and is defined
as,
\be
\lambda_d = 2.47 {\rm km} \frac{E}{GeV} \frac{{\rm eV}^2}{\alpha}
\label{ld}
\ee

We see that the neutrino survival probability (\ref{pmumudo}) depends on
the decay constant $\alpha$ and the mass squared difference \dm between the 
states that mix, apart from the mixing angle $\theta$. Hence one may 
consider either $\alpha=0$, which would give 
pure $\numu-\nutau$ oscillations of 
stable neutrinos or $\alpha$ non-zero, which would corresponds to the 
case of decay along with oscillations. The latter may again be 
subdivided into two cases depending on the value of \dm:  
(a) $\dm  >0.1$ eV$^2$ and (b) \dm unconstrained. We present below the 
results for the \chisq fits to all the three above mentioned cases. 
The data used for $R$ and $Y$ is shown in Table \ref{skrydata} in 
chapter 3. 

\section{Results of the $\chi^2$ analysis}
\subsection{Two-Generation $\numu-\nutau$ Oscillations}

For the two flavor $\numu-\nutau$ oscillations with the \chisq 
defined as in eq. (\ref{chiry}) of chapter 3, 
the $\chi^2_{min}$ that we get for the 848 day data 
is 1.21 with the best-fit values as
\dm = 0.003 $eV^2$ and \st = 1.0. With six data points and two parameters 
this provides a good fit to the data being allowed at 87.64\%.
If we use the 535 day data then
the $\chi^2_{min}$ that we get is 4.25 with the best-fit values as
\dm = 0.005 $eV^2$ and \st = 1.0, the g.o.f being 37.32\%.
Thus the fit becomes much better with the 848 day data with
no significant change in the best-fit values.
Though we have used a different procedure of data fitting, our results
agree well with that obtained by the SK
collaboration\footnote{The best-fit values that the
SK collaboration had got for the
848 day data are \cite{kate5}
\dm = 0.003 eV$^2$, \st = 0.995 and $\chi^2$ = 55.4 for 67 d.o.f.
This corresponds to a g.o.f of 84.33\%.}.

\subsection{Neutrino Decay with $\Delta m^2 > 0.1$ eV$^2$}

If the unstable component in the $\nu_\mu$ state decays to some other state
with which it mixes then bounds from $K$ decays imply 
$\Delta m^2 > 0.1$ eV$^2$ \cite{pak25}. 
In this case the $\cos(2 \pi L/\lambda_{osc})$ term
averages to zero and the probability becomes
\be
P_{\numu\numu} = \sin^4\theta + \cos^4\theta \exp(-4 \pi L/\lambda_d) \,.
\label{pmumu}
\ee
\begin{figure}[t]
\vskip -3.5cm
      \centerline{\psfig{file=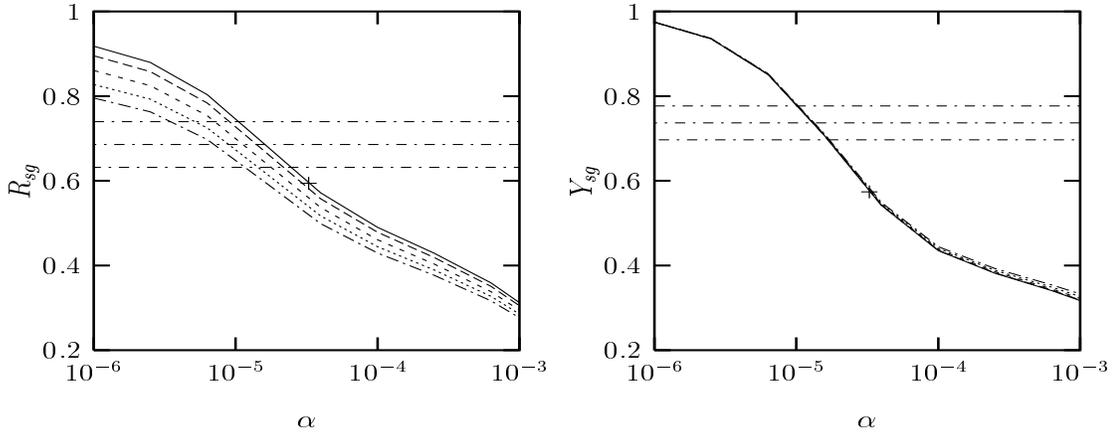,width=8.0in,height=9.0in}}
\vskip -14cm
    \caption[$R_{sg}$ and $Y_{sg}$ vs. $\alpha$ for \dm $> 0.1$ eV$^2$]
  {\label{r1}
The variation of R and Y with $\alpha$ for the sub-GeV
neutrinos (denoted by the subscript sg) assuming neutrino decay
with \dm $> 0.1 eV^2$. The curves
are drawn at fixed values of $\sin^2\theta$=0.03 (solid line),
$\sin^2\theta$=0.04 (long dashed line), $\sin^2\theta$=0.06
(short dashed line), $\sin^2\theta$=0.08 (dotted line)
and $\sin^2\theta$=0.1 (long dashed-dotted line).
The short dashed-dotted lines give the SK 848 day results within a
$\pm 1\sigma$ band. Also shown by a cross are the R and
Y at the best-fit point.}
\end{figure}
In figs. \ref{r1} and \ref{y1} 
we show the variation of $R$ and $Y$ with $\alpha$ for
various values of $\sin^2 \theta$ for the sub-GeV and multi-GeV cases. 
For higher values of $\alpha$, the decay length $\lambda_d$ 
given  by eq. (\ref{ld}) is low and the exponential term in the survival
probability is less implying that more number of neutrinos decay 
and hence  $R$ is
low.  As $\alpha$ decreases the decay length increases and the number 
of decaying neutrinos decreases, increasing $R$. 
For very low values of $\alpha$ 
the exponential term goes to 1, the neutrinos do not get the time to 
decay so that the 
probability becomes $1-\frac{1}{2}\sin^2 2\theta$ and remains constant 
thereafter for all lower values of $\alpha$. 
This is to be contrasted with the $\nu_\mu-\nu_\tau$ oscillation case 
where in the no oscillation limit the $\sin^2 (\pi L/\lambda_{osc})$ term
$\rightarrow$ 0 and the survival probability $\rightarrow$ 1.  
For multi-GeV neutrinos since the energy is higher the
$\lambda_d$ is higher and the no decay limit is reached for a 
larger value of $\alpha$ as compared to
the sub-GeV case. This explains why the multi-GeV curves become flatter 
at a higher $\alpha$.
The behavior of the up-down asymmetry parameter is also completely
different from the only oscillation case \cite{yasudafig5}. 
In particular the plateau obtained for a range of
\dm which was considered as a characteristic prediction for
up-down asymmetries is missing here.
For the decay case even for $\alpha$ as high as 0.001 $eV^2$,
the decay length $\lambda_d = 2500~(E$/GeV) km so that the 
exponential term is 1,  
there is almost no decay for the downward neutrinos 
and the survival probability is $P = 1 - \frac{1}{2}\sin^2
2\theta$ while the upward going neutrinos have some decay and so $Y$ is 
less than 1. As $\alpha$ decreases, the $\lambda_d$ increases, and
the fraction of upward going neutrinos decaying decreases and this increases 
$Y$. For very small values of $\alpha$ even the upward neutrinos do not
decay and $Y \rightarrow$ 1 being independent of $\theta$. 
\begin{figure}[t]
\vskip -3.5cm
      \centerline{\psfig{file=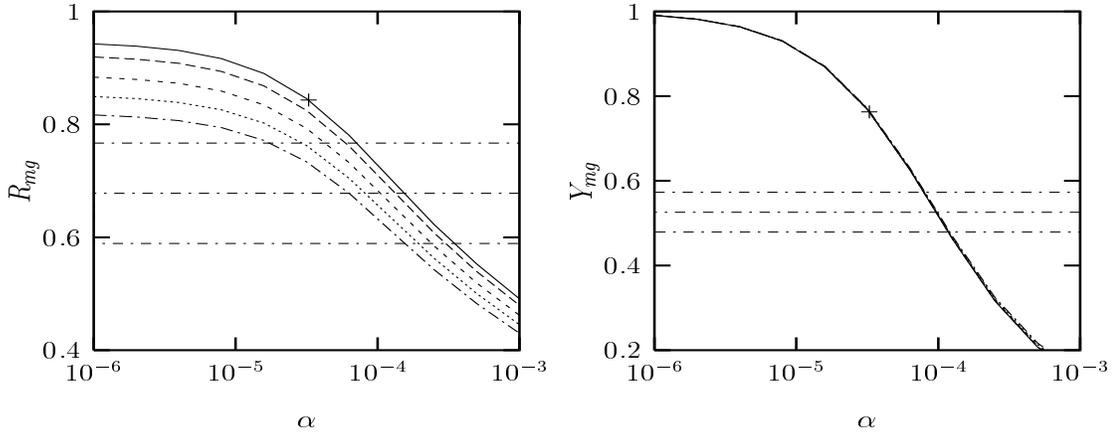,width=8.0in,height=9.0in}} 
\vskip -14cm
    \caption[$R_{mg}$ and $Y_{mg}$ vs. $\alpha$ for \dm $> 0.1$ eV$^2$]
  {\label{y1}
Same as in fig. \ref{r1} but for multi-GeV neutrinos.}
\end{figure}

We also perform a $\chi^2$ analysis of the data calculating the 
``th" quantities in (\ref{chiry}) for this scenario. 
For the 848 day data the 
best-fit values that we get are $\alpha = 0.33 \times 10^{-4}$ in
$eV^2$ 
and $\sin^2\theta = 0.03$ with a $\chi^2_{\rm{min}}$ of 49.16.
For 4 degrees of freedom this solution is ruled out at 100\%.  
The best-fit values for the 535 day of data 
that we get are $\alpha = 0.28 \times 10^{-4}$ in
$eV^2$ 
and $\sin^2\theta = 0.08$ with a $\chi^2_{\rm{min}}$ of 31.71.
For 4 degrees of freedom this solution is ruled out at 99.99\% 
\cite{lisidecay5}.
Thus the fit becomes worse with the 848 day data as compared to the
535 day data. 
We have marked the $R$ and $Y$ corresponding to the 
best-fit value of the parameters $\alpha$ and $\sin^2\theta$ in figs. 
\ref{r1} and \ref{y1}. 
It can be seen that the best-fit value of R for the sub-GeV 
neutrinos is just below and that for the multi-GeV neutrinos is just above 
the $\pm 1\sigma$ allowed band of the SK 848 day of data. The up-down 
asymmetry parameter $Y$ is quite low for the sub-GeV neutrinos and 
extremely high for the multi-GeV neutrinos as compared to that allowed 
by the data. The fig. \ref{r1} shows that for the sub-GeV neutrinos the data 
demands a lower value of $\alpha$ while from fig. \ref{y1} 
we see that the multi-GeV 
neutrinos need a much higher $\alpha$ to explain the SK data.
It is 
not possible to get an $\alpha$ that can satisfy both the sub-GeV and the
multi-GeV SK data, particularly its zenith angle distribution.   
In this 
scenario, decay for the sub-GeV upward neutrinos is more than that for the 
multi-GeV upward neutrinos (downward neutrinos do not decay much) and as a 
result $Y$ for sub-GeV is lower than the $Y$ for multi-GeV, a fact not 
supported by the data. 
Since the 848 day data needs even lesser depletion of the sub-GeV flux 
as compared to the multi-GeV flux, 
the fit gets worse.  


\subsection{Neutrino Decay with $\Delta m^2$ unconstrained}

In this section we present the results of our $\chi^2$-analysis removing
the constraint on $\Delta m^2$. 
This case corresponds to the unstable neutrino state decaying to some sterile
state with which it does not mix \cite{pak5}.
The probability will be still given by eq. (\ref{pmumudo}).

\begin{figure}[t]
\vskip -3.5cm
      \centerline{\psfig{file=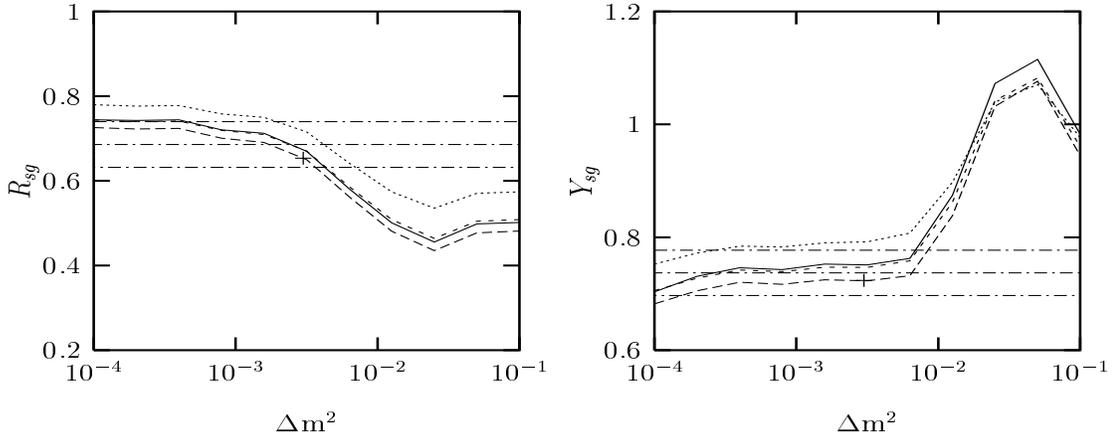,width=8.0in,height=9.0in}} 
\vskip -14cm
    \caption[$R_{sg}$ and $Y_{sg}$ vs. \dm for fixed $\alpha$]
  {\label{r2}
The variation of R and Y with \dm for the sub-GeV
neutrinos (denoted by the subscript sg)
assuming neutrino decay with \dm unconstrained. In these curves
the $\alpha$ is fixed at its best-fit value of $0.3\times 10^{-5} eV^2$.
The curves are drawn at fixed values of $\sin^2\theta$=0.7 (dotted line),
$\sin^2\theta$=0.6 (short dashed line) and $\sin^2\theta$=0.5 (long
dashed line). The solid lines give the curves for the best-fit value
($\sin^2\theta = 0.5$)
of the $\numu-\nutau$  oscillation case. The dotted-dashed lines give the
SK 848 day results within a $\pm 1\sigma$ band. Also shown are
the R and Y at the best-fit point.}
\end{figure}

\begin{figure}[t]
\vskip -3.5cm
      \centerline{\psfig{file=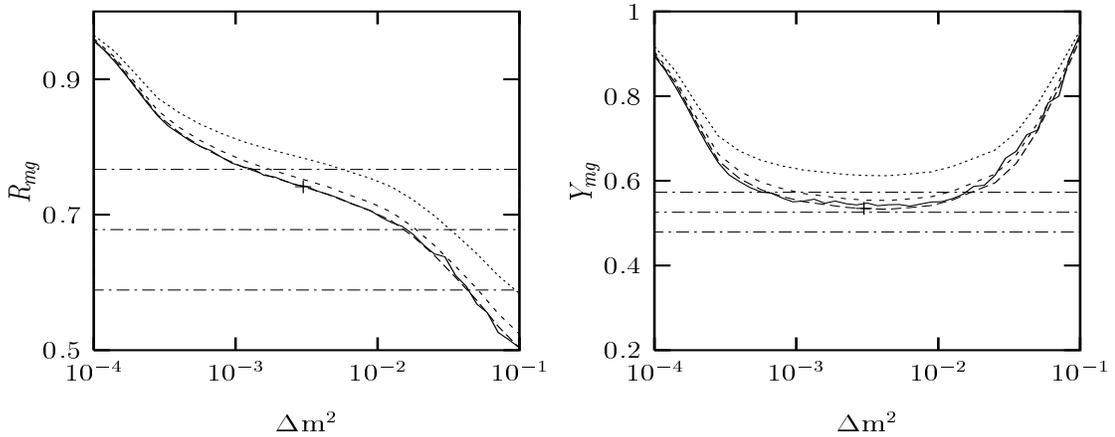,width=8.0in,height=9.0in}} 
\vskip -14cm
    \caption[$R_{mg}$ and $Y_{mg}$ vs. \dm at fixed $\alpha$]
  {\label{y2}
Same as in fig. \ref{r2} but for multi-GeV neutrinos.}
\end{figure}

In fig. \ref{r2} and \ref{y2} 
we plot the R vs. \dm and Y vs. \dm for the sub-GeV and
multi-GeV data for $\alpha$ = 0.3 $\times 10^{-5} eV^2$ 
(which is the best-fit 
value we get for the 848 day data) and  
compare with the curve obtained for the best-fit 
value of $\sin^2 \theta$ (=0.5) 
for the only oscillation case (solid line). 
For the best-fit value of $\alpha$ that we get, the downward neutrinos do
not have time to decay while the upward neutrinos undergo very little decay. 
Thus the curves are very similar in nature to the only oscillation
curves.
In the sub-GeV case (fig. \ref{r2}), for 
high values of \dm around 0.1 $eV^2$ both upward and downward
neutrinos undergo \dm independent average oscillations and 
R stays more or less constant with \dm. 
For the upward going neutrinos in addition to average oscillation there
is little amount of decay as well and hence Y $\sim$ $N_{up}/N_{down}$
is $\stackrel{<}{\sim}$ 1. 
As \dm decreases to about 0.05 $eV^2$  the oscillation wavelength increases 
--  for upward neutrinos it is still average oscillation but for the downward 
neutrinos, 
the $\cos (2\pi L/\lambda_{osc})$ term becomes negative which corresponds to 
maximum oscillation effect
and the survival probability 
of these neutrinos decreases, and hence R decreases;
while the upward neutrinos continue to decay and oscillate at the same rate   
and $Y$ becomes greater than 1. 
As \dm decreases further, the downward neutrino oscillation wavelength becomes
greater than the distance traversed and they are converted 
less and less and thus R increases and Y decreases.
Below \dm = 0.001 $eV^2$ the downward neutrinos stop oscillating completely 
while for the upward neutrinos the $\cos (2\pi L/ \lambda_{osc})$ 
term goes to 1, and R and Y no longer 
vary much with \dm.

For the multi-GeV case (fig. \ref{y2}) the oscillation wavelength is more than
the sub-GeV case and for \dm around 0.1 $eV^2$ 
the $\cos (2 \pi L/\lambda_{osc})$ 
term stays close to 1 for the downward neutrinos; while 
the upward neutrinos undergo average oscillations and slight decay and
Y is less than 1. As \dm decreases the
downward neutrinos oscillate even less and 
the upward neutrinos also start departing from average
oscillations and hence $R$ increases and $Y$ decreases. 
Around 0.01 $eV^2$ the downward neutrinos stop oscillation while for 
upward neutrinos the oscillation effect is maximum ($\lambda \sim L/2$) and 
the $cos (2\pi L/ \lambda_{osc})$ term is $\sim$ -1
and $Y$ stays constant with \dm.  
As \dm decreases further the upward neutrino oscillation
wavelength increases and they oscillate less in number making 
both R and Y approach 1 for \dm around 0.0001 $eV^2$. 
For multi-GeV neutrinos the decay term contributes even
less as compared to the sub-GeV case.  
 
\begin{table}[b]
\begin{center}
\begin{tabular}{||c|c|c||} \hline\hline
{\rule[-3mm]{0mm}{8mm}
{Quantity}} & {$\alpha = 0.3 \times 10^{-5}$ eV$^2$} & {$\alpha = 0.0$
eV$^2$} \\ \hline\hline 
{$R^{sg}$} & {0.085} & {0.021} \\ 
{$Y^{sg}_{\mu}$} & {0.011} & {0.033} \\ 
{$R^{mg}$} & {0.48} & {0.56} \\ 
{$Y^{mg}_\mu$} & {0.014} & {0.073} \\ 
{$Y^{sg}_e$} & {0.344} & {0.344} \\ 
{$Y^{mg}_e$} & {0.176} & {0.176} \\ \hline\hline
\end{tabular}
\end{center}
\caption[Individual contributions to $\chi^2_{min}$]
{\label{tabchicont}
The various contributions to the ${\chi^2}_{\rm{min}}$ at the
best-fit value of $\alpha$ and at $\alpha$=0.0}
\end{table}

We perform a $\chi^2$ minimization in the three parameters 
\dm, \st and $\alpha$.
The best-fit values that we get for the 848 day data are 
$\Delta m^2 = 0.003 eV^2$, \st = 1.0 and 
$\alpha = 0.3 \times 10^{-5} eV^2$. 
The $\chi^2$ minimum that we get is 1.11 which is an acceptable fit being 
allowed at 77.46\%.
For the 535 day data 
the best-fit values that we get are $\Delta m^2 = 0.002 eV^2$, \st = 0.87 and 
$\alpha = 0.0023 eV^2$ with a $\chi^2_{\rm {min}}$ of 4.14 which is 
allowed at 24.67\%.
Thus compared to the 535 day data, the fit improves immensely and 
the best-fit shifts towards the 
oscillation limit, the best-fit value of the decay constant $\alpha$ 
being much lower now. 
It is to be noted however,  
that the best-fit in this model does not come out to be
$\alpha = 0.0$, {\it viz} the only oscillation limit. 
In Table \ref{tabchicont} 
we give the contributions to $\chi^2$ from the $R$'s and $Y$'s
at the best-fit value of $\alpha$ and for the $\alpha$ = 0.0 case.  

\begin{figure}[t]
      \centerline{\psfig{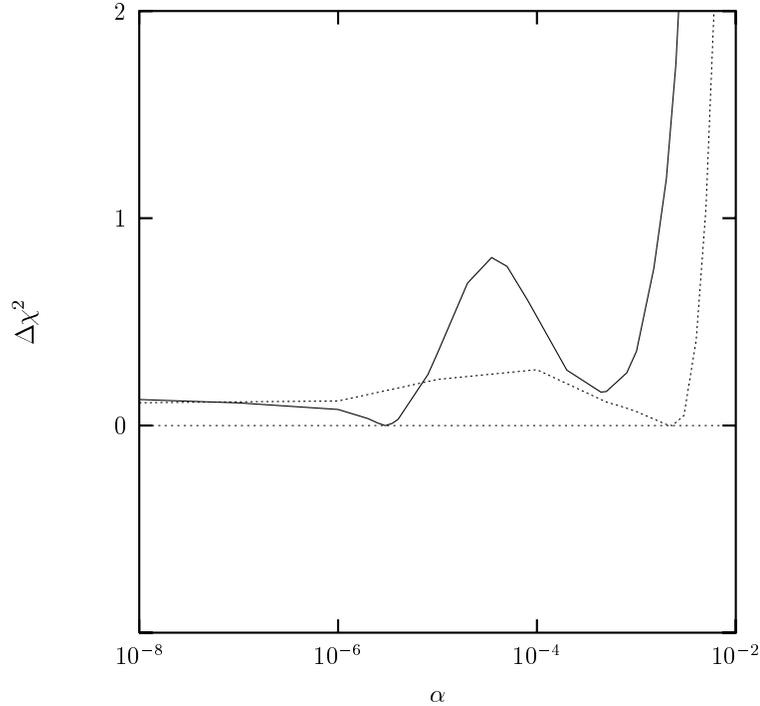}}
\vskip -1cm
    \caption[$\Delta \chi^2$ vs. $\alpha$ for osc+decay]
  {\label{deltachi}
The $\Delta \chi^2 = \chi^2 - \chi^2_{\min}$ vs. $\alpha$
with \dm and \st unconstrained for the 848 day (solid line) and
535 day data (dotted line).}
\end{figure}

Thus from the contributions to $\chi^2$ 
we see that for the best- fit case 
there is improvement for the multi-GeV $R$ and $Y$ as 
compared to the $\alpha = 0.0$ case. The $\chi^2$ for sub-GeV $Y$ also 
improves. 
In fig. \ref{deltachi} 
we plot $\Delta \chi^2 = \chi^2 - \chi^2_{\min}$ vs. $\alpha$ with
\dm and \st unconstrained.  
There are two distinct minima in this curve -- one for lower 
values and another at higher values of $\alpha$. 
The best-fit \dm in both cases is $\sim$ 0.001 $eV^2$.
In this model there are two competing processes -- oscillation and decay.
For lower values of $\alpha$ the decay length is greater than the
the oscillation wavelength and oscillation dominates. 
The decay term exp$(-\alpha L/E)$ is close to 1 and does not vary much with
the zenith distance $L$. As $\alpha$ increases the exponential term 
starts varying very sharply with $L$ and the variation is much more sharp
for the sub-GeV as compared to multi-GeV. 
This behavior is inconsistent with the data and that is why one gets a peak
in $\Delta \chi^2$ for higher $\alpha$. As $\alpha$ increases further
the exp$(-\alpha L/E)$ term goes to zero
for the upward neutrinos and there is complete decay of these
neutrinos while the downward neutrinos do not decay, the exponential
term still being 1. 
Whenever the exponential term is 0 or 1 for the upward neutrinos, 
the wrong energy dependence of this term
does not spoil the fit and  
these scenarios can give good fit to the data. 
Even though fig. \ref{deltachi} 
shows that the data allows a wide range of $\alpha$, 
we get the two distinct minima in the $\Delta \chi^2$ vs. $\alpha$ curve 
for high and low $\alpha$ values, for both the 535 day (dotted line) 
and 848 day (solid line) data. 
But while the 848 day data prefers the lower $\alpha$ limit, the 535 day 
data gives a better fit for the high $\alpha$ limit. The reason behind 
this is that for the 848 day data the R is much higher than for the 
535 day data. Hence the 848 day data prefers lower $\alpha$ and hence 
lower suppression. 

\begin{figure}[t]
      \centerline{\psfig{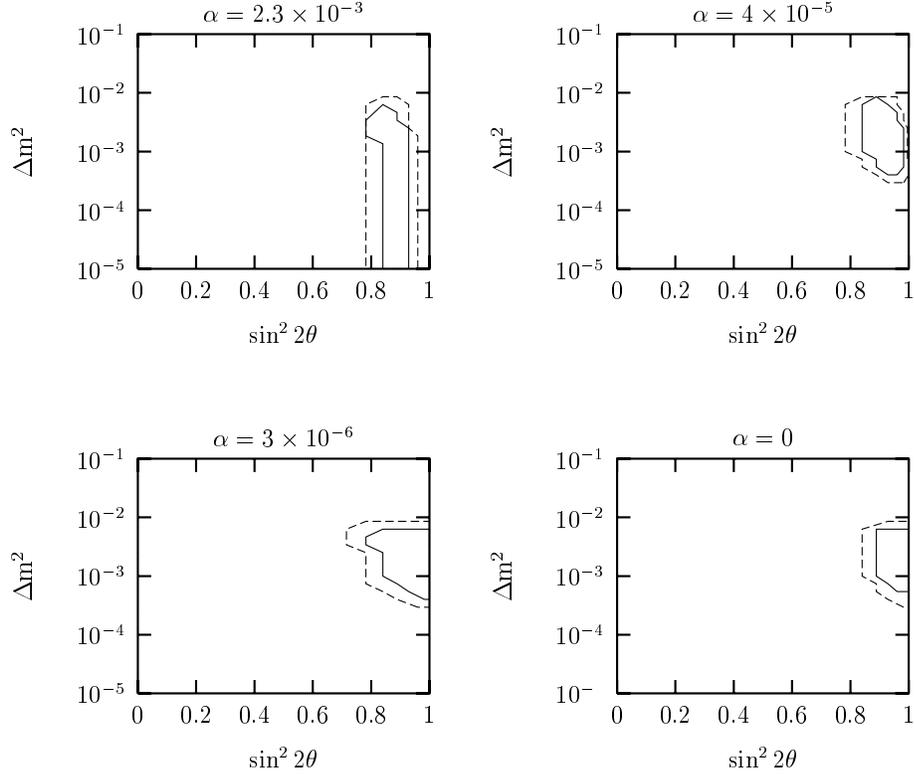}}
    \caption[Contour in $\dm-\st$ plane]
  {\label{cont1}
The allowed parameter region for the 848 day data
in the \dm-\st plane for 4
different values of $\alpha$ shown at the top of each panel. The
solid and the dashed lines correspond to the area allowed at 90\% C.L.
and 99\% C.L. respectively.}
\end{figure}

\begin{figure}[t]
      \centerline{\psfig{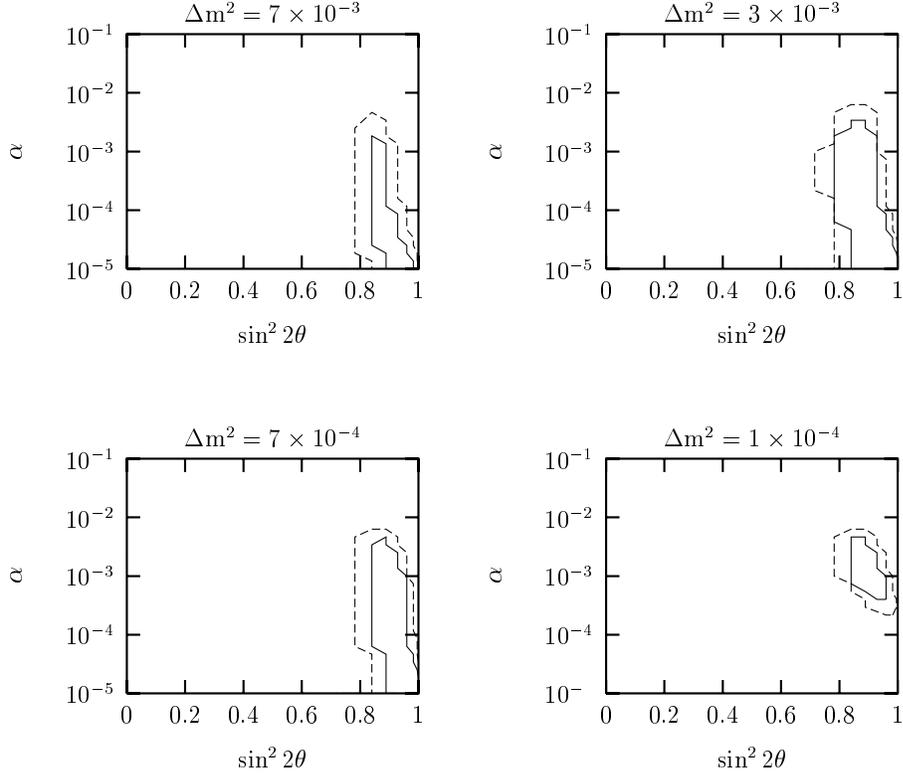}}
    \caption[Contour in $\alpha-\st$ plane]
  {\label{cont2}
The allowed parameter region for the 848 day data
in the $\alpha$-\st plane
for 4 different values of \dm shown at the top of each panel.
The solid and the dashed lines correspond to the area allowed at
90\% C.L. and 99\% C.L. respectively.}
\end{figure}

In fig. \ref{cont1} we show the 90\% and 99\% C.L. 
allowed parameter region in the
\dm- \st plane for
a range of values of the parameter $\alpha$. 
In fig. \ref{cont2} we show the 90\% and 99\% C.L. contours in the 
$\alpha$ - \st plane fixing
$\Delta m^2$ at different  values. 
These contours are obtained from the definition 
$\chi^2  \leq {\chi^2}_{\min} + \Delta \chi^2$, with 
$\Delta \chi^2$ = 6.25 and 15.5 for the three parameter case 
for 90\% and 99\% C.L. respectively. 
The bottom left panel in fig. \ref{cont1} 
is for the best-fit value of $\alpha$.
For high $\alpha$ (the top left panel)  
no lower limit is obtained on
$\Delta m^2$, because even if \dm becomes so low so that there is no
oscillation the complete decay of upward neutrinos can explain their
depletion. 
As we decrease $\alpha$ the allowed parameter region shrinks and finally for
$\alpha=0$ we get the two parameter limit modulo the small difference in the
C.L. definitions for the two and three parameter cases. The upper right 
panel of fig. \ref{cont2} corresponds to the best-fit value of \dm. 
For very low $\alpha$, even though there is 
no decay, we still have oscillations and that ensures that when \dm is large 
enough there is no lower bound on $\alpha$ as evident in the 
fig. \ref{cont2}. For 
\dm$=10^{-4} eV^2$ the neutrinos stop oscillating and hence we get a lower 
bound on $\alpha$ beyond which the depletion in the neutrino flux is not 
enough to explain the data.  

\section{Comparisons and Discussions}

\begin{figure}[t]
      \centerline{\psfig{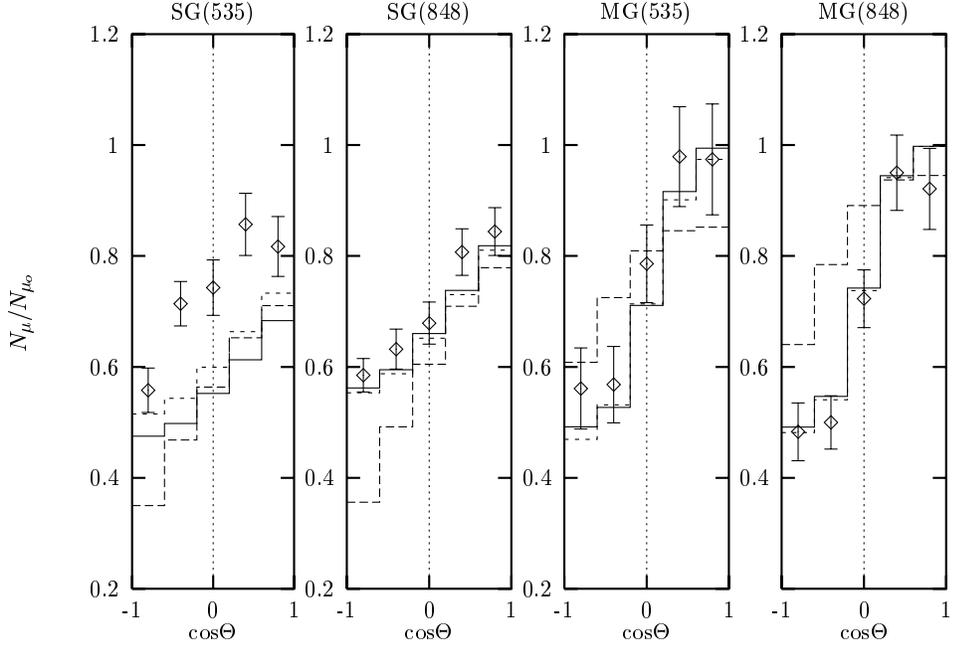}}
\vskip -2cm
    \caption[Zenith angle distribution of the atmospheric events]
  {\label{zendist}
The sub-GeV and multi-GeV $\mu$ event distributions vs.
zenith angle for the various scenarios
considered. $N_\mu$ is the number of $\mu$ events as given by eq. 
(\ref{rate}) 
and $N_{\mu 0}$ is the corresponding number with survival probability 1.
The panels labelled SG(535) and MG(535) give the histograms for the sub-GeV
and multi-GeV 535 day data respectively,
while the SG(848) and MG(848) give the corresponding
histograms for the 848 day data. For the both the sets
the solid line corresponds to the best-fit $\numu-\nutau$
oscillation solution, the long dashed line is for the best-fit
for scenario (a) and the short dashed
line for the best-fit
for scenario (b).
Also shown are the SK $\mu$ event
distributions with $\pm 1\sigma$ error bars for both the sets.}
\end{figure}

In fig. \ref{zendist} we show 
the histogram of the muon event distributions for the 
sub-GeV
and multi-GeV data under the assumptions of $\nu_\mu-\nu_\tau$ oscillation,
and the two scenarios of neutrino decay for the best-fit values of the 
parameters both for the 535 and the 848 day of data. 
From the figures it is clearly seen that the
scenario (a) (big dashed line) ($\dm > 0.1 eV^2$) does not fit the data well
there being
too much suppression for the sub-GeV upward going neutrinos and too less
suppression for the multi-GeV upward going neutrinos.
The scenario (b) (\dm unconstrained, small dashed line), however,  
reproduces the
event distributions well. However with the 848 day data the sub-GeV events are
reproduced better as compared to the 535 day data and the quality of the
fit improves.

The neutrino decay is an interesting idea as
it can preferentially suppress the upward $\nu_\mu$ flux and can cause
some up-down asymmetry in the atmospheric neutrino data.
However the intrinsic defect in the decay term $\exp(-\alpha L/E)$ is that
one has more decay for lower energy neutrinos than for the higher energy ones.
Thus neutrino decay by itself fails to reproduce the observed
data \cite{lipari5}.
If  however one considers the most general case of neutrinos with
non-zero mixing then there are three factors which control
the situation
\begin{itemize}
\item the decay constant $\alpha$ which determines the decay rate
\item the mixing angle $\theta$ which determines the proportion
of neutrinos decaying and mixing with the other flavour
\item the \dm which determines if there are oscillations as well
\end{itemize}

If the heavier state decays to a state with which it mixes then
\dm has to be $> 0.1 eV^2$  because of bounds coming from $K$ decays 
\cite{pak25}. 
The best-fit value of $\alpha$ that one gets is $0.33\times 10^{-4} eV^2$ 
with the 848 day SK data. At this value of $\alpha$ the  
$e^{-\alpha L/E}$ term tends to 1  
for the downward going neutrinos signifying
that they 
do not decay much. The survival probability  
goes to ($1 - \frac{1}{2} \sin^2 2 \theta $) which is
just the average oscillation probability.
In order to suppress this average oscillation  
the best-fit value of $\sin^2 \theta$ comes out to be small in this picture.
For the upward going neutrinos, in scenario (a),
there  will be both decay and average
oscillations. If one had only average oscillation then the probability would
have stayed constant for a fixed value of the mixing angle $\theta$. 
But because of the exponential decay term the survival probability 
drops very sharply as we go towards $\cos\Theta=-1.0$.
The drop and hence the decay is more for lower energy neutrinos. 
As a result the sub-GeV flux gets more depleted than the multi-GeV flux, 
a fact not supported by the data. In fact the  
848 day data requires the sub-GeV flux to be even less suppressed 
than the multi-GeV flux as compared to the 535 day data and the fit 
worsens with the 848 day data.  
The small mixing signifies that the
$\nu_\mu$ has a large fraction of the unstable component
$\nu_2$ (see eq. (\ref{nm})).
Hence the constant $\alpha$ comes out to be low so that
the decay rate is less  to compensate this. 
However even at the best-fit $\alpha$ of 0.33 $\times 10^{-4} eV^2$  
the survival probability in the bin with $\cos \Theta$ between $-1.0$ 
to $-0.6$ 
comes out to be 0.15 for E=1 GeV, much lower than the value of
$\sim$ 0.5 as required by the data. 
Thus scenario (a) fails to explain the upward going neutrino data properly
because of two main reasons
\begin{itemize}
\item{$\theta$ is low in order to suppress the average oscillations of the downward neutrinos}
\item{the energy dependence of the exponential decay term is
in conflict with the data}
\end{itemize}

In the scenario (b), in addition to mixing with $\nu_\tau$,
the unstable component in $\nu_\mu$ decays to some 
sterile state with which it does not mix. 
In this case there is no restriction on \dm and it enters the $\chi^2$ fit as
an independent parameter. We find that: 
\begin{itemize}
\item{ The best-fit \dm does not come out naturally to be in the
\dm independent average oscillation  regime 
of $>$ 0.1 $eV^2$, rather  
it is  $0.003 eV^2$.}
\item{The best-fit value of the decay constant $\alpha = 0.3 \times 10^{-5} eV^2$ implying that the decay rate is small so that the mixing angle is 
maximal ($\sin^2 \theta = 0.5$).} 
\item{Large values of $\alpha$ giving complete decay of upward neutrinos 
are also allowed with a high C.L. In fact with 535 day data the best-fit
was in this region.} 
\item{
The 
best-fit value of the decay constant $\alpha$ is non-zero 
signifying that a little amount of decay combined with \dm dependent 
oscillations gives a better fit to the data.}
\end{itemize} 
At the best-fit values of the parameters 
there is no oscillation of the downward neutrinos so that the 
$\cos (2 \pi L/\lambda_{osc})$ term goes to 1. The 
decay term also goes to 1 signifying that there is not 
much decay either for the downward neutrinos and 
the survival probability is $\approx$ 1 
without requiring the mixing angle to be low. 
On the other hand for the upward neutrinos there are oscillations as well as  
little amount of decay. The sub-GeV  
upward neutrinos have smaller oscillation wavelength and they are close to the  
average oscillation limit (survival probability $\sim$ 0.5) 
while for the multi-GeV neutrinos the oscillation wavelength is such that 
one has maximum oscillations
and the survival probability is less than 0.5. Thus this scenario
reproduces the correct energy dependence of the suppression -- namely 
sub-GeV is suppressed less as compared to multi-GeV neutrinos. 
The best-fit value of 
$\alpha$ being even smaller now than the scenario (a) the 
decay term $e^{-\alpha L/E}$  does not vary very
sharply with the zenith distance $L$ or the energy $E$ so that
its wrong energy dependence does not spoil the fit. 

The conversion probability of $\numu$ to $\nutau$ is given by eq.
(\ref{ntb}). 
The value of $P_{\numu\nutau}$ integrated over the energy and
the zenith angle, for $\alpha=0.3\times 10^{-5} eV^2$ 
(the best fit for scenario (b)) is 0.33
for sub-GeV and 0.26 for multi-GeV. 
For $\alpha=0.44\times 10^{-3} eV^2$ (the second minima in the
$\Delta \chi^2$ vs. $\alpha$ curve) 
the corresponding numbers are 0.21 and 0.15, while
for the only $\numu-\nutau$ oscillation case, the
corresponding values are 0.37 and 0.26 respectively. The values of \dm and
\st for all the cases are taken as $0.003 eV^2$ and 1.0 respectively.

The fig. \ref{zendist} shows that the 
zenith angle dependence of the scenario (b) is
almost similar to the case of $\nu_\mu - \nu_\tau$ oscillation. 
But the two cases are very different in principle.
For the oscillation case a larger 
$\theta$ implies a larger conversion whereas in scenario (b)
a larger $\theta$ means the fraction of the unstable component is less
in $\nu_\mu$ and the depletion is less. 
If one compares the conversion probability as given by eq.(\ref{ntb}) 
with the one for the $\nu_\mu - \nu_\tau$ oscillation case, then the 
scenario (b)
considered in this chapter would have smaller number of $\nu_\tau$s
in the resultant flux at the detector, especially for the larger values 
of $\alpha$ which are still allowed by the data and the two cases might be 
distinguished when 
one has enough statistics to detect $\tau$ appearance in Super-Kamiokande 
\cite{sk12895,hall5} or from neutral current events \cite{smirnov5}.

In this chapter we have followed the procedure of data fitting as done in
\cite{yasuda5}. Thus we use the ratios for which the common systematic
errors get canceled out. 
Strictly speaking one should use the absolute number of events and
include all the correlations between bins and $e$-like and $\mu$-like
events. But the best-fit points and the allowed regions are not
expected to change significantly. 
We have compared the scenarios of neutrino oscillation and
decay with the same definition of $\chi^2$ and for this purpose of
comparison neglecting the correlation matrix will not make much
difference.
Apart from the statistical analysis we have given plots of $R$ and $Y$
for various values of the parameters. The allowed parameter ranges
from these plots are consistent with what we get from our statistical
analysis. The histograms that we have plotted are 
also independent of our definition of $\chi^2$.  We have checked that if  
we estimate the allowed ranges from the histograms these are 
consistent with what we get from our definition of $\chi^2$.
Thus we agree with the observation in ref. \cite{yasuda5} that although
this method of data fitting is approximate it works well. 


In \cite{pakvasa25} 
the authors have considered another scenario where $\dm < 0.0001$
eV$^2$ and have obtained good fits to the SK data. They have also
discussed in details the neutrino decay models and the consequences
of such models for astrophysics and cosmology.

%% file: chapter7.tex
\chapter{Massive Neutrinos in Supernova}

The core of a massive star $(M\ge 8M_\odot)$ starts collapsing
once it runs out of nuclear fuel.
The collapse continues to densities beyond the nuclear
matter density after which a bouncing of the infalling matter takes place
leading to supernova explosion and the formation of a protoneutron star 
\cite{bethe6}.
Only a small fraction of the huge gravitational energy released in the
process
goes into the explosion and all the rest of the energy is carried away by
neutrinos and antineutrinos of all three flavors. 
About $10^{58}$ neutrinos, in all three flavors
carrying a few times $10^{53}$ ergs of energy are released in a type II
supernova. The luminosities of all the neutrino species are almost same 
while the average energies are approximately 
11 MeV for $\nue$, 16 MeV for $\anue$ and the 
average energy of $\numu$ and $\nutau$ and their antiparticles is 
25 MeV \cite{rprocess}. 
These neutrinos for galactic supernova events can be
detected by the current water \chr detectors, the Super-Kamiokande
(SK) and the Sudbury Neutrino Observatory (SNO) \cite{bkg16}. 
In contrast to the solar, the atmospheric
as well as the accelerator/reactor neutrinos 
where one has neutrino flux of a single flavor at the
source, postbounce supernova neutrinos (antineutrinos)
start from the source in
all three flavors but with $\nu_{\mu}/\nu_{\tau}$
($\anumu/\anutau$) having average energies
more than that for $\nu_e (\anue)$ and it is an interesting
problem to study whether
their flux and their signal at the terrestrial $\nu$ detectors
get appreciably altered in reaching the earth if
neutrinos do oscillate. 

The detection of the SN1987A neutrinos by the water
\chr detectors at Kamioka \cite{kamsn} and IMB \cite{imbsn}
settled many important issues in
the subject of type II supernova theory. The observation of neutrinos
from any future
galactic supernova event will answer the remaining
questions regarding the understanding of the supernova mechanisms. A
galactic supernova event will also bring in a lot of information
on neutrino mass, which of late, has been an issue of much discussion. 
The effect of
neutrino mass can show up in the observed neutrino
signal in these detectors in two ways,
\begin{itemize}
\item by causing delay in the time of flight measurements
\item by modifying the neutrino spectra through neutrino flavor mixing
\end {itemize}

Massive neutrinos travel with speed less than the speed of light and
for typical galactic supernova distances $\sim$ 10 kpc,
even a small mass results in a measurable delay in the arrival time
of the neutrino. Many different analyses have been performed before
to give bounds on the neutrino mass by looking at this delay in the 
arrival time of the massive neutrinos (\cite{bv16,bv26} and references
therein). Neutrino oscillations on the other hand convert the more
energetic $\numu/\nutau (\anumu/\anutau)$ into $\nue (\anue)$ thereby
hardening the resultant $\nue (\anue)$ energy spectra and
hence enhancing their signal at the detector \cite{kp6,bkg26,qf6,ds6,mat6}.
In this chapter we give quantitative predictions for 
the number of neutrino events coming from a typical
type II supernova at a distance of 10kpc in both SNO and SK and
show how the number of events for each detection process would change
in case oscillations do take place.
We study in detail the effect of neutrino mass 
and mixing 
on the total number of events recorded in the detector, the distortion 
of the neutrino spectra due to oscillations, the effect of delay due 
to neutrino mass on the time response of the signal and the 
effect of oscillations on the delay due to mass recorded at the detector.

The water \chr detectors detect neutrinos through various charged and 
neutral current processes. 
The differential number of neutrino events at the detector for a
given reaction process is
\begin{equation}
\frac{d^2 S}{dEdt}=\sum_i\frac{n}{4\pi D^2}
N_{\nu_i}(t)f_{\nu_i}(E) \sigma(E)
\epsilon(E)
\label{sig}
\end{equation}
where $i$ runs over the neutrino species concerned,
$N_{\nu_i}(t) = L_{\nu_i}(t)/\langle E_{\nu_i}(t)\rangle$,
are the number of neutrinos produced at the source
where $L_{\nu_i}(t)$ is the neutrino luminosity and
$\langle E_{\nu_i}(t) \rangle$ is the average energy,
$\sigma (E)$ is the reaction cross-section for the neutrino with
the target particle, $D$ is the distance of the neutrino source
from the detector (taken as 10kpc),
$n$ is the number of detector particles for the reaction considered
and $f_{\nu_i} (E)$ is the energy spectrum for the neutrino species
involved,
while $\epsilon(E)$ is the detector efficiency as a function of the
neutrino energy.
By integrating out the energy from eq.(\ref{sig}) we get the time
dependence of the various reactions at the detector. To get the total
numbers both integrations over energy and time has to be done.

In section 7.1 we use the eq. (\ref{sig}) to estimate the signal that a
future galactic supernova event would register in SK and SNO, using 
the luminosities and average energies from a realistic $20 M_\odot$ 
supernova model. We consider a scheme of neutrino mass and mixing 
such that one has almost pure vacuum oscillations and study its effect on 
the neutrino spectrum and hence on the signal at the detector. 
We next consider in section 7.2 a neutrino mass spectrum where 
one of the neutrino masses is in the eV range and we study the 
effect of delay in the arrival time on the neutrino signal for 
both with and without the presence of neutrino flavor mixing. We end 
this chapter by drawing the main conclusions in section 7.3.

\section{Effect of Neutrino Oscillations in Vacuum}

There have been various attempts before to estimate the effect of
non-zero neutrino mass and mixing on the expected neutrino signal
from a galactic supernova.
Matter enhanced resonant flavor conversion has been observed to have
a large effect on the $\nue$ signal \cite{kp6,bkg26,qf6,ds6,mat6,akh6}. 
The $\anue$
events of course remain unchanged in this case. With vacuum
oscillations we can expect an increase in both the $\nue$ and $\anue$
signal.
Burrows {\it et al.} \cite{bkg26} have considered for SNO, the effect of
vacuum oscillations as well and have found that
with two-flavors the effect of vacuum oscillations on the signal is
small, using their model predictions for the
different $\nu$ luminosities.

We have considered a three-generation mixing scheme and have
calculated the effect of neutrino oscillations on the signal
from a 20 $M_\odot$ supernova model developed recently \cite{totani6} 
by Totani {\it et al.} 
based on the hydrodynamic code developed by Wilson and Mayle. 
For the neutrino luminosities and average energies we 
use the model predictions from \cite{totani6}. 
Though in their paper Totani {\it et al.}
observe that the neutrino spectrum is not a pure black body, but we
as a first approximation use a Fermi-Dirac spectrum for the neutrinos,
characterized by the $\nu$ temperature alone
for simplicity. The effect of a
chemical potential is to cut the 
high energy tail of the neutrino spectrum
and we also study its effect on the the $\nu$ signal and on the
enhancement of the signal when oscillations are introduced. 

\noindent
For the mass and mixing parameters we consider two scenarios 

$\bullet$ {\bf Scenario 1:} 
First we do our calculations for the threefold maximal mixing model
\cite{smirmax6,kimkimmax6,harmax6,footmax6} 
consistent with the solar ($\Delta m_{12}^2 \sim 10^{-10}$ eV$^2$) and the
atmospheric neutrino data ($\Delta m_{23}^2 \approx \Delta m_{13}^2  
\sim 10^{-3}$ eV$^2$).
For $\Delta m^2 \sim 10^{-3}$ eV$^2$
normally we expect matter enhanced resonance in the supernova. But
for the particular case of maximal mixing it has been shown before,
both numerically \cite{harrison6} and analytically \cite{bilenky6}, that
there are no discernable matter effects in the resultant neutrino spectrum
on earth. Though the arguments in both these previous papers are for
solar neutrinos, extension to the case of supernova neutrinos is
straightforward. 
Hence we are concerned with vacuum
oscillations only. Since the oscillation 
wavelengths (cf. eq. (\ref{wv})) corresponding to both the mass scales 
are much smaller compared to the Earth--supernova distance $L$, the 
oscillation probabilities reduce to \cite{kimkimmax6,harmax6}
\be
P_{\nue \nue}=\frac{1}{3}
\label{prb1}
\ee
\be
P_{\numu \nue}+P_{\nutau \nue}=1-P_{\nue \nue}
\ee
We call this case 1.

$\bullet$ {\bf Scenario 2:} Here we set 
$\Delta m_{12}^2 \sim 10^{-18}$ eV$^2$
for which $\lambda \sim L$ and hence the oscillation effects are 
observable in the neutrino spectrum. The other mass range in 
kept in the solar vacuum oscillation regime 
$\Delta m_{13}^2 \approx \Delta m_{23}^2  \sim 10^{-10}$ eV$^2$. 
For this case the oscillations due to $\Delta m_{13}^2$ and 
$\Delta m_{23}^2$ are averaged out as the neutrinos travel to 
Earth but those due to $\Delta m_{12}^2$ survive. In this 
scenario again there is no matter effects and one has vacuum 
oscillations. The transition and survival probabilities in 
this case are
\be
P_{\nue \nue}=1-\sin^2 2\theta_{12}\cos^4 \theta_{13}\sin^2
\frac{\pi L}{\lambda_{12}} - \frac{1}{2} \sin^2 2\theta_{13}
\label{prb2}
\ee
\be
P_{\numu \nue}+P_{\nutau \nue}=1-P_{\nue \nue}
\ee
We use the for the mixing matrix $U=R_{23}R_{13}R_{12}$, where 
$R_{ij}$ have been defined before in chapter 5.  
For $\theta_{13}$ we consider two sets of
values allowed by the solar $\nu$ data. We have done our calculations
for $\sin^2 2\theta_{13}$ = 1.0 and
$\sin^2 2\theta_{13}$ = 0.75. The
first set is called Case 2a while the second is called Case 2b.
Since nothing constrains $\Delta m_{12}^2$ in this scenario
we can vary $\theta_{12}$ and study its effect on the $\nu$ signal.
We have tabulated our results for $\sin^2 2\theta_{12}$ = 1.0 since
it gives the maximum increase in the signal from the no oscillation
value.

The corresponding expressions for the antineutrinos will be identical.
We note that because the energy spectra of the $\numu$ and $\nutau$
are identical, we do not need to distinguish them and keep the
combination
$P_{\numu \nue}+P_{\nutau \nue}$. We have made here a three-generation
analysis where all the three neutrino flavors are active. Hence if both
the solar $\nu$ problem and the atmospheric $\nu$ anomaly require $\nu$
oscillation solutions, then in the {\bf scenario 2},
the atmospheric data has to be reproduced by $\numu-\nu_{\rm sterile}$
oscillations\footnote{The pure $\numu-\nu_{\rm sterile}$ solution to the 
atmospheric neutrino problem is now disfavored from the SK 
atmospheric data \cite{skatmsterile6}. 
However schemes in which $\numu$ can oscillate 
into a combination of active and sterile species are still allowed 
\cite{fl4gen6}.}. 
We are interested in this scenario as only
with neutrinos from a supernova can one probe very small
mass square differences $\sim 10^{-18}$ eV$^2$.
To find the number of events with oscillations we will have to fold the
expression (\ref{sig}) with the expressions for survival and
transition probabilities for the neutrinos
for all the cases considered.

\begin{table}[htbp]
    \begin{center}
        \begin{tabular}{||c||c||c|c|c||} \hline\hline
 & signal
& \multicolumn{3}{c||}{signal with oscillation}\\
\cline{3-5}
reaction & without &
{\bf scenario 1} &
\multicolumn{2}{c||}{\bf scenario 2}\\ \cline{3-5}
& oscillation & Case 1
& Case2a & Case2b
\\ \hline\hline
{$\nu_e+d\rightarrow p+p+e^-$} & {78} & {155}
& {150} & {153}\\ 
{$\bar\nu_e +d\rightarrow n+n+e^+$} & {93} & {136}
& {133} & {135}\\ 
{$\nu_x+d\rightarrow n+p+\nu_x$} & {455} & {455} & {455} &
{455} \\ 
{$\bar\nu_e +p\rightarrow n+e^+$} & {263} & {330}
& {326} & {329}\\ 
{$\nu_e +e^- \rightarrow \nu_e +e^-$}
& {4.68} & {5.68} & {5.61} & {5.66} \\ 
{$\bar\nu_e+e^- \rightarrow \bar\nu_e+e^-$}
& {1.54} & {1.77} & {1.76} & {1.77} \\ 
{$\nu_{\mu,\tau}(\bar\nu_{\mu,\tau}) +e^- \rightarrow
\nu_{\mu,\tau}(\bar\nu_{\mu,\tau}) +e^-$}
& {3.87} & {3.55} & {3.50} & {3.53} \\ 
{$\nu_e +^{16}O \rightarrow e^- +^{16}F$} & {1.13} & {14.58}
& {13.78} & {14.45} \\ 
{$\bar\nu_e + ^{16}O\rightarrow e^+ + ^{16}N$} & {4.57} & {10.62}
& {10.23} & {10.53} \\ 
{$\nu_x +^{16}O \rightarrow \nu_x +\gamma +X$} & {13.6} & {13.6} &
{13.6} & {13.6} \\ \hline\hline
\end{tabular}
      \end{center}
      \caption[Signal from a galactic supernova with vacuum oscillations]
{\label{votab}
          The expected number of neutrino events for a 1 kton
water \chr detector (${\rm H_2O}$ or ${\rm D_2O}$).} 
\end{table}

In Table \ref{votab} 
we report the calculated number of expected events for the 
main reactions in $\rm H_2O$ and $\rm D_2O$. 
Column 2 of Table \ref{votab} 
gives the expected numbers for the 
supernova model under consideration when 
the neutrino masses are assumed to be zero. Column 3,4,5 give the 
corresponding numbers for the two scenarios of neutrino mixing 
that we have considered (see Table \ref{votab} for details). All the numbers 
tabulated have been calculated for 1 kton of detector mass. To get 
the actual numbers we have to multiply these numbers with the 
relevant fiducial mass of the detector. The efficiency of both the 
detectors (SNO and SK) is taken to be 1 \cite{bv16,bv26,totani6}. 
The energy threshold is taken to be 5 MeV for both SK \cite{bv16} and 
SNO \cite{bv26}. The energy threshold of SNO in the recently declared 
solar neutrino results is 6.75 MeV \cite{sno6}. 
For the cross-section of the $(\nue-d), (\anue-d), 
(\nu_x-d)$ 
and $(\anue-p)$ reactions we refer to \cite{burrows6}. 
The cross-section of the $(\nue(\anue)-e^-)$ and $(\nu_x-e^-)$  
scattering has been taken from \cite{kolb6} while the neutral current 
$(\nu_x-^{16}O)$ scattering cross-section is taken from \cite{bv16}. 
For the 
$^{16}O(\nue,e^-)^{16}F$ and $^{16}O(\anue,e^+)^{16}N$ reactions we 
refer to \cite{haxton6} where we have used the cross-sections 
for the detector with perfect efficiency.  
From a comparison of the 
predicted numbers in Table \ref{votab}, 
it is evident that neutrino oscillations 
play a significant role in supernova neutrino detection. 
For the neutral current sector the number of 
events remain unchanged as the interaction is flavor blind.

The 32 kton of pure water in SK detects neutrinos 
primarily through the capture of $\anue$ on 
protons ($\anue p \rightarrow n e^+$) and $(\nue(\anue)-e^-)$ scattering. 
The energy threshold for $^{16}O(\nue,e^-)^{16}F$ is 15.4 MeV and 
that for $^{16}O(\anue,e^+)^{16}N$ is 11.4 MeV, hence these reactions   
are important only for very high energy neutrinos. 
The typical average energies of $\nue$ and $\anue$ from a type II 
supernova is 
about 11 MeV and 16 MeV respectively, so we do not expect significant 
contribution from these two reactions. This is evident from Table \ref{votab} 
where the $^{16}O$ events are only 2.1\% of the total charge 
current signal at SK. 
As a result of mixing the mu and tau neutrinos 
and antineutrinos oscillate (with average energy $\sim$ 25 MeV) into 
$\nue$ and $\anue$ during their flight from the galactic 
supernova to the 
detector resulting in higher energy $\nue$ and $\anue$ 
and the number of $^{16}O$ 
events are increased appreciably (for Case 1 $(\nue-^{16}O)$ events go 
up by 13 times) so that after oscillations 
they are 7\% (Case 1) of the 
total charge current events at SK. 
The effect of oscillations on the ($\anue$-p) capture is to enhance 
the expected signal by about 25\% (Case 1).  
In all previous studies where the effect 
of MSW transition on the neutrino signal has been studied \cite{akh6,qf6}, 
there is no enhancement in the number of expected events for the ($\anue$-p) 
sector while we do get a significant change in the expected 
signal with vacuum oscillations. 
For the ($\nue(\anue) - e^-$) scattering the effect of 
oscillation is very small. 

\begin{figure}[t]
    \centerline{\psfig{file=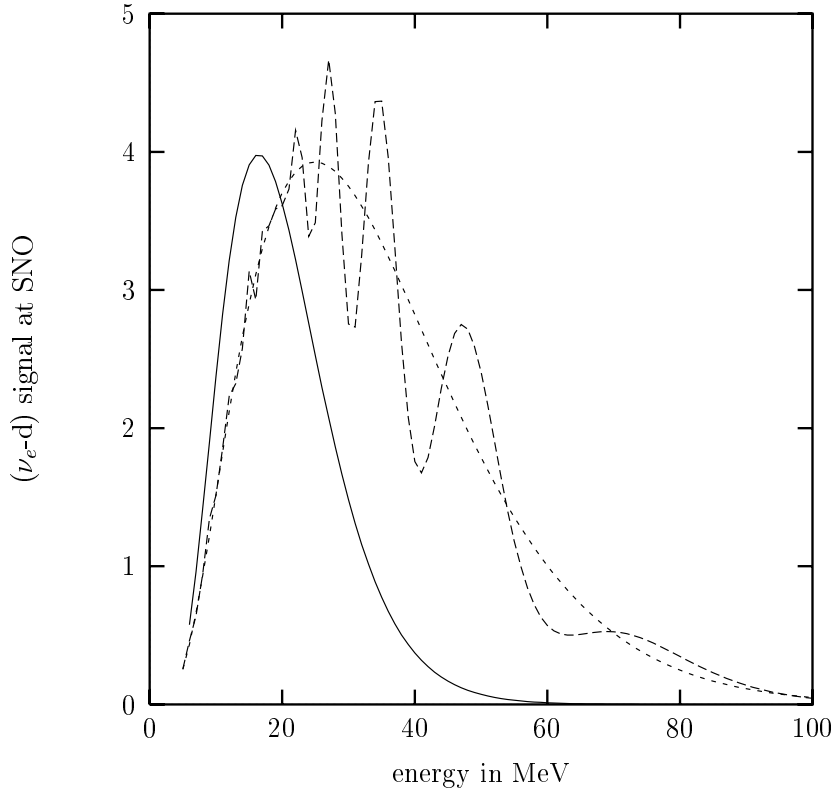,width=5.0in}}
\vskip -1cm
   \caption[The supernova ($\nu_e-d$) signal at SNO vs neutrino energy]
{\label{nued6}
The ($\nue-d$) signal at SNO vs neutrino energy
without (solid line) and with oscillations for
the Case 1 (short dashed line) and Case 2b (long dashed line).}
\end{figure}

The SNO is the world's first heavy water detector made of 1 kton of 
pure $\rm D_2O$ surrounded by ultra pure $\rm H_2O$. 
We find about 99\% increase in $(\nue-d)$ events and about 46\% increase 
in $(\anue-d)$ events for the Case 1. 
From the column 2 of Table \ref{votab} 
we can see that there are more $(\anue-d)$ than $(\nue-d)$ 
events even though there are more $\nue$ than $\anue$ coming 
from the supernova. This is  because the reaction cross-section $\sigma \sim 
E^{2.3}$ and the $\anue$ spectrum is harder than the $\nue$ 
spectrum. This also results in a greater enhancement due to oscillations 
for the $(\nue-d)$ events, as the difference between the energies of the 
$\nue$ and $\numu(\nutau)$ is greater than those between $\anue$ and 
$\anumu(\anutau)$ and hence the effect on the $\nue$ events is more. 
As a result after oscillations are switched on the number of $(\nue-d)$ 
events supersede the $(\anue-d)$ events. 
We observe a similar effect for the $^{16}O$ events, where the $\anue$ 
signal without oscillations is more than the $\nue$ signal, 
while the effect of oscillations 
is more for the latter. The effect is more magnified in this case due to 
the very strong energy dependence of the reaction cross-section and 
also due to the fact that the energy threshold for $(\anue-^{16}O)$ 
event is lower than for the $(\nue-^{16}O)$ event. 
In fig. \ref{nued6} we plot the signal due to the ($\nue-d$) events as a 
function of neutrino energy, 
without oscillations and with oscillations for the Case 1 and 
Case 2b. All the features mentioned are clearly seen. The plot 
for the Case 2b clearly shows oscillations. 

\begin{figure}[t]
    \centerline{\psfig{file=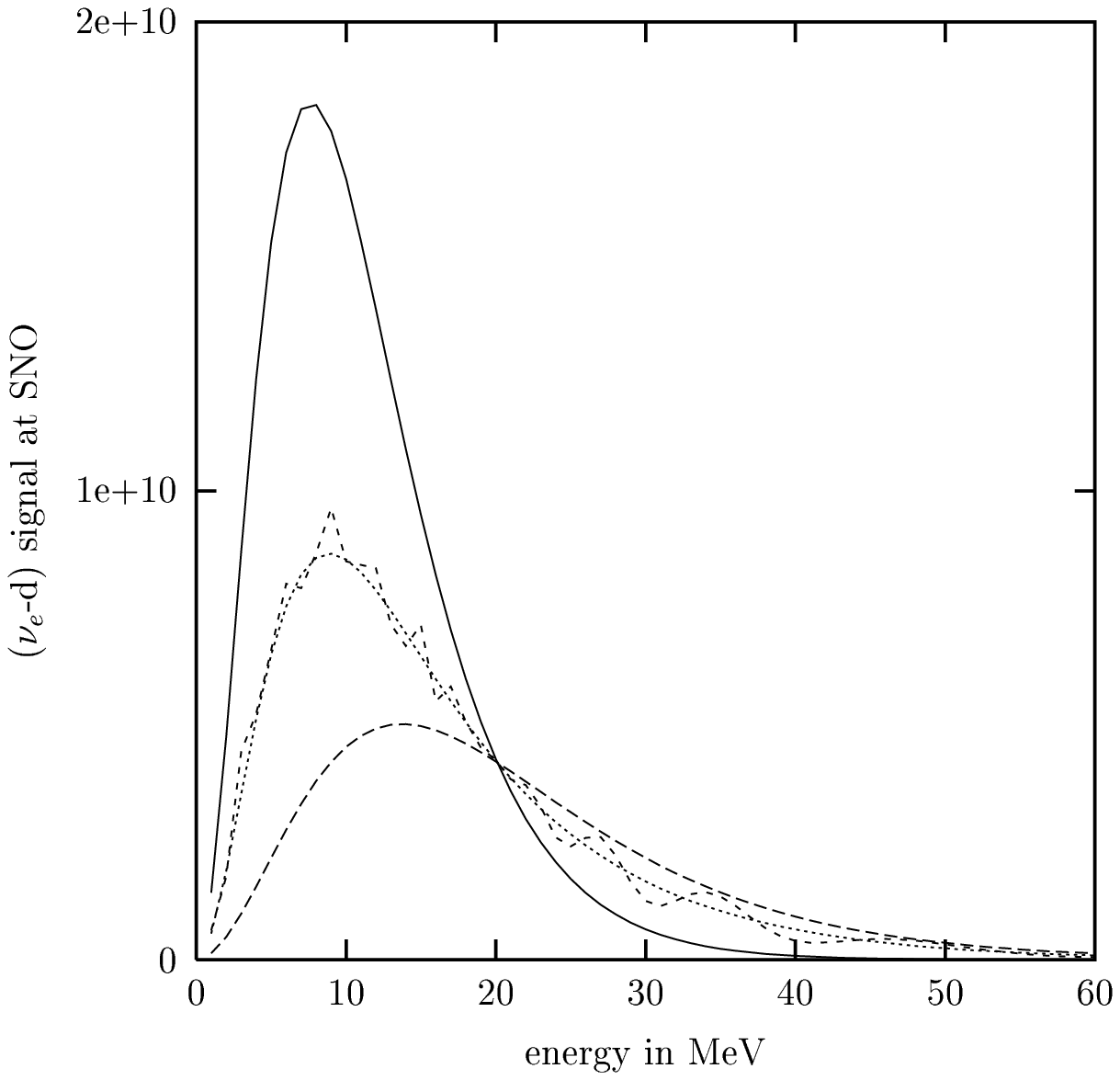,width=5.0in}}
\vskip -1cm
    \caption[The supernova neutrino fluence with and without oscillation]
{\label{nuflu}The cumulative $\nue$ fluence as a function of the neutrino
energy without (solid line) 
and with oscillations for the Case 1 (dotted line) and Case 2b 
(short dashed line). Also
shown is the $\numu$ fluence (long dashed line) for comparison.}
\end{figure}

In fig. \ref{nuflu} we plot the cumulative fluence of the $\nue$ 
coming from the supernova at 10 kpc without oscillations and with 
oscillations for Case 1 and Case 2b. It is seen that the result of 
oscillation in fact is to reduce the total number of $\nue$. Yet 
as seen from Table \ref{votab}, we have 
obtained significant increase in the $(\nue-d)$ events and 
the $(\nue -^{16}O)$ events. The solution to 
this apparent anomaly lies in the fact that the 
cross-section of these reactions 
are strongly energy dependent. As a result of oscillations the $\nue$ flux 
though depleted in number, gets enriched in high energy neutrinos. 
It is these higher energy neutrinos which enhance the $\nu$ signal at the 
detector. This also explains the difference in the degree of enhancement 
for the different processes. For the $(\nue-d)$ and $(\nue - ^{16}O)$ 
events, especially for the latter, the effect is huge while for the 
$(\nue-e^-)$ scattering it is negligible as its reaction cross-section 
is only linearly proportional to the neutrino energy.  
Due to their high energy dependent cross-sections  
the $^{16}O(\nue,e^-)^{16}F$ events turn out to be 
extremely sensitive to oscillations. 
A similar argument holds true for the case of the antineutrinos, only 
here the effect of oscillations is less than in the case for the 
neutrinos as the difference between the energies of the $\anue$ and 
$\anumu/\anutau$ is comparatively less as discussed earlier. 

For the {scenario 2} we have studied the effect of the mixing 
angles on the signal. For a fixed $\theta_{13}$ the effect of 
oscillations is enhanced if we raise $\theta_{12}$. The effect of 
$\theta_{13}$ is more subtle. The effect of oscillations increase 
with $\theta_{13}$ initially and then 
decrease. 
We have also checked the effect of a chemical potential 
$\mu$ on the neutrino signal. A non-zero $\mu$ cuts the high energy 
tail of the neutrino signal as a result of which the total signal goes 
down for both with and without oscillations, the effect being greater 
for the more energy sensitive reactions. 

With the supernova model of Totani {\it et al.} \cite{totani6}, we 
have obtained oscillation effects in the expected $\nu$ signal which 
are significantly larger than those obtained by Burrows {\it et al.} 
\cite{bkg16,bkg26}. In the model that Burrows {\it et al.} use in their 
study, the $\nu$ luminosities $L_\nu$ are more than those for 
Totani {\it et al.} model, but the average energy is much smaller, 
particularly for the $\anue$ and $\nu_{\mu,\tau}(\bar\nu_{\mu,\tau})$. 
Hence their $\numu$ spectra lacks in high energy neutrinos which 
results in almost negligible effect of oscillations in their case. 
Again in the model of Burrows {\it et al.} 
the average energies decrease with time while in the model of Totani 
{\it et al.} not only the average energies but also the difference between 
the average energies of $\nue(\anue)$ and $\nu_{\mu,\tau}(\bar\nu_{\mu,\tau})$ 
increases with time. The effect of all these is to magnify the effect of 
oscillations in our case.

\section{The Effect of Delay in the Time of Flight}

In the previous section we studied the effects
of neutrino
flavor oscillations on the supernova neutrino spectrum and the number of
charged current events at the detector using a realistic supernova model.
In this section we study the neutral current signal
as a function of time
in the water Cerenkov detectors, for a mass range of the neutrinos
where both the phenomenon of delay and flavor conversion are operative.
That the time response of the event rate in the detector is
modified if the neutrinos have mass alone and hence delay
is a well known feature \cite{bv16,bv26}.
Here we stress the point that since
neutrino flavor conversions change the energy spectra
of the neutrinos, and since
the time delay of the massive neutrinos is energy dependent,
the time dependence of the event rate at the detector
is altered appreciably in the presence of mixing.
We suggest various variables which act as tools for
measuring this change in the time response curve of the neutral
current events and in differentiating the cases of (a) massless
neutrinos (b) neutrinos with mass but no mixing and (c) neutrinos
with mass as well as mixing. In particular
we study the ratio of the charged current to neutral current ratio R(t),
as a function of time in the SNO detector and show that the
change in the value and the shape of R(t) due to flavor mixing
cannot be emulated by
uncertainties. We also study other variables like the normalized
$n$-th energy moments of the neutral current events and the ratio of
charged to the neutral current $n$-th moments as important diagnostic tools
in filtering out the effects of neutrino mass and mixing.

For the neutrino luminosities and average energies, though it is best to
use a numerical supernova model,
but for simplicity, we will here use a profile of the
neutrino luminosities and temperatures which have general agreement
with most supernova models. We take
the total supernova energy radiated in neutrinos to be 3 $\times 10^{53}$
ergs. This luminosity, which is almost the same
for all the neutrino species, has a fast rise over a period of 0.1 sec
followed by a slow fall over several seconds in most
supernova models. We use a luminosity that
has a rise in 0.1 sec using one side of the Gaussian with $\sigma$ = 0.03
and then an exponential decay with time constant $\tau$ = 3 sec for
all the flavors \cite{bv16,bv26}.

The average energies associated with the $\nue, \anue ~\rm{and}~ \numu$ (the
$\numu, \anumu, \nutau ~\rm{and}~ \anutau$ have the same energy spectra)
are 11 MeV, 16 MeV and 25 MeV respectively in most numerical models.
We take these average energies and consider them to be
constant in time. We have also checked our calculations with
time dependent average energies and estimated its effect.
The neutrino spectrum is taken to be a pure Fermi-Dirac
distribution characterized by the
neutrino temperature alone.

If the neutrinos are massless then the time response of
their signal at the detector
reflect just the time dependence of their luminosity function at the
source, which is the same for all the three
flavors and hence the same for the charged current
and neutral current reactions. If neutrinos have mass $\sim$ eV then
they pick up a measurable delay during their course of flight from the
supernova to the earth.
For a neutrino of mass m (in eV) and energy E (in MeV), the delay
(in sec) in traveling a distance D (in 10 kpc) is
\begin{equation}
\Delta t(E) = 0.515{(m/E)}^2 D
\label{deltime}
\end{equation}
where we have neglected all the small higher order terms.
The time response curve then has contributions
from both the luminosity and the mass.
We will now consider a scheme of neutrino masses such that
$\Delta m_{12}^2 \sim 10^{-5}$ eV$^2$ consistent with the LMA MSW 
solution of the solar
neutrino problem and $\Delta m_{13}^2 \approx \Delta
m_{23}^2 \sim 1-10^4$ eV$^2$.
For $\dm \sim 10^{-5}$ eV$^2$ 
the effect of Earth matter can show up in the neutrino signal 
\cite{snearth6} but we neglect it here for simplicity. 
The neutrino mass
model considered here is one of several, given for the purpose
of illustration only.  In this scheme the atmospheric neutrino anomaly
will have to be explained by the $\numu-\nu_{\rm sterile}$ oscillation mode.
The mass range for the neutrinos as the hot component of hot plus cold
dark matter scenario in cosmology is a few eV only \cite{hdm}, which
will conflict with the higher values in the range of
$m_{\nu_3} = 1-100$ eV that we consider here if $\nu_3$ is stable.
Hence, we assume that the $\nu_3$
state is unstable but with a large enough life time so that it is does
not conflict with the observations of SN 1987A \cite{pbpal6}
(even though SN1987A
observations did not correspond to any $\nutau$ event, one can put limits
on the $\nu_3/\bar{\nu_3}$ lifetime as the $\nue/\anue$ state is a mixture
of all the three mass eigenstates) and is
also consistent with Big Bang Nucleosynthesis. In fact, from the
ref. \cite{bv16,bv26} we know that using the time delay technique, the
SK and SNO can be used to probe neutrino masses down to only 50 eV and
30 eV respectively. Hence we have presented all our results for a
particular representative value of $m_{\nu_3} = 40$ eV.
There have been proposals in the past for an unstable neutrino with
mass $\sim 30$ eV and lifetime $\sim 10^{23}$ s \cite{sciama}.
Since direct kinematical measurements
give $m_\nue < 5$ eV \cite{nue}, we have taken the $\nue$ to be
massless and
the charged current events experience no change.
But since the $\nutau(\anutau)$ pick up a
detectable time delay (for the mass spectrum of the neutrinos that we
consider here, the $\numu(\anumu)$ do not have
measurable time delay), the expression for the neutral current events gets
modified to,
\begin{eqnarray}
\frac{dS_{nc}^d}{dt}&=&\frac{n}{4\pi D^2} \int dE \sigma (E)
\{N_\nue(t) f_\nue (E) + N_\anue (t)f_\anue (E) +
N_{\numu}(t)f_{\numu}(E) +
\nonumber\\
&+& N_{\anumu}(t)f_{\anumu}(E) +N_{\nutau}(t-\Delta t(E))f_{\nutau}(E) +
N_{\anutau}(t-\Delta t(E))f_{\anutau}(E)\}
\label{del1}
\end{eqnarray}
where $dS_{nc}^d/dt$ denotes the neutral current $(nc)$ event rate with
delay $(d)$.
Delay therefore distorts the neutral current event
rate vs. time curve. By doing a $\chi^2$
analysis of this shape distortion one can put limits on the $\nutau$
mass \cite{bv16,bv26}.

We next consider the neutrinos to have flavor mixing as well.
The mixing angle $\sin^2\theta_{12}$ can be constrained from the
solar neutrino data (cf. chapter 3)
while for
$\sin^2\theta_{13}$ there is no experimental data to fall back upon,
but from r-process considerations in the `hot bubble" of the
supernova, one can restrict $\sin^2\theta_{13} \sim 10^{-6}$
\cite{rprocess,qf6}.
In this scenario there will be first a matter enhanced
$\nue-\nutau$ resonance in the mantle of the supernova followed by a
$\nue-\numu$ resonance in the envelope.
The MSW mechanism in the supernova for the neutrino mass scheme that we
consider here is discussed in details in ref.
\cite{qf6}.
As the average energy of the $\numu/\nutau$ is greater
than the average energy of the $\nue$, neutrino flavor mixing
modifies their energy spectrum.
Hence as pointed out in the previous section 
the $\nue$ flux though depleted in number, gets enriched
in high energy neutrinos and since the detection
cross-sections are strongly energy dependent, this results in the
enhancement of the charged current signal.
The total number of events in SNO, integrated over time
in this scenario 
are given in Table \ref{mswtab}.
In the third column of Table \ref{mswtab} (marked A) we 
show the number of events for 
galactic supernova neutrinos with 
luminosities and average energies considered in this section while 
the last column (marked B) gives the number of events with complete flavor
oscillations ($P_{\nue\nue}=0$).  
Of course since the $\anue$
do not have any conversion here, the $\anue$ signal remains unaltered.
Also as the neutral current
reactions are flavor blind, the total neutral current signal remains
unchanged. But whether the time response curve of the neutral current
signal remains unchanged in presence of mixing, in addition to delay,
is an interesting question.

\begin{table}[t]
    \begin{center}
\begin{tabular}{||c| c| c|c||} \hline\hline
&&A&B\\\hline
reactions&
{$\nu_e+d\rightarrow p+p+e^-$} & { 75 } & {239} \\ 
in&
{$\bar\nu_e +d\rightarrow n+n+e^+$} & {91} & {91} \\ 
1 kton&
{$\nu_i+d\rightarrow n+p+\nu_i$} & {544} & {544} \\ 
$\rm{D_2O}$&
{$\nu_e +e^- \rightarrow \nu_e +e^-$} & {4} & {6} \\ 
&
{$\bar\nu_e+e^- \rightarrow \bar\nu_e+e^-$}
& {1} & {1} \\ 
&
{$\nu_{\mu,\tau}(\bar\nu_{\mu,\tau}) +e^- \rightarrow
\nu_{\mu,\tau}(\bar\nu_{\mu,\tau}) +e^-$}
& {4} & {3} \\ 
&
{$\nu_e +^{16}O \rightarrow e^- +^{16}F$} & {1} & {55} \\ 
&
{$\bar\nu_e + ^{16}O\rightarrow e^+ + ^{16}N$} & {4} & {4} \\ 
&
{$\nu_i +^{16}O \rightarrow \nu_i +\gamma +X$} & {21} & {21} \\ \hline

reactions& 
{$\bar\nu_e +p\rightarrow n+e^+$} & {357} & {357} \\ 
in &
{$\nu_e +e^- \rightarrow \nu_e +e^-$} & {6} & {9} \\ 
1.4 kton&
{$\bar\nu_e+e^- \rightarrow \bar\nu_e+e^-$} & {2} & {2} \\ 
$\rm{H_2O}$ &
{$\nu_{\mu,\tau}(\bar\nu_{\mu,\tau}) +e^- \rightarrow
\nu_{\mu,\tau}(\bar\nu_{\mu,\tau}) +e^-$} & {6} & {5} \\ 
&
{$\nu_e +^{16}O \rightarrow e^- +^{16}F$} & {2} & {86} \\ 
&
{$\bar\nu_e + ^{16}O\rightarrow e^+ + ^{16}N$} & {6} & {6} \\ 
&
{$\nu_i +^{16}O \rightarrow \nu_i +\gamma +X$} & {33} & {33} \\ 
\hline\hline
\end{tabular}
\end{center}
\caption[Signal from a galactic supernova for complete conversion]
{\label{mswtab}
The expected number of neutrino events in SNO. To get
the number of events in SK, one has to scale the number of events in
$\rm{H_2O}$ given here to its fiducial mass of 32 kton.
The column A corresponds to massless neutrinos,
column B to neutrinos with complete flavor conversion
The $\nu_i$ here refers to all the six neutrino species.}
\end{table}
\vskip 20pt

If the neutrinos have mass as well as mixing, then the neutrinos are
produced in their flavor eigenstate, but they travel in their mass
eigenstate. The neutrino mass eigenstates
will travel with different speeds depending on their mass and will
arrive at the detector at different times.
For the scenario that
we are considering only $\nu_3$ and $\bar\nu_3$ will be delayed.
Hence to take this
delay in arrival time into account, the eq.(\ref{del1}) has to be
rewritten in terms of the mass eigenstates.
It can be shown that expression for the neutral current event rate
in terms of the mass eigenstates is,
\begin{eqnarray}
\frac{dS_{nc}^{do}}{dt}&=&\frac{n}{4\pi D^2} \int dE \sigma (E)
\{N_{\nu_1}(t) f_{\nu_1} (E)+N_{\bar\nu_1}(t)f_{\bar\nu_1}(E)+
N_{\nu_2}(t)f_{\nu_2}(E)
\nonumber\\
&+& N_{\bar\nu_2}(t)f_{\bar\nu_2}(E)+
N_{\nu_3}(t-\Delta t(E))f_{\nu_3}(E)+
N_{\bar\nu_3}(t-\Delta t(E))f_{\bar\nu_3}(E)\}
\end{eqnarray}
where $N_{\nu_i}$ is the $\nu_i$ flux at the source. If the neutrinos
are produced at densities much higher than their resonance densities,
all the mixings in matter are highly suppressed, and the neutrinos
are produced almost entirely in their mass eigenstates. For the three
generation case that we are considering, $\nue \approx \nu_3$,
$\numu \approx \nu_1$ and $\nutau \approx \nu_2$. For the antineutrinos
on the other hand, at the
point of production in the supernova $\anue \approx \bar\nu_1$,
$\anumu \approx \bar\nu_2$ and $\anutau \approx \bar\nu_3$.
Hence the above expression for the neutral current event rate in the
presence of delay and mixing can be written as,
\begin{eqnarray}
\frac{dS_{nc}^{do}}{dt}&\!\!\!=\!\!\!&\frac{n}{4\pi D^2} \int dE \sigma (E)
\{ N_\numu (t) f_\numu(E) + N_\anue(t)f_\anue(E)
+N_\nutau (t) f_\nutau(E)+ 
\nonumber\\
&\!\!\!+\!\!\!&N_\anumu (t) f_\anumu(E) \!+ 
N_\nue(t\!\!-\!\!\Delta t(E))f_\nue (E)\!+\!
N_{\anutau}(t\!\!-\!\!\Delta t(E))f_{\anutau}(E)\}
\label{do1}
\end{eqnarray}
Note that the above expression does not depend on the neutrino conversion
probability as the neutral current interaction is flavor blind.

\begin{figure}[t]
      \centerline{\psfig{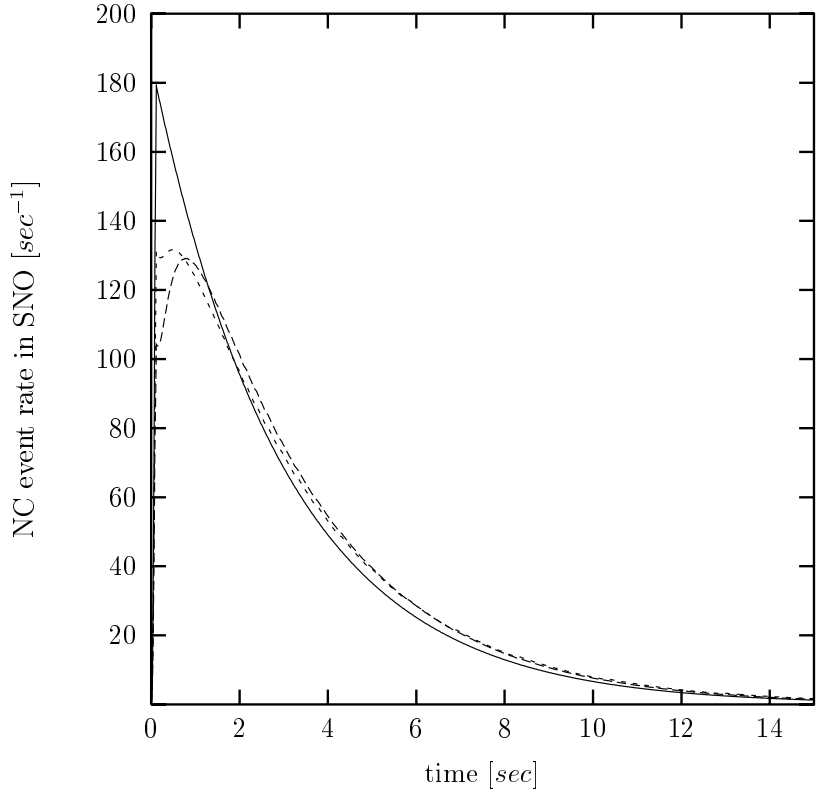}}
\vskip -1cm
\caption[Time response of the neutral current events]
{\label{time}
The neutral current event rate as a function of time in $\rm D_2O$
in SNO. The solid line corresponds to the case of
massless neutrinos, the long dashed line to neutrinos with only mass but
no mixing, while the short dashed line gives the event rate for neutrinos
with mass as well as flavor mixing.}
\end{figure}

In fig. \ref{time} we 
have plotted the neutral current event rate for the reaction
($\nu_i + d \rightarrow n+p+\nu_i$, where $\nu_i$ stands for all the 6 neutrino 
species) as a function of time for massless
neutrinos along with the cases for mass but no mixing (eq.(\ref{del1}))
and mass along with mixing (eq.(\ref{do1})).
The figure looks similar for the other
neutral current reactions as well, apart from a constant normalization factor
depending on the total number of events for the process concerned. The
curves corresponding to the massive neutrinos have been given for
$m_{\nu_3} = 40$ eV. 
As expected, the shape of the neutral current event rate changes due 
to the delay of massive $\nutau$. 
Since the delay given by eq.(\ref{deltime}) depends quadratically
on the neutrino mass, the distortion is more for larger masses 
\cite{bv2}.
But the noteworthy point is that 
the presence of mixing further distorts the rate vs. time curve. 
The reason for this distortion can be traced to the fact that 
the time delay $\propto 1/E^2$. As the energy
spectrum of the neutrinos change due to flavor mixing, 
the resultant delay is also modified and
this in turn alters the neutral current event rate as a function of time.
In fact the flavor conversion in the supernova results in 
de-energising the $\numu/\nutau$ 
spectrum and hence the delay given by eq.(\ref{deltime}) 
should increase. As larger delay caused
by larger mass results in further lowering of the neutral current event
rate vs. time curve for early times, one would normally expect that the
enhanced delay as a result of neutrino flavor conversion would have a
similar effect. But the fig. \ref{time} 
shows that during the first second, the curve
corresponding to delay with mixing is higher than the one with only time
delay. This at first sight seems unexpected. But then one realizes
that while the flavor conversion reduces the average energy of the massive
$\nutau$ increasing its delay and hence
depleting its signal at early times, it energizes the 
massless and hence undelayed $\nue$ beam, which
is detected with full strength. Therefore, while for no mixing the $\nutau$
gave the larger fraction of the signal, 
for the case with mixing it is the $\nue$ that
assume the more dominant role, and so even though the $\nutau$ arrive 
more delayed compared to the case without mixing, 
the delay effect is diluted 
due to the enhancement of the $\nue$ fraction 
and the depletion of the $\nutau$ fraction of the neutral current events.  
We have also checked that although it may seem that the curve
with delay and mixing can be simulated by another curve with delay alone
but with smaller mass, the actual shape of the two curves would still be
different. This difference in shape though may not be statistically
significant and hence one may not be able to see the effect of 
mixing in the time delay of the neutrinos  
just by looking at the time response of 
the neutral current event rate 
in the present water Cerenkov detectors. We therefore look for 
various other variables which can be studied to compliment this. 

\begin{figure}[t]
      \centerline{\psfig{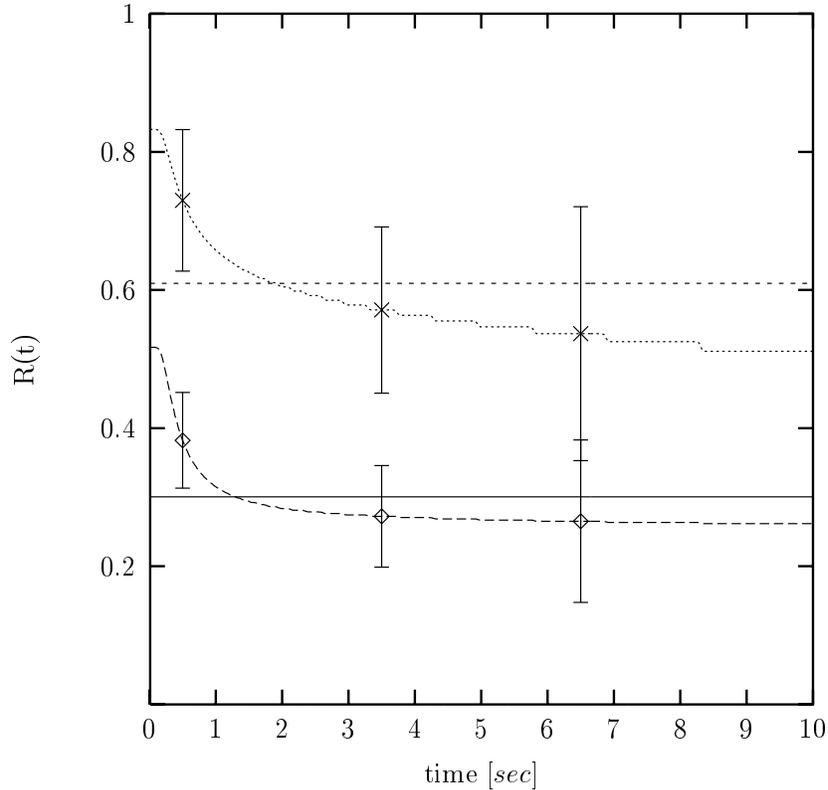}}
\vskip -1cm
\caption[Ratio of CC to NC events as a function of time]
{\label{ccnc6}
The ratio R(t) of the total charged current to neutral current
event rate in SNO versus time. The solid line is for massless
neutrinos, the short dashed line for neutrinos with complete
flavor conversion but
no delay, the long dashed line for neutrinos with only delay and no flavor
conversion and the dotted line is for neutrinos with both delay and
complete flavor conversion.
Also shown are the $\pm 1\sigma$ statistical errors for
delay with and without mixing in the $1^{st}$, $4^{th}$ and the
$7^{th}$ time bins.}
\end{figure}

One such variable which carries information about both the neutrino mass and
their mixing is R(t), the ratio of charged to neutral current event rate as
a function of time. 
In fig. \ref{ccnc6} 
we give the ratio R(t) of the total charged current to the
neutral current event rate in $\rm{D_2 O}$ in 
SNO as a function of time. Plotted are the
ratios (i) without mass, (ii) with only mixing, (iii) with delay
but zero mixing and (iv) with delay
and flavor mixing. The differences in the behavior of R(t) for the four
different cases are clearly visible. For no mass R(t)=0.3 and since the
time dependence of both the charged current and neutral current reaction
rates are the same, their ratio is constant in time. 
As the presence of mixing enhances the charged current signal keeping the 
neutral current events unaltered, R(t) goes up to 0.61 for only mixing,  
remaining constant
in time, again due to the same reason. With the introduction of delay the
ratio becomes a function of time as the neutral current reaction now has
an extra time dependence coming from the mass. At early times as the
$\nutau$ get delayed the neutral current event rate drops increasing
R(t). These delayed $\nutau$s arrive later and hence R(t) falls at
large times.
This feature can be seen for both the curves with and without mixing.
The curve for only delay starts at R(t)=0.52 at t=0 s and falls to about
R(t)=0.26 at t=10 s. For the delay with mixing case the corresponding
values of R(t) are 0.83 and 0.51 at t=0 and 10 s respectively.
The important point is that the curves with and without mixing are
clearly distinguishable and should allow one to differentiate between the
two cases of only delay and delay with neutrino flavor conversion.
 
\begin{figure}[t]
      \centerline{\psfig{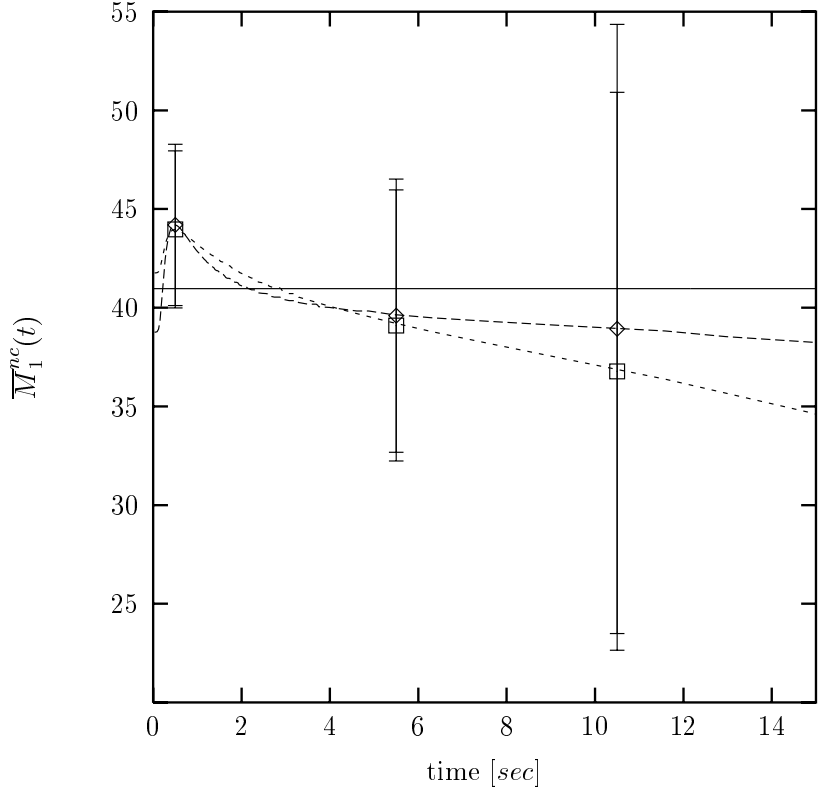}}
\vskip -1cm
\caption[Moments of the neutral current events vs. time]
{\label{mom}
The 1st normalized energy moment of the neutral current
events in SNO $\overline M_1^{nc}(t)$ versus time. The solid line
corresponds to the case of
massless neutrinos, the long dashed line to neutrinos with only mass but
no mixing, while the short dashed line gives the event rate for neutrinos
with mass as well as flavor conversion.
Also shown are the $\pm 1\sigma$ statistical errors for
delay with and without mixing in the $1^{st}$, $6^{th}$ and the
$11^{th}$ time bins.}
\end{figure}

In order to
substantiate our claim that the two scenarios of only delay and delay with
mixing are distinguishable in SNO, we divide the time into bins of
size 1 second. The number of events in each bin is then used to estimate
the $\pm 1\sigma$ statistical error in the ratio R(t)
in each bin and these are then plotted in fig. \ref{ccnc6} 
for the typical time bin 
numbers 1, 4 and 7. 
From the figure we see that the two cases of delay, with 
and without mixing, are
certainly statistically distinguishable in SNO for the first
6 seconds. 

We next focus our attention on $M_n^{nc}(t)$, 
the neutral current $n$-th moments of the neutrino energy 
distributions \cite{dmarc} 
observed at the detector, defined as
\begin{equation}
M_n^{nc}(t)=\int\frac{d^2S}{dEdt} E^n dE
\end{equation}
while the corresponding normalized moments are given by
\begin{equation}
\overline M_n^{nc}(t)=\frac{M_n^{nc}(t)}{M_0^{nc}(t)}
\end{equation}
We have shown the behavior of the 1st normalized moment 
$\overline M_1^{nc}(t)$ in fig. \ref{mom} as 
a function of time in SNO. For massless neutrinos, the 
$\overline M_1^{nc}$ has a 
value 40.97, constant in time, as this is again a ratio and hence 
the time dependence gets canceled out as in the case of R(t). 
For the case where the $\nutau$ is massive 
and hence delayed, it assumes a time dependence. 
Since the delay $\propto 1/E^2$ and since the neutrinos are produced 
at the source with an energy distribution, 
hence at each instant the lower 
energy $\nutau$ will be delayed more than the higher energy $\nutau$. 
Therefore $\overline M_1^{nc}(t)$, which gives the energy centroid of the 
neutral current event distribution in $\rm{D_2O}$, 
starts from a low value 38.76 at t=0 s as all the $\nutau$ are delayed, 
rises sharply as the higher energy neutrinos arrive first 
and then falls slowly as the lower energy delayed $\nutau$ 
start arriving. If the $\nutau$ are allowed to mix with the $\nu_e$, 
then they are de-energized and the above mentioned effect is 
further enhanced. 
To make an estimate of whether SNO would be able to distinguish the 
three cases discussed above, we compute the $\pm 1 \sigma$ statistical 
errors in the $1^{st}$ normalized moment for the two scenarios 
of delay, with and without mixing, and show them for the 
$1^{st}$, $6^{th}$ and $11^{th}$ bins. We see that the errors 
involved are large enough to completely wash out the differences 
between the energy moments with and without neutrino mass and mixing. 
Hence the normalized energy moments fail to probe neutrino mass and mixing 
as at early times we don't see much difference between the different 
cases considered, while at late times the number of events become 
very small so that the error in $M_0^{nc}(t)$ becomes huge, increasing 
the error in $\overline M_1^{nc}(t)$.  

\begin{figure}[t]
      \centerline{\psfig{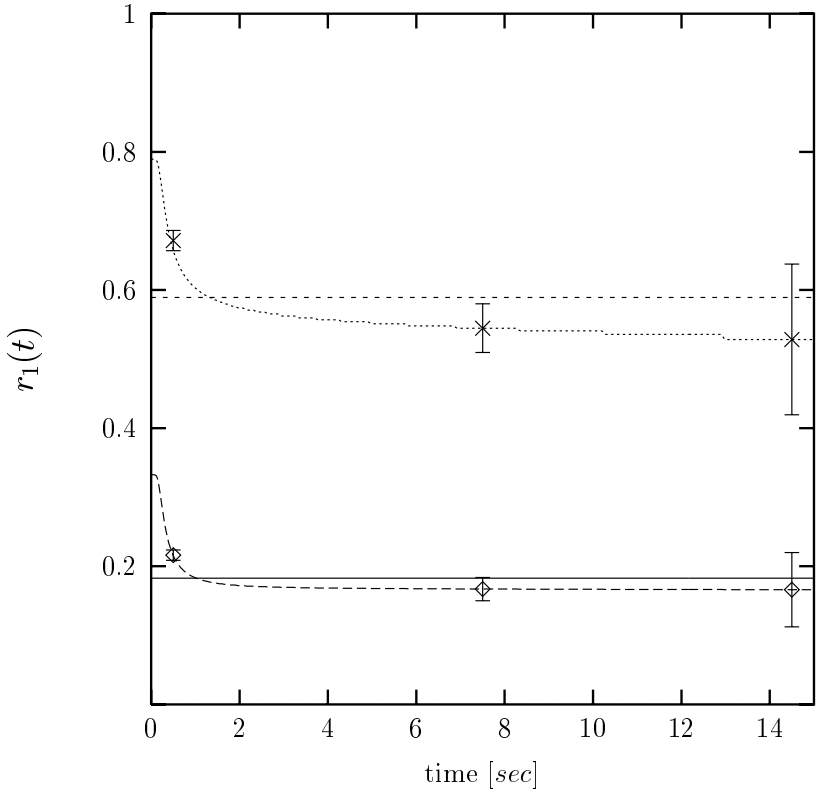}}
\vskip -1cm
\caption[Ratio of CC to NC moments vs. time]
{\label{rmom}
The variation of $r_1(t)$ with time in SNO.
The solid line is for massless
neutrinos, the short dashed line for neutrinos with complete
flavor conversion but
no delay, the long dashed line for neutrinos with only delay and no flavor
conversion and the dotted line is for neutrinos with both delay and
complete flavor conversion.
Also shown are the $\pm 1\sigma$ statistical errors for
delay with and without mixing in the $1^{st}$, $8^{th}$ and the
$15^{th}$ time bins.}
\end{figure}

The variable that can be a useful probe 
for differentiating the case for delay 
with mixing from the case for delay without mixing is
the ratio of the unnormalized moment of the charged to neutral 
current events 
\begin{equation}
r_n(t)= \frac{M_n^{cc}(t)}{M_n^{nc}(t)}
\end{equation}
We present in fig. \ref{rmom}, for SNO, 
the $r_n(t)$ vs. time plots (for n=1) for the cases of 
(a) massless neutrinos (b) with mixing but no delay (c) with 
delay but no mixing and (d) with delay as well as mixing. 
Since this is a ratio, the supernova flux uncertainties get 
canceled out to a large extent and since the unnormalized moments 
have smaller statistical errors, this is a better variable than the 
normalized moments to observe the signatures of neutrino mixing. In the 
figure we have shown the $\pm 1 \sigma$ statistical errors in $r_1(t)$ 
for the two cases of delay alone and delay with mixing, for the 
$1^{st}$, $8^{th}$ and $15^{th}$ bins in time, and the two cases are 
clearly distinguishable in SNO for early as well as late times. 
Note that $r_1(t)$ is different from the ratio R(t) as it gives 
information about the ratio of the energy centroids of the 
charged current and neutral current distributions as a function 
of time, while the latter gives only the ratio of the number of events 
as a function of time.  

\begin{figure}[t]
      \centerline{\psfig{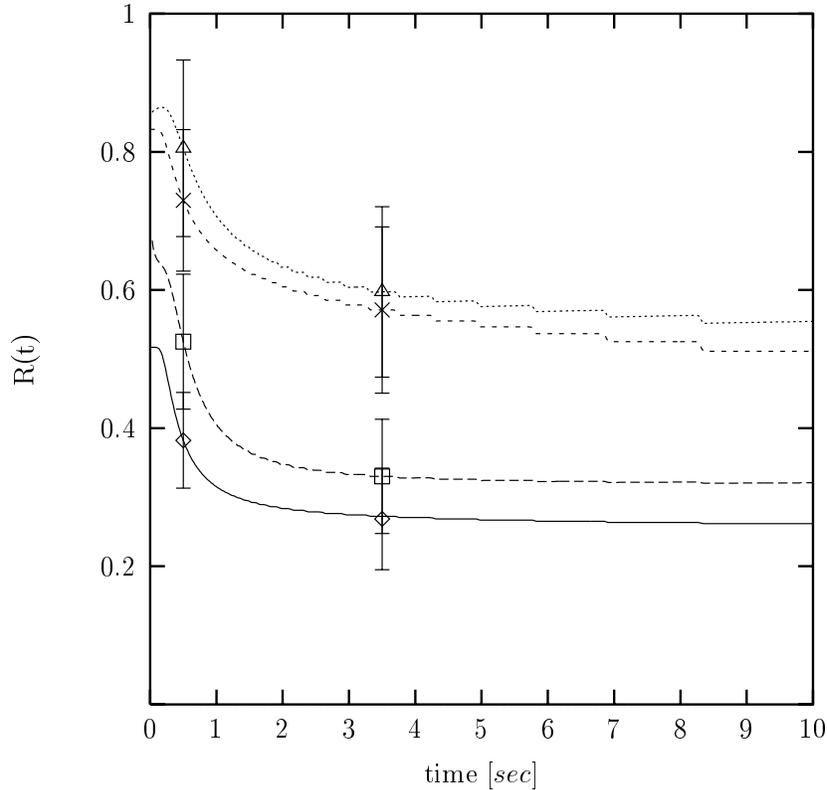}}
\vskip -1cm
\caption[Effect of supernova uncertainties]
{\label{rtime}
The ratio R(t) in SNO for the two cases of fixed
and time dependent neutrino temperatures. The solid line and the long
dashed line give R(t) for the cases of fixed temperatures
and varying temperatures respectively for only delay,
while the short dashed line and the dotted line
give the corresponding R(t) for delay with mixing.
We have also given the $\pm 1\sigma$ statistical errors
in the $1^{st}$ and the $4^{th}$ time bin,
for the both the curves for fixed and time dependent temperatures.}
\end{figure}

The advantage of using ratios 
is that, they are not only sensitive to the mass and
mixing parameters but are also almost insensitive to the details of 
supernova models. 
Since they are a ratio they are almost
independent of the luminosity and depend only
on some function of the ratio of neutrino temperatures. 
All the calculations presented so far are for fixed neutrino 
temperatures. In 
order to show that the time dependence of the neutrino temperatures 
does not alter our conclusions much, we present our analysis
with time dependent neutrino temperatures. We take
\begin{equation}
T_\nue = 0.16\log t + 3.58,~~
T_\anue= 1.63 \log t + 5.15,~~
T_\numu= 2.24 \log t + 6.93
\end{equation}
These forms for the neutrino temperatures follow from fits to the results of
the numerical supernova model given in Totani {\it et al.} \cite{totani6} 
which we had used in the previous section. In
fig. \ref{rtime} we compare the ratio R(t) for the cases of delay and
delay with mixing for the two cases
of fixed temperatures and the time dependent temperatures. 
It is clear from the figure that
that the time dependence of the neutrino temperatures does not have
much effect on the time dependence of the ratio of the
charged current to neutral current rates. In fact the two curves
corresponding to fixed and time dependent temperatures, fall within
$\pm 1\sigma$ statistical errorbars for both the cases of only delay 
and delay with mixing.

\section{Summary}

In conclusion, we have shown that with the model of Totani {\it et al.}
even with vacuum oscillations we obtain appreciable enhancement in
the expected $\nu$ signal in SNO and SK even though the number of
neutrinos arriving at the detector from the supernova goes down. In
contrast to the case where we have MSW resonance in the supernova, with
vacuum oscillations we get enhancement for both $\nue$ as well as $\anue$
events. If we have a galactic supernova event in the near future and if we
get a distortion in the neutrino spectrum and an enhancement in the signal,
for both $\nue$ as well as $\anue$ then that would indicate vacuum
neutrino oscillations.

We have shown that even though neutrino flavor mixing cannot alter the 
total neutral current signal in the detector - the neutral current 
interaction being flavor blind, it can have a non-trivial impact on 
the delay of massive neutrinos, which alters the neutral current event 
rate as a function of time. The neutral current event rate though 
does not depend on the neutrino conversion probability. 
In order to study the effect of neutrino mass and mixing we have suggested  
various variables. Of the different variables that we have presented here, 
the ratio of the charged to neutral current event rate R(t), can show 
the effect of mixing during the first few seconds, while the 
charged to neutral 
current ratio of the energy moments 
are useful diagnostic tools for all times. 
These variables are not just sensitive 
to flavor mixing and time delay, they are also insensitive to 
supernova model uncertainties and hence are excellent tools to study 
the effect of flavor mixing on the time delay of massive supernova 
neutrinos.  
Though we have considered a mass spectrum for the neutrinos 
where only the $\nu_3$ have a measurable delay but 
the model considered is one of many 
and one can easily 
extend the above formalism to include 
more general classes of neutrino models. 

%% file: chapter8.tex
\chapter{Conclusions}

\section{Summary}

In this thesis we have explored the signatures of neutrino mass and 
mixing in the solar and atmospheric neutrino experiments. We have 
derived bounds on the mixing parameters through elaborate \chisq 
analyses. We have studied the effect of the neutrino mass and 
mixing on supernova neutrino detection. 

In chapter 2 we have presented the basic aspects of neutrino oscillations, 
both in vacuum and in matter. 

In chapter 3 we have discussed the experimental status of the solar 
neutrino problem, the atmospheric neutrino anomaly and the 
accelerator/reactor neutrino experiments. 
We performed comprehensive statistical analysis of the solar and 
atmospheric neutrino data in terms of two flavor oscillations, 
identified the best-fit solutions and presented the 
allowed area in the neutrino parameter space. For the solar neutrino 
problem we have presented our results in Table \ref{totchi}. 
The LMA MSW solution emerges as the best-fit solution with 
\chisqmin = 33.42 which for 39 degrees of freedom is allowed at 72.18\%. 
The other solutions with large mixing angles also give good  
fit. 
The C.L. contours for the global analysis of the rates and rates+spectrum 
data are displayed in fig. \ref{contsolar} where we 
note that after the inclusion on the SNO CC results the SMA solution 
fails to appear even at $3\sigma$. 
The $\nue-\nu_{\rm sterile}$ option is largely disfavored. 
For the atmospheric neutrino case we defined two methods of \chisq 
analysis and discussed their merits and demerits. We performed the 
bin-by-bin analysis of the 1144 day SK contained events 
in terms of pure $\numu-\nutau$ oscillations 
and presented the results. The 90\% and 99\% C.L. allowed zones in the 
parameter space are displayed in fig. \ref{c2atm}. 
We gave an outline of the experimental bounds on the mixing parameters 
from the most stringent accelerator/reactor data. 

In chapter 4 we continued our discussions on the solar neutrino problem 
and explored the viability of the energy independent scenario 
in explaining the global solar data. 
We identified regions of the parameter space 
where the survival probability is within $10\%$ of the energy 
independent survival probability and called these the 
quasi-energy independent regions.   
Allowing for modest 
renormalizations for the Cl data and the BPB00 \br flux we studied 
the comparative fit for the energy independent solution {\it vis a vis} 
the MSW solutions. We showed that these renormalizations 
enlarge the allowed large mixing angle MSW regions 
and most of these enlarged 
regions overlap with the quasi-energy independent region. 
We briefly commented on the potential of some of the future experiments 
in distinguishing the energy independent scenario from the MSW solutions. 

In chapter 5 we did a three-generation oscillation
analysis of the 1144 day (SK)
atmospheric neutrino data going beyond the one mass scale dominance (OMSD)
approximation. We fixed $\Delta_{12} = \Delta_{13}$ ($\Delta_{LSND}$)
in the range eV$^2$ as allowed by the results from
LSND and other accelerator and reactor experiments on neutrino
oscillation and kept $\Delta_{23}$ ($\Delta_{ATM}$) and the three mixing
angles as free parameters.
We incorporated the matter effects, indicated some new allowed regions
with small $\Delta_{23}$ ($< 10^{-4}$ eV$^2$) and $\sin^2 2\theta_{23}$
close to 0 and discussed the differences with the
two-generation and OMSD pictures. In our scenario,
the oscillation probabilities for the accelerator and reactor neutrinos
involve only two of the mixing angles $\theta_{12}$ and $\theta_{13}$
and one mass scale. But the atmospheric neutrino oscillation is in general
governed by both mass scales and all the  three mixing
angles. The higher mass scale gives rise to \dm independent average
oscillations for atmospheric neutrinos and does not enter the $\chi^2$
analysis as an independent parameter. The $\Delta_{23}$ and the three
mixing angles on the other hand appear as independent parameters in
the $\chi^2$ analysis and the best-fit values of these are determined
from an analysis of a) the SK data, b) the SK and CHOOZ data.
The allowed values of the mixing angles $\theta_{12}$ and $\theta_{13}$
from the above analysis are compared with the constraints from all
accelerator and reactor experiments including the latest results
from LSND and KARMEN2.
Implications for future long baseline experiments are discussed. 

In chapter 6 we did a detailed $\chi^2$-analysis of the
848 day SK atmospheric neutrino data
under the assumptions of $\nu_\mu - \nu_\tau$
oscillation and neutrino
decay. For the latter we took the most general case of neutrinos
with non-zero mixing and considered the possibilities of
the unstable component in $\nu_\mu$ decaying
to a state with which it mixes (scenario (a)) and to a sterile state
with which it does not mix (scenario (b)). In the first case
$\Delta m^2$ (mass squared difference between the two mass states
which mix) has
to be $>$ 0.1 $eV^2$ from constraints on $K$ decays while for the
second case $\Delta m^2$ can be unconstrained.
For case (a) \dm does not enter
the $\chi^2$-analysis while in case (b) it enters the
$\chi^2$-analysis as an independent parameter.
In scenario (a) there is \dm averaged oscillation in addition to decay
and this gets ruled out at 100\% by the SK data.
Scenario (b) on the other hand gives a reasonably good
fit to the data for \dm  $\sim 0.001 ~eV^2$. We discussed the 
possibility of differentiating this latter scenario from the 
pure $\numu-\nutau$ scenario with more statistics on $\nutau$ appearance in 
SK. 

Neutrinos and antineutrinos of all three flavors
are emitted during the post bounce phase of a core
collapse supernova with $\numu/\nutau(\anumu/\anutau)$ having average energies
more than that of $\nue(\anue)$. These neutrinos can be detected by the 
earth bound detector like the SK and SNO. 
In chapter 7 we made realistic predictions 
for the observed signal in the detector. We studied the 
effect of three flavor oscillations in vacuum on the resultant neutrino 
spectra, in the framework of 
two different mass spectrum and presented our results. 
We showed that even though neutrino oscillations result in a depletion in the
number of $\nue$ and $\anue$ coming from the supernova, the actual signals
at the detectors are appreciable enhanced.
In particular we found a huge 
enhancement in the $^{16}O$ charged current rate in the water \chr 
detectors due to oscillations. We next considered a mass spectrum where 
one has detectable delay in the time of flight of the massive $\nutau$ 
and studied its effect. We showed that even though 
the neutral current interaction is flavor blind,
and hence neutrino flavor mixing cannot alter the total neutral current
signal in the detector, it can have a non-trivial impact on
the delay of massive neutrinos and alters the neutral current event
rate as a function of time. We have suggested various variables of the
neutral and charged current events that can be used to study this effect.
In particular the ratio of charged to neutral current events can
be used at early times while the
ratio of the energy moments for the charged
to the neutral current events can form useful diagnostic tools even
at late times to study neutrino mass and mixing.

\section{Future Prospects}

A pertinent question at this point is the potential of the future 
experiments in building upon the current information about the 
mass and mixing scenario. They should also look towards confirming 
the presence of neutrino oscillations in the deficit of the 
solar and the atmospheric neutrinos. Though neutrino oscillations 
is the favored solution for both the problems, there is no conclusive 
evidence in its favor. For the atmospheric problem one should be 
able to see the periodicity, or in other words the dips of the 
oscillations in the resultant neutrino beam. Likewise for the solar 
neutrinos the future experiments should be designed to give 
smoking gun evidence for the oscillation scenario and should 
have the strength to distinguish between the various allowed solutions. 

The MONOLITH experiment in Gran Sasso, Italy is a magnetized tracking 
Calorimeter which will detect atmospheric neutrinos. This has the 
sensitivity in L/E to detect unambiguously the dip in 
survival probability predicted by the $\numu-\nutau$ oscillation 
hypothesis. 
The other planned experiment which has the potential 
to observe the oscillations in the atmospheric neutrino signal  
is the ring imaging \chr detector AQUA RICH also in Italy. 

Among the solar neutrino experiments the most promising upcoming 
detector is the BOREXINO in Gran Sasso, Italy. 
The BOREXINO will detect the monoenergetic $^7Be$ neutrinos via 
$\nue-e$ scattering. This detector will be able to put strong bounds 
on the already disfavored SMA solution which predicts a 
very small 
survival probability for the \ber neutrinos. 
BOREXINO also expects to see large day-night asymmetry in the 
LOW region and can put constraints there. 
LENS 
also in Gran Sasso will be able to 
detect $pp$, \ber, $pep$ and \br neutrinos and hopes to 
compliment the results of BOREXINO. The other low energy detectors 
are HERON, HELLAZ and the Lithium-Berrylium detector.  

The Kamland in Kamioka, Japan will detect both solar neutrinos as 
well as $\anue$ from ten different reactors all across Japan 
with various baselines within 350 km. Using these known fluxes 
from the different reactors with known baselines, it 
is expected to scan the entire allowed LMA zone, which is 
currently the favored solution to the solar neutrino problem. 
The other exciting experiment is MINOS, which with a baseline 
of 730 km from Fermilab to Soudan will use the $\numu$ 
disappearance 
channel to explore almost the entire area allowed by the  
atmospheric neutrino data. The other experiment probing this 
region of the neutrino parameter space is K2K in Japan. It has 
already given positive signal for oscillations. 

The MiniBooNE experiment at Fermilab 
will probe the oscillations of muon 
neutrinos to electron neutrinos and it is expected that within one year 
of data taking it will have the sensitivity to either 
confirm or refute the LSND claim of having seen oscillations. 
If MiniBooNE does see oscillations then there are plans to upgrade 
it to BooNE which will have better sensitivity and hence will determine 
the mass and mixing parameters more accurately. 

In conclusion, the large mixing angle solutions, the LMA, LOW and 
the vacuum oscillation solutions are the favored solution to the 
solar neutrino problem with the LMA giving the best-fit. 
The $\numu-\nutau$ oscillations give an excellent description 
of the SK atmospheric neutrino data, which has confirmed the 
existence of neutrino mass. 
However one still needs smoking 
gun evidence to establish neutrino oscillations, particularly 
for the solar neutrinos. The future experiments should be able to 
throw more light on this question and 
are also expected to put severe constraints on the allowed 
values of the mass and mixing parameters. 

%% file: list.tex
\begin{plist}

\noindent The papers included in this thesis are marked with a star ($\star$). 

\begin{enumerate}

\item {\bf ($\star$) 
Energy independent solution to the solar neutrino anomaly
including the SNO data}\\
{\it S. Choubey, S. Goswami and D.P. Roy},
preprint hep-ph/0109017.

\item {\bf Impact of the first sno results on neutrino mass and mixing}\\
{\it A. Bandyopadhyay, S. Choubey, S. Goswami and K. Kar},
preprint hep-ph/0106264, Phys. Lett. {\bf B} (in press). 

\item {\bf Global oscillation analysis of solar neutrino data with
helioseismically constrained fluxes}\\
{\it S. Choubey, S. Goswami, K. Kar, H.M. Antia and S.M. Chitre},
preprint hep-ph/0106168, Phys. Rev. {\bf D64}, (in press).

\item {\bf ($\star$) Reviving the 
energy independent suppression of the solar neutrino}\\
{\it S. Choubey, S. Goswami, N. Gupta and D.P. Roy},
preprint hep-ph/0103318, Phys. Rev. {\bf D64}, 053002 (2001).

\item {\bf MSW mediated neutrino decay and the solar neutrino problem}\\
{\it A. Bandopadhyay, S. Choubey and S. Goswami},
preprint hep-ph/0101273, Phys. Rev. {\bf D63}, 113019 (2001).

\item {\bf Status of the neutrino decay solution to the solar neutrino
problem}\\
{\it S. Choubey, S. Goswami and D. Majumdar},
preprint hep-ph/0004193, Phys. Lett. {\bf B484}, 73 (2000).

\item {\bf ($\star$)
A three generation oscillation analysis of the Super-Kamiokande
atmospheric neutrino data beyond one mass scale dominance approximation}\\
{\it S. Choubey, S. Goswami and K. Kar},
preprint hep-ph/0004100, Astropart. Phys. (in press).

\item {\bf ($\star$) 
Effect of flavor mixing on the delay of massive supernova
neutrinos}\\
{\it S. Choubey and K. Kar},
preprint hep-ph/0003256, Phys. Lett. {\bf B479}, 402 (2000).

\item {\bf Detection of massive supernova neutrinos}\\
{\it S. Choubey},
Indian J. Physics {\bf 73B(6)}, 977 (1999).

\item {\bf ($\star$) Is Neutrino decay really ruled out as a solution to the
atmospheric neutrino problem from Super-Kamiokande data}\\
{\it S. Choubey and S. Goswami},
preprint hep-ph/9904257, Astropart. Phys. {\bf 14}, 67 (2000).

\item {\bf ($\star$) Effect of flavour oscillations on the detection of
supernova neutrinos}\\
{\it S. choubey, D. Majumdar and K. Kar},
preprint hep-ph/9809424, J. Phys. {\bf G25}, 1001 (1999).

\item {\bf Spectral distribution studies with a modified Kuo-Brown
interaction in the upper half of the fp shell}\\
{\it S. Choubey, K. Kar, J.M.G. Gomez and V.R. Manfredi},
preprint nucl-th/9807004, Phys. Rev. {\bf C58}, 597 (1998).

\end{enumerate}

\end{plist}